\documentclass[dvipdfm,final]{pittetd}
\usewithpatch{graphicx}
\usepackage{amsmath,amsthm}
\usepackage{graphicx}
\usepackage{dcolumn}
\usepackage{bm}

\InputIfFileExists{amsmatch.pit}{%
            \ClassInfo{pittetd}{Patch for amsmatch loaded}}{%
            \ClassInfo{pittetd}{No patch found for amsmatch}} \InputIfFileExists{amsthm.pit}{%
            \ClassInfo{pittetd}{Patch for amsthm loaded}}{%
            \ClassInfo{pittetd}{No patch found for amsthm}}
\title[A pittetd-thesis sample]{Study of meson properties in quark models}
\author{Olga Lakhina}
\degree{B.S., Omsk State University, 2001}
\date{October 19th 2006}
\subject{Elementary Particle Theory}
\school{Department of Physics}
\begin{document}
\maketitle
%
\committeemember{Eric Swanson, Assistant Professor}
\committeemember{Daniel Boyanovsky, Professor}
\committeemember{Vladimir Savinov, Associate Professor}
\committeemember{Adam Leibovich, Assistant Professor}
\committeemember{Charles Jones, Lecturer/Advisor for B.S. programs}
\school{Physics Department} \makecommittee
\begin{abstract}
The main motivation is to investigate meson properties in the quark
model to understand the model applicability and generate possible
improvements. Certain modifications to the model are suggested which
have been inspired by fundamental QCD properties (such as running
coupling or spin dependence of strong interactions). These
modifications expand the limits of applicability of the constituent
quark model and illustrate its weaknesses and strengths. The meson
properties studied include meson spectra, decay constants,
electromagnetic and electroweak form-factors and radiative
transitions. The results are compared to the experimental data,
lattice gauge theory calculations and other approaches.
Modifications to the quark model suggested in this dissertation lead
to a very good agreement with available experimental data and
lattice results.
\end{abstract}
\tableofcontents
\listoftables                      
\listoffigures                     
\preface

I would like to thank my Research Advisor, Dr. Eric Swanson, whose
support and guidance made my thesis work possible. He has been
actively interested in my work and has always been available to
advise me. I am very grateful for his patience, motivation,
enthusiasm, and immense knowledge in Elementary Particle Physics
that, taken together, make him a great mentor.

I thank my parents, Nadegda and Vladimir, for being wonderful
parents, and my brothers, Anton and Dmitriy, for being good friends
to me.
%
%
\chapter{Introduction: Quantum Chromodynamics and its~properties}

Quantum Chromodynamics (QCD) was proposed in the 1970s as a theory
of the strong interactions. It was widely accepted after the
discovery of asymptotic freedom in 1973 as it offered a satisfying
explanation to some of the puzzling experimental results at the
time.

However, understanding of the strong interactions is far from
complete. One of the open problems is the difficulty to explain much
of the experimental data on the particle properties from the first
principles. Building models, which capture the most important
features of strong QCD, is one way to resolve this problem.

The main motivation for the present dissertation is to investigate
meson properties in the quark model to understand the model
applicability and generate possible improvements. Certain
modifications to the model are suggested which have been inspired by
fundamental QCD properties (such as running coupling or spin
dependence of strong interactions). These modifications expand the
limits of applicability of the constituent quark model and
illustrate its weaknesses and strengths.

The meson properties studied include meson spectra, decay constants,
electromagnetic and electroweak form-factors and radiative
transitions. The results are compared to the experimental data,
lattice gauge theory calculations and other approaches.
Modifications to the quark model suggested in this dissertation lead
to a very good agreement with available experimental data and
lattice results.

In the next section of the introduction, different approaches to the
problem of strong QCD are discussed. After that, the most important
properties of QCD are described, including asymptotic freedom,
confinement and chiral symmetry breaking. The quark models studied
here are introduced and the theory necessary for understanding our
methods is explained in Chapter \ref{theory}. Our results are
presented and discussed in Chapter \ref{spectrum} (Spectroscopy),
Chapter \ref{dc} (Meson decay constants), Chapter \ref{ff}
(Form-factors),
 Chapter \ref{gg_chapter} (Gamma-gamma decays), and Chapter \ref{RadTrans}
(Radiative transitions). Chapter \ref{Conclusions} gives conclusions
and an outlook for the future.

\section{Overview}

Year after year, QCD continues to succeed in explaining the physics
of strong interactions, and no contradictions between this theory
and experiment have been found yet. QCD is especially successful in
the ultraviolet region, for which firm methods from the first
principles have been developed, and some nontrivial and unexpected
properties of QCD have been well understood and confirmed
experimentally (such as scaling violations in deep inelastic
scattering).

However, properties of medium and low energy QCD still present
challenges to particle physicists and remain to be understood. For
instance, a rigorous proof is still lacking that QCD works as a
microscopic theory of strong interactions that give rise to the
macroscopic properties of chiral symmetry breaking and quark
confinement. The main problem is that perturbation theory (which
proved to be very useful for high energy region) is not applicable
at low energy scales, and no other analytical methods have been
developed so far. The situation is well described by the 2004 Nobel
Laureate David J. Gross (who received the prize for the discovery of
asymptotic freedom together with F. Wilczek and H. D. Politzer).
Gross said in 1998 \cite{Gross}:

\emph{At large distances however perturbation theory was useless. In
fact, even today after nineteen years of study we still lack
reliable, analytic tools for treating this region of QCD. This
remains one of the most important, and woefully neglected, areas of
theoretical particle physics.}

The only reliable method of studying the physical properties of low
energy QCD is the unquenched lattice formulation of gauge theory.
Unfortunately, the numerical integrations needed in this approach
are extremely computationally expensive. Even with the use of
efficient Monte Carlo methods, approximations must be done in order
to obtain results with the computational technology of today.
However, unquenched lattice gauge theory calculations are appearing
and have already made an impact. They are still preliminary, but a
good understanding exists on the sources of error, and plans are in
place to address them.

The only other way to proceed is to invent models that capture the
most important features of strong QCD. A great variety of models
have been developed during 30 years of QCD. Among them are quenched
lattice gauge theory, the Dyson-Schwinger formalism, constituent
quark models, light cone QCD, and various effective field theories
(heavy quark effective field theory, chiral perturbation theory and
other theories).

\subsection{Quenched lattice gauge theory}

The lattice formulation of gauge theory was proposed in 1974 by
Wilson \cite{Wilson} (and independently by Polyakov \cite{Polyakov}
and Wegner \cite{Wegner}). They realized how to implement the
continuous SU(3) gauge symmetry of QCD and that lattice field theory
provided a non-perturbative definition of the functional integral.
The basic idea was to replace continuous finite volume spacetime
with a discrete lattice. From a theoretical point of view, the
lattice and finite volume provide gauge-invariant ultraviolet and
infrared cutoffs, respectively. A great advantage of the lattice
formulation of gauge theory is that the strong coupling limit is
particularly simple and exhibits confinement \cite{Wilson}.
Moreover, the lattice approach can be formulated numerically using
Monte Carlo techniques. This approach is in principle only limited
by computer power, and much progress has been made since the first
quantitative results emerged in 1981 \cite{Creutz}. However,
numerous uncertainties arise in moving the idealized problem of
mathematical physics to a practical problem of computational
physics. For instance, an uncontrolled systematic effect of many
lattice calculations has historically been the quenched
approximation, in which one ignores the effects related to particle
creation and annihilation so the contribution from the closed quark
loops is neglected. It is hard to estimate the associated error, and
only in isolated cases can one argue that it is a subdominant error.
As a result, it is very difficult to describe light meson properties
from the lattice formulation in quenched approximation. Certainly,
additional analysis in other models is needed to make any firm
conclusions about quenched lattice QCD results.

\subsection{Dyson-Schwinger formalism}

One of the techniques that has been quite successful in explaining
light hadron properties is based on Dyson-Schwinger equations (DSEs)
derived from QCD. The set of DSEs is an infinite number of coupled
integral equations; a simultaneous, self-consistent solution of the
complete set is equivalent to a solution of the theory. In practice,
the complete solution of DSEs is not possible for QCD. Therefore one
employs a truncation scheme by solving only the equations important
to the problem under consideration and making assumptions for the
solutions of other equations. Both the truncation scheme and the
assumptions have to respect the symmetries of the theory, which
could be achieved by incorporating Ward-Takahashi identities. One
important advantage of this model is that it is Poincar\'{e}
covariant and directly connected to the underlying theory and its
symmetries. In particular, chiral symmetry and its dynamical
breaking have been successfully studied in this model. A good review
of this approach can be found in \cite{Maris}. Unfortunately, heavy
and heavy-light mesons are more difficult to investigate using this
approach, as a lot can be learned from the available experimental
data for these states. This is opposite to the naive quark models,
which work surprisingly well for heavy mesons but have problems
describing light particles.

\subsection{Quark model}

The quark model of hadrons was first introduced in 1964 by Gell-Mann
\cite{GM} and, independently, by Zweig \cite{Zweig}. At a time the
field theory formulation of strong interactions was disfavored and
many eminent physicists advocated abandoning it altogether. As Lev
Landau wrote in 1960 \cite{Landau}:

\emph{Almost 30 years ago Peierls and myself had noticed that in the
region of relativistic quantum theory no quantities concerning
interacting particles can be measured, and the only observable
quantities are the momenta and polarizations of freely moving
particles. Therefore if we do not want to introduce unobservables we
may introduce in the theory as fundamental quantities only the
scattering amplitudes.}

\emph{The $\psi$ operators which contain unobservable information
must disappear from the theory and, since a Hamiltonian can be built
only from $\psi$ operators, we are driven to the conclusion that the
Hamiltonian method for strong interaction is dead and must be
buried, although of course with deserved honour.}

Until the discovery of asymptotic freedom it was not considered
proper to use field theory without apologies. Even in their paper
describing the original ideas on the quark gluon gauge theory, which
was later named QCD, Gell-Mann and Fritzsch wrote \cite{FGM}:

\emph{For more than a decade, we particle theorists have been
squeezing predictions out of a mathematical field theory model of
the hadrons that we don't fully believe - a model containing a
triple of spin 1/2 fields coupled universally to a neutral spin 1
field, that of the 'gluon'....}

\emph{....Let us end by emphasizing our main point, that it may well
be possible to construct an explicit theory of hadrons, based on
quarks and some kind of glue, treated as fictitious, but with enough
physical properties abstracted and applied to real hadrons to
constitute a complete theory. Since the entities we start with are
fictitious, there is no need for any conflict with the bootstrap or
conventional dual model point of view.}

Today almost no one seriously doubts the existence of quarks as
physical elementary particles, even though they have never been
observed experimentally in isolation. It is believed that the
dynamics of the gluon sector of QCD contrives to eliminate free
quark states from the spectrum. In principle, the possibility of
observing free quarks and gluons exists at extremely high
temperature and density, in a phase of QCD called the quark-gluon
plasma (QGP). Experiments at CERN's Super Proton Synchrotron first
tried to create the QGP in the 1980s and 1990s, and they may have
been partially successful. Currently, experiments at Brookhaven
National Laboratory's Relativistic Heavy Ion Collider (RHIC) are
continuing this effort. CERN's new experiment, ALICE, will start
soon (around 2007-2008) at the Large Hadron Collider (LHC).

In a nonrelativistic constituent quark model, one ignores the
dynamical effects of gluon fields on the hadron structure and
properties. Quarks are considered as nonrelativistic objects
interacting via an instantaneous adiabatic potential provided by
gluons. One model of the potential which proves to be rather
successful in describing the heavy meson spectrum is the `Coulomb +
linear potential'. In the weak-coupling limit (at small distances),
this is a Coulomb potential with an asymptotically free coupling
constant. The strong coupling limit (large distances), on the other
hand, gives a linear potential which confines color.

The quark model has been used to study the low-lying hadron spectrum
with a remarkable success. Moreover, as is demonstrated in the
present dissertation, it is also able to describe and predict other
meson properties, for example those relevant to transitions, and
could be applied to different types of mesons, from light to
heavy-light and heavy.

However, there exist a number of phenomena for which gluon dynamics
could be important, such as the existence of hybrid mesons and
baryons suggested by QCD. Hybrid hadrons, in addition to static
quarks and antiquarks, consist of excited gluon fields. These states
can be studied on the lattice or in modified quark models and give
important insights on the phenomenon of confinement.

Another model that is based on the potential quark model, but with
significant modifications, is a `Coulomb Gauge model', which is
described in section \ref{CoulombGauge} of this dissertation. The
model consists of a truncation of QCD to a set of diagrams which
capture the infrared dynamics of the theory. The efficiency of the
truncation is enhanced through the use of quasiparticle degrees of
freedom. In addition, the random phase approximation could be used
to obtain mesons. This many-body truncation is sufficiently powerful
to generate Goldstone bosons and has the advantage of being a
relativistic truncation of QCD.

All models have been designed to reproduce certain QCD properties
and have their limits. Therefore, it is quite important to
understand when and why a model can be considered reliable.

Certainly, as we apply some model to investigate new effects and
properties, that are different from what it was designed for,
necessary changes and adjustments have to be made to reproduce
experimental data. The process of improving the model can teach us a
great deal about QCD properties and show us which aspects of it are
crucial for describing certain effects and which can be neglected.

In the next few sections of the introduction the most important
properties of QCD are described, including asymptotic freedom,
confinement and chiral symmetry breaking.

\section{Quarks, color, and asymptotic freedom}

In the 1960s a growing number of new particles was being discovered,
and it became clear that they could not all be elementary.
Physicists were looking for the theory to explain this phenomenon.
Gell-Mann and Zweig provided a simple idea which solved the problem
- they proposed that all mesons consisted of a quark and an
antiquark and all baryons consisted of three quarks. It is now
widely accepted that quarks come in six flavors: \emph{u} (up),
\emph{d} (down), \emph{s} (strange), \emph{c} (charm), \emph{b}
(bottom) and \emph{t} (top), and carry fractional electric charge
(up, charm and top quarks have charge $+\frac{2}{3}e$, and down,
strange and bottom have charge $-\frac{1}{3}e$). Quarks also have
another property called color charge which was introduced in 1964 by
Greenberg \cite{Greenberg}, and in 1965 by Han and Nambu
\cite{HanNambu}. Quarks and antiquarks combine together to form
hadrons in such a way that all observed hadrons are color neutral
and carry integer electric charge. Quarks are fermions and have spin
$s=\frac{1}{2}$.

By analogy with Quantum Electrodynamics (QED), in which photons are
the carriers of the electromagnetic field, particles called
\emph{gluons} carry the strong force as they are exchanged by
colored particles. The important difference of QCD is that gluons
also carry color charge and therefore can interact with each other.
This leads to the fact that gluons in the system behave in such a
way as to increase the magnitude of an applied external color field
as the distance increases. Quarks being fermions have the opposite
effect on the external field - they partially cancel it out at any
finite distance (screening of the color charge occurs much as the
screening of the electric charge by electrons happens in QED). The
composite effect of the quarks and gluons on the vacuum polarization
depends on the number of quark flavors and colors. In QCD, for 6
quark flavors and 3 colors, the anti-screening of gluons overcomes
the screening due to the quarks and leads to the emergence of
interesting phenomenon called \emph{asymptotic freedom}. The name of
the phenomena suggests its meaning -- at short distances (high
energies) strong interacting particles behave as if they are
asymptotically free (effective coupling is very small).

Asymptotic freedom was introduced in 1973 by Gross and Wilczek
\cite{Gross:1973id} and Politzer \cite{Politzer:1973fx} in an effort
to explain rather puzzling deep inelastic scattering experiments
performed at SLAC and MIT. In these experiments, a hydrogen target
was hit with a 20 GeV electron beam and the scattering rate was
measured for large deflection angles (hard scattering). This
experiment was very similar to Rutherford's famous experiment, where
the gold target was hit by alpha particles and the rate of particles
scattered with a large angle was measured.

Hard scattering corresponds to a high momentum transfer between the
electrons and protons in the target, so detecting a large rate would
mean that the structure of the proton is similar to that of an
elementary particle. Because the hypothesis at the time was that the
hadrons were loose clouds of constituents, like jelly, relatively
low rates were expected. However, not only was a high rate for hard
electron scattering detected, but also only in rare cases did a
single proton emerge from the process. Instead, an electromagnetic
impulse shattered the proton and produced a system with a large
number of hadrons. It looked like the proton behaved like an
elementary particle in electromagnetic processes, but as a complex
softly bound system for strong interaction processes.

The explanation for this phenomenon was offered by Bjorken
\cite{Bjorken:1969ja} and Feynman \cite{Feynman:1969ej}. They
introduced the \emph{parton model}, which assumes that the proton is
a loosely bound system of a small number of constituents called
\emph{partons} that are unable to transfer large momenta through
strong interactions. These constituents included electrically
charged quarks and antiquarks and possibly some other neutral
particles. The idea was that when a quark (or antiquark) in a proton
was hit by an electron, they could interact electromagnetically and
the quark was knocked out of the proton. The remainder of the proton
then experienced a soft momentum transfer from the knocked out quark
and materialized as a jet of hadrons.

This model imposes a strong constraint on the behavior of the deep
inelastic scattering cross section, called \emph{Bjorken scaling}.
The physical meaning of Bjorken scaling is basically the statement
that the structure of the proton looks the same to an
electromagnetic probe independently of the energy of the system, so
the strong interaction between the constituents of the proton can be
ignored.

\begin{figure}[h!]
\begin{center}
\includegraphics[trim=2cm 1cm 0cm -2cm, width=10cm]{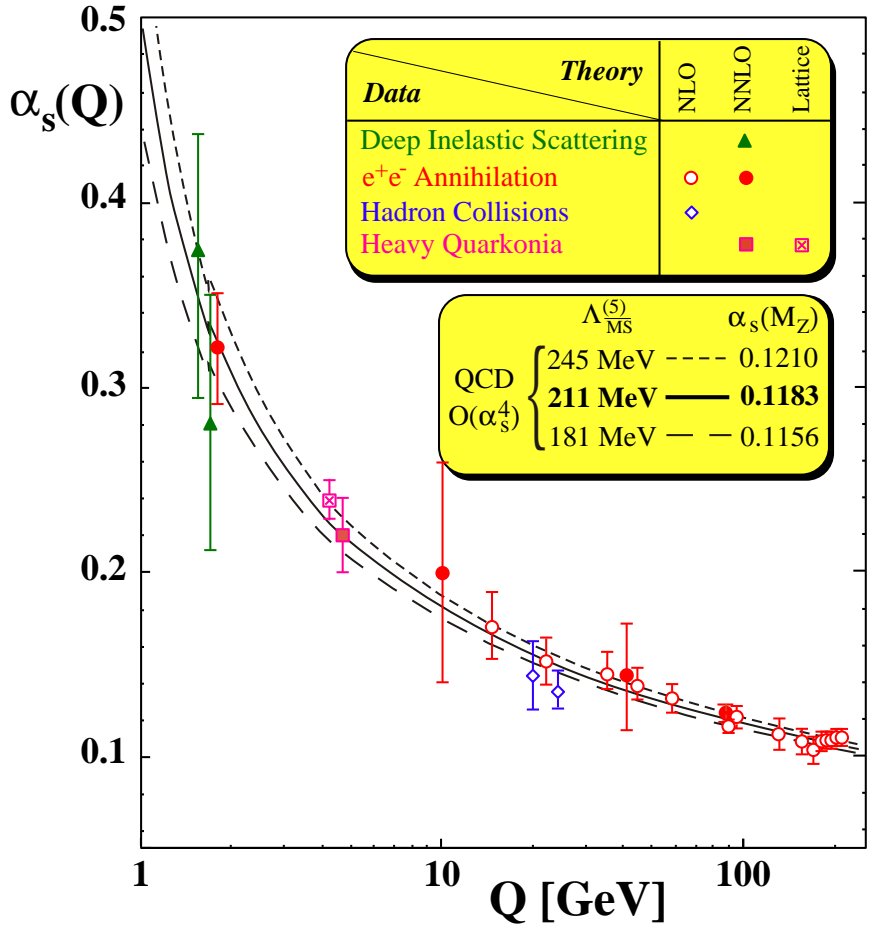}
\caption{Summary of measurements of the running coupling of strong
interactions $\alpha_s(Q)$ and its dependence on the energy scale
\cite{Bethke:2002rv}.}\label{RunCoupl}
\end{center}
\end{figure}

However, the deep inelastic scattering experiments showed slight
deviation from Bjorken scaling, suggesting that the coupling of
strong interactions was still not zero at any finite momentum
transfer. This fit perfectly with the predictions of dependence of
running coupling on an energy scale calculated from the
renormalization group approach by Gross, Wilczek and Politzer.
Later, more experiments were performed that confirmed this result.
The dependence of the coupling on the energy scale and the
experimental data are demonstrated in Fig. \ref{RunCoupl}.

Asymptotic freedom turned out to be a very useful property for
studying high energy QCD. It allows one to treat the coupling
constant perturbatively for sufficiently small distances and
therefore calculate physical properties under consideration in a
systematic and controlled manner.

The property of confinement is another interesting QCD phenomenon,
it is discussed in the next section.

\section{Confinement}

Confinement is an important property of the strong interaction that
is widely accepted and incorporated into any model claiming to
imitate strong QCD. Being an essentially nonperturbative phenomenon,
confinement still lacks a rigorous explanation from first principles
despite more than 30 years of investigation.

Quark confinement is often defined as the absence of isolated quarks
in nature as they have never been experimentally observed. Searches
for free quarks normally focus on free particles with fractional
electrical charge. But the observation of a particle with fractional
charge does not necessarily mean that a free quark has been
observed. For instance, there might exist heavy colored scalar
particles that can form bound states with quarks producing massive
states with fractional electric charge \cite{Greensite,SWconf}.

Another definition of confinement is the physics phenomenon that
color-charged particles cannot be isolated. But this confuses
confinement with color screening, and also works for spontaneously
broken gauge theories which are not supposed to exhibit confinement.

One can try to define confinement by its physical properties, for
instance, the long range linear potential between quarks. However,
this requirement is only reasonable for infinitely heavy quarks.
When two quarks with finite masses become separated, at some point
it becomes more energetically favorable for a new quark/anti-quark
pair to be created out of the vacuum than to allow the quarks to
separate further.

\begin{figure}[h!]
\begin{center}
\includegraphics[width=10cm]{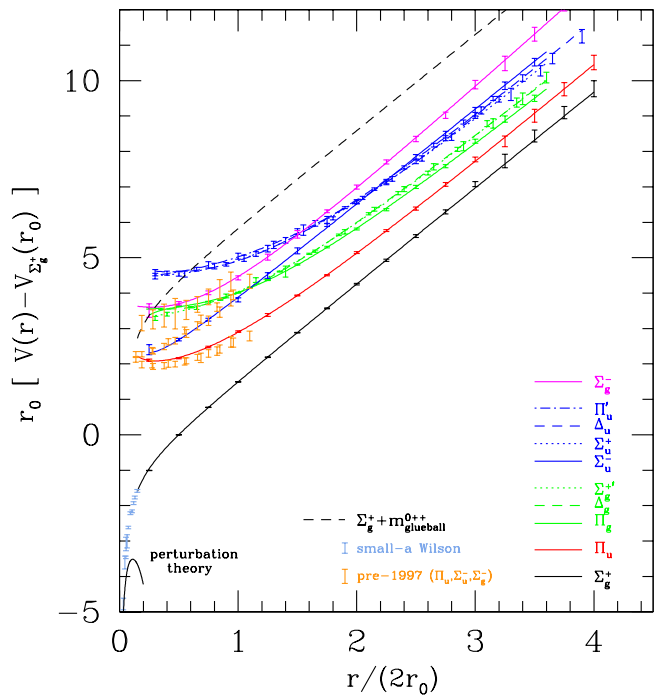}
\caption{Wilson loop measurements of various static quark potentials
\cite{Juge:1997nd}.}\label{potl}
\end{center}
\end{figure}

The lattice gauge approach has its own definition of confinement.
Field theory is said to exhibit confinement if the interaction
potential between quark and antiquark in this theory (which
corresponds to the Wilson loop calculated on the lattice) has
asymptotic linear behavior at large distances. Wilson loop
measurements of various static quark potentials in the QCD vacuum
are presented in Fig. \ref{potl}. The lowest curve corresponds to
the ground state of the gluonic field in the quark-antiquark system
(meson) while higher curves correspond to the excited gluonic field
(possibly hybrid states). One can see that for large distances all
the potentials show linear behavior (confinement).

\begin{figure}[h!]
\begin{center}
\includegraphics[trim=0cm 13.3cm 1cm 0cm, width=8cm]{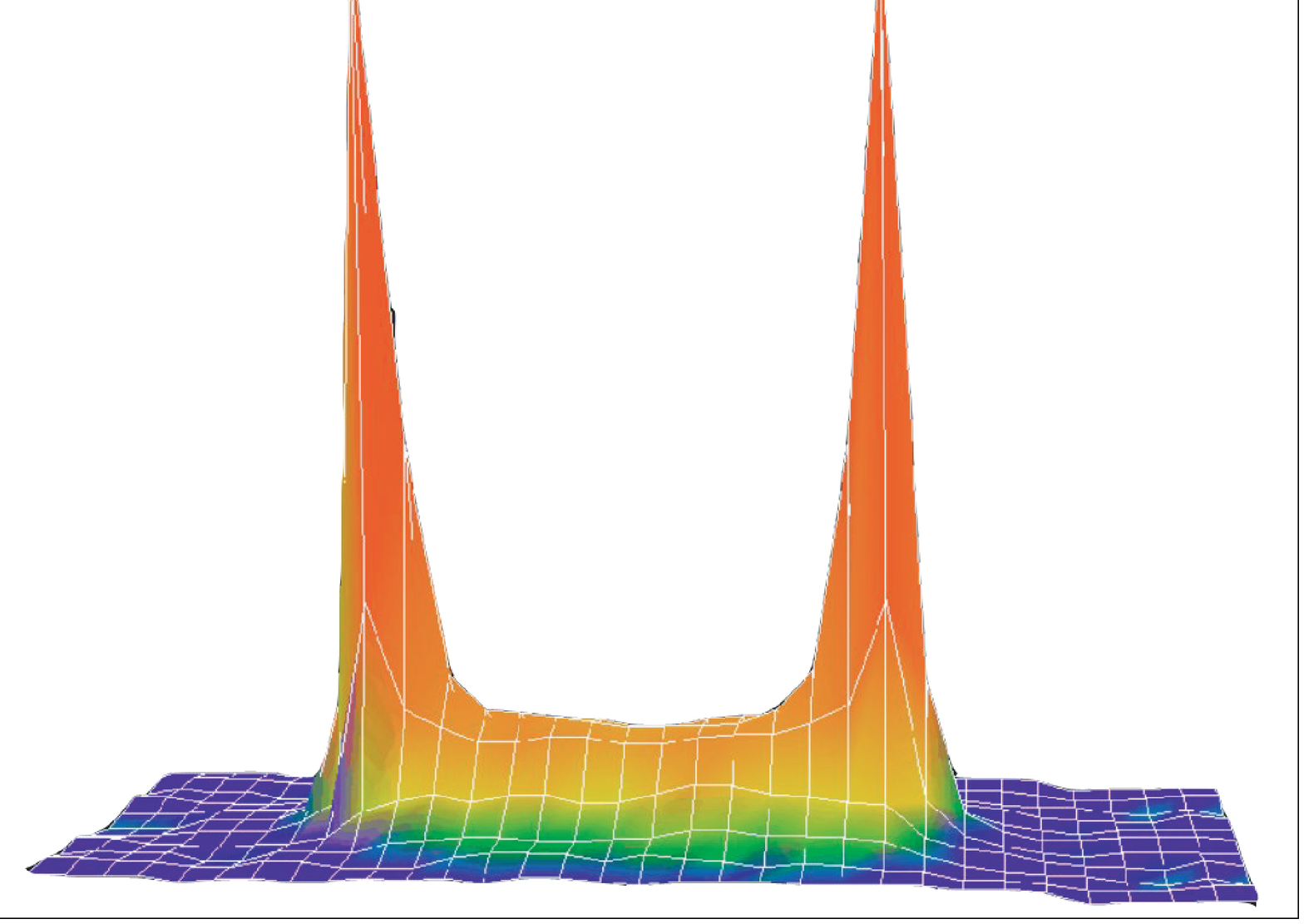}
\includegraphics[trim=0cm 0cm 0cm 0cm, width=7.5cm]{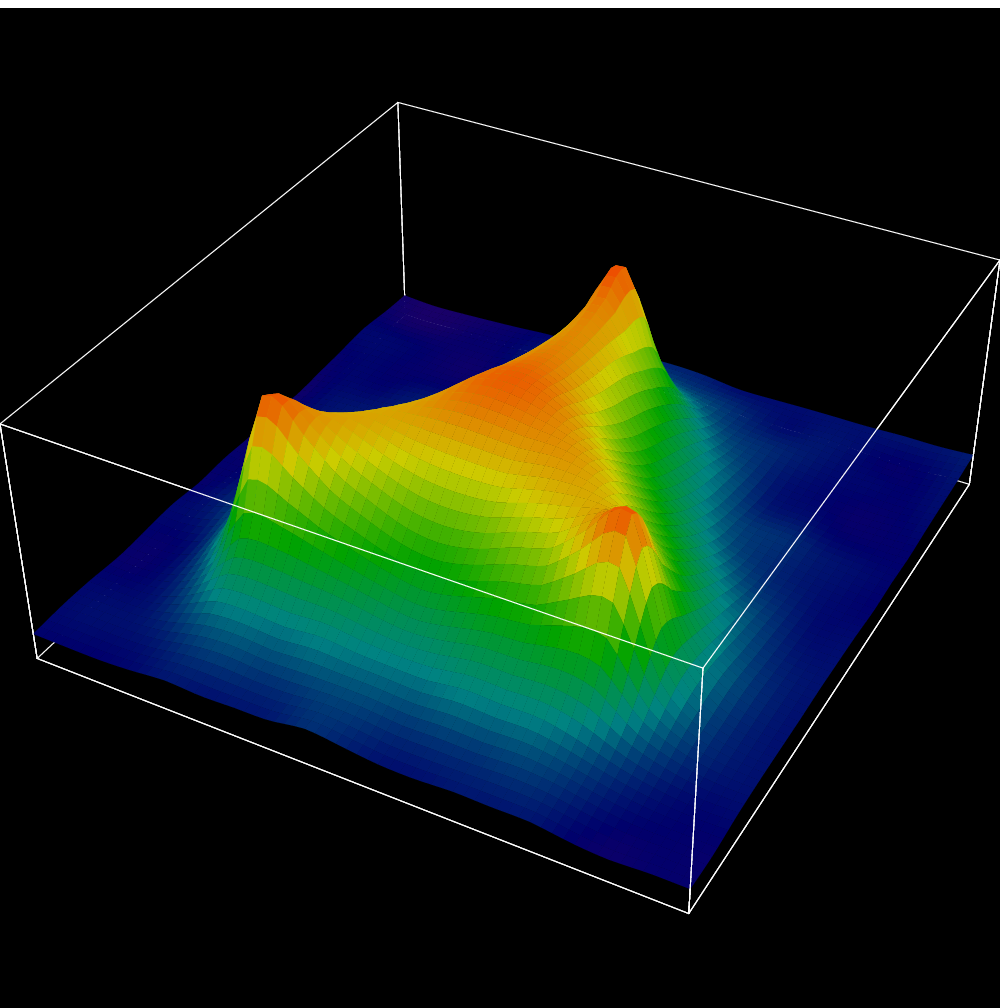}
\caption{Mesonic \cite{Bali} and baryonic \cite{Ichie:2002dy} flux
tubes.}\label{fluxtube}
\end{center}
\end{figure}

Gluonic fields can be visualized with the help of the plots of the
action or gluonic field density made on the lattice (see Fig.
\ref{fluxtube} for meson (left) and baryon (right)). They clearly
show that the quarks in a hadron are sources of color electric flux
and that flux is trapped in a flux tube connecting the quarks. The
formation of the flux tube is related to the self-interaction of
gluons via their color charge. There exists a possibility that a
gluonic field can be excited and, by interacting with quarks,
produce mesons with exotic quantum numbers. Studying the spectrum of
the exotic mesons one can learn a great deal about the structure of
gluonic degrees of freedom and the confinement.


Since QCD is a gauge theory, it might be convenient to choose a
specific gauge to study the particular property of the theory, such
as confinement. It has been shown that the confinement of color
charge could be easily understood in minimal Coulomb gauge, while,
for instance, in Landau gauge the mechanism of this phenomenon is
rather mysterious \cite{GOZ}. In minimal Coulomb gauge the 0-0
component of the gluon propagator,
\begin{equation}
    D_{00}(x,t)=V_{coul}(x)\delta(t)+non\!-\!instantaneous,
\end{equation}
has an instantaneous part, $V_{coul}(r)$, that is long range and
confining and couples universally to all color-charge. The data of
numerical study \cite{GOZ1} are consistent with a linearly rising
potential, $V_{coul}(r)\sim \sigma_{coul} r$, and a Coulomb string
tension that is larger than the phenomenological string tension,
$\sigma_{coul} > \sigma$. Moreover, the 3-dimensionally transverse
physical components of the gluon propagator,
\begin{equation}
    D_{ij}(x,t)=\langle A_i(x,t)A_j(0,0) \rangle,
\end{equation}
are short range, corresponding to the absence of gluons from the
physical spectrum. This property makes Coulomb gauge especially
convenient to study nonperturbative QCD. More details on a study of
confinement in Coulomb gauge can be found in \cite{CoulGaugeConf}.
The first serious look at Coulomb gauge and the problem of
confinement there was in the paper by Szczepaniak and Swanson
\cite{SS}.

Every theory of confinement aims at explaining the linear rise of
the static quark potential, which is suggested by the linearity of
meson Regge trajectories. However, this phenomenon has a number of
other interesting properties that a satisfactory theory of
confinement is obligated to explain, one of them being Casimir
scaling. Casimir scaling \cite{DelDebbio:1995gc} refers to the fact
that there is an intermediate range of distances where the string
tension of static sources in color representation r is approximately
proportional to the quadratic Casimir of the representation; i.e.
\begin{equation}
    \sigma_r=\frac{C^2_r}{C^2_F}\sigma_F,
\end{equation}
where the subscript F refers to the fundamental representation. This
behavior was first suggested in Ref. \cite{Ambjorn:1984dp}. The term
`Casimir scaling' was introduced much later, in Ref.
\cite{DelDebbio:1995gc}, where it was emphasized that this behavior
poses a serious challenge to some prevailing ideas about
confinement.

\begin{figure}[h!]
\begin{center}
\includegraphics[width=10cm]{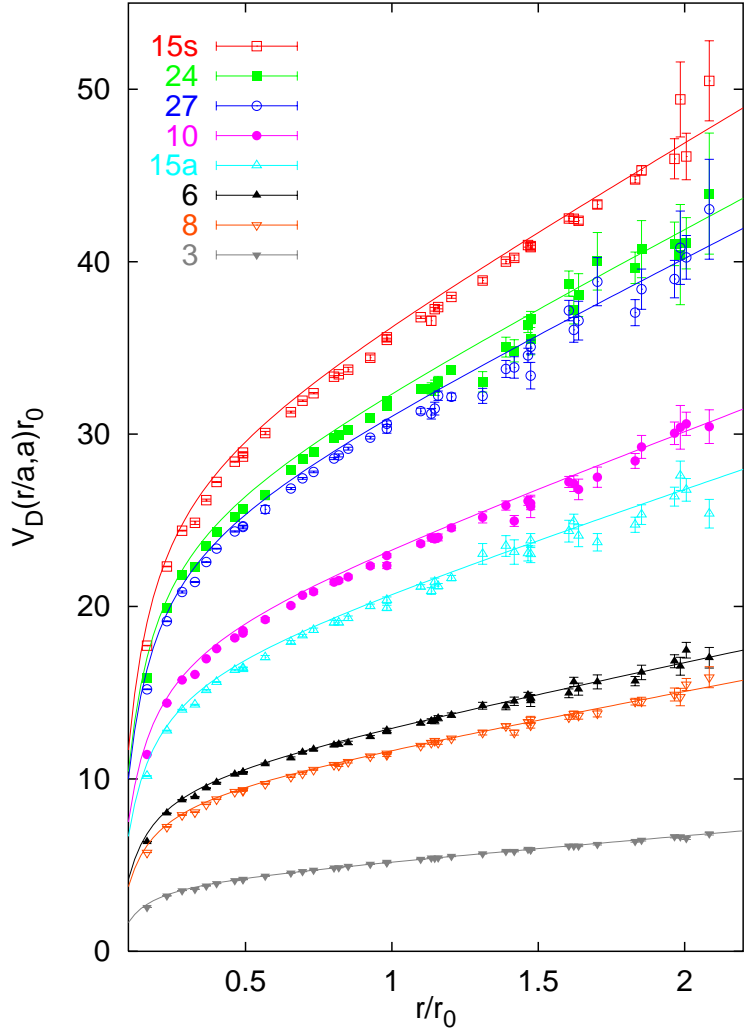}
\caption{Casimir scaling of confinement
\cite{Bali:2000un}.}\label{Casimir}
\end{center}
\end{figure}

Figure \ref{Casimir} shows in a compelling way the property of
Casimir scaling of confinement. The figure was obtained by measuring
the Wilson loop for sources in various representations of SU(3). The
interaction between color triplets is the lowest surface in the
figure and forms the template for the others. In the figure one sees
higher surfaces with sources in the 8, 6, $15_A$, 10, 27, 24, and
$15_S$ representations. The curves are obtained by multiplying a fit
to the lowest (fundamental representation) surface by the quadratic
Casimir, $C^2_r=\langle r|T^aT^a|r\rangle$ divided by $C^2_F$. The
quadratic Casimir is given by $(p^2+q^2+pq)/3+p+q$ where (p,q) is
the Dynkin index of the representation. The agreement is remarkable
and is a strong indication that the color structure of confinement
may be modelled as
\begin{equation}
    \int \bar{\psi}T^a \psi...\bar{\psi}T^a \psi
\end{equation}
where the ellipsis represents Lorentz and spatial dependence.

Chiral symmetry breaking is another interesting QCD property, it is
discussed in the next section.

\section{Chiral symmetry breaking}

In quantum field theory, chiral symmetry is a possible symmetry of
the Lagrangian under which the left-handed and right-handed parts of
Dirac fields transform independently. QCD Lagrangian has an
approximate flavor chiral symmetry $SU_L(N_f)\times SU_R(N_f)$ due
to the relative smallness of the masses of up, down and strange
quarks. This approximate symmetry is dynamically broken to $SU(N_f)$
and leads to the appearance of $(N_f^2-1)$ Goldstone bosons in the
theory (which are pseudoscalar mesons for QCD). Since chiral
symmetry is not exact (explicitly broken by small but nonzero quark
masses), Goldstone bosons in QCD are not massless but relatively
light. The actual masses of these mesons can in principle be
obtained in chiral perturbation theory through an expansion in the
(small) actual masses of the quarks.

The mechanism of dynamical chiral symmetry breaking is closely
related to the structure of the vacuum. In QCD, quarks and
antiquarks are strongly attracted to each other, therefore if these
quarks are massless, the energy cost of the pair creation from the
vacuum is small. So we expect that QCD vacuum contains
quark-antiquark condensates with the vacuum quantum numbers (zero
total momentum and angular momentum). It means that the condensates
have nonzero chiral charge, pairing left-handed quarks with the
antiparticles of right-handed quarks. It leads to the nonzero vacuum
expectation value for the scalar operator
\begin{equation}
    \langle0|\bar{Q}Q|0\rangle=\langle0|\bar{Q}_LQ_R+\bar{Q}_RQ_L|0\rangle \not{\!\!=} 0.
\end{equation}

The expectation value signals that the vacuum mixes the two quark
helicities. This allows massless quarks to acquire effective mass as
they move through the vacuum. Inside quark-antiquark bound states,
quarks appear to move if they are massive, even though they have
zero bare mass (in the Lagrangian).

Dynamical chiral symmetry breaking is impossible in perturbation
theory because at every finite order in perturbation theory the
self-energy of the particle is proportional to its renormalized
mass. So if one starts with a chirally symmetric theory then one
will also end up with a chirally symmetric theory, if using
perturbative approaches. Therefore dynamical chiral symmetry
breaking has to be studied using nonperturbative methods.

\begin{figure}[h!]
\begin{center}
\includegraphics[trim=0cm 12cm 0cm 0cm, width=10cm]{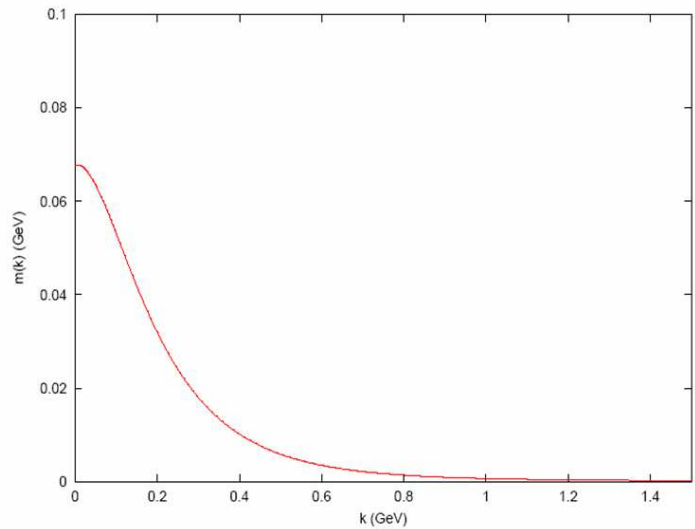}
\caption{Dependence of the dynamical quark mass on the momentum
calculated in Coulomb gauge model introduced in section
\ref{CoulombGauge}.}\label{DynMass}
\end{center}
\end{figure}

In the many-body approach dynamical chiral symmetry breaking and
momentum-dependent mass generation of elementary excitations can be
described by the \emph{Gap Equation} (an example of a gap equation
will be presented in section \ref{CoulombGauge}). The Gap Equation
allows one to calculate the mass function of the particle which is
momentum-dependent. The mass function of the quark calculated in
this approach is presented in the Fig. \ref{DynMass}. One can see
that the dynamical quark mass is large in the infrared and
suppressed in the ultraviolet, this result is not possible in weakly
interacting theories.

Another useful tool to study dynamical chiral symmetry breaking is
the method based on Dyson-Schwinger equations. In fact, the simplest
Dyson-Schwinger equation is the gap equation for the dressed quark
propagator. By solving this equation one would obtain the mass
function of the quark (dependence of the quark mass on the
momentum). This equation cannot be solved exactly however since it
is one of the equations of the self-consistent set of infinite
number of coupled nonlinear integral equations. Truncation schemes
appropriate to this problem have been found and the
momentum-dependence of the quark mass has been calculated. It is in
excellent agreement with lattice gauge theory calculations.

In the next chapter of the present dissertation we introduce the
quark models studied and explain our methods.

\chapter{THEORY}\label{theory}
The study of the meson sector has attracted much attention, with a
great variety of different models. The fundamental reason is that it
is a very good laboratory for exploring the nonperturbative QCD
regime. `Composed of a quark and an antiquark', a meson is the
simplest nontrivial system that can be used to test basic QCD
properties. In particular, the meson spectra can be reasonably
understood in non-relativistic or semi-relativistic models with
simple or sophisticated versions of the funnel potential, containing
a long-range confining term plus a short-range Coulomb-type term
coming from one-gluon exchange \cite{DeRujula:1975ge,BGS}.

Energies are not very stringent observables and to test more deeply
the wave functions, one needs to rely on more sensitive observables.
Electromagnetic properties, such as decay constants or form factors
can be employed. In that case the transition operator is precisely
known. On the other hand, one can also study hadronic transitions
occurring through the strong interaction; this kind of transition is
able to explain the decay of a meson into several mesons, or
baryon-antibaryon, or other more complicated channels. The
hadronization process is quite difficult to understand and model in
terms of basic QCD. One reason is that, contrary to the
electromagnetic case, the transition operator is not defined
precisely.

\section{Quark models of hadron structure}

\subsection{Nonrelativistic Potential Quark Model}\label{NPQM}

In the nonrelativistic potential quark model the meson is
approximated to be a bound state of interacting quark and antiquark.
The meson state for such a system is:
\begin{eqnarray}\label{meson}
|M\rangle&=&\sqrt{2E_P}\sum_{c\bar{c}s\bar{s}f\bar{f}M_SM_L}
\frac{\delta_{c\bar{c}}}{\sqrt{3}} \langle JM|LM_LSM_S \rangle
\chi^{SM_S}_{s\bar{s}}\, \Xi^{II_z}_{f\bar{f}}\nonumber\\
&&\times\int\frac{d^3kd^3\bar{k}}{(2\pi)^3}\Phi\left(\frac{m_{\bar{q}}\vec{k}-m_q\vec{\bar{k}}}{m_q+m_{\bar{q}}}\right)
\frac{\delta^{(3)}(\vec{k}+\vec{\bar{k}}-\vec{P})}{\sqrt{2E_k}\sqrt{2E_{\bar{k}}}}|\vec{k},\vec{\bar{k}}\rangle
\end{eqnarray}
where $\vec{P}$ is the meson momentum, $S$, $L$ and $J$ are the
meson spin, orbital and angular momenta with projections $M_S$,
$M_L$ and $M$. $X^{SM_S}_{s\bar{s}}$ is the spin wave function of
the meson, it depends on spin projections of quark and antiquark $s$
and $\bar{s}$ and also on the meson spin and its projection.
$\Xi^{II_z}_{f\bar{f}}$ is the flavor wave function and it depends
on the flavors of the quark and antiquark $f$ and $\bar{f}$ and on
the meson isospin $I$ and its projection $I_z$. $\Phi$ is the
spatial wave function, it depends on the momenta $k$ and $\bar{k}$
of quark and antiquark with masses $m_q$ and $m_{\bar{q}}$.

In the nonrelativistic approximation the mesonic wave function is
the eigenfunction of a Schrodinger equation:
\begin{equation}
\hat{H}\Psi=E\Psi,
\end{equation}
and the Hamiltonian for the system is:
\begin{equation}
H=K+V(r),
\end{equation}
where $K$ is the nonrelativistic kinetic energy and $V(r)$ is the
potential energy.

Several phenomenological models for the interaction potential exist.
The simplest one is a spherical harmonic oscillator potential. It is
a rather crude approximation and doesn't give good description of
the meson properties, for example it can't distinguish between two
mesons with different spins. But it allows analytical calculations
for most of the meson properties and easy Fourier transformations of
the wave functions, so it is useful as a simple estimate of some
physical quantities of interest.

Another variation of the nonrelativistic potential model is ISGW
\cite{ISGW}, which is based on SHO potential model but with an
artificial factor $\kappa$ introduced so that
$|\vec{q}|\rightarrow|\vec{q}|/\kappa$. The $\kappa$ factor was
added to achieve better agreement with the experimental data for the
pion form-factor and certain heavy quark transitions.

A more realistic model of the potential is Coulomb+linear+hyperfine
interaction model:
\begin{equation}\label{vcEq}
V(r)=-\frac{4}{3}\frac{\alpha_C}{r}+br+C+\frac{32\alpha_H\sigma^3
e^{-r^2\sigma^2}}{9m_1m_2\pi^{1/2}}\,\vec{S}_1\cdot\vec{S}_2.
\end{equation}
The strengths of the Coulomb and hyperfine interactions are taken as
separate parameters. Perturbative gluon exchange implies that
$\alpha_C = \alpha_H$ and we find that the fits prefer the near
equality of these parameters.

The Coulomb term corresponds to the quark interaction due to the one
gluon exchange and dominates at short range. The linear term
describes confinement. The hyperfine term is spin-dependent and
makes it possible to distinguish between mesons of different spins.
This potential has 3 parameters ($\alpha$, $\beta$ and $\sigma$),
and together with the mass of the quarks they could be adjusted to
describe the properties of the mesons (for examples the masses of
several meson ground states). After the parameters have been
adjusted, calculations of other meson properties could be done and
compared to the experimental data to see how the model works. Also
predictions of the physical properties, potentially observable in
the future, could be made.

As will be described in the next chapter, the observables that we
consider require a weaker ultraviolet interaction than that of Eq.
\ref{vcEq}. We therefore introduce a running coupling that recovers
the perturbative coupling of QCD but saturates at a phenomenological
value at low momenta:
\begin{equation}
\alpha_C \to
\alpha_C(k)=\frac{4\pi}{\beta_0\log\left({e^{\frac{4\pi}{\beta_0\alpha_0}}+\frac{k^2}{\Lambda^2}}\right)}
\label{alRunEq}
\end{equation}
where $k^2=|\vec{k}|^2$ is the square of the three-momentum
transfer, $\beta_0=11-2N_f/3=9$, $N_f$ is the number of flavors
taken to be 3. One can identify the parameter $\Lambda$ with
$\Lambda_{QCD}$ because $\alpha_C(k)$ approaches the one loop
running constant of QCD. However, this parameter will also be fit to
experimental data in the following (nevertheless, the resulting
preferred value is reassuringly close to expectations). Parameters
and details of the fit are presented in the Chapter \ref{dc}.

\begin{figure}[h!]
\begin{center}
\includegraphics[trim=-0.5cm 0cm -0.5cm 0cm, width=4.5cm]{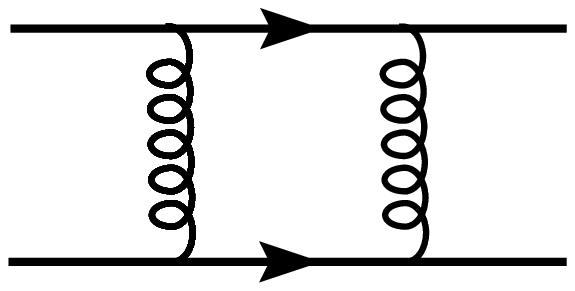}
\includegraphics[trim=-0.5cm 0cm -0.5cm 0cm, width=4.5cm]{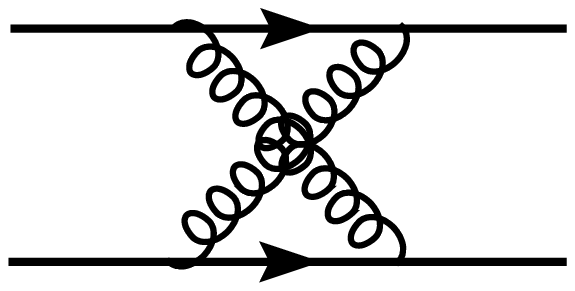}
\includegraphics[trim=-0.5cm 0cm -0.5cm 0cm, width=4.5cm]{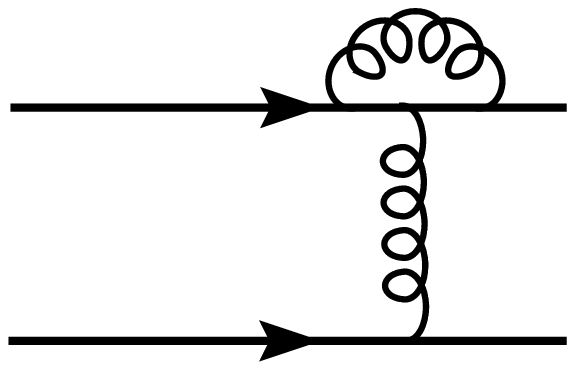}
\includegraphics[trim=-0.5cm 0cm -0.5cm 0cm, width=4.5cm]{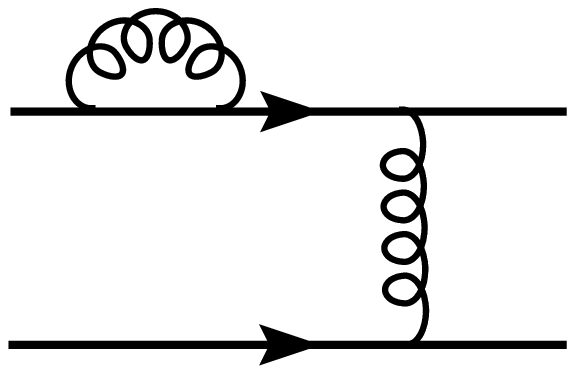}
\includegraphics[trim=-0.5cm 0cm -0.5cm 0cm, width=4.5cm]{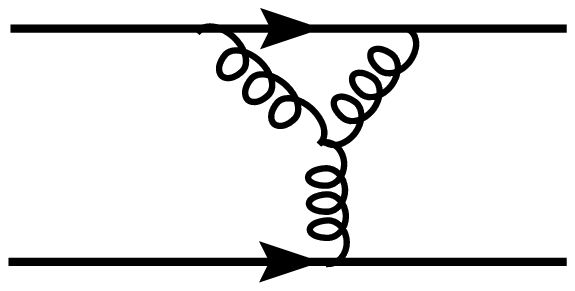}
\caption{One-loop diagrams of the heavy quark
interaction.}\label{OneLoopDiags}
\end{center}
\end{figure}

Potential of Eq. \ref{vcEq} cannot explain P-wave mass splittings
induced by spin-dependent interactions, which are due to spin-orbit
and tensor terms. A common model of spin-dependence is based on the
Breit-Fermi reduction of the one-gluon-exchange interaction
supplemented with the spin-dependence due to a scalar current
confinement interaction. The general form of this potential has been
computed by Eichten and Feinberg\cite{EF} at tree level using Wilson
loop methodology. The result is parameterized in terms of four
nonperturbative matrix elements, $V_i$, which can be determined by
electric and magnetic field insertions on quark lines in the Wilson
loop. Subsequently, Pantaleone, Tye, and Ng\cite{PTN} performed in a
one-loop computation of the heavy quark interaction and showed that
a fifth interaction, $V_5$ is present in the case of unequal quark
masses. The diagrams that have been calculated in addition to the
tree level diagram are presented in Fig. \ref{OneLoopDiags}.

The net result is a quark-antiquark interaction that can be written
as:

\begin{eqnarray}
V_{q\bar q}=V_{conf}+V_{SD}
\end{eqnarray}
where $V_{conf}$ is the standard Coulomb+linear scalar form:
\begin{equation}
V_{conf}(r)=-\frac{4}{3}\frac{\alpha_s}{r}+br
\end{equation}
and

\begin{eqnarray}
V_{SD}(r) &=& \left( {\bm{\sigma}_q \over 4 m_q^2} +
{\bm{\sigma}_{\bar q} \over 4 m_{\bar q}^2} \right)\cdot {\bf L}
\left( {1\over r} {d V_{conf} \over d r} + {2 \over r} {d V_1 \over
d r} \right) + \left( {\bm{\sigma}_{\bar q} +
        \bm{\sigma}_q \over 2 m_q m_{\bar q}} \right)\cdot {\bf L}
        \left( {1 \over r} {d V_2 \over d r} \right) \nonumber \\
&&+ {1 \over 12 m_q m_{\bar q}}\Big( 3 \bm{\sigma}_q \cdot \hat {\bf
r} \,
 \bm{\sigma}_{\bar q}\cdot \hat {\bf r} -   \bm{\sigma}_q\cdot
 \bm{\sigma}_{\bar q} \Big) V_3
+ {1 \over 12 m_q m_{\bar q}} \bm{\sigma}_q \cdot \bm{\sigma}_{\bar
q}
V_4 \nonumber \\
&&+ {1\over 2}\left[ \left( {\bm{\sigma}_q \over m_q^2} -
{\bm{\sigma}_{\bar q}\over m_{\bar q}^2}\right)\cdot {\bf L} +
\left({\bm{\sigma}_q - \bm{\sigma}_{\bar q}\over m_q m_{\bar
q}}\right)\cdot {\bf L} \right] V_5. \label{VSD}
\end{eqnarray}
Here ${\bf L} = {\bf L}_q = - {\bf L_{\bar q}}$, $r=|{\bf r}|= |{\bf
r}_q - {\bf r}_{\bar q}|$ is the ${\bar Q Q}$ separation and the
$V_i=V_i(m_q,m_{\bar q}; r)$ are the Wilson loop matrix elements
discussed above. The explicit expressions for $V_i$'s can be found
in the section \ref{OpenCharm} of the present dissertation.

The first four $V_i$ are order $\alpha_s$ in perturbation theory,
while $V_5$ is order $\alpha_s^2$; for this reason $V_5$ has been
ignored by quark modelers. For example, the analysis of Cahn and
Jackson\cite{CJ} only considers $V_1$ -- $V_4$. In practice this is
acceptable (as we show later) {\it except in the case of unequal
quark masses}, where the additional spin-orbit interaction can play
an important role.

\subsection{Relativistic Many-Body Approach in Coulomb Gauge}\label{CoulombGauge}

The canonical nonrelativistic quark model relies on a potential
description of quark dynamics and therefore neglects many-body
effects in QCD. Related to this is the question of the reliability
of nonrelativistic approximations, the importance of hadronic
decays, and the chiral nature of the pion. The latter two phenomena
depend on the behavior of nonperturbative glue and as such are
crucial to the development of robust models of QCD and to
understanding soft gluodynamics. Certainly, one expects that
gluodynamics will make its presence felt with increasing insistence
as experiments probe higher excitations in the spectrum. Similarly
the chiral nature of the pion cannot be understood in a fixed
particle number formalism. This additional complexity is the reason
so few models attempt to derive the chiral properties of the pion.
This is an unfortunate situation since the pion is central to much
of hadronic and nuclear physics.

To make progress one must either resort to numerical experiments or
construct models which are closer to QCD. One such model is based on
the QCD Hamiltonian in Coulomb gauge
\cite{SSJC,Cotanch,Felipe,RSSJC}.

In this approach the exact QCD Hamiltonian in the Coulomb gauge is
modeled by an effective, confining Hamiltonian, that is relativistic
with quark field operators and current quark masses. However, before
approximately diagonalizing H, a similarity transformation is
implemented to a new quasiparticle basis having a dressed, but
unknown constituent mass. As described later, this transformation
entails a rotation that mixes the bare quark creation and
annihilation operators. By then performing a variational calculation
to minimize the ground state (vacuum) energy, a specific mixing
angle and corresponding quasiparticle mass is selected. In this
fashion chiral symmetry is dynamically broken and a non-trivial
vacuum with quark condensates emerges. This treatment is precisely
analogous to the Bardeen, Cooper, and Schrieffer (BCS) description
of a superconducting metal as a coherent vacuum state of interacting
quasiparticles combining to form condensates (Cooper pairs). Excited
states (mesons) can then be represented as quasiparticle excitations
using standard many-body techniques, for example Tamm-Dancoff (TDA)
or random phase approximation (RPA) methods.

There are several reasons for choosing the Coulomb gauge framework.
As discussed by Zwanziger \cite{Zw}, the Hamiltonian is
renormalizable in this gauge and, equally as important, the Gribov
problem ($\Delta \cdot A = 0$ does not uniquely specify the gauge)
can be resolved (see Refs. \cite{ZW,RSSJC} for further discussion).
Related, there are no spurious gluon degrees of freedom since only
transverse gluons enter. This ensures all Hilbert vectors have
positive normalizations which is essential for using variational
techniques that have been widely successful in atomic, molecular and
condensed matter physics. Second, an advantage of Coulomb gauge is
the appearance of an instantaneous potential.

By introducing a potential $K^{(0)}$, the QCD Coulomb gauge
Hamiltonian \cite{RSSJC} for the quark sector can be replaced by an
effective Hamiltonian
\begin{equation}
H=\int d\vec{x} \bar{\Psi}^{\dag}\left(\vec{x}\right)
\left(-i\vec{\alpha}\cdot\vec{\Delta}+\beta m\right)
\bar{\Psi}\left(\vec{x}\right) +\frac{1}{2}\int d\vec{x} d\vec{y}
\rho^a\left(\vec{x}\right) K^{(0)}\left(|\vec{x}-\vec{y}| \right)
\rho^a\left(\vec{y}\right),
\end{equation}
where $\Psi$, $m$ and
$\rho^a(\vec{x})=\Psi^{\dag}\left(\vec{x}\right)T^a\Psi\left(\vec{x}\right)$
are the current (bare) quark field, mass and color density,
respectively. For notational ease the flavor subscript is omitted
(same H for each flavor) and the color index runs $a = 1...8$.

$K^{(0)}$ is defined as the vacuum expectation value of the
instantaneous non-Abelian Coulomb interaction. The procedure for
calculating $K^{(0)}$ is described in \cite{SS}. The solution is
well approximated by the following expression:
\begin{eqnarray}
K^{(0)}(k)=\frac{12.25}{k^2}\left\{
\begin{array}{lc}
\left(\frac{m_g}{k}\right)^{1.93} & k<m_g,\\
0.6588\log{(k^2/m^2_g+0.82)}^{-0.62} \log{(k^2/m^2_g+1.41)}^{-0.80}
& k>m_g.
\end{array}\right.
\end{eqnarray}

To find the meson wave function, equation $H\Psi=E\Psi$ has to be
solved as accurately as possible. First the ground state has to be
studied, and the Bogoliubov-Valatin, or BCS, transformation is
introduced.

The plane wave, spinor expansion for the quark field operator is:
\begin{equation}
\Psi(\vec{x})=\sum_{c\lambda} \int \frac{d\vec{k}}{(2\pi)^3}
\left[u_{c\lambda}(\vec{k})b_{c\lambda}(\vec{k})+v_{c\lambda}(-\vec{k})d^{\dag}_{c\lambda}(-\vec{k})\right]
e^{i\vec{k}\cdot\vec{x}}
\end{equation}
with free particle, anti-particle spinors $u_{c\lambda}$,
$v_{c\lambda}$ and bare creation, annihilation operators
$b_{c\lambda}$, $d_{c\lambda}$ for current quarks, respectively.
Here the spin state (helicity) is denoted by $\lambda$ and color
index by $c=1,2,3$ (which is hereafter suppressed). Because $\Psi$
could be expanded in terms of any complete basis, a new
quasiparticle basis may equally well be used:
\begin{equation}
\Psi(\vec{x})=\sum_{\lambda} \int \frac{d\vec{k}}{(2\pi)^3}
\left[U_{\lambda}(\vec{k})B_{\lambda}(\vec{k})+V_{\lambda}(-\vec{k})D^{\dag}_{\lambda}(-\vec{k})\right]
e^{i\vec{k}\cdot\vec{x}}
\end{equation}
entailing quasiparticle spinors $U_{\lambda}$, $V_{\lambda}$ and
operators $B_{\lambda}$,$D_{\lambda}$. The Hamiltonian is equivalent
in either basis and the two are related by a similarity
(Bogoliubov-Valatin or BCS) transformation. The transformation
between operators is given by the rotation
\begin{eqnarray}
B_{\lambda}(\vec{k})=\cos{\frac{\theta_k}{2}}b_{\lambda}(\vec{k})-\lambda\sin{\frac{\theta_k}{2}}d^{\dag}_{\lambda}(-\vec{k}),\nonumber\\
D_{\lambda}(-\vec{k})=\cos{\frac{\theta_k}{2}}d_{\lambda}(-\vec{k})+\lambda\sin{\frac{\theta_k}{2}}b^{\dag}_{\lambda}(\vec{k}),
\end{eqnarray}
involving the BCS angle $\theta_k=\theta(k)$. Similarly the rotated
quasiparticle spinors are
\begin{eqnarray}
U_{\lambda}(\vec{k})=\cos{\frac{\theta_k}{2}}u_{\lambda}(\vec{k})-\lambda\sin{\frac{\theta_k}{2}}v_{\lambda}(-\vec{k})=\frac{1}{\sqrt{2}}\left[
\begin{array}{c}
\sqrt{1+\sin{\phi(\vec{k})}}\,\,\,\chi_{\lambda}\\
\sqrt{1-\sin{\phi(\vec{k})}}\,\,\,\vec{\sigma}\cdot\hat{k}\,\,\,\chi_{\lambda}
\end{array}\right],\nonumber\\
V_{\lambda}(-\vec{k})=\cos{\frac{\theta_k}{2}}v_{\lambda}(-\vec{k})+\lambda\sin{\frac{\theta_k}{2}}u_{\lambda}(\vec{k})=\frac{1}{\sqrt{2}}\left[
\begin{array}{c}
-\sqrt{1-\sin{\phi(\vec{k})}}\,\,\,\vec{\sigma}\cdot\hat{k}\,\,\,\chi_{\lambda}\\
\sqrt{1+\sin{\phi(\vec{k})}}\,\,\,\chi_{\lambda}
\end{array}\right],
\end{eqnarray}
where $\chi_{\lambda}$ is the standard two-dimensional Pauli spinor.
The gap angle, $\phi_k=\phi(k)$, has also been introduced, which is
related to the BCS angle, $\theta/2$, by $\phi=\theta+\alpha$ where
$\alpha$ is the current, or perturbative, mass angle satisfying
$\sin{\alpha}=m/E_k$ with $E_k=\sqrt{m^2+k^2}$. Hence
\begin{eqnarray}
\sin{\phi_k}=\frac{m}{E_k}\cos{\theta_k}+\frac{k}{E_k}\sin{\theta_k},\nonumber\\
\cos{\phi_k}=\frac{k}{E_k}\cos{\theta_k}-\frac{m}{E_k}\sin{\theta_k}.
\end{eqnarray}
Similarly, the perturbative, trivial vacuum, defined by
$b_{\lambda}|0\rangle=d_{\lambda}|0\rangle=0$, is related to the
quasiparticle vacuum,
$B_{\lambda}|\Omega\rangle=D_{\lambda}|\Omega\rangle=0$, by the
transformation
\begin{equation}
|\Omega\rangle=exp\left(-\sum_{\lambda}\int\frac{d\vec{k}}{(2\pi)^3}\lambda\tan{\frac{\theta_k}{2}}
b^{\dag}_{\lambda}(\vec{k})d^{\dag}_{\lambda}(-\vec{k})\right)|0\rangle.
\end{equation}

Here $\Omega$ is so called BCS vacuum (later we introduce the RPA
vacuum labeled $|\Omega_{RPA}\rangle$ which is required to obtain a
massless pion). Expanding the exponential and noting that the form
of the operator $b^{\dag}d^{\dag}$ is designed to create a current
quark/antiquark pair with the vacuum quantum numbers, clearly
exhibits the BCS vacuum as a coherent state of quark/antiquark
excitations (Cooper pairs) representing $^{2S+1}L_J=\,^3P_0$
condensates. One can regard $tan{\frac{\theta_k}{2}}$ as the
momentum wavefunction of the pair in the center of momentum system.

An approximate ground state for our effective Hamiltonian could be
found by minimizing the BCS vacuum expectation,
$\langle\Omega|H|\Omega\rangle$. It could be done variationally
using the gap angle, $\phi_k$, which leads to the gap equation,
$\delta\langle\Omega|H|\Omega\rangle=0$. After considerable
mathematical reduction, the nonlinear integral gap equation follows
\begin{equation}
k\sin{\phi_k}-m\cos{\phi_k}=\frac{2}{3}\int\frac{d\vec{q}}{(2\pi)^3}K^{(0)}\left(|\vec{k}-\vec{q}|\right)
\left[\sin{\phi_k}\cos{\phi_q}\hat{k}\cdot\hat{q}-\sin{\phi_q}\cos{\phi_k}\right].
\end{equation}

This gap equation is to be solved for the unknown Bogoliubov angle,
which then specifies the quark vacuum and the quark field mode
expansion via spinors. Comparing the quark spinor to the canonical
spinor permits a simple interpretation of the Bogoliubov angle
through the relationship $\mu(k)=k\tan{\phi_k}$ where $\mu(k)$ may
be interpreted as a dynamical momentum-dependent quark mass.
Similarly $\mu(0)$ may be interpreted as a constituent quark mass.

The numerical solution for the dynamical quark mass is very
accurately represented by the functional form
\begin{equation}
\mu(k)=\sigma K^{(0)}(k) \left(1-e^{-M/(\sigma K^{(0)})(k)} \right)
\end{equation}
where M is a 'constituent' quark mass and $\sigma$ is a parameter
related to the quark condensate. Notice that this form approaches
the constituent mass for small momenta and $\sigma K^{(0)}$ for
large momenta.

With explicit expressions for the quark interaction and the
dynamical quark mass the mesonic bound states can now be obtained.
The definitions of the meson creation operators in TDA and RPA
approximations are (see $\S 59$ of Ref. \cite{FW}, also
\cite{Ring,Mattueck}):
\begin{eqnarray}
Q^{\dag}_M(TDA)&=&\sum_{\gamma\delta} \int\frac{d^3k}{(2\pi)^3}
\psi_{\gamma\delta}(\vec{k})
B^{\dag}_{\gamma}(\vec{k})D^{\dag}_{\delta}(-\vec{k}),\\
Q^{\dag}_M(RPA)&=&\sum_{\gamma\delta} \int\frac{d^3k}{(2\pi)^3}
\left[\psi^{\dag}_{\gamma\delta}(\vec{k}) B^{\dag}_{\gamma}(\vec{k})
D^{\dag}_{\delta}(-\vec{k}) -\psi^-_{\gamma\delta}(\vec{k})
B_{\gamma}(\vec{k}) D_{\delta}(-\vec{k})\right]
\end{eqnarray}
with $B$ and $D$ being the quasiparticle operators. It is worthwhile
recalling that the RPA method is equivalent to the Bethe-Salpeter
approach with instantaneous interactions \cite{Resag}.

 A meson is then represented by the Fock space expansion:
\begin{eqnarray}
|M_{TDA}\rangle&=&Q^{\dag}_M(TDA)|\Omega\rangle,\\
|M_{RPA}\rangle&=&Q^{\dag}_M(RPA)|\Omega_{RPA}\rangle.
\end{eqnarray}

Here $\Omega_{RPA}$ is RPA vacuum, it has both fermion (two
quasiparticles or Cooper pairs) and boson (four quasiparticles or
meson pairs) correlations.

To derive the TDA and RPA equations of motion we project the
Hamiltonian equation onto the truncated Fock sector. It gives:
\begin{eqnarray}
&&\langle M_{TDA}|[H,Q^{\dag}_M(TDA)]|\Omega
\rangle=(E_M-E_0)\langle M_{TDA}|Q^{\dag}_M(TDA)|\Omega
\rangle,\label{TDA_eq}\\
&&\langle M_{RPA}|[H,Q^{\dag}_M(RPA)]|\Omega_{RPA}
\rangle=(E_M-E_0)\langle M_{RPA}|Q^{\dag}_M(RPA)|\Omega_{RPA}
\rangle\label{RPA_eq}
\end{eqnarray}

In TDA (\ref{TDA_eq}) generates an integral equation for the meson
wave function $\psi(\vec{k})$, and in RPA (\ref{RPA_eq}) generates
two coupled nonlinear integral equations for two wave functions
$\psi^{\dag}(\vec{k})$ and $\psi^-(\vec{k})$.

The RPA and TDA equations include self energy terms (denoted
$\Sigma$) for each quark line and these must be renormalized. In the
zero quark mass case renormalization of the TDA or RPA equations
proceeds in the same way as for the quark gap equation. In fact, the
renormalization of these equations is consistent and one may show
that a finite gap equation implies a finite RPA or TDA equation.
This feature remains true in the massive case. The RPA equation in
the pion channel reads:
\begin{eqnarray}\label{RPA_pion}
(E_{\pi}-E_{BCS})\psi^{\dag}(k)&=&2\left[m\sin{\phi_k}+k\cos{\phi_k}+\Sigma(k)\right]\psi^{\dag}(k)-\nonumber\\
&&-\frac{C_F}{2}\int\frac{q^2dq}{(2\pi)^3}\Big[V_0(k,q)(1+\sin{\phi_k}\sin{\phi_q})+V_1(k,p)\cos{\phi_k}\cos{\phi_q}\Big]\psi^{\dag}(q)-\nonumber\\
&&-\frac{C_F}{2}\int\frac{q^2dq}{(2\pi)^3}\Big[V_0(k,q)(1-\sin{\phi_k}\sin{\phi_q})-V_1(k,p)\cos{\phi_k}\cos{\phi_q}\Big]\psi^{-}(q),\nonumber\\
\end{eqnarray}
where
\begin{equation}
V_L(k,q)=2\pi\int d(\hat{q}\cdot\hat{k})
K^{(0)}\left(|\vec{q}-\vec{k}|\right) P_L(\hat{q}\cdot\hat{k}).
\end{equation}

A similar equation for $\psi^{-}$ holds with $(+\rightarrow-)$ and
$E\rightarrow-E$. The wavefunctions $\psi^{\pm}$ represent forward
and backward moving components of the many-body wavefunction and the
pion itself is a collective excitation with infinitely many
constituent quarks in the Fock space expansion. These two coupled
nonlinear integral equation could be solved numerically to obtain
meson spectrum and wave functions.

TDA equation may be obtained from the RPA equation (\ref{RPA_pion})
by neglecting the backward wave function $\psi^{-}$. The spectrum in
the random phase and Tamm-Dancoff approximations has been computed
\cite{Norbert} and it has been confirmed that the pion is massless
in the chiral limit. It was also found that the Tamm-Dancoff
approximation yields results very close to the RPA for all states
except the pion. All other mesons have nearly identical RPA and TDA
masses. The complete hidden flavor meson spectrum in the
Tamm-Dancoff approximation is given by the following equations.
\begin{equation}
E\psi_{PC}(k)=2\left[m\sin{\phi_k}+k\cos{\phi_k}+\Sigma(k)\right]\psi_{PC}(k)-\frac{C_F}{2}\int
\frac{q^2 dq}{(2\pi)^3} K^{PC}_J(k,q)\psi_{PC}(q)
\end{equation}
with
\begin{equation}
\Sigma(k)=\frac{C_F}{2}\int\frac{q^2 dq}{(2\pi)^3}
\left(V_0\sin{\phi_k}\sin{\phi_q}+V_1\cos{\phi_k}\cos{\phi_q}\right)
\end{equation}
and where $\psi$ is the meson radial wavefunction in momentum space.
Note that the imaginary part of the self-energy $Im(\Sigma)=0$, this
follows from the fact that the quark-antiquark interaction is
instantaneous in the Coulomb gauge.

The kernel $K_J$ in the potential term depends on the meson quantum
numbers, $J^{PC}$. In the following possible values for the parity
or charge conjugation eigenvalues are denoted by $(J)=+$ if $J$ is
even and $-$ if $J$ is odd. These interaction kernels have been
derived in the quark helicity basis (see for example Ref.
\cite{Norbert}).

\begin{itemize}
  \item $0^{++}$\\
   \begin{equation}
    K(p,k)=V_0\cos{\phi_p}\cos{\phi_k}+V_1\left(1+\sin{\phi_p}\sin{\phi_k}\right)
   \end{equation}
  \item $J^{(J+1)(J)}\,\,\,\,[^1J_J, J\ge 0]$\\
   \begin{equation}
    K_J(p,k)=V_J\left(1+\sin{\phi_p}\sin{\phi_k}\right)
    +\left(V_{J-1}\frac{J}{2J+1}+V_{J+1}\frac{J+1}{2J+1}\right)\cos{\phi_p}\cos{\phi_k}
   \end{equation}
  \item $J^{(J+1)(J+1)}\,\,\,\,[^3J_J, J\ge 1]$\\
   \begin{equation}
    K_J(p,k)=V_J\left(1+\sin{\phi_p}\sin{\phi_k}\right)
    +\left(V_{J-1}\frac{J+1}{2J+1}+V_{J+1}\frac{J}{2J+1}\right)\cos{\phi_p}\cos{\phi_k}
   \end{equation}
  \item $J^{(J)(J)}\,\,\,\,[^3(J-1)_J, ^3(J+1)_J, J\ge 1]$
   \begin{eqnarray}
    K_{11}(p,k)&=&V_J\cos{\phi_p}\cos{\phi_k}
    +\left(V_{J-1}\frac{J}{2J+1}+V_{J+1}\frac{J+1}{2J+1}\right)
    \left(1+\sin{\phi_p}\sin{\phi_k}\right)\nonumber\\
    K_{22}(p,k)&=&V_J\cos{\phi_p}\cos{\phi_k}
    +\left(V_{J-1}\frac{J+1}{2J+1}+V_{J+1}\frac{J}{2J+1}\right)
    \left(1+\sin{\phi_p}\sin{\phi_k}\right)\nonumber\\
    K_{12}(p,k)&=&\left(V_{J-1}-V_{J+1}\right)\frac{\sqrt{J(J+1)}}{2J+1}
    \left(\sin{\phi_k}+\sin{\phi_p}\right)
   \end{eqnarray}
\end{itemize}

\section{Strong decays}\label{3P0} 

The decay of a meson into two mesons is the simplest example of a
strong decay. The decay of a baryon into a meson and a baryon has
also been extensively studied. Even in those particularly simple
decays, various models have been proposed to explain the mechanism.
Among them, let us cite the naive $SU(6)_W$ model \cite{SU6W}, the
elementary meson-emission model
\cite{Mitra,Faiman,Sakurai,Moorhouse,Koniuk} (in which one emitted
meson is considered as an elementary particle coupled to the quark),
the $^3S_1$ model \cite{Furui,Alcock} (in which a quark-antiquark
pair is created from the gluon emitted by a quark of the original
meson), the flux-tube model \cite{FluxTube} and the $^3P_0$ model
(in which a quark-antiquark pair is created from the vacuum)
\cite{Micu,Yaouanc3P0,Chaichian,ABS}.

This last model ($^3P_0$) is especially attractive because it can
provide the gross features of various transitions with only one
parameter, the constant corresponding to the creation vertex. This
property is of course an oversimplification because there is no
serious foundation for a creation vertex independent of the momenta
of the created quarks. Even in the $^3P_0$ model, the form of the
vertex is essentially unknown.

This phenomenological model of hadron decays was developed in the
1970s by LeYaouanc et al, \cite{Yaouanc3P0}, which assumes, as
suggested earlier by Micu in \cite{Micu}, that during a hadron decay
a $q\bar{q}$ pair is produced from the vacuum with vacuum quantum
numbers, $J^{PC} = 0^{++}$. Since this corresponds to a $^3P_0$
$q\bar{q}$ state, this is now generally referred to as the $^3P_0$
decay model. The $^3P_0$ pair production Hamiltonian for the decay
of a $q\bar{q}$ meson A to mesons B + C is usually written in a
rather complicated form with explicit wavefunctions \cite{3P0sum},
which in the conventions of Geiger and Swanson \cite{GS} (to within
an irrelevant overall phase) is
\begin{eqnarray}\label{Ham3P0}
\langle BC|H_I |A\rangle = \gamma \int\int
\frac{d^3rd^3y}{(2\pi)^{3/2}} e^{\frac{i}{2}\vec{P}_B\cdot\vec{r}}
\Psi_A\left(\vec{r}\right) \langle \vec{\sigma}
\rangle_{q\bar{q}}\cdot \left(
i\vec{\Delta}_B+i\vec{\Delta}_C\right. && \!\!\!\!\!\!\!\!\left.
+\vec{P}_B \right) \Psi^*_B\left( \frac{\vec{r}}{2}+\vec{y} \right)
\Psi^*_C\left( \frac{\vec{r}}{2}-\vec{y} \right)\nonumber\\
&&\delta\left( \vec{P}_A-\vec{P}_B-\vec{P}_C \right)
\end{eqnarray}
for all quark and antiquark masses equal. The strength $\gamma$ of
the decay interaction is regarded as a free constant and is fitted
to data \cite{gamma}.

Studies of hadron decays using this model have been concerned almost
exclusively with numerical predictions, and have not led to any
fundamental modifications. Recent studies have considered changes in
the spatial dependence of the pair production amplitude as a
function of quark coordinates \cite{3P0sum} but the fundamental
decay mechanism is usually not addressed; this is widely believed to
be a nonperturbative process, involving "flux tube breaking".

\section{Electromagnetic and electroweak transitions}

Since the operator of electromagnetic and electroweak transitions is
very well known, studying these processes for hadrons could provide
us with valuable information on the hadron structure. Still these
transitions are complicated enough, so that simplifying
approximations are typically in use. In this section, different
types of electromagnetic and electroweak transitions are described,
and approaches to study them are explained.

\subsection{Decay constants}\label{dcSection}

Leptonic decay constants are a simple probe of the short distance
structure of hadrons and therefore are a useful observable for
testing quark dynamics in this regime. Decay constants are computed
by equating their field theoretic definition with the analogous
quark model definition. This identification is rigorously valid in
the nonrelativistic and weak binding limits where quark model state
vectors form good representations of the Lorentz
group\cite{ISGW,HI}. The task at hand is to determine the
reliability of the computation away from these limits.

The method is illustrated with the vector meson decay constant
$f_V$,  which is defined by

\begin{equation}
m_V \, f_V \, \epsilon^{\mu}= \langle
0|\bar{\Psi}(0)\gamma^{\mu}\Psi(0)|V\rangle \label{VdecayEq}
\end{equation}
where $m_V$ is the vector meson mass and $\epsilon^{\mu}$ is its
polarization vector. Note that the vector current is locally
conserved for the physical vector meson.

The decay constant is computed in the conceptual weak binding and
nonrelativistic limit of the quark model and is assumed to be
accurate away from these limits. One thus employs the quark model
state:

\begin{equation}
|V(P)\rangle = \sqrt{\frac{2 E_P}{N_c}} \chi^{SM_S}_{s\bar{s}}\,
\int\frac{d^3k\,
d^3\bar{k}}{(2\pi)^3}\Phi\left(\frac{m_{\bar{q}}\vec{k}-m_q\vec{\bar{k}}}{m_{\bar{q}}+m_q}\right)
\delta^{(3)}(\vec{k}+\vec{\bar{k}}-\vec{P}) b^\dag_{ks}
d^\dag_{\bar{k}\bar s}|0\rangle, \label{VmesonEq}
\end{equation}
where $m_q$ and $m_{\bar{q}}$ are the masses of quark and antiquark
with momenta $\vec{k}$ and $\vec{\bar{k}}$ accordingly, $\vec{P}$ is
the vector meson momentum. The decay constant is obtained by
computing the spatial matrix element of the current in the vector
center of mass frame (the temporal component is trivial) and yields
\begin{equation}
f_V=\sqrt{\frac{N_c}{m_V}}\int\frac{d^3k}{(2\pi)^3} \Phi(\vec{k})
\sqrt{1+\frac{m_q}{E_k}} \sqrt{1+\frac{m_{\bar{q}}}{E_{\bar{k}}}}
\left(1+\frac{k^2}{3(E_k+m_q)(E_{\bar{k}}+m_{\bar{q}})}\right).
\label{relfVEq}
\end{equation}
The nonrelativistic limit is proportional to the meson wave function
at the origin
\begin{equation}
f_V
= 2\sqrt{\frac{N_c}{m_V}} \tilde{\Phi}(r=0); \label{nonrelfVEq}
\end{equation}
which recovers the well-known result of van Royen and
Weisskopf\cite{vRW}.

Decay constant for vector mesons with quark and antiquark of the
same flavor could be determined from the experimental data for the
decay $V\rightarrow e^{+}e^{-}$. In this process the vector meson
first converts into the photon and then photon becomes the
electron-positron pair. The amplitude of this process is then:
\begin{equation}
A_{s_1s_2}=\langle
e^{+}e^{-}|\bar{\Psi}\gamma^{\mu}\Psi|0\rangle\langle
0|\bar{\Psi}\gamma_{\mu}\Psi|V\rangle =\frac{e^2}{q^2}Q
\bar{u}_{s_1}(p_1)\gamma^{\mu}v_{s_2}(p_2)f_Vm_V\epsilon_{\mu}
\end{equation}
where $p_1$ and $p_2$ are the momenta, $s_1$ and $s_2$ are spins of
the electron and positron, Q is the quark charge (in units of $e$),
$q$ is the photon momentum.

Then the squared amplitude summed over the electron and positron
spins and averaged over vector meson polarizations is:
\begin{eqnarray}
|A|^2=\frac{1}{3}\sum_{s_1s_2}|A_{s_1s_2}|^2&=&\frac{e^4}{3q^4} Q^2
f^2_V m^2_V \epsilon_{\mu} \epsilon^*_{\nu}
\sum_{s_1s_2}\left[\bar{u}_{s_1}(p_1)\gamma^{\mu}v_{s_2}(p_2) \bar{v}_{s_2}(p_2)\gamma^{\nu}u_{s_1}(p_1)\right]\nonumber\\
&=&\frac{e^4}{3q^4} Q^2 f^2_V m^2_V (-g_{\mu\nu}) Tr[(\rlap{/}{p_2}-m_e)\gamma^{\nu}(\rlap{/}{p_1}+m_e)\gamma^{\mu}]\nonumber\\
&=&-\frac{4e^4}{3q^4} Q^2 f^2_V m^2_V g_{\mu\nu} [p^{\nu}_2
p^{\mu}_1 + p^{\nu}_1 p^{\mu}_2-g^{\mu\nu}(p_1\cdot p_2+ m^2_e)]
\end{eqnarray}

Since in this process the masses of electron and positron are much
smaller than their momenta, we can neglect $m_e$. Then:
\begin{equation}
|A|^2=\frac{8e^4}{3q^4}Q^2f^2_Vm^2_V(p_1\cdot p_2)
\end{equation}

From the momentum conservation law $q=p_1+p_2$ so
\begin{equation}
(p_1+p_2)^2=p^2_1+p^2_2+2p_1p_2=2m^2_e+2p_1p_2\approx 2p_1p_2=q^2
\end{equation}
and then $p_1p_2=q^2/2$ so
\begin{equation}
|A|^2=\frac{4e^4}{3q^2}Q^2f^2_Vm^2_V=\frac{4}{3}e^4Q^2f^2_V
\end{equation}
because in the meson rest frame $q=m_V$.

Now we can calculate the decay rate of this process:
\begin{equation}
\Gamma_{V\rightarrow
e^{+}e^{-}}=\frac{1}{2m_V}\int\frac{d\Omega_{cm}}{32\pi^2}\left(\frac{2|\vec{p}_1|}{E_{cm}}\right)|A|^2
\end{equation}

Here $E_{cm}$ is the energy of the final state in its rest frame.
Since $m_e\approx 0$ then $E_{cm}=2|\vec{p}_1|$ and then the decay
rate is:
\begin{equation}
\Gamma_{V\rightarrow e^{+}e^{-}}=\frac{e^4Q^2f^2_V}{12\pi
m_V}=\frac{4\pi\alpha^2}{3}\frac{Q^2f^2_V}{m_V}
\end{equation}
and the decay constant is:
\begin{equation}
f_V=\left(\frac{3m_V\Gamma_{V\rightarrow
e^{+}e^{-}}}{4\pi\alpha^2Q^2}\right)^{1/2}
\end{equation}

That gives the following results for the existing vector mesons:
\begin{eqnarray}
f_{\rho}=217 MeV & (Q=\frac{1}{\sqrt{2}})\nonumber\\
f_{\phi}=229 MeV & (Q=-\frac{1}{3})\nonumber\\
f_{J/\psi}=411 MeV & (Q=\frac{2}{3})\nonumber\\
f_{\Upsilon}=704 MeV & (Q=-\frac{1}{3})
\end{eqnarray}

Similar results hold for other mesons that couple to electroweak
currents. A summary of the results for a variety of models and the
discussion are presented in Chapter \ref{dc}. The expressions used
to compute the table entries and the data used to extract the
experimental decay constants are collected in Appendix
\ref{DecayConstantsApp}. \clearpage

\subsection{Impulse approximation}\label{imp}

\begin{figure}[h!]
\begin{center}
\includegraphics[trim=1cm 20cm 1cm -1cm,width=10cm]{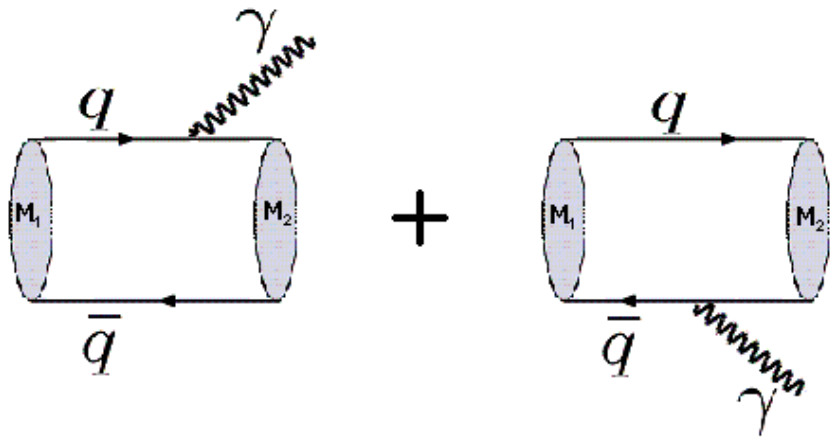}
\caption{Impulse approximation diagrams.}\label{Impulse}
\end{center}
\end{figure}

The impulse approximation is widely used in studies of meson
transitions and form-factors. In this approximation the possibility
of quark-antiquark pair creation from the vacuum is neglected. The
interaction of the external current with the meson is the sum of its
coupling to quark and antiquark as illustrated in figure
\ref{Impulse}. In the diagrams, $M_1$ and $M_2$ are the initial and
final state mesons (bound states of quark $q$ and antiquark
$\bar{q}$, which are represented by lines with arrows). In this
section, our approach to the calculations of the form-factors and
radiative transition decay rates in the impulse approximation of the
quark model is presented.

Form factors are a powerful determinant of internal hadronic
structure because the external current momentum serves as a probe
scale. And of course, different currents are sensitive to different
properties of the hadron, so it is useful to study the form-factors
when tuning and testing models.

The technique used to compute the form factors is illustrated by
considering the inelastic pseudoscalar electromagnetic matrix
element $\langle P_2 | J^\mu | P_1\rangle$, where $P$ refers to a
pseudoscalar meson. The most general Lorentz covariant decomposition
of this matrix element is

\begin{equation}
\langle P_2(p_2)|\bar{\Psi}(0)\gamma^{\mu}\Psi(0)|P_1(p_1)\rangle =
f(Q^2)\left((p_2+p_1)^{\mu}-\frac{M^2_2-M^2_1}{q^2}(p_2-p_1)^{\mu}\right)
\end{equation}
where conservation of the vector current has been used to eliminate
a possible second invariant. The argument of the form factor is
chosen to be $Q^2 = - (p_2-p_1)_\mu (p_2-p_1)^\mu$.

Now the matrix element on the left could be calculated in some
model, for example in the quark model, and then the result for the
form-factor $f(Q^2)$ could be compared to the experimental data (if
available).

In the impulse approximation, using the temporal component of the
vector current and computing in the rest frame of the initial meson
yields
\begin{eqnarray}\label{ffFullEq}
f_{sq}(Q^2)&=&\frac{\sqrt{M_1E_2}}{(E_2+M_1)-\frac{M^2_2-M^2_1}{q^2}(E_2-M_1)}\\
&\times&
\int \frac{d^3k}{(2\pi)^3} \Phi(\vec{k})
\Phi^*\left(\vec{k}+\vec{q}\frac{\bar{m}_2}{m_2+\bar{m}_2}\right)
\sqrt{1+\frac{m_1}{E_k}}\sqrt{1+\frac{m_2}{E_{k+q}}}
\left(1+\frac{(\vec{k}+\vec{q})\cdot\vec{k}}
{(E_k+m_1)(E_{k+q}+m_2)}\right).\nonumber
\end{eqnarray}
The pseudoscalars are assumed to have valence quark masses $m_1,
\bar m_1$ and $m_2, \bar m_2$ for $P_1$ and $P_2$ respectively. The
masses of the mesons are labeled $M_1$ and $M_2$. The single quark
elastic form factor can be obtained by setting $m_1= \bar m_1 = m_2
= \bar m_2$ and $M_1 = M_2$. In the nonrelativistic limit Eq.
\ref{ffFullEq} reduces to the simple expression:

%

\begin{equation}
f_{sq}(Q^2)= \int \frac{d^3k}{(2\pi)^3} \Phi(\vec{k})
\Phi^*\left(\vec{k}+\frac{\vec{q}}{2}\right). \label{ff0Eq}
\end{equation}
In this case it is easy to see the normalization condition
$f_{sq}(\vec q =0) = 1$. This is also true for the relativistic
elastic single quark form factor of Eq. \ref{ffFullEq}.

To calculate the decay rate of the radiative transition
$M_1\rightarrow \gamma M_2$ we need to know the electromagnetic
matrix element $\langle M_2|J^{\mu}|M_1\rangle$ at $q^2=0$, where
$M_1$ and $M_2$ are the initial and final meson states. In the
impulse approximation using the vector component of the current we
have:
\begin{eqnarray}\label{oper_n}
\vec{A}_{em}&\equiv& \langle M_2|\vec{J}|M_1\rangle=eQ_q\langle
M_2|\vec{J}^{(q)}|M_1\rangle
+eQ_{\bar{q}}\langle M_2|\vec{J}^{(\bar{q})}|M_1\rangle,\\
\vec{J}^{(q)}&=&-u^\dag_{s_2}({k_2})\,{\vec{\alpha}}\,u_{s_1}(k_1),\nonumber\\
\vec{J}^{(\bar{q})}&=&-v^\dag_{\bar{s}_1}(\bar{k}_1)\,{\vec{\alpha}}\,v_{\bar{s}_2}(\bar{k}_2).\nonumber
\end{eqnarray}
where ${\bf{k}}_1,\bar{\bf{k}}_1,s_1$ and $\bar{s}_1$ are the
momenta and spins of the quark and antiquark of the initial state
meson, and $\bf{k}_2,\bar{\bf{k}}_2,s_2,\bar{s}_2$ are the
corresponding momenta and spins of the final state meson. $Q_q$ and
$Q_{\bar{q}}$ are the quark and antiquark charges. The two terms of
(\ref{oper_n}) corresponds to the quark and antiquark
electromagnetic interactions.

It is very common to consider quark and antiquark being
nonrelativistic when studying radiative transition. We investigate
the validity of this approximation by comparing two cases: taking
the full relativistic expressions for quark and antiquark spinors
and then comparing our results to those calculated with the
nonrelativistic approximation. We find considerable differences for
the decay rates, even for heavy mesons, as will be shown in Chapter
\ref{RadTrans}, and conclude that quarks should be treated
relativistically.

We illustrate the technique used to study radiative transitions for
the nonrelativistic approximation of the quark spinors. The
treatment of the case with full relativistic expressions for spinors
is completely analogous, except for much more complicated
expressions for the matrix elements. The study of full relativistic
case have been performed numerically.

In the rest frame of the initial state meson we have:
\begin{equation}\label{Aem}
\vec{A}_{em}=\frac{eQ_q}{2m_q}\, \Big{\langle} M_2
\Big|(2\vec{k}+\vec{q})\delta_{s_1s_2}-i
\vec{q}\times\vec{\sigma}_{21} \Big|M_1\Big{\rangle}+
\frac{eQ_{\bar{q}}}{2m_{\bar{q}}}\, \Big{\langle} M_2
\Big|(2\vec{k}-\vec{q})\delta_{\bar{s}_1\bar{s}_2}-i
\vec{q}\times\vec{\sigma}_{\bar{1}\bar{2}}
 \Big|M_1\Big{\rangle},
\end{equation}
where $\vec{\sigma}_{21}=\chi^{\dag}_{s_2}\vec{\sigma}\chi_{s_1}$
and
$\vec{\sigma}_{\bar{1}\bar{2}}=\tilde{\chi}^{\dag}_{\bar{s}_1}\vec{\sigma}\tilde{\chi}_{\bar{s}_2}$,
here $\chi_{s}$ is the Dirac spinor for quark or antiquark.

Radiative transitions are usually said to be either of electric or
magnetic type depending on the dominating term in multipole
expansion of the amplitude. If the initial and final state mesons
have different spins but same angular momentum then the transition
is magnetic, and the contribution of the terms proportional to
$\delta_{s_1s_2}$ or $\delta_{\bar{s}_1\bar{s}_2}$ in expression
\ref{Aem} is zero. The example of the magnetic transition is the
vector to pseudoscalar meson transition $^3S_1\rightarrow\,^1S_0\,
\gamma$.

If the initial and final states have different angular momentum then
the transition is electric and all the terms in \ref{Aem} contribute
to the amplitude. An example of the electric transition is P-wave
state to the vector meson state transition $^3P_0\rightarrow\,^3S_1
\gamma$.

Very often when considering electric transitions the second term in
the square brackets of (\ref{Aem}) is ignored, which is called the
dipole approximation, and also the limit $\vec{q}\rightarrow 0$ is
taken, which corresponds to the long-wavelength approximation. In
this case the expression for the amplitude of E1 transitions is very
simple:
\begin{equation}\label{AemApprox}
\vec{A}_{em}=\langle
M_2|\vec{k}|M_1\rangle\left[\frac{eQ_q}{m_q}+\frac{eQ_{\bar{q}}}{m_{\bar{q}}}\right]
\delta_{s_1s_2}\delta_{\bar{s}_1\bar{s}_2}
\end{equation}
Using Siegert's theorem one can write $i[H,r]=2p/m_q$ and then the
transition amplitude is proportional to the matrix element $ \langle
M_2|r|M_1\rangle$.

The technique described in the previous paragraph is the usual way
to calculate radiative transitions in the literature. We have tested
the validity of the approximations typically made. In particular, we
have taken into account all terms in the operator of the equation
(\ref{Aem}) and we have not made the zero recoil approximation.
Comparing the results of our calculations to the results with usual
approximations in Chapter \ref{RadTrans} we find significant
differences and conclude that it is important to treat radiative
transitions carefully in the most possible general way.

Matrix elements in (\ref{Aem}) could be calculated using the quark
model meson state (\ref{meson}). For example, for vector meson to
pseudoscalar meson transition $V\rightarrow P\gamma$ in the
nonrelativistic approximation for the quark spinors it is:
\begin{eqnarray}\label{AemNR}
\vec{A}_{em}(\!\!\!\!&V& \!\rightarrow P\gamma)=-2i\left(\vec{q}\times\vec{\epsilon}_V\right)\sqrt{M_1E_2}\times\\
&\times &\left[\frac{eQ_q}{2m_q}\int\frac{d^3\vec{k}}{(2\pi)^3}
\Phi^*_2\left({\vec{k}}+\frac{m_{\bar{q}}}{m_q+m_{\bar{q}}}\vec{q}\right)
\Phi_1(\vec{k})
+\frac{eQ_{\bar{q}}}{2m_{\bar{q}}}\int\frac{d^3\vec{k}}{(2\pi)^3}
\Phi^*_2\left(\vec{k}-\frac{m_q}{m_q+m_{\bar{q}}}\vec{q}\right)\Phi_1(\vec{k})\right],\nonumber
\end{eqnarray}
where $\vec{\epsilon}_V$ is the polarization vector of the vector
meson, $\Phi(k)_{(1,2)}$ are the spatial wave function of initial
and final state mesons in the momentum space, $M_1$ is the mass of
the initial state meson and $E_2$ is the energy of the final state
meson.

As an approximation to the meson wave function, spherical harmonic
oscillator wave functions are widely in use. This approximation
greatly simplifies the calculations, and most of the quantities of
interest could be calculated analytically. Thus we conclude that it
is reasonably good for the crude estimation of the ground state
meson wave function and main features of matrix elements but for
qualitative studies realistic meson wave functions should be
employed. Another use of this approximation is testing the numerical
methods which then could be applied to the more complex cases.

The SHO spatial wave function for vector and pseudoscalar mesons is:
\begin{equation}
\Phi(k)=\left(\frac{4\pi}{\beta^2}\right)^{3/4} e^{-k^2/2\beta^2},
\end{equation}
and then the amplitude (\ref{AemNR}) for $V\rightarrow P\gamma$ in
the SHO nonrelativistic approximation is:
\begin{equation}
\vec{A}_{em}=-2i\left(\vec{q}\times\vec{\epsilon}_V\right)\sqrt{M_1E_2}\left[\frac{eQ_q}{2m_q}e^{-{\mu^2q^2}/4\beta^2}
+\frac{eQ_{\bar{q}}}{2m_{\bar{q}}}e^{-{\bar{\mu}^2q^2}/4\beta^2}\right],
\end{equation}
where
\begin{equation}
\mu=\frac{m_{\bar{q}}}{m_q+m_{\bar{q}}}\,\,\,\,\,\,\,\,\,and\,\,\,\,\,\,\,\,\,\bar{\mu}=\frac{m_q}{m_q+m_{\bar{q}}}.
\end{equation}

In the special case of $m_q=m_{\bar{q}}$ we have $\mu=\bar{\mu}=1/2$
and:
\begin{equation}\label{impSHO}
\vec{A}_{em}=-i\left(\vec{q}\times\vec{\epsilon}_V\right)\sqrt{M_1E_2}
\frac{eQ_q+eQ_{\bar{q}}}{m_q} e^{-{q^2}/16\beta^2}.
\end{equation}

The decay rate for a radiative transition is:
\begin{equation}
\Gamma(M_2\rightarrow
M_1\gamma)=\frac{1}{32\pi^2}\frac{1}{2J_1+1}\int
d\Omega_{\hat{q}}\frac{|\vec{q}|}{M^2_1}\sum_{\lambda,V}|\vec{\epsilon}^{\,*}_{\lambda}\cdot\vec{A}_{em}|^2.
\end{equation}

In our example for $V\rightarrow P\gamma$ using SHO wave functions
the decay rate is:
\begin{equation}
\Gamma(V\rightarrow
P\gamma)=\frac{\alpha}{3}\frac{E_2|\vec{q}|^3}{M_1}\left[\frac{Q_q}{m_q}e^{-{\mu^2q^2}/4\beta^2}
+\frac{Q_{\bar{q}}}{m_{\bar{q}}}e^{-{\bar{\mu}^2q^2}/4\beta^2}\right]^2,
\end{equation}

The same approach could be used for any other meson radiative
transitions. The results of our calculations for a variety of
models, discussion of the effects of approximations described above
and comparison to the experiment are presented in Chapter
\ref{RadTrans}.

\subsection{Higher order diagrams}\label{HigherOrder}

Higher order diagrams take into account the possibility of
quark-antiquark pair appearing from the vacuum. Studying these
diagrams is important as they might give significant contribution to
the impulse approximation since there is no small parameter
associated with the quark-antiquark pair creation in low energy QCD.

\begin{figure}[h!]
\begin{center}
\includegraphics[trim=0 26cm 0 0,width=10cm]{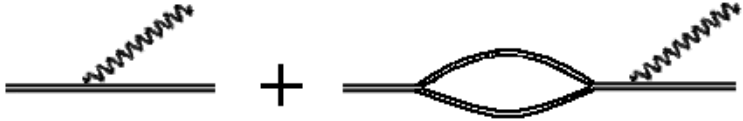}
\caption{Higher order diagrams in Cornell
model.}\label{CornellModelRad}
\end{center}
\end{figure}

\begin{figure}[h!]
\begin{center}
\includegraphics[trim=0 23cm 0 -4cm,width=6cm]{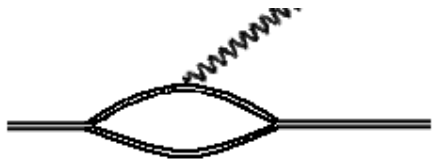}
\caption{Higher order diagram not taken into account in Cornell
model.}\label{CornellModelRad1}
\end{center}
\end{figure}

One way to introduce higher order diagrams was developed by Cornell
group \cite{Cornell}, the so called `Cornell' model. In this model
the mesonic state is described as a superposition of a naive
quark-antiquark state and all possible decay channels of a naive
state into two other mesons. There are two diagrams contributing to
the radiative transition, shown in Fig. \ref{CornellModelRad}.
Mesons in the Cornell model diagrams are represented by double line.
The first diagram corresponds to the impulse approximation and the
second diagram is higher order. However, for this model to be
consistent, coupling of the electromagnetic current to the products
of the decay in the intermediate state should also be taken into
account, for example, diagram shown in Fig. \ref{CornellModelRad1}
should be considered. These kinds of diagrams have been neglected in
\cite{Cornell}.

\begin{figure}[h!]
\begin{center}
\includegraphics[trim=6cm 18cm -2cm 0,width=3.8cm]{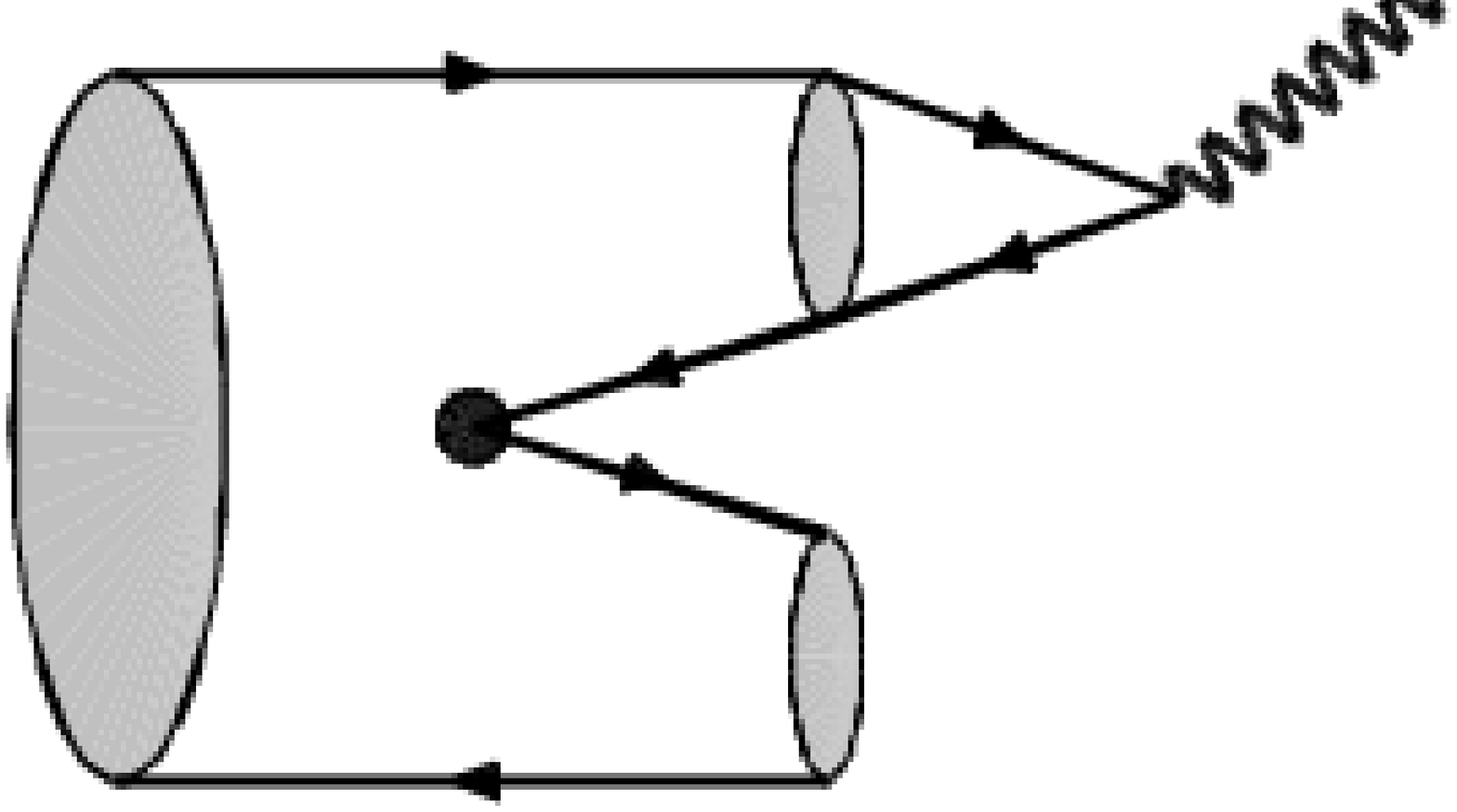}
\includegraphics[trim=-5cm 18cm 6cm 0,width=3.5cm]{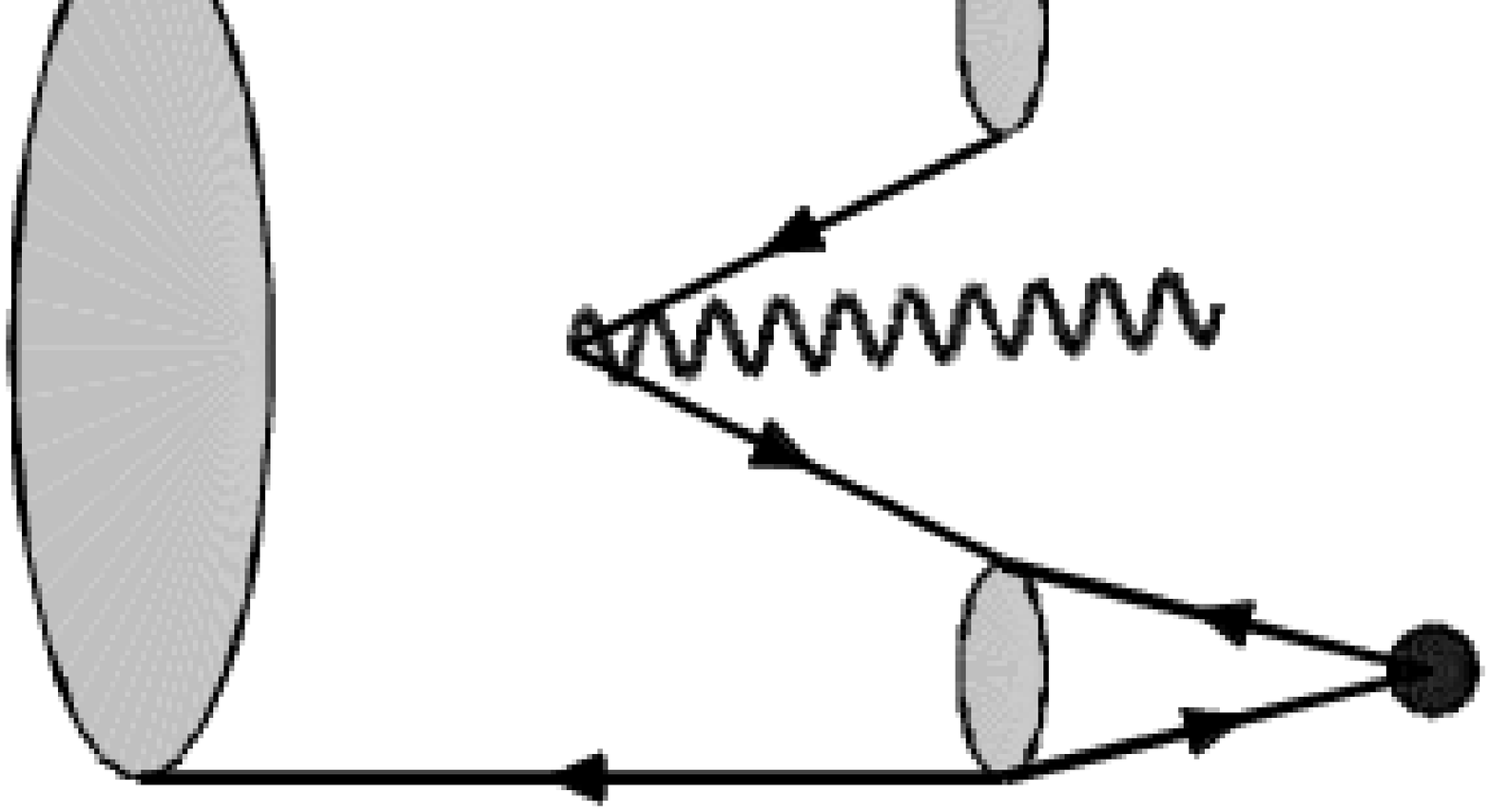}
\caption{Higher order diagrams in the bound state time ordered
perturbation theory.}\label{HOfig}
\end{center}
\end{figure}

We offer a different way of describing higher order diagrams in
radiative transitions. In our approach, we use $^3P_0$ model to
describe the quark-antiquark pair creation ($^3P_0$ model is
explained in section \ref{3P0}) and then employ the bound state time
ordered perturbation theory to obtain higher order diagrams. There
are two diagrams which contribute to the transition in addition to
the impulse approximation, they are shown in Fig. \ref{HOfig}. When
calculating the diagrams in the quark model all possible
intermediate bound states have to be summed over. Details of the
calculations and our estimations of these diagrams are presented in
Chapter \ref{RadTrans}.

\subsection{Gamma-gamma transitions}\label{gg_theory}

Two-photon decays of mesons are of considerable interest as a search
mode, a probe of internal structure, and as a test of
nonperturbative QCD modeling. An illustration of the importance of
the latter point is the recent realization that the usual
factorization approach to orthopositronium (and its extensions to
QCD) decay violates low energy theorems\cite{ps}.

It has been traditional to compute decays such as $Ps \to
\gamma\gamma$ by assuming factorization between soft bound state
dynamics and hard rescattering into photons\cite{pw}. This
approximation is valid when the photon energy is much greater than
the binding energy $E_B \sim m\alpha^2$. This is a difficult
condition to satisfy in the case of QCD where $\alpha \to \alpha_s
\sim 1$. Nevertheless, this approach has been adopted to inclusive
strong decays of mesons\cite{ap,brodsky,kwong} and has been
extensively applied to two-photon decays of quarkonia\cite{bc}.

The application of naive factorization to orthopositronium decay (or
$M \to ggg$, $\gamma gg$ in QCD) leads to a differential decay rate
that scales as $E_\gamma$ for small photon energies\cite{op} -- at
odds with the $E_\gamma^3$ behavior required by gauge invariance and
analyticity (this is Low's theorem\cite{low}). The contradiction can
be traced to the scale dependence of the choice of relevant states
and can be resolved with a careful NRQED analysis\cite{mr}. For
example, a parapositronium-photon intermediate state can be
important in orthopositronium decay at low energy. Other attempts to
address the problem by treating the binding energy nonperturbatively
can be found in Refs. \cite{smith,ni}.

\begin{figure}[h]
\begin{center}
\includegraphics[trim=0 0 0 -0.5cm,angle=0,width=5cm]{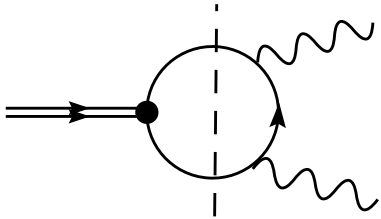}
\caption{Naive Factorization in Positronium Decay.}
\label{posDecayFig}
\end{center}
\end{figure}

Naive factorization is equivalent to making a vertical cut through
the loop diagram representing $Ps \to n\gamma$\cite{smith} (see Fig.
\ref{posDecayFig}). Of course this ignores cuts across photon
vertices that correspond to the neglected intermediate states
mentioned above. In view of this, a possible improvement is to
assume that pseudoscalar meson decay to two photons occurs via an
intermediate vector meson followed by a vector meson dominance
transition to a photon. This approach was indeed suggested long ago
by van Royen and Weisskopf\cite{vRW} who made simple estimates of
the rates for $\pi^0\to \gamma\gamma$ and $\eta \to \gamma\gamma$.
This proposal is also in accord with time ordered perturbation
theory applied to QCD in Coulomb gauge, where intermediate bound
states created by instantaneous gluon exchange must be summed over.

Finally, one expects that an effective description should work for
sufficiently low momentum photons. The effective Lagrangian for
pseudoscalar decay can be written as

\begin{equation}
{\cal L} = g\int \eta F^{\mu\nu}\tilde F_{\mu\nu}
\end{equation}
leading to the prediction $\Gamma(\eta\to \gamma\gamma) \propto g^2
m_\eta^3$. Since this scaling with respect to the pseudoscalar mass
appears to be experimentally satisfied for $\pi$, $\eta$, $\eta'$
mesons, Isgur {\it et al.} inserted an {\it ad hoc} dependence of
$m_\eta^3$ in their quark model computations\cite{HI,GI}. While
perhaps of practical use, this approach is not theoretically
justified and calls into doubt the utility of the quark model in
this context. Indeed simple quark model computations of the
amplitude of Fig. \ref{posDecayFig} are not dependent on binding
energies and can only depend on kinematic quantities such as quark
masses.

In view of the discussion above, we chose to abandon the
factorization approach and compute two-photon charmonium decays in
the quark model in bound state time ordered perturbation theory.
This has the effect of saturating the intermediate state with all
possible vectors, thereby bringing in binding energies, a nontrivial
dependence on the pseudoscalar mass, and incorporating oblique cuts
in the loop diagram.

Details of our calculations and the results are presented in Chapter
\ref{gg_chapter}.


\subsection{Meson transitions in `Coulomb gauge model'}\label{RPA}

As was described in section \ref{CoulombGauge}, a relativistic
many-body approach in Coulomb gauge (`Coulomb gauge model') is a
richer model of hadron structure than the nonrelativistic potential
model. It can explain some fundamental properties of QCD, such as
chiral symmetry breaking and dependence of the quark mass on the
energy scale, in a fully relativistic way. Until now only meson
spectra have been calculated in this model, and the agreement with
the experiment is impressive. However, for testing and improving the
model, other meson properties should be investigated.

For Coulomb gauge model the same approach to the calculation of the
meson properties could be used as for the nonrelativistic potential
quark model, the main difference being the spatial meson wave
functions. As was explained in section \ref{CoulombGauge} in order
to calculate spatial meson wave function in RPA approximation we
need to solve the system of two nonlinear coupled integral
equations. After that the formulas from appendices
\ref{DecayConstantsApp} and \ref{ffApp} could be used to calculate
form-factors, decay constants and radiative transitions.

The only practical exception of the statement above is the study of
pion properties. In RPA approximation the wave function of each
meson is a superposition of the forward and backward propagating
components. The backward propagating component is negligible for all
the mesons, except pion. In the pion case this leads to a change in
the wave function normalization and has a considerable effect on the
pion properties. As an example, our results for radiative transition
decay rates involving pion will be presented in Chapter
\ref{RadTrans}. They have much better agreement with the experiment
in Coulomb gauge model than in the nonrelativistic potential model.

\clearpage

\chapter{SPECTROSCOPY}\label{spectrum}

New spectroscopy from the B factories and the advent of CLEO-c and
the BES upgrade have led to a resurgence of interest in charmonia.
Among the new developments are the discovery of the $\eta_c'$ and
$h_c$ mesons and the observation of the enigmatic $X(3872)$ and
$Y(4260)$ states at Belle\cite{HHreview}.

BaBar's discovery of the $D_s(2317)$ state\cite{BabarDs} generated
strong interest in heavy meson spectroscopy --  chiefly due to its
surprisingly low mass with respect to expectations. These
expectations are based on quark models or lattice gauge theory.
Unfortunately,  at present large lattice systematic errors do not
allow a determination of the $D_s$ mass with  a precision better
than several hundred MeV. And, although quark models appear to be
exceptionally accurate in describing charmonia, they are less
constrained by experiment and on a weaker theoretical footing in the
open charm sector. It is therefore imperative to examine reasonable
alternative descriptions of the open charm sector.

The $D_s(2317)$ was produced in $e^+e^-$ scattering and discovered
in the isospin violating $D_s\pi$ decay mode in $K\bar K \pi\pi$ and
$K\bar K \pi\pi\pi$ mass distributions. Its width is less than 10
MeV and it is likely that the quantum numbers are $J^P =
0^+$\cite{HHreview}. Finally, if the $D_s\pi^0$ mode dominates the
width of the $D_s(2317)$ then the measured product of branching
ratios\cite{belleDs}
\begin{equation}
Br(B^0 \to D_{s}(2317)K) \cdot Br(D_{s}(2317) \to D_s \pi^0) = (4.4
\pm 0.8 \pm 1.1)\cdot 10^{-5}
\end{equation}
implies that $Br(B \to D_{s}(2317)K) \approx Br(B \to D_s K)$,
consistent with the $D_s(2317)$ being a canonical $0^+$ $c\bar s$
meson.

In view of this, Cahn and Jackson have examined the feasibility of
describing the masses and decay widths of the low lying $D$ and
$D_s$ states within the constituent quark model\cite{CJ}. They
assume a standard spin-dependent structure for the quark-antiquark
interaction (see below) and allow general vector and scalar
potentials. Their conclusion is that it is very difficult to
describe the data in this scenario.

Indeed, the $D_s(2317)$ lies some 160 MeV below most model
predictions (see Ref.\cite{HHreview} for a summary), leading to
speculation that the state could be a $DK$ molecule\cite{BCL} or a
tetraquark\cite{tetra}. Such speculation  is supported by the
isospin violating discovery mode of the $D_s(2317)$ and the
proximity of the S-wave $DK$ threshold at 2358-2367 MeV.

Although these proposals have several attractive features, it is
important to exhaust possible canonical $c\bar s$ descriptions of
the $D_s(2317)$ before resorting to more exotic models.
In section \ref{OpenCharm} we propose a simple modification to the
standard vector Coulomb+scalar linear quark potential model that
maintains good agreement with the charmonium spectrum and agrees
remarkably well with the $D$ and $D_s$ spectra. Possible
experimental tests of this scenario are discussed.

Below the results of our study of charmonium, bottomonium and open
charm spectroscopy are presented and discussed.

\section{Charmonium}

We adopt the standard practice of describing charmonia with
nonrelativistic kinematics, a central confining potential, and order
$v^2/c^2$ spin-dependent interactions. Thus $H =  2m +
P_{rel}^2/2\mu + V_C + V_{SD}$ where

\begin{equation}
V_C(r) = -\frac{4}{3}\frac{\alpha_C}{r} + br,
\end{equation}
and
\begin{equation}
V_{SD}(r) = \frac{32\alpha_H\pi e^{-k^2/4\sigma^2}}{9 m_q^2 }
\vec{S}_q \cdot \vec{S}_{\bar q}  +
\Big(\frac{2\alpha_s}{r^3}-\frac{b}{2r}\Big)\frac{1}{m_q^2}\vec L
\cdot \vec S + \frac{4\alpha_s}{m_q^2 r^3} T, \label{vsdEq}
\end{equation}
where $3T = 3\hat r \cdot S_q \hat r \cdot \vec S_{\bar q} - \vec
S_q \cdot \vec S_{\bar q}$. The strengths of the Coulomb and
hyperfine interactions have been taken as separate parameters.
Perturbative gluon exchange implies that $\alpha_C = \alpha_H$ and
we find that the fits prefer the near equality of these parameters.
The variation of this model, as described in section \ref{NPQM},
includes running coupling \ref{alRunEq}.

The resulting low lying spectra are presented in Table
\ref{spectrumTab}. The first column presents the results of the
`BGS'  model\cite{BGS}, which was tuned to the available charmonium
spectrum. Parameters are: $m_c = 1.4794$ GeV, $\alpha_c = \alpha_H =
0.5461$, $\sigma = 1.0946$ GeV, and $b = 0.1425$ GeV$^2$. No
constant is included.

The second and third columns, labeled BGS+log, makes the replacement
of Eq. \ref{alRunEq}; the parameters have not been retuned. One sees
that the $J/\psi$ and $\eta_c$ masses have been raised somewhat and
that the splitting has been reduced to 80 MeV. Heavier states have
only been slightly shifted. It is possible to fit the $J/\psi$ and
$\eta_c$ masses by adjusting parameters, however this tends to ruin
the agreement of the model with the excited states. We therefore
choose to compare the BGS and BGS+log models without any further
adjustment to the parameters. A comparison with other models and
lattice gauge theory can be found in Ref. \cite{HHreview}.

\addtolength{\topmargin}{-1.5cm} \addtolength{\footskip}{2cm}

\begin{table}[h]
\caption{\label{mass} Spectrum of $c\bar{c}$ mesons (GeV).}
\begin{center}
\begin{tabular}{c|cccc}
\hline
state & BGS & BGS log & BGS log & experiment \\
      &     & $\Lambda = 0.25$ GeV & $\Lambda = 0.4$ GeV & \\
\hline \hline
$\eta_c(1^1S_0)$    &2.981 &3.088  &3.052  &2.979\\
$\eta_c(2^1S_0)$    &3.625 &3.669  &3.655  &3.638\\
$\eta_c(3^1S_0)$    &4.032 &4.067  &4.057  &-    \\
$\eta_c(4^1S_0)$    &4.364 &4.398  &4.391  &-    \\
$\eta_{c2}(1^1D_2)$ &3.799 &3.803  &3.800  &-    \\
$\eta_{c2}(2^1D_2)$ &4.155 &4.158  &4.156  &-    \\
$J/\psi(1^3S_1)$    &3.089 &3.168  &3.139  &3.097\\
$\psi(2^3S_1)$      &3.666 &3.707  &3.694  &3.686\\
$\psi(3^3S_1)$      &4.060 &4.094  &4.085  &4.040\\
$\psi(4^3S_1)$      &4.386 &4.420  &4.412  &4.415\\
$\psi(1^3D_1)$      &3.785 &3.789  &3.786  &3.770\\
$\psi(2^3D_1)$      &4.139 &4.143  &4.141  &4.159\\
$\psi_2(1^3D_2)$    &3.800 &3.804  &3.801  &-    \\
$\psi_2(2^3D_2)$    &4.156 &4.159  &4.157  &-    \\
$\psi_3(1^3D_3)$    &3.806 &3.809  &3.807  &-    \\
$\psi_3(2^3D_3)$    &4.164 &4.167  &4.165  &-    \\
$\chi_{c0}(1^3P_0)$    &3.425 &3.448  &3.435  &3.415\\
$\chi_{c0}(2^3P_0)$    &3.851 &3.870  &3.861  &-    \\
$\chi_{c0}(3^3P_0)$    &4.197 &4.214  &4.207  &-    \\
$\chi_{c1}(1^3P_1)$    &3.505 &3.520  &3.511  &3.511\\
$\chi_{c1}(2^3P_1)$    &3.923 &3.934  &3.928  &-    \\
$\chi_{c1}(3^3P_1)$    &4.265 &4.275  &4.270  &-    \\
$\chi_{c2}(1^3P_2)$    &3.556 &3.564  &3.558  &3.556\\
$\chi_{c2}(2^3P_2)$    &3.970 &3.976  &3.972  &-    \\
$\chi_{c2}(3^3P_2)$    &4.311 &4.316  &4.313  &-    \\
$h_c(1^1P_1)$       &3.524 &3.536  &3.529  &-    \\
$h_c(2^1P_1)$       &3.941 &3.950  &3.945  &-    \\
$h_c(3^1P_1)$       &4.283 &4.291  &4.287  &-    \\
\hline \hline
\end{tabular}
\end{center}
\label{spectrumTab}
\end{table}

%

Meson spectrum is not a particularly robust test of model
reliability because it only probes gross features of the
wavefunction. Alternatively, observables such as strong and
electroweak decays and production processes probe different
wavefunction momentum scales. For example, decay constants are short
distance observables while strong and radiative transitions test
intermediate scales. Thus the latter do not add much new information
unless the transition occurs far from the zero recoil point. In this
case the properties of boosted wavefunctions and higher momentum
components become important. Production processes can provide
information on the short distance behavior of the wavefunctions
since much experimental data is available. Unfortunately, the
underlying mechanisms at work are still under debate, even for
$J/\psi$ and $\psi'$ \cite{Lansberg:2006dh}.

\clearpage \addtolength{\topmargin}{1.5cm}
\addtolength{\footskip}{-2cm}

\section{Bottomonium}\label{bb-bar}

The bottomonium parameters were obtained by fitting the potential
model of Eqs. \ref{vcEq} and \ref{vsdEq} (C+L) to the known
bottomonium spectrum. The results are $m_b = 4.75$ GeV, $\alpha_C =
\alpha_H =0.35$, $b=0.19$ GeV$^2$, and $\sigma=0.897$ GeV. All the
calculations have been performed as for charmonia.
\begin{table}[!h]
\caption{Bottomonium Spectrum (GeV).} \label{BSpectrum}
\begin{center}
\begin{tabular}{c|cccc}
\hline
Meson & C+L & C+L log  & C+L log & PDG \\
      &     & $\Lambda = 0.4$ GeV & $\Lambda = 0.25$ GeV & \\
\hline \hline
$\eta_b$                       &9.448   &9.490   &9.516    &\\
$\eta_b'$                      &10.006  &10.023  &10.033   &\\
$\eta_b''$                     &10.352  &10.365  &10.372   &\\
$\Upsilon$                     &9.459   &9.500   &9.525    &$9.4603\pm 0.00026$\\
$\Upsilon'$                    &10.009  &10.026  &10.036   &$10.02326\pm 0.00031$\\
$\Upsilon''$                   &10.354  &10.367  &10.374   &$10.3552\pm 0.0005$\\
$\chi_{b0}$                       &9.871   &9.873   &9.879    &$9.8599\pm 0.001$\\
$\chi_{b0}'$                      &10.232  &10.235  &10.239   &$10.2321\pm 0.0006$\\
$\chi_{b0}''$                     &10.522  &10.525  &10.529   &\\
$\chi_{b1}$                       &9.897   &9.900   &9.904    &$9.8927\pm 0.0006$\\
$\chi_{b1}'$                      &10.255  &10.257  &10.260   &$10.2552\pm 0.0005$\\
$\chi_{b1}''$                     &10.544  &10.546  &10.548   &\\
$\chi_{b2}$                       &9.916   &9.917   &9.921    &$9.9126\pm 0.0005$\\
$\chi_{b2}'$                      &10.271  &10.272  &10.275   &$10.2685\pm 0.0004$\\
$\chi_{b2}''$                     &10.559  &10.560  &10.563   &\\
\hline \hline
\end{tabular}
\end{center}
\end{table}

Second and third columns correspond to the model with logarithmic
dependence of running coupling \ref{alRunEq}. The parameters of the
potential have not been refitted. One can see that, as for
charmonium, introducing the running coupling has a small effect on
the excited states while considerably shifting ground state masses
of $\eta_b$ and $\Upsilon$.

\clearpage

\section{Spectroscopy of Open Charm States}\label{OpenCharm}

The spectra we seek to explain are summarized in Table
\ref{DDsSpectraTab}.  Unfortunately, the masses of the $D_0$
(labeled $a$) and $D_1'$ (labeled $b$) are poorly determined. Belle
have observed\cite{belleD0} the $D_0$ in $B$ decays, and claim a
mass of $2308  \pm 17 \pm 32$ MeV with a width of $\Gamma = 276 \pm
21 \pm 18 \pm 60$ MeV, while FOCUS\cite{focusD0} find $2407 \pm 21
\pm 35$ MeV with a width $\Gamma = 240 \pm 55 \pm 59$ MeV. While
some authors choose to average these values, we regard them as
incompatible and consider the cases separately below. Finally, there
is an older mass determination from Belle\cite{belleOld} of $2290
\pm 22 \pm 20$ MeV with a width of $\Gamma = 305 \pm 30 \pm 25$. The
$D_1'$ has been seen in $B$ decays to $D\pi\pi$ and $D^*\pi\pi$ by
Belle \cite{belleD0}. A Breit-Wigner fit yields a mass of $2427 \pm
26\pm 20 \pm 15$ MeV and a width of $384^{+107}_{-90} \pm 24 \pm 70$
MeV. Alternatively, a preliminary report from CLEO\cite{cleoD1}
cites a mass of $2461^{+41}_{-34} \pm 10 \pm 32$ MeV and a width of
$290^{+101}_{-79} \pm 26 \pm 36$ MeV. Finally, FOCUS \cite{focusD1}
obtain a lower neutral $D_1'$ mass of $2407 \pm 21 \pm 35$ MeV.
Other masses in Table \ref{DDsSpectraTab} are obtained from the PDG
compilation\cite{PDG}.


\begin{table}[h]
\caption{Low Lying $D$ and $D_s$ Spectra} \label{DDsSpectraTab}
\begin{tabular}{l|cccccc}
\hline \hline
$J^P$ & $0^-$ &  $1^-$ & $0^+$  &  $1^+$  &  $1^+$  & $2^+$ \\
\hline
$D$ & $1869.3 \pm 0.5$ & $2010.0 \pm 0.5$ & a  & b & $2422.2 \pm 1.8$  & $2459 \pm 4$ \\
$D_s$ & $1968.5 \pm 0.6$ & $2112.4 \pm 0.7$ &  $2317.4 \pm 0.9$ & $2459.3 \pm 1.3$ &  $2535.35 \pm 0.34$ & $2572.4 \pm 1.5$ \\
\hline
\end{tabular}
\end{table}

In addition to the unexpectedly low mass of the $D_s(2317)$, the
$D_s(2460)$ is also somewhat below predictions (Godfrey and Isgur,
for example,  predict a mass of 2530 MeV\cite{GI}). It is possible
that an analogous situation holds in the $D$ spectrum, depending on
the mass of the $D_0$. The quark model explanation of these states
rests on P-wave mass splittings induced by spin-dependent
interactions.

Here we propose to take the spin-dependence of Eq. \ref{VSD}
seriously and examine its effect on low-lying heavy-light mesons.
Our model can be described in terms of vector and scalar kernels
defined by

\begin{equation}
V_{conf} = V + S
\end{equation}
where $V = -4\alpha_s/ 3 r$ is the vector kernel and $S = br$ is the
scalar kernel, and by the order  $\alpha_s^2$ contributions to the
$V_i$, denoted by $\delta V_i$. Expressions for the matrix elements
of the spin-dependent interaction are then

\begin{eqnarray}
V_1 &=& -S + \delta V_1 \\
V_2 &=& V + \delta V_2 \\
V_3 &=& V'/r - V'' + \delta V_3 \\
V_4 &=& 2 \nabla^2 V + \delta V_4\\
V_5 &=& \delta V_5
\end{eqnarray}
Explicitly,

\begin{eqnarray}
V_1(m_q,m_{\bar{q}},r)&=&-br-C_F\frac{1}{2r}\frac{\alpha_s^2}{\pi}
\left(C_F-C_A \left(\ln{\left[(m_qm_{\bar{q}})^{1/2}r\right]}+\gamma_E\right)\right)\nonumber\\
V_2(m_q,m_{\bar{q}},r)&=&-\frac{1}{r}C_F\alpha_s\left[1+\frac{\alpha_s}{\pi}
\left[\frac{b_0}{2}[\ln{(\mu r)}+\gamma_E]+\frac{5}{12}b_0-\frac{2}{3}C_A+\right.\right.\nonumber\\
&&+\left.\left.\frac{1}{2}\left(C_F-C_A \left(\ln{\left[(m_qm_{\bar{q}})^{1/2}r\right]}+\gamma_E\right)\right)\right]\right]\nonumber\\
V_3(m_q,m_{\bar{q}},r)&=&\frac{1}{r^3}3C_F\alpha_s\left[1+\frac{\alpha_s}{\pi}
\left[\frac{b_0}{2}[\ln{(\mu r)}+\gamma_E-\frac{4}{3}]+\frac{5}{12}b_0-\frac{2}{3}C_A+\right.\right.\nonumber\\
&&+\left.\left.\frac{1}{2}\left(C_A+2C_F-2C_A \left(\ln{\left[(m_qm_{\bar{q}})^{1/2}r\right]}+\gamma_E-\frac{4}{3}\right)\right)\right]\right]\nonumber\\
V_4(m_q,m_{\bar{q}},r)&=&\frac{32\alpha_s\sigma^3 e^{-\sigma^2 r^2}}{3\sqrt{\pi}}\nonumber\\
V_5(m_q,m_{\bar{q}},r)&=&\frac{1}{4r^3}C_FC_A\frac{\alpha_s^2}{\pi}\ln{\frac{m_{\bar{q}}}{m_q}}
\label{Vmodel}
\end{eqnarray}
where $C_F=4/3$, $C_A=3$, $b_0=9$, $\gamma_E=0.5772$, the scale
$\mu$ has been set to 1 GeV.

The hyperfine interaction (proportional to $V_4$) contains a delta
function in configuration space and is normally `smeared' to make it
nonperturbatively tractable. For this reason we choose not to
include $\delta V_4$ in the model definition of Eq. \ref{Vmodel}. In
following, the hyperfine interaction ($V_4$) have been included in
the meson wave function calculations and the remaining
spin-dependent terms are treated as mass shifts using leading-order
perturbation theory.

We have confirmed that the additional features do not ruin previous
agreement with, for example, the charmonium spectrum. For example,
Ref. \cite{BGS} obtains very good agreement with experiment for
parameters $m_c = 1.4794$ GeV, $\alpha_s = 0.5461$, $b = 0.1425$
GeV$^2$, and $\sigma = 1.0946$ GeV. Employing the model of Eqn.
\ref{Vmodel} worsens the agreement with experiment, but the original
good fit is recovered upon slightly modifying parameters (the refit
parameters are $m_c = 1.57$ GeV, $\alpha_s = 0.52$, $b = 0.15$
GeV$^2$, and $\sigma = 1.3$ GeV).

\begin{table}[h!]
\caption{Model Parameters} \label{ParamsTab}
\begin{center}
\begin{tabular}{l|ccccc}
\hline \hline
model & $\alpha_s$ & $b$ (GeV$^2$) & $\sigma$ (GeV) & $m_c$ (GeV) & $C$ (GeV) \\
\hline
low & 0.46 & 0.145 & 1.20 & 1.40 & -0.298 \\
avg & 0.50 & 0.140 & 1.17 & 1.43 & -0.275 \\
high & 0.53 & 0.135 & 1.13 & 1.45 & -0.254 \\
\hline
\end{tabular}
\end{center}
\end{table}

The low lying  $c\bar s$ and $c \bar u$ states are fit reasonably
well with the parameters labeled `avg' in Table \ref{ParamsTab}.
Predicted masses are given in Table \ref{DTab}. Parameters labeled
`low' in Table \ref{ParamsTab} fit the $D$ mesons very well, whereas
those labeled `high' fit the known $D_s$ mesons well. It is thus
reassuring that these parameter sets are reasonably similar to each
other and to the refit charmonium parameters. (Note that constant
shifts in each flavor sector are fit to the relevant pseudoscalar
masses.)

The predicted $D_{s0}$ mass is 2341 MeV, 140 MeV lower than the
prediction of Godfrey and Isgur and only 24 MeV higher than
experiment. We remark that the best fit to the $D$ spectrum predicts
a mass of 2287 MeV for the $D_0$ meson, in good agreement with the
preliminary Belle measurement of 2290 MeV, 21 MeV below the current
Belle mass, and in disagreement with the FOCUS mass of 2407 MeV.

The average error in the predicted P-wave masses is less than 1\%.
It thus appears likely that the simple modification to the
spin-dependent quark interaction is capable of describing
heavy-light mesons with reasonable accuracy.

\begin{table}[h!]
\caption{Low Lying Charm Meson Spectra (GeV)} \label{DTab}
\begin{center}
\begin{tabular}{l|cccccc}
\hline \hline
flavor & $0^-$ & $1^-$ & $0^+$ & $1^+$ & $1^+$ & $2^+$ \\
\hline
$D$ & 1.869 & 2.017 & 2.260 & 2.406 & 2.445 & 2.493 \\
\hline
$D_s$ & 1.968  & 2.105 & 2.341 & 2.475 & 2.514 & 2.563 \\
\hline
\end{tabular}
\end{center}
\end{table}

\begin{figure}[h!]
\includegraphics[width=5.5 true cm, angle=270]{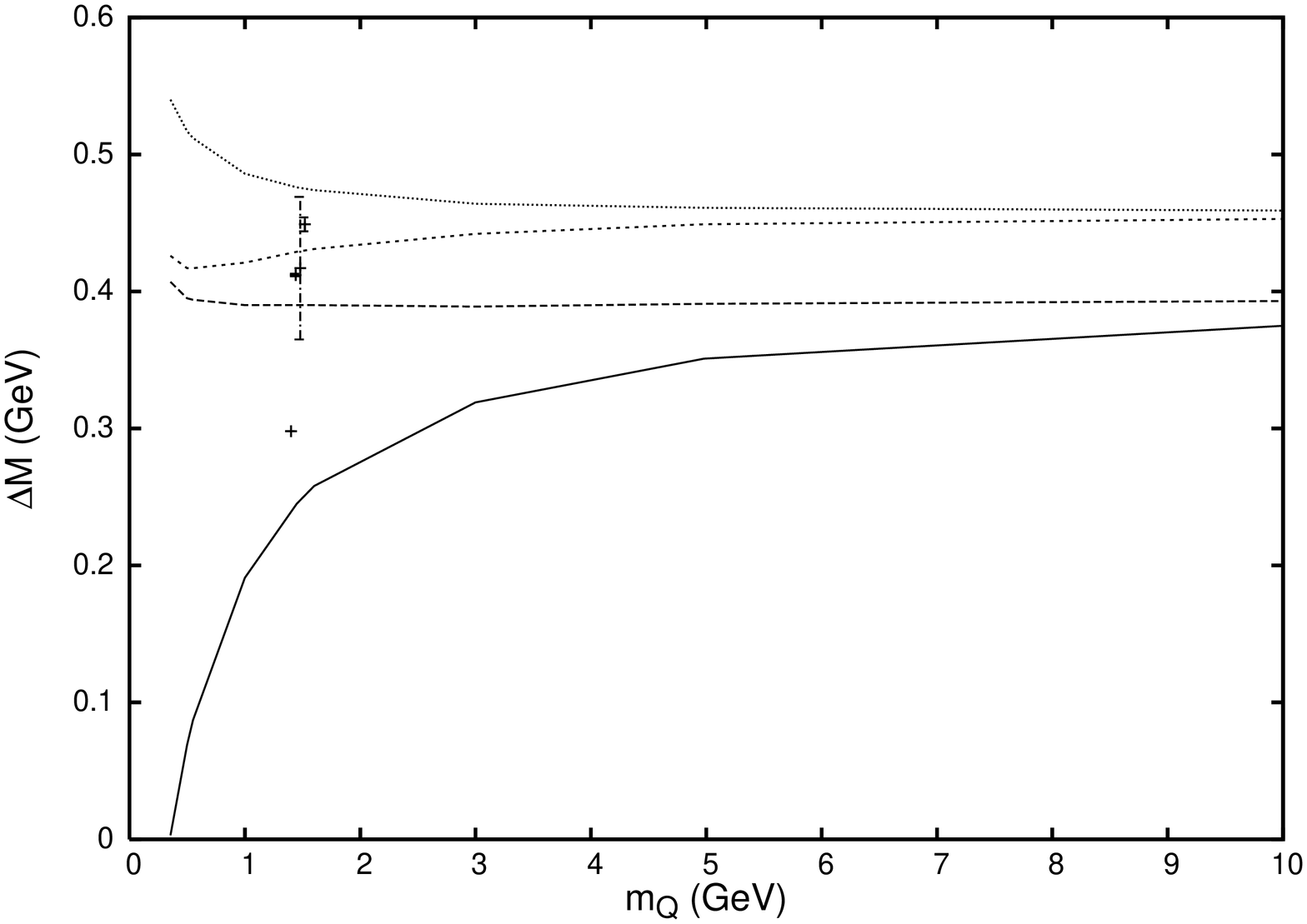}
\hskip 1 true cm
\includegraphics[width=5.5 true cm, angle=270]{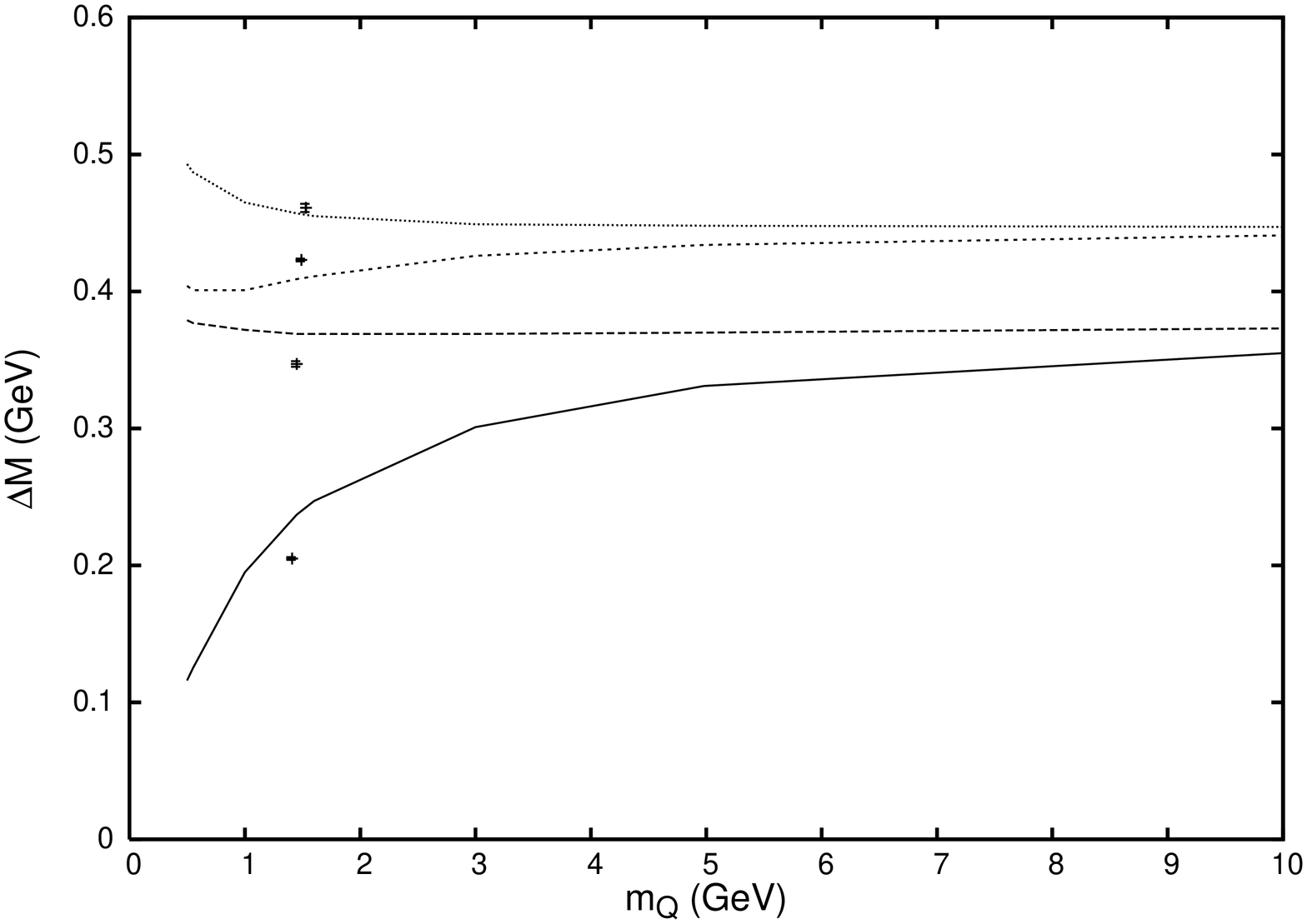}
\includegraphics[width=5.5 true cm, angle=270]{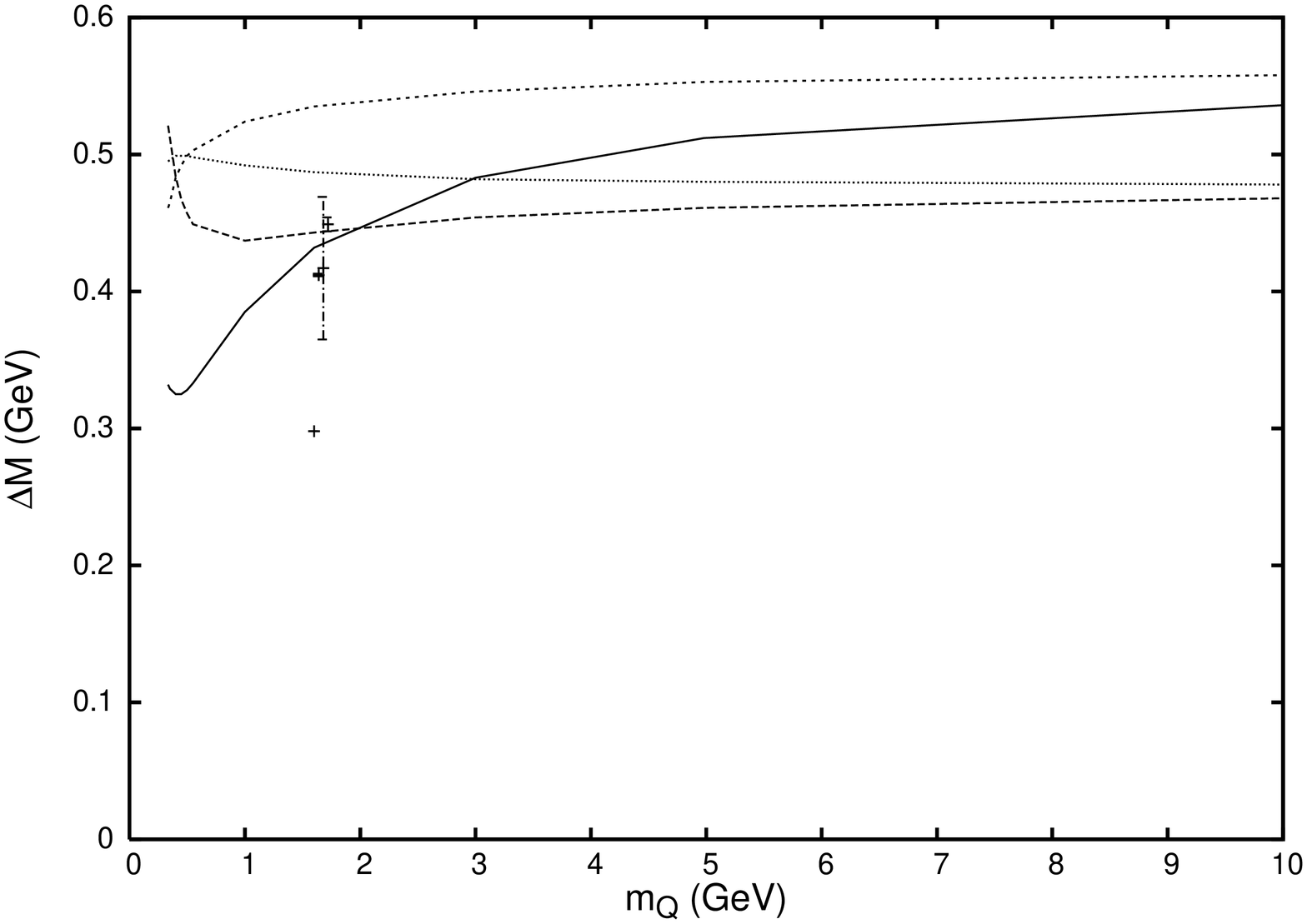}
\hskip 1 true cm
\includegraphics[width=5.5 true cm, angle=270]{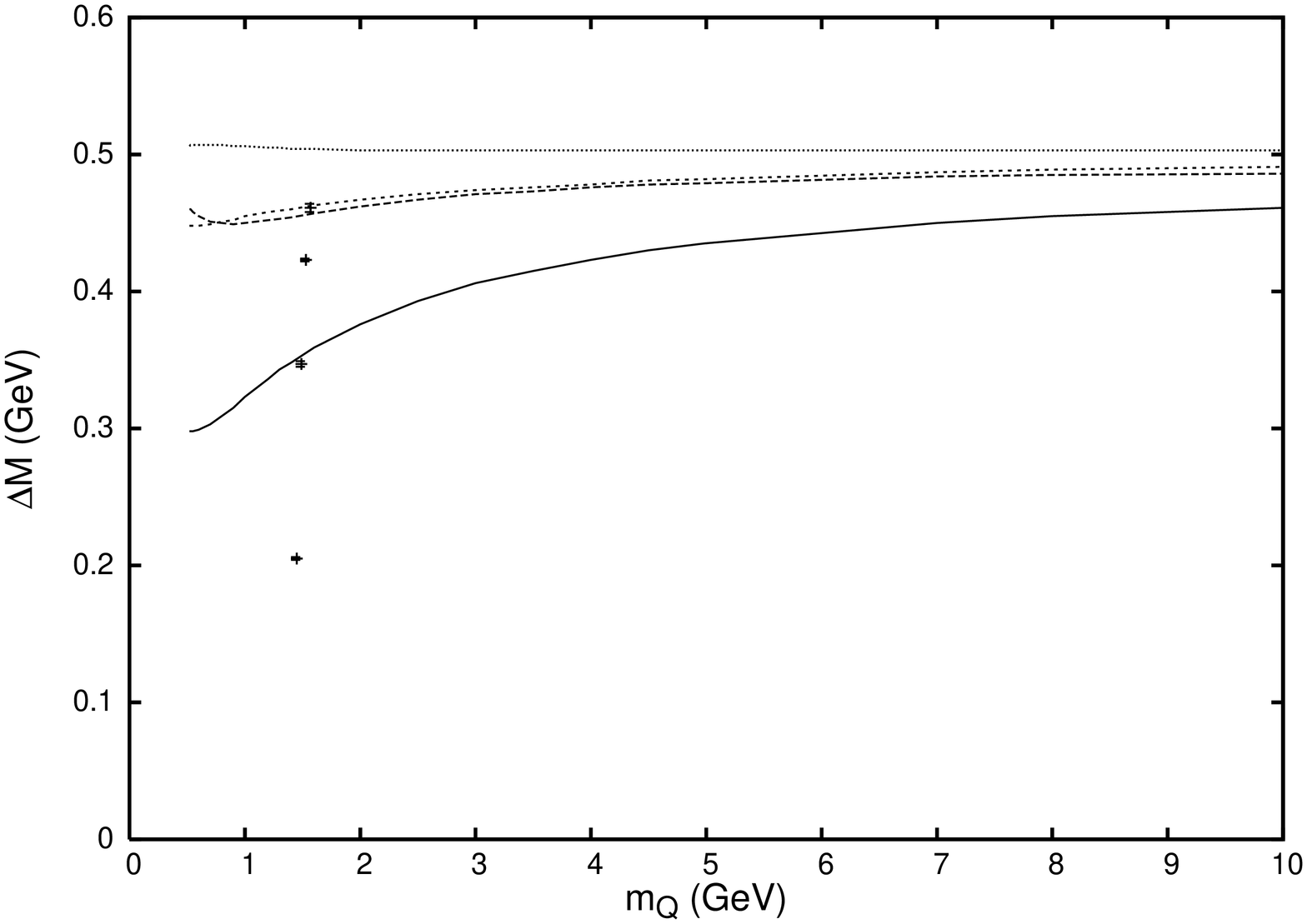}
\caption{M(P-wave) - M(vector) as a Function of the Heavy Quark
Mass. $D$ System (left); $D_s$ System (right).} \label{massesFig}
\end{figure}

We examine the new model in more detail by computing P-wave meson
masses (with respect to the ground state vector) as a function of
the heavy quark mass. Results for $Q\bar u$ and $Q\bar s$ systems
are displayed in Fig. \ref{massesFig}. One sees a very slow approach
to the expected heavy quark doublet structure. Level ordering ($D_2
> D_1', D_1 > D_0$) is maintained for all heavy quark masses. This
is not the case in the canonical quark model, and ruins the
agreement with experiment at scales near the charm quark mass.  It
is intriguing that the scalar-vector mass difference gets very small
for light $Q$ masses, raising the possibility that the enigmatic
$a_0$ and $f_0$ mesons may simply be $q\bar q$ states.

Finally, one obtains $M(h_c) > M(\chi_{c1})$ in one-loop and
traditional models, in agreement with experiment. However,
experimentally $M(f_1) - M(h_1)\approx 100$ MeV and $M(a_1) - M(b_1)
\approx 0$ MeV, indicating that the $^3P_1$ state is heavier than
(or nearly degenerate with) the $^1P_1$ light meson state. Thus the
sign of the combination of tensor and spin-orbit terms that drives
this splitting must change when going from charm quark to light
quark masses. This change is approximately correctly reproduced in
the traditional model (lower left panel of Fig.\ref{massesFig}). The
one-loop model does not reproduce the desired cross over, although
it does come close, and manipulating model parameters can probably
reproduce this behavior. We do not pursue this here since the focus
is on heavy-light mesons.

\clearpage

\subsection{Mixing Angles and Radiative Decays}

The lack of charge conjugation symmetry implies that two nearby low
lying axial vector states exist (generically denoted as $D_1$ and
$D_1'$ in the following). The mixing angle between these states can
be computed and compared to experiment (with the help of additional
model assumptions). We define the mixing angle via the relations:
\begin{eqnarray}
&|D_1\rangle  &= +\cos(\phi) |^1P_1\rangle + \sin(\phi) | ^3P_1\rangle \nonumber \\
&|D_1'\rangle  &= -\sin(\phi) |^1P_1\rangle + \cos(\phi) |
^3P_1\rangle. \label{D1Eqn}
\end{eqnarray}
In the following, we choose to define the $D_1'$ as the heavier
axial state in the heavy quark limit. In this limit a particular
mixing angle follows from the quark mass dependence of the
spin-orbit and tensor terms, $\phi_{HQ}=-54.7^o (35.3^o)$, if the
expectation of the heavy-quark spin-orbit interaction is positive
(negative). It is often assumed that the heavy quark mixing angle
holds for charmed mesons.

\begin{figure}[h]
\includegraphics[width=5.5 true cm, angle=270]{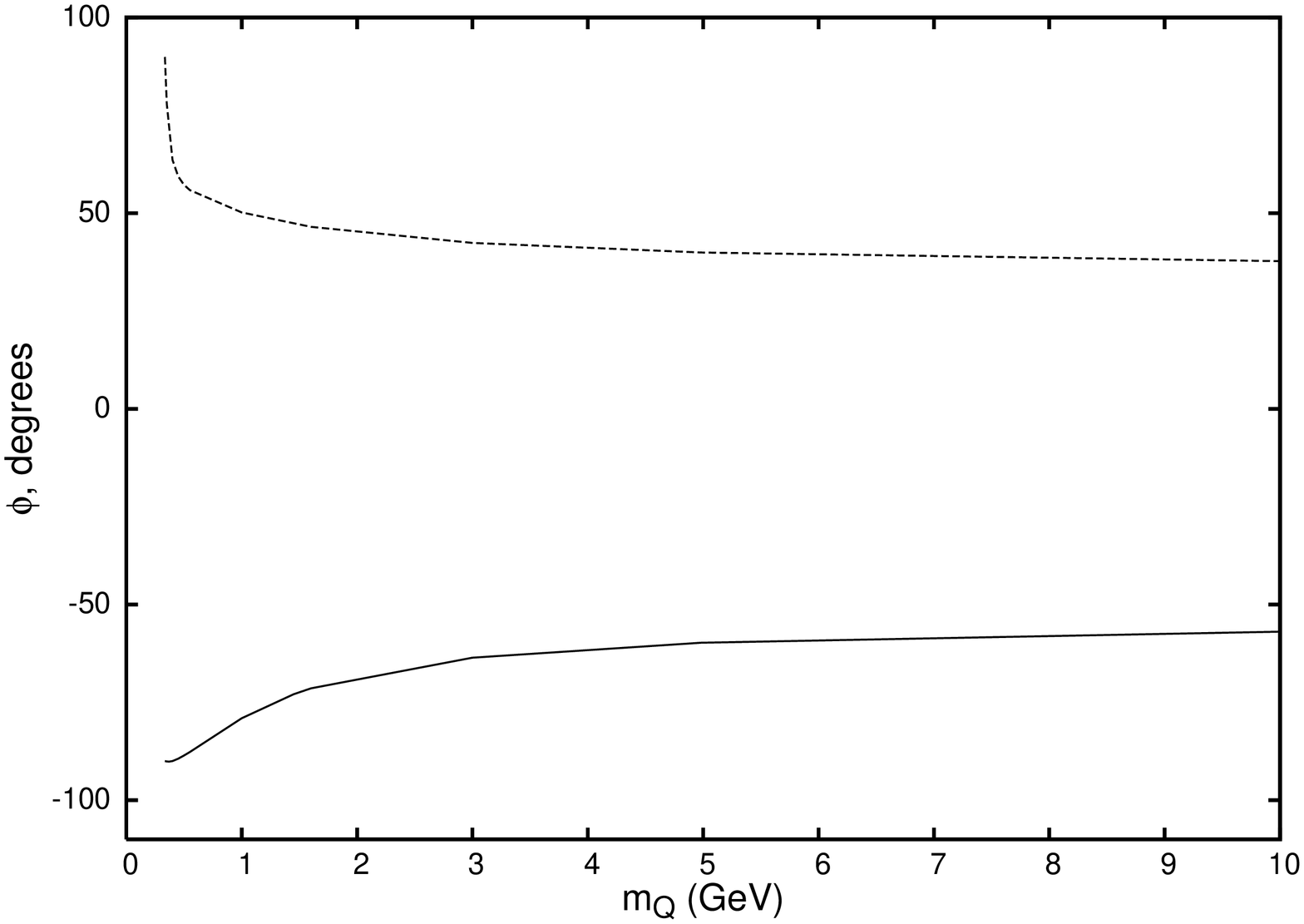}
\hskip 1 true cm
\includegraphics[width=5.5 true cm, angle=270]{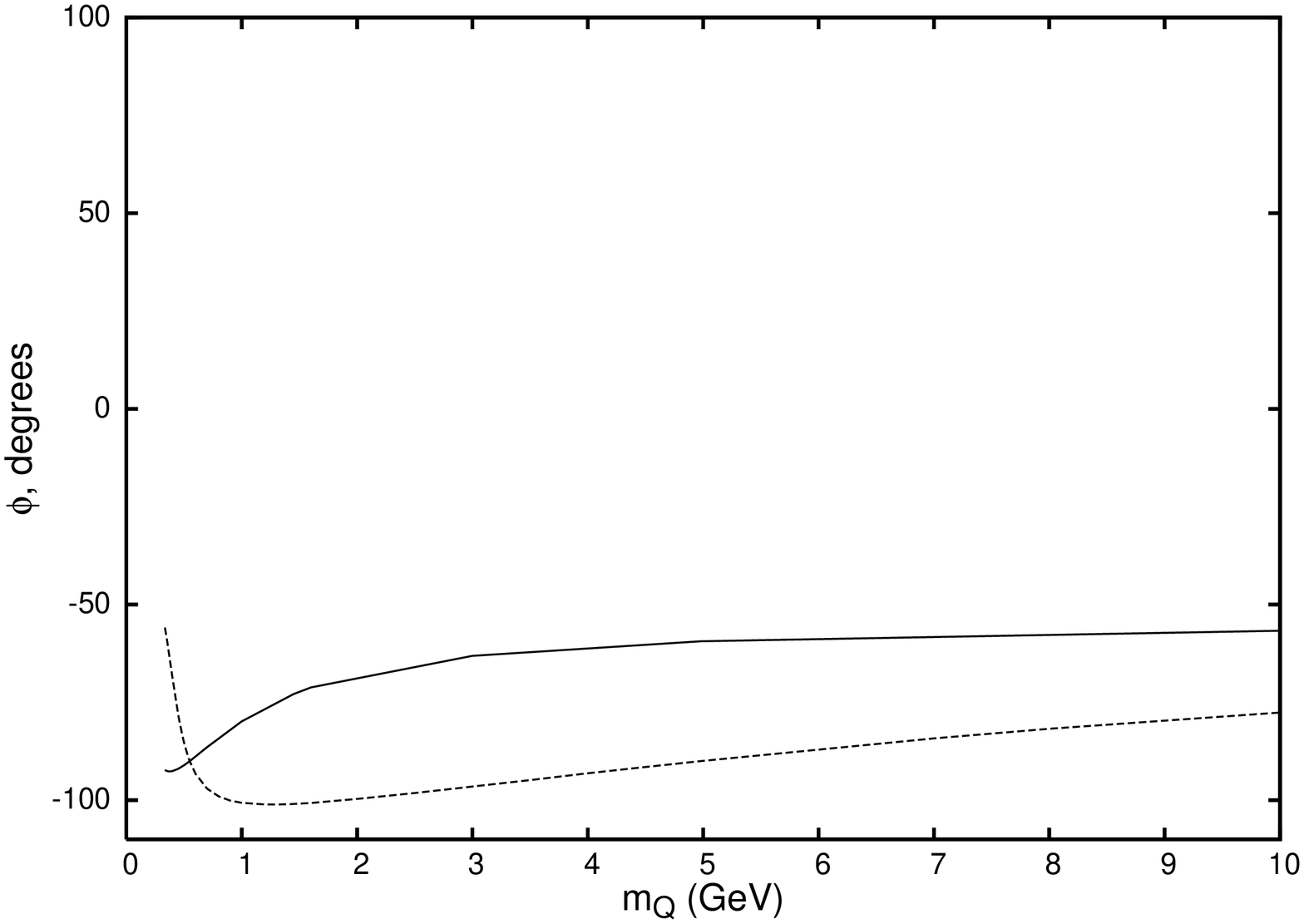}
\caption{$D$ (left) and $D_{s1}$ (right) Mixing Angles. The
traditional model is given by the dashed line; the extended model is
the solid line.} \label{mixingAngleFig}
\end{figure}

Fig. \ref{mixingAngleFig} shows the dependence of the mixing angle
on the heavy quark mass for $Q\bar u$ and $Q \bar s$ mesons for the
traditional and extended models. The effect of the one-loop terms is
dramatic: for the $Q\bar u$ system the relevant spin-orbit matrix
element changes sign, causing the heavy quark limit to switch from
$35.3^o$ to $-54.7^o$. Alternatively, both models approach $-54.7^o$
in the $Q \bar s$ system. There is strong deviation from the heavy
quark limit in both cases: $\phi(D_s)\approx\phi(D)\approx -70^o$.
This result is not close to the heavy quark limit (which is
approached very slowly) – indeed it is reasonably close to the
unmixed limit of $\pm 90^o$!

Mixing angles can be measured with the aid of strong or radiative
decays. For example, the $D_1'$ is a relatively narrow state,
$\Gamma(D_1')=20.4±1.7$ MeV, while the $D_1$ is very broad. This
phenomenon is expected in the heavy quark limit of the $^3P_0$ and
Cornell strong decay models \cite{HHreview}, \cite{cs}, \cite{gk}.
Unfortunately, it is difficult to exploit these widths to measure
the mixing angle because strong decay models are rather imprecise.

Radiative decays are possibly more accurate probes of mixing angles
because the decay vertex is established and the impulse
approximation has a long history of success. Table \ref{RadTransTab}
presents the results of two computations of radiative decays of $D$
and $D_s$ mesons. Meson wavefunctions are computed with `average'
parameters, as above. Transition matrix elements are evaluated in
the impulse approximation and full recoil is allowed. The column
labeled `nonrel' reports transition matrix elements computed in the
nonrelativistic limit, while the column labeled `rel' contains
results obtained with the full spinor structure at the photon
vertex.

The nonrelativistic results can differ substantially from those of
Refs. \cite{cs,Godfrey} because those computations were made in the
zero recoil limit where an E1 transition, for example, is diagonal
in spin. Thus the decay $D_1 \to D^* \gamma$ can only proceed via
the $^3P_1$ component of the $D_1$. Alternatively, the computations
made here are at nonzero recoil and hence permit both components of
the $D_1$ to contribute to this decay. The table entries indicate
that nonzero recoil effects can be surprisingly large.

Further complicating the analysis is the large difference seen
between the nonrelativistic and relativistic models (see, eg,
$D^{+*}\to \gamma D^+$). This unfortunate circumstance is due to
differing signs between the heavy and light quark impulse
approximation subamplitudes. Employing the full quark spinors leaves
the heavy quark subamplitude largely unchanged, whereas the light
quark subamplitude becomes larger, thereby reducing the full
amplitude. The effect appears to be at odds with the only available
experimental datum ($D^* \to D \gamma$).  Clearly it would be useful
to measure as many radiative transitions as possible in these
sectors to better evaluate the efficacy of these (and other) models.
Once the decay model reliability has been established, ratios such
as $\Gamma(D_{1} \to \gamma D^*)/\Gamma(D_1' \to \gamma D^*)$ and
$\Gamma(D_1 \to \gamma D)/\Gamma(D_1\to \gamma D)$ will help
determine the $D_1$ mixing  angle.

\clearpage \addtolength{\oddsidemargin}{-1.5cm}
\begin{table}[h]
\caption{Meson Radiative Decay rates (keV).} \label{RadTransTab}
\begin{tabular}{|c|c|c|c|c|}
\hline \hline
mode & $q_\gamma$ (MeV) & nonrel & rel & expt \\
\hline
$D^{+*}\rightarrow\gamma D^+$       &136 &1.38                          &0.08                                &$1.5\pm0.5$\\
$D^{0*}\rightarrow\gamma D^0$       &137 &32.2                          &13.3                                &$<800$ \\
$D_0^+\rightarrow\gamma D^{*+}$     &361 &76.0                          &7.55                                & \\
$D_0^0\rightarrow\gamma D^{*0}$     &326 &1182                          &506.                                & \\
$D_1^+\rightarrow\gamma D^{*+}$     &381 &$(6.34s)^2+(3.22s+5.9c)^2$    &$(2.00s-0.13c)^2+(0.13s+4.23c)^2$   & \\
$D_1^0\rightarrow\gamma D^{*0}$     &380 &$(27.05s)^2+(19.33s+9.63c)^2$ &$(17.65s-0.15c)^2+(12.28s+6.01c)^2$ & \\
$D_1'^+\rightarrow\gamma D^{*+}$    &381 &$(6.34c)^2+(-3.22c+5.9s)^2$   &$(2.00c+0.13s)^2+(-0.13c+4.23s)^2$  & \\
$D_1'^0\rightarrow\gamma D^{*0}$    &384 &$(27.26c)^2+(19.35c-9.83s)^2$ &$(17.78c+0.15s)^2+(12.29c-6.13s)^2$ & \\
$D_1^+\rightarrow\gamma D^+$        &494 &$(5.49s+4.75c)^2$             &$(4.17s-0.60c)^2$               & \\
$D_1^0\rightarrow\gamma D^0$        &493 &$(8.78s+31.42c)^2$            &$(5.56s+18.78c)^2$              & \\
$D_1'^+\rightarrow\gamma D^+$       &494 &$(-5.49c+4.75s)^2$            &$(4.17c-0.60s)^2$               & \\
$D_1'^0\rightarrow\gamma D^0$       &498 &$(-8.90c+31.41s)^2$           &$(-5.62c+18.78s)^2$             & \\
$D_2^+\rightarrow\gamma D^{*+}$     &413 &15.0                          &6.49                            & \\
$D_2^0\rightarrow\gamma D^{*0}$     &412 &517                           &206                             & \\
\hline
$D_s^*\rightarrow\gamma D_s$        &139 &0.20                         &0.00                               & \\
$D_{s0}\rightarrow\gamma D_s^*$     &196 &6.85                         &0.16                               & \\
$D_{s1}\rightarrow\gamma D_s^*$     &322 &$(1.84s)^2+(0.99s+2.39c)^2$  &$(0.18s-0.07c)^2+(-0.44s+2.13c)^2$ & \\
$D_{s1}'\rightarrow\gamma D_s^*$    &388 &$(2.13c)^2+(-0.87c+3.62s)^2$ &$(0.24c-0.10s)^2+(0.64c+3.19s)^2$  & \\
$D_{s1}\rightarrow\gamma D_s$       &441 &$(2.68s+1.37c)^2$            &$(2.55s-1.21c)^2$                  & \\
$D_{s1}'\rightarrow\gamma D_s$      &503 &$(3.54c-1.12s)^2$            &$(3.33c+1.52s)^2$                  & \\
$D_{s2}\rightarrow\gamma D_s^*$     &420 &1.98                         &3.94                               & \\
\hline
\end{tabular}
\end{table}
\clearpage \addtolength{\oddsidemargin}{1.5cm}

\subsection{Discussion and Conclusions} 

A popular model of the $D_s$ mesons is based on an effective
lagrangian description of mesonic fields in the chiral and heavy
quark limits\cite{chiral}. Deviations from these limits induce mass
splittings which imply that the axial--vector and
scalar-pseudoscalar mass differences are the same. Since the premise
of this idea has been questioned in Refs. \cite{HHreview, Bicudo},
it is of interest to consider this mass difference in the present
model. Splittings for the three parameter sets considered above are
shown in Table \ref{ChiralSplittingsTab}. Evidently, the chiral
multiplet relationship holds to a very good approximation in both
the $D$ and $D_s$ sectors and is robust against variations in the
model parameters.

Nevertheless, the near equivalence of these mass differences must be
regarded as an accident. Indeed, the $B$ spectra given in Table
\ref{BTab} clearly indicate that this relationship no longer holds.
It would  thus be of interest to find P-wave open bottom mesons
(especially scalars). These data will distinguish chiral multiplet
models from the model presented here and from more traditional
constituent quark models. For example, Godfrey and Isgur claim that
the $B_0$ meson lies between 5760 and 5800 MeV; the $B_{s0}$ mass is
5840-5880 MeV, and the $B_{c0}$ mass is 6730-6770 MeV. Of these, our
$B_{s0}$ mass is predicted to be 65-105 MeV lower than the
Godfrey-Isgur mass.

\begin{table}[h]
\caption{Chiral Multiplet Splittings (MeV).}
\label{ChiralSplittingsTab}
\begin{center}
\begin{tabular}{l|cc}
\hline \hline
params & $M(1^+(1/2^+)) - M(1^-)$ & $M(0^+) - M(0^-)$ \\
\hline
$D$ low & 411  &  412 \\
$D$ avg & 391  & 389 \\
$D$ high & 366 & 368 \\
\hline
$D_s$ low & 384 & 380 \\
$D_s$ avg & 373 & 370 \\
$D_s$ high & 349 & 346 \\
\hline
\end{tabular}
\end{center}
\end{table}

\begin{table}[h]
\caption{Low Lying Bottom Meson Masses (MeV).} \label{BTab}
\begin{center}
\begin{tabular}{l|cccccc}
\hline \hline
flavor & $0^-$ & $1^-$ & $0^+$ & $1^+$ & $1^+$ & $2^+$ \\
\hline
$B$ & 5279 & 5322 & 5730 & 5752 & 5753 & 5759 \\
expt & 5279 & 5325 &  --  & $5724\pm 4 \pm 7$ &  --   &  $5748 \pm 12$ \\
\hline
$B_s$ & 5370 & 5416 & 5776 & 5803 & 5843 & 5852 \\
expt & 53696 & 54166 &  --  &  --   &  --   &  --  \\
\hline
$B_c$ & 6286 & 6333 & 6711 & 6746 & 6781 & 6797 \\
expt & 6286 & -- &  --  &  --   &  --   &  --  \\
\hline
\end{tabular}
\end{center}
\end{table}

The bottom flavored meson spectra of Table \ref{BTab} have been
obtained with the `average' extended model parameters and $m_b =
4.98$ GeV. As with the open charm spectra, a flavor-dependent
constant was fit to each pseudoscalar. The second row reports
recently measured P-wave B meson masses \cite{D0}; these are in
reasonable agreement with the predictions of the first row.

When these results are (perhaps incorrectly) extrapolated to light
quark masses, light scalar mesons are possible. Thus a simple $q
\bar q$ interpretation of the enigmatic $a_0$ and $f_0$ mesons
becomes feasible.

Finally, the work presented here may explain the difficulty in
accurately computing the mass of the $D_{s0}$ in lattice
simulations. If the extended quark model is correct, it implies that
important mass and spin-dependent interactions are present in the
one-loop level one-gluon-exchange quark interaction. It is possible
that current lattice computations are not sufficiently sensitive to
the ultraviolet behavior of QCD to capture this physics. The problem
is exacerbated by the nearby, and presumably strongly coupled, DK
continuum; which requires simulations sensitive to the infrared
behavior of QCD. Thus heavy-light mesons probe a range of QCD scales
and make an ideal laboratory for improving our understanding of the
strong interaction. \clearpage

\section{$D_s(2860)$ and $D_s(2690)$}
BaBar have recently announced the discovery of a new $D_s$ state
seen in $e^+e^-$ collisions decaying to $K^-\pi^+ K^+$,
$K^-\pi^+\pi^0K^+$  ($D^0K^+$), or $D^+K^0_S$\cite{palano}. The
Breit-Wigner mass of the new state is

\begin{equation}
M(D_{sJ}(2860)) = 2856.6 \pm 1.5 \pm 5.0 \ {\rm Mev}
\end{equation}
and the width is
\begin{equation}
\Gamma(D_{sJ}(2860)) =  48 \pm 7 \pm 10\ {\rm MeV}.
\end{equation}
The signal has a significance greater than 5 $\sigma$ in the $D^0$
channels and 2.8 $\sigma$ in the $D^+$ channel. There is no evidence
of the $D_{sJ}(2860)$ in the $D^*K$ decay mode\cite{palano} or the
$D_s \eta$ mode\cite{bill}.

There is, furthermore, structure in the $DK$ channel near 2700 MeV
that yields Breit-Wigner parameters of

\begin{equation}
M(D_{sJ}(2690)) = 2688 \pm 4 \pm 2 \ {\rm MeV}
\end{equation}
and
\begin{equation}
\Gamma(D_{sJ}(2690)) = 112 \pm 7 \pm 36 \ {\rm MeV}.
\end{equation}
The significance of the signal was not stated.

The discovery of these states is particularly germane to the
structure of the $D_s(2317)$. For example, the low mass and isospin
violating decay mode, $D_s\pi^0$, of the $D_s(2317)$ imply that the
state could be a $DK$ molecule\cite{BCL}. If this is the case, the
$D_{sJ}(2690)$ could be a supernumerary scalar $c\bar s$ state.
Alternatively, the $D_s(2317)$ could be the ground state scalar
$c\bar s$ state and the new $D_{sJ}$'s could be canonical radial
excitations. Clearly, constructing a viable global model of all the
$D_s$ states is important to developing a solid understanding of
this enigmatic sector\cite{HHreview}.

Previous efforts to understand the new BaBar states have argued that
the $D_{sJ}(2860)$ is a scalar $c \bar s$ state predicted at 2850
MeV in a coupled channel model\cite{vBR} or that it is a $J^P = 3^-$
$c \bar s$ state\cite{CFN}.

Here we pursue a simple model that assumes that all of the known
$D_s$ states are dominated by simple $c\bar s$ quark content. It is
known that this is difficult to achieve in the `standard'
constituent quark model with $O(\alpha_s)$ spin-dependent mass
shifts because the $D_{s0}(2317)$ is much lighter than typical
predictions (for example, Godfrey and Isgur obtain a $D_{s0}$ mass
of 2480 MeV\cite{GI}). An essential feature in such phenomenology
has been the assumption of two static potentials: a Lorentz scalar
confining potential and a short range Coulombic vector potential.
Following the discovery of the $D_s(2317)$, Cahn and
Jackson\cite{CJ} analyzed the $D_s$ states with a scalar potential
$S$, whose shape they allowed to be arbitrary, while retaining a
vector potential $V$ that they assumed to be Coulombic. In the limit
that the mass $m_2 \gg m_1$ this enabled the spin dependent
potential applicable to P-states to take the form

\begin{equation}
V_{SD} = \lambda L\cdot S_1 + 4\tau L\cdot S_2 + \tau S_{12}
\end{equation}
(see the discussion around Eq. 1 of \cite{CJ} for details). For
$\lambda \gg \tau$ a reasonable description of the masses could be
obtained though a consistent picture of $D_s, D$ spectroscopies and
decays remained a problem. As the authors noted, ``the ansatz taken
for the potentials $V$ and $S$ may not be as simple as assumed". The
more general form  \cite{gromes} is

\begin{equation}
V_{SD} = \lambda L\cdot S_1 + 4\tau L\cdot S_2 + \mu S_{12}
\end{equation}
 only in the particular case of a Coulomb potential need $\mu = \tau$\cite{gromes}. Direct channel
couplings (such as to $DK$ and $D^*K$ thresholds\cite{BCL,Rupp})
will induce effective potentials that allow the above more general
form. Similarly, higher order gluon exchange effects in pQCD will
also. Indeed,
the full spin-dependent structure expected at order $\alpha_s^2$ in
QCD has been computed\cite{PTN} and reveals that an additional
spin-orbit contribution to the spin-dependent interaction exists
when quark masses are not equal. When these are incorporated in a
constituent quark model there can be significant mass shifts leading
to a lowered mass for the $D_{s0}$ consistent with the
$D_{s0}(2317)$\cite{ls}. Here we apply this model to the recently
discovered $D_s$ states.

\subsection{Canonical $c\bar s$ States}

Predictions of the new model in the $D_s$ sector are summarized in
Table \ref{spectrumTab} (the `high' parameters of  Table
\ref{ParamsTab} are employed).

\begin{table}[!h]
\caption{$D_s$ Spectrum.} \label{spectrumTab}
\begin{center}
\begin{tabular}{l|cc}
\hline \hline
state                          & mass (GeV) & expt\protect\cite{PDG06} (GeV) \\
\hline
$D_s(1^1S_0)$                  &1.968       &1.968                           \\
$D_s(2^1S_0)$                  &2.637       &                                \\
$D_s(3^1S_0)$                  &3.097       &                                \\
$D^*_s(1^3S_1)$                &2.112       &2.112                           \\
$D^*_s(2^3S_1)$                &2.711       &2.688?                          \\
$D^*_s(3^3S_1)$                &3.153       &                                \\
$D_s(1^3D_1)$                  &2.784       &                                \\
$D_{s0}(1^3P_0)$               &2.329       &2.317                           \\
$D_{s0}(2^3P_0)$               &2.817       &2.857?                          \\
$D_{s0}(3^3P_0)$               &3.219       &                                \\
$D_{s1}(1P)$                   &2.474       &2.459                           \\
$D_{s1}(2P)$                   &2.940       &                                \\
$D_{s1}(3P)$                   &3.332       &                                \\
$D'_{s1}(1P)$                  &2.526       &2.535                           \\
$D'_{s1}(2P)$                  &2.995       &                                \\
$D'_{s1}(3P)$                  &3.389       &                                \\
$D_{s2}(1^3P_2)$               &2.577       &2.573                           \\
$D_{s2}(2^3P_2)$               &3.041       &                                \\
$D_{s2}(3^3P_2)$               &3.431       &                                \\
\hline
\end{tabular}
\end{center}
\end{table}

Since the $D_{sJ}(2690)$ and $D_{sJ}(2860)$ decay to two
pseudoscalars, their quantum numbers are $J^P = 0^+$, $1^-$, $2^+$,
etc. Given the known states\cite{PDG06} and that the energy gap for
radial excitation is hundreds of MeV, on almost model independent
grounds the only possibility for a $D_{sJ}(2690)$ is an excited
vector. Table \ref{spectrumTab} shows that the $D_{sJ}(2690)$ can
most naturally be identified with the excited vector $D_{s}^*(2S)$;
the D-wave vector is predicted to be somewhat too high at 2784 MeV
though mixing between these two basis states may be expected. For
the $D_{sJ}(2860)$, Table \ref{spectrumTab} indicates that this is
consistent with the radially excited
 scalar state $D_{s0}(2P)$. It appears that the $D_{s2}(2P)$ is too heavy to form a viable
identification.


\subsection{Decay Properties}

Mass spectra alone are insufficient to classify states. Their
production and decay properties also need to be compared with model
expectations. For example, strong decay widths can be computed with
the quark model wavefunctions and the strong decay vertex of the
$^3P_0$ model. An extensive application of the model to heavy-light
mesons is presented in Ref. \cite{cs}. Here we focus on the new
BaBar states with the results given in Table \ref{StrongDecayTab}.

\begin{table}[h!]
\caption{Strong Partial Widths for Candidate $D_s$ States.}
\label{StrongDecayTab}
\begin{center}
\begin{tabular}{l|cc}
\hline \hline
state (mass) & decay mode & partial width (MeV) \\
\hline
$D_s^*(2S)(2688)$ & $DK$   & 22 \\
            & $D^*K$ & 78 \\
            & $D_s\eta$ & 1 \\
            & $D_s^* \eta$ & 2 \\
            & total & 103 \\
\hline
$D_{s0}(2P)(2857)$ & $DK$  & 80 \\
            & $D_s\eta$ & 10 \\
            & total &  90 \\
\hline
$D_{s2}(2P)(2857)$ &  $DK$  & 3  \\
            &  $D_s\eta$ & 0 \\
            & $D^*K$ &  18 \\
            & $DK^*$ & 12 \\
            & total & 33 \\
\hline
$D_{s2}(2P)(3041)$ &  $DK$  & 1  \\
            &  $D_s\eta$ & 0 \\
            & $D^*K$ &  6 \\
            & $DK^*$ & 47 \\
            & $D^*K^*$ & 76 \\
            & total & 130 \\
\hline
\end{tabular}
\end{center}
\end{table}

The total width of the $D_s^*(2S)$ agrees very well with the
measured width of the $D_{sJ}(2690)$ ($112 \pm 37$ MeV), lending
support to this identification. No signal in $D_s\eta$ is seen or
expected, whereas the predicted  large $D^*K$ partial width implies
that this state should be visible in this decay mode. The data in
$D^{*0}(K) \to D^0\pi^0(K)$ do not support this contention; however,
the modes $D^{*+}(K) \to D^0\gamma (K)$ and $D^{*+}(K) \to D^+\pi^0
(K)$ show indications of a broad structure near 2700
MeV\cite{palano}. There is the possibility that $1^3D_1$ mixing with
$2^3S_1$ shift the mass down by 30 MeV to that observed and also
suppress the $D^*K$ mode. For a specific illustration, take the
model masses for the  $2^3S_1$ as 2.71GeV and $1^3D_1$ as 2.78 GeV.
A simple mixing matrix then yields a solution for the physical
states with masses 2.69 GeV and its predicted heavy partner at
around 2.81 GeV with eigenstates

\begin{eqnarray}
|D_s^*(2690)\rangle &\approx& \frac{1}{\sqrt{5}}(- 2 |1S\rangle + 1 |1D\rangle) \nonumber \\
|D_s^*(2810)\rangle &\approx& \frac{1}{\sqrt{5}}(|1S\rangle + 2
|1D\rangle )
\end{eqnarray}
and hence a mixing angle consistent with -0.5 radians.

\begin{figure}[h!]
\includegraphics[width=5 true cm, angle=270]{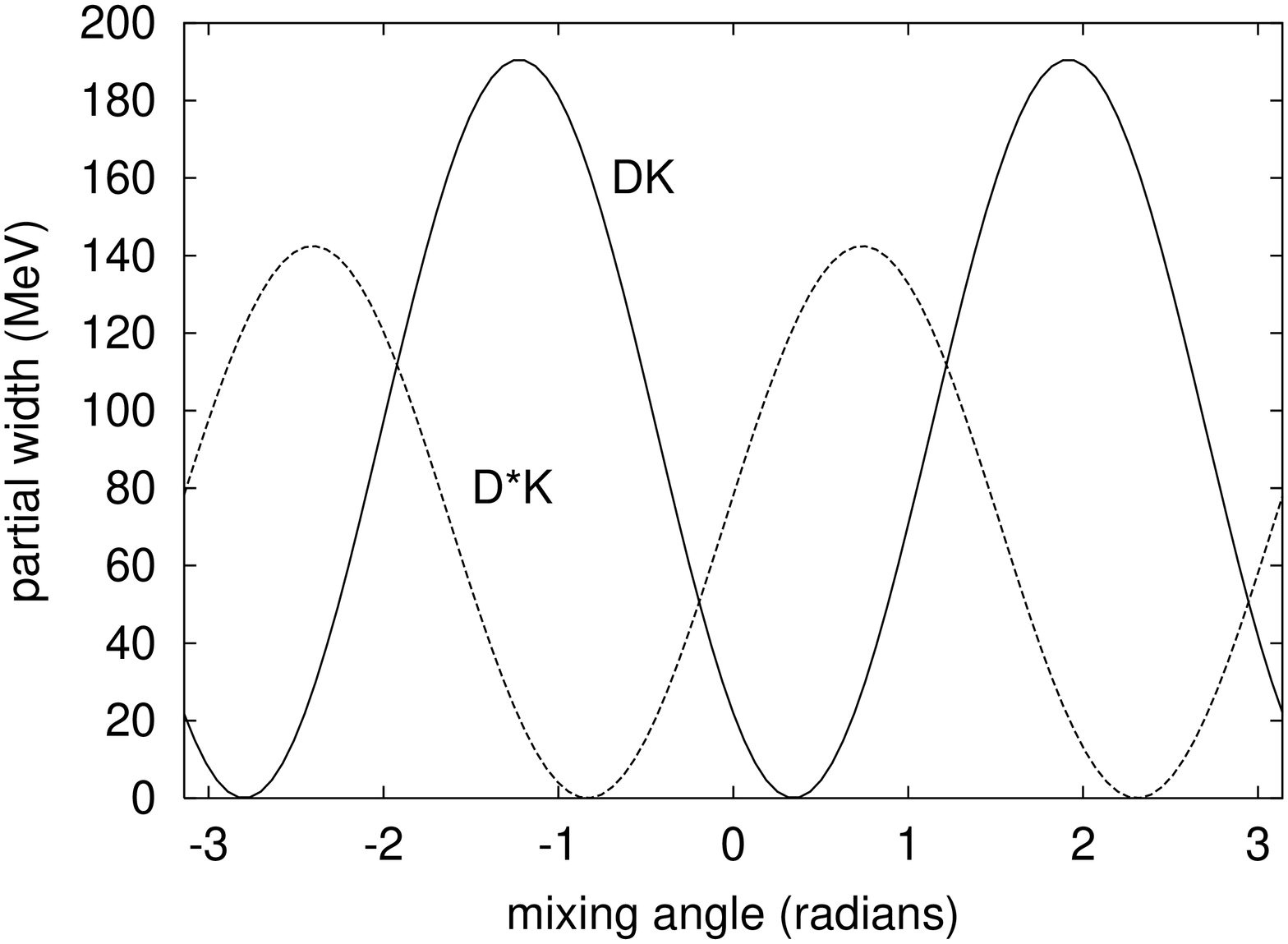}
\hskip 1 true cm
\includegraphics[width=5 true cm, angle=270]{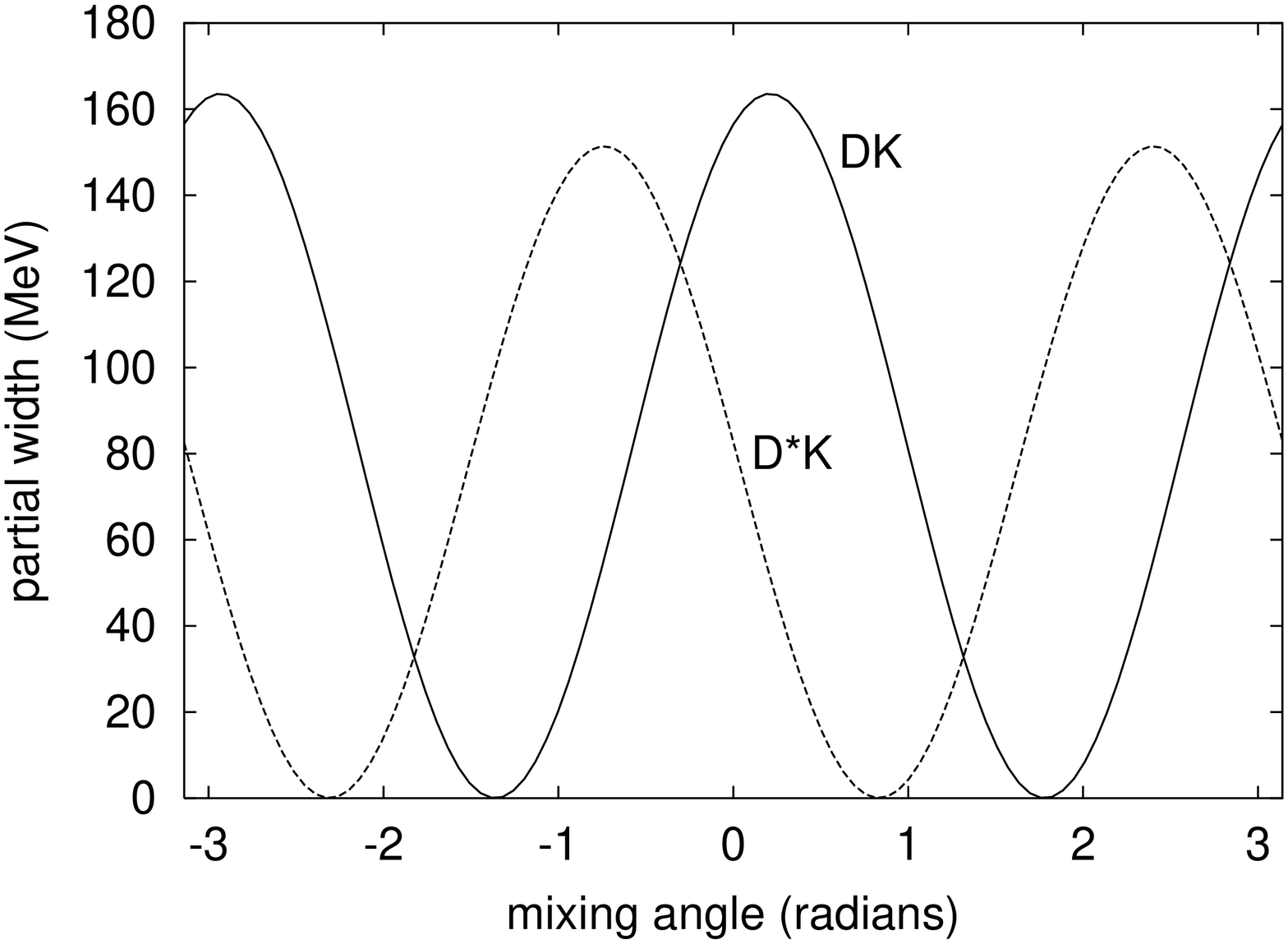}
\caption{$DK$ and $D^*K$ Partial Widths vs. Mixing Angle. Low vector
(left); high vector (right).} \label{mixedFig}
\end{figure}

The results of an explicit computation in the $^3P_0$ model are
shown in  Fig. \ref{mixedFig}. One sees that a mixing angle of
approximately -0.5 radians suppress the $D^*K$ decay mode of the low
vector (with mass set to 2688 MeV) and have a total width of
approximately 110 MeV, in agreement with the data. The orthogonal
state would then have a mass around 2.81 GeV and has a significant
branching ratio to both $DK$ and $D^*K$, albeit with a broad width,
greater than 200 MeV.

In summary, if the $D_{sJ}(2690)$ is confirmed as vector resonance,
then signals in the $D^*K$ channel are expected, either in the low
lying state (if the mixing is weak) or in a higher vector near 2.8
GeV.

For the $D_{sJ}(2860)$, the $D_{s2}(2P)$ assignment is further
disfavored. At either its model mass of 3041 MeV or at 2860 MeV the
$DK$ mode is radically suppressed, due to the $D$-wave barrier
factor. BaBar see their $D_{sJ}(2860)$ signal in $DK$ and do not
observe it in the $D^*K$  decay mode, making the $D_{s2}(2P)$
assignment unlikely.

By contrast, the properties of $D_{sJ}(2860)$ are consistent with
those predicted for the $D_{s0}(2P)$. Within the accuracy typical of
the $^3P_0$ model for S-wave decays, the total width is in accord
with the prediction that the $D_{s0}(2P)$ total width is less than
that of the excited vectors, and qualitatively in accord with the
measured $48 \pm 12$ MeV.


\clearpage

\subsection{Radiative Transitions}

The meson assignments made here can be tested further by measuring
radiative transitions for these states. Predictions made with the
impulse approximation, with and without nonrelativistic reduction of
quark spinors, are presented in Table \ref{RadTransTab}.

\begin{table}[h]
\caption{$D_s$ E1 Radiative Transitions (keV).}
\label{RadTransTab}
\begin{center}
\begin{tabular}{l|ccc}
\hline \hline
decay mode (mass)& $q_\gamma$ (MeV) & Non Rel rate  & Rel Rate\\
\hline
$D_s^*(2S)(2688) \to D_{s0} \gamma$    & 345   & 12.7 & 4.6 \\
$D_s^*(1D)(2784) \to D_{s0} \gamma$    & 428   & 116 & 82  \\
$D_{s0}(2P)(2857) \to D_s^* \gamma$    & 648   & 13  & 0.4  \\
$D_{s2}(2P)(3041) \to D_s^* \gamma$    & 787   & 6.8  & 1.9 \\
\hline
\end{tabular}
\end{center}
\end{table}

\subsection{Production}

The production of the radially excited $D_{s0}$ in $B$ decays can be
estimated with ISGW and other formalisms\cite{ISGW,chris}. Since
vector and scalar $c\bar s$ states can be produced directly from the
$W$ current, the decays $B \to D_s^*(2S) D_{(J)}$ or $D_{s0}(2P)
D_{(J)}$ serve as a viable source excited $D_s$ states.
Computationally, the only differences from ground state $D_s$
production are kinematics and the excited $D_s$ decay constants.

Production systematics can reveal structural information. For
example, the decay $B^0 \to D_s^+D^-$ goes via $W$ emission with a
rate proportional to $V_{bc}V_{cs}$, while $W$ exchange gives rise
to $B^0 \to D_s^- K^+ \sim V_{bc}V_{ud}$ and $B^0 \to D_s^+K^- \sim
V_{cd}V_{bu}$.  $W$ exchange is suppressed compared to $W$ emission,
thus the expected hierarchy of rates is

\begin{equation}
\Gamma(B^0 \to D_s^+D^-) \gg \Gamma(B^0 \to D_s^- K^+) \gg
\Gamma(B^0 \to D_s^+K^-).
\end{equation}
This suppression of $W$ exchange is confirmed by the
data\cite{PDG06} with $BR(B^0 \to D_s^+ D^-) = (6.5\pm2.1)\times
10^{-3}$ and $BR(B^0 \to D_s^- K^+) = (3.1\pm0.8)\times 10^{-5}$.
The decay to $D_s^+ K^-$ has not been observed.

It is therefore intriguing that the observed rate for $B^0 \to
D_s(2317)^+ K^-$ ($(4.3\pm1.5)\times 10^{-5}$) is comparable to $B^0
\to D_s^- K^+$.  Assuming accurate data, one must conclude either
that this simple reasoning is wrong, the $D_s(2317)^- K^+$ mode will
be found to be large, or the $D_s(2317)$ is an unusual state.
Searching for the process $B^0 \to D_s(2317)^- K^+$ is clearly of
great interest.

With the previous warning in mind, we proceed to analyze the
production of excited $D_s$ states in a variety of models. Rates
with decay constants set to 1 MeV for $D_s(2317)$ and $D_s(2860)$
production assuming that they are simple $c\bar s$ scalar and
excited scalar states are presented in Table
\ref{table:ScalarsComparison}.

Unfortunately, decay constants cannot be accurately computed at this
time. We have evaluated ratios of decay constants assuming a simple
harmonic oscillator quark model, a Coulomb+linear+hyperfine quark
model, and a relativized quark model. The resulting ratio for scalar
mesons fall in the range ${f_{D_s(2860)} \over f_{D_s(2317)}}
\approx 0.8 - 1.4$. The final estimates of the production of excited
scalar $D_s$ mesons in $B$ decays are thus

\begin{equation}
{B \to D_s(2860) D \over B \to D_s(2317)D} = 0.5 - 2
\end{equation}
and
\begin{equation}
{B \to D_s(2860) D^* \over B \to D_s(2317)D^*} = 0.3 - 1.3.
\end{equation}

\begin{table}[h]
\begin{center}
\begin{tabular}{c|cccc}
\hline {Decay Mode} & {ISGW} & {HQET - Luo \&
Rosner}\protect{\cite{Rosner:HQET}} &
{Pole}\protect{\cite{Rosner:HQET}} & {HQET - Colangelo}\protect{\cite{Colangelo:2003sa}} \\
\hline \hline $D_s(2317) D$ & $2.78 \times 10^{-7}$ & $1.95 \times
10^{-7}$ & $1.91 \times 10^{-7}$ & $2.24
\times 10^{-7}$ \\
$D_s(2317) D^*$ & $1.06 \times 10^{-7}$ & $8.82 \times 10^{-8}$ &
$8.79 \times 10^{-8}$ & $1.23
\times 10^{-7}$ \\
$D_s(2860) D$ & $2.09 \times 10^{-7}$ & $1.72 \times 10^{-7}$ &
$1.66 \times 10^{-7}$ & $1.83
\times 10^{-7}$ \\
$D_s(2860) D^*$ & $4.57 \times 10^{-8}$ & $3.61 \times 10^{-8}$ &
$3.55 \times 10^{-8}$ & $4.66
\times 10^{-8}$ \\
\hline
\end{tabular}
\end{center}
\caption{Branching ratios to scalars in different models with decay
constants set to 1 MeV} \label{table:ScalarsComparison}
\end{table}


A similar analysis for vector $D_s^*$ production is presented in
Table \ref{table:VectorsComparison}.

\begin{table}[h]
\begin{center}
\begin{tabular}{c|cccc}
\hline {Decay Mode} & {ISGW} & {HQET - Luo \&
Rosner}\protect{\cite{Rosner:HQET}} &
{Pole}\protect{\cite{Rosner:HQET}} & {HQET - Colangelo}\protect{\cite{Colangelo:2003sa}} \\
\hline \hline $D_s^* D$ & $ 1.97\times 10^{-7}$ & $ 1.33\times
10^{-7}$ & $ 1.32\times 10^{-7}$ & $ 1.57\times
10^{-7}$  \\
$D_s^* D^*$ & $ 4.20\times 10^{-7}$ & $ 3.22\times 10^{-7}$ & $
3.23\times 10^{-7}$ & $ 4.52\times
10^{-7}$  \\
$D_s(2690) D$ & $ 1.01\times 10^{-7}$ & $ 8.06\times 10^{-8}$ & $
7.77\times 10^{-8}$ & $
8.79\times 10^{-8}$  \\
$D_s(2690) D^*$ & $ 4.66\times 10^{-7}$ & $ 3.55\times 10^{-7}$ & $
3.49\times 10^{-7}$ & $
4.65\times 10^{-7}$  \\
\hline
\end{tabular}
\end{center}
\caption{Branching ratios to vectors in different models with decay
constants set to 1 MeV} \label{table:VectorsComparison}
\end{table}

Estimating vector decay constant ratios as above yields
${f_{D_s(2690)} \over f_{D_s^*}} \approx 0.7 - 1.1$. Finally,
predicted ratios of excited vector production are

\begin{equation}
{B \to D_s(2860) D \over B \to D_s^*(2110)D} = 0.3-0.7
\label{vecRatEq}
\end{equation}
and
\begin{equation}
{B \to D_s(2860) D^* \over B \to D_s^*(2110)D^*} = 0.5-1.3.
\end{equation}
We note that Eqn. \ref{vecRatEq} agrees well with the earlier
prediction of Close and Swanson\cite{cs}.

\subsection{Summary and Conclusions} 

Given the controversial nature of the $D_s(2317)$, establishing a
consistent picture of the entire $D_s$ spectrum is very important.
The new states claimed by BaBar can be useful in this regard. We
have argued that the six known $D_s$ and two new states can be
described in terms of a constituent quark model with novel
spin-dependent interactions. Predicted strong decay properties of
these states appear to agree with experiment.

Perhaps the most important tasks at present are (i) discovering the
$D_{s2}(2P)$ state, (ii) searching for resonances in $D^*K$ and
$DK^*$ up to 3100 MeV, (iii) analyzing the angular dependence of the
$DK$ final state in $D_{sJ}(2860)$ decay, (iv) assessing whether the
$D_{sJ}(2690)$ appears in the $D^*K$ channel, (v) searching for
these states in $B \to D_{sJ}D^{(*)}$ with branching ratios of $\sim
10^{-3}$.


\subsection{Postscript: Belle discovery}

Subsequent to these calculations, Belle\cite{newBelle} has reported
a vector state whose mass, width, and possibly production rate and
decay characteristics are consistent with our predictions.
Specifically, their measured mass and total width are $M = 2715 \pm
11 ^{+11}_{-14}$ MeV and $\Gamma = 115 \pm 20 ^{+36}_{-32}$ MeV, in
remarkable agreement with our predictions. The specific parameters
we have used in our analysis are contained within their
uncertainties.

Belle\cite{newBelle} find the new state in $B$ decays, which we have
proposed as a likely source. They report $Br(B \to
\bar{D^0}D^*_{s}(2700))\times Br(D^*_{s}(2700) \to D^0K^+) = (7.2
\pm 1.2 ^{+1.0}_{-2.9})\cdot 10^{-4}$. When compared to the
production of the ground state vector\cite{PDG06} which is $Br(B \to
\bar{D^0}D^*_{s}(2112)) = (7.2 \pm 2.6)\cdot 10^{-3}$, the ratio of
production rates in $B$ decay is then
${\cal{O}}(0.1)/Br(D^*_{s}(2700) \to D^0K^+)$. From our Table
\ref{StrongDecayTab}, and assuming flavor symmetry for the strong
decay, we predict that $Br(D^*_{s}(2690) \to D^0K^+) \sim 10\%$ ,
which within the uncertainties will apply also to the Belle state.
Thus the absolute production rate, within the large uncertainties,
appears to be consistent with that predicted in Section 4. If the
central value of the Belle mass is a true guide, then a significant
branching ratio in $D^*K$ would be expected (Table
\ref{StrongDecayTab} and Fig \ref{mixedFig}). The orthogonal vector
state would then be dominantly 1D at 2.78 GeV, but hard to produce
in $B$ decays.  These statements depend on the dynamics underlying
$2S$-$1D$ mixing, which is poorly understood. It is therefore very
useful that $B$ decay systematics and the strength of the $D^*K$
decay channel in the excited vector $D_s$ mesons can probe this
dynamics.


Searching for this state in the other advocated modes, and improving
the uncertainties, now offers a significant test of the dynamics
discussed here.

\clearpage

\chapter{DECAY CONSTANTS}\label{dc}

Decay constants describe the simplest electroweak transitions, where
a meson couples directly to the photon or $W$ boson. They are often
used in more complicated calculations, for example, nonleptonic
decays, gamma-gamma transitions or higher order diagrams for
radiative transitions, so it is important to know them with good
accuracy. Decay constants for some mesons could be determined from
experimental data, for example from $e^{+}e^{-}$ decays for
quarkonium. Comparison to the experimental data makes it possible to
test the meson wave function in different models and help us
understand our models better.

In this chapter the results for the decay constants calculated in
the nonrelativistic potential quark model with the variety of the
potentials are presented and discussed.

\section{Charmonium}
Results for the decay constants of charmonium states are shown in
Table \ref{dcTab}. In the following we will demonstrate that
agreement with experimental charmonium decay constants requires a
weakening of the short range quark interaction with respect to the
standard Coulomb interaction. This weakening is in accord with the
running coupling of perturbative QCD and eliminates the need for an
artificial energy dependence that was introduced by Godfrey and
Isgur\cite{GI} to fit experimental decay constants.

Since our results depend substantially on the parameters of the
potential used, the global study of this dependence is required
before any firm conclusions can be made. We performed this study of
five experimentally observable quantities by varying the parameters
of the potential and minimizing the deviations between the
calculated and experimental values. These five quantities are: the
masses of the first two excited vector meson states relative to the
mass of the ground state vector meson; the spin average mass of the
scalar ($0^{++}$), axial vector ($1^{++}$) and tensor ($2^{++}$)
meson states relative to the mass of the ground state vector meson;
the decay constants of the ground and the first excited vector meson
states.

We found that for usual `Coulomb+linear+hyperfine' potential no set
of parameters exist that could reproduce the values of all of the
five quantities better than $10\%$. However, with the introduction
of the logarithmic dependence of running coupling, all five
calculated quantities are not further than $5\%$  from the
experimental values. We also found that BGS parameters \cite{BGS}
are very close to the best fit parameters for all of the five
quantities, so we use BGS parameters for all our calculations. Our
results for the relative differences of the calculated meson
properties from the experimental data are presented in Table
\ref{dc_param} (the full spinor structure has been used for
calculation of our results in this Table). It is very hard to obtain
the decay constant of the second excited vector meson state close to
the experimental value for both potentials (`BGS' and `BGS log') as
illustrated in the same Table.

\begin{table}[!h]
\caption{Relative differences between the calculated and
experimental values in \%.}\label{dc_param}
\begin{center}
\begin{tabular}{c|ccc}

\hline

&BGS Rel &BGS log &BGS log\\
&&$\Lambda=0.4$ GeV &$\Lambda=0.25$ GeV\\
\hline \hline
$m_{\psi'}-m_{\psi}$     &-2.04 &-5.77 &-8.49 \\
$m_{\psi''}-m_{\psi}$    &2.97  &0.32  &-1.80 \\
$m_{\chi}-m_{\psi}$      &1.64  &-8.96 &-14.02\\
$f_{\psi}$               &32.6  &2.92  &-4.38 \\
$f_{\psi'}$              &33.0  &9.68  &5.02  \\
$f_{\psi''}$             &115.5 &53.4  &48.3  \\
\hline \hline
\end{tabular}
\end{center}
\end{table}

\begin{table}[!h]
\caption{Charmonium Decay Constants (MeV).} \label{dcTab}
\begin{tabular}{c|cccccc}
\hline
Meson & BGS NonRel & BGS Rel & BGS log & BGS log & lattice & experiment \\
      &            &         &  $\Lambda= 0.4$ GeV & $\Lambda = 0.25$ GeV & & \\
\hline \hline
$\eta_c$                       &795    &493   &424   &402      &$429\pm 4\pm 25$  &$335\pm 75$\\
$\eta'_c$                      &477    &260   &243   &240      &$56\pm 21\pm 3$   &\\
$\eta''_c$                     &400    &205   &194   &193      &                  &\\
$J/\psi$                       &615    &545   &423   &393      &$399\pm 4$        &$411\pm 7$\\
$\psi(2S)$                     &431    &371   &306   &293      &$143\pm 81$       &$279 \pm 8$\\
$\psi(3S)$                     &375    &318   &267   &258      &                  &$174\pm 18$\\
$\chi_{c1}$                    &239    &165   &155    &149       &                  &\\
$\chi'_{c1}$                   &262    &167   &157   &152      &                  &\\
$\chi''_{c1}$                  &273    &164   &155   &151      &                  &\\
\hline
\end{tabular}
\end{table}

The second column of Table \ref{dcTab} shows results of the
nonrelativistic computation (Eq.\ref{nonrelfVEq}) with wavefunctions
determined in the Coulomb+linear model with BGS parameters
\cite{BGS}.  A clear trend is evident as all predictions are
approximately a factor of two larger than experiment (column seven).
Using the full spinor structure (column three) improves agreement
with experiment substantially, but still yields predictions which
are roughly 30\% too large.  At this stage the lack of agreement
must be ascribed to strong dynamics, and this motivated the running
coupling model specified above. The fourth and fifth columns give
the results obtained from this model. It is apparent that the
softening of the short range Coulomb potential induced by the
running coupling brings the predictions into very good agreement
with experiment.

Column six lists the quenched lattice gauge computations of Ref.
\cite{JLlatt}. The agreement with experiment is noteworthy; however,
the predictions for the $\eta_c'$ and $\psi(2S)$ decay constants
are much smaller than those of the quark model (and experiment in
the case of the $\psi(2S)$). It is possible that this is due to
excited state contamination in the computation of the mesonic
correlators.

The good agreement between model and experiment has been obtained
with a straightforward application of the quark model. This stands
in contrast to the methods adopted in Ref. \cite{GI} where the
authors insert arbitrary factors of $m/E(k)$ in the integrand in
order to obtain agreement with experiment (the extra factors of
$m/E$ serve to weaken the integrand, approximating the effect of the
running coupling used here).

It is very difficult to obtain a value for $f_{\psi(3S)}$ that is as
small as experiment. Assuming that the experimental value is
reliable it is possible that this difficulty points to serious
problems in the quark model. A simple mechanism for diminishing the
decay constant is via S-D wave mixing, because the D-wave decouples
from the vector current in the nonrelativistic limit. This mixing
can be generated by the tensor interaction of Eq. \ref{vsdEq};
however, computations yield amplitude reductions of order 2\% -- too
small to explain the effect. Note that S-D mixing can also be
created by transitions to virtual meson-meson pairs. Unfortunately,
evaluating this requires a reliable model of strong  Fock sector
mixing and we do not pursue this here.

A similar discussion holds for the $e^+e^-$ width of the
$\psi(3770)$. Namely, the large decay constant $f_{\psi(3770)} = 99
\pm 20$ MeV can perhaps be explained by mixing with nearby S-wave
states. Again, the computed effect due to the tensor interaction is
an order of magnitude too small and one is forced to look elsewhere
(such as loop effects) for an explanation.

Attempts to compute Lorentz scalars such as decay constants or form
factors in a noncovariant framework are necessarily ambiguous.  As
stated above, the results of a computation in the nonrelativistic
quark model are only guaranteed to be consistent in the weak binding
limit. However the accuracy of the quark model can be estimated by
examining the decay constant dependence on model assumptions. For
example, an elementary aspect of covariance is that a single decay
constant describes the vector (for example) decay amplitude in all
frames and for all four-momenta. Thus the decay constant computed
from the temporal and spatial components of the matrix element
$\langle 0 | J_\mu|V\rangle$ should be equal. As pointed out above,
setting $\mu=0$ yields the trivial result $0=0$ in the vector rest
frame. However, away from the rest frame one obtains the result
\begin{eqnarray}
f_V = \sqrt{N_c E(P)} \int {d^3 k \over (2\pi)^3} &&\!\!\!\!\!\!\!\!\Phi(k;P) {1 \over \sqrt{E(k+P/2)}\sqrt{E(k-P/2)}}\times\nonumber\\
&&\times\frac{1}{2}\left(
\frac{\sqrt{E(k+P/2)+m}}{\sqrt{E(k-P/2)+m}} +
                  \frac{\sqrt{E(k-P/2)+m}}{\sqrt{E(k+P/2)+m}}  \right)
\end{eqnarray}
or, in the nonrelativistic limit
\begin{equation}
f_V = {\sqrt{N_c M_V}\over m} \tilde\Phi(0).
\end{equation}
One sees that covariance is recovered in the weak binding limit
where the constituent quark model is formally valid.

\begin{figure}[h]
\includegraphics[angle=-90,width=15cm]{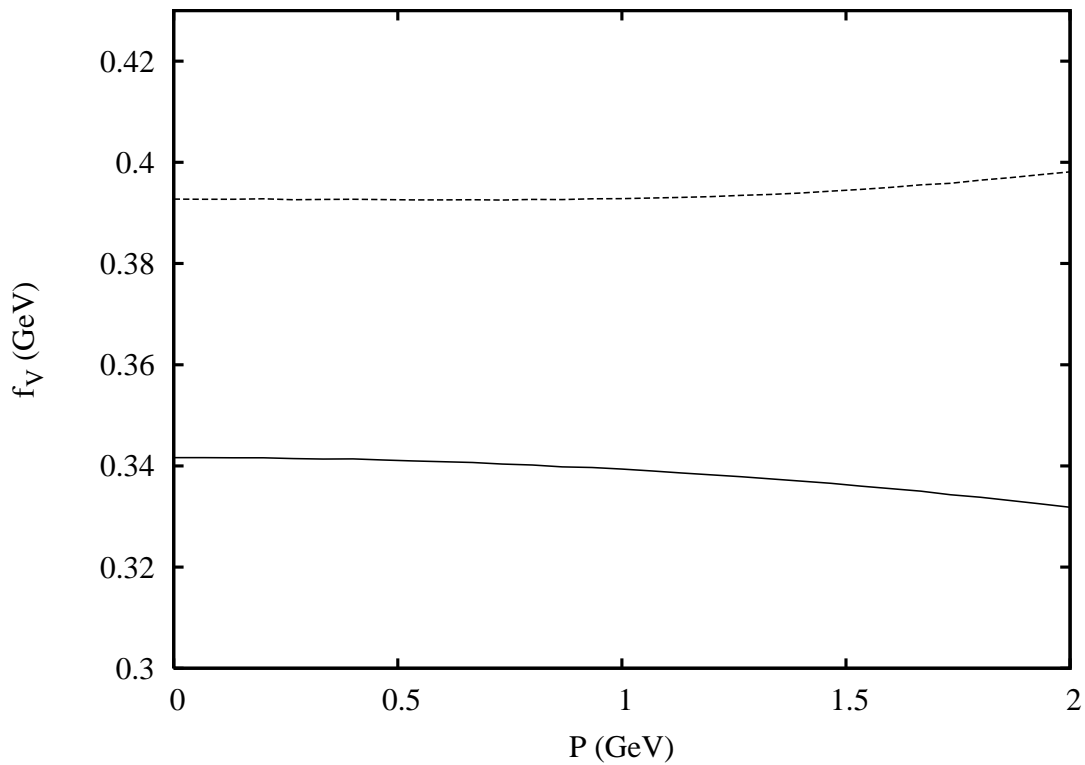}
\caption{Temporal (top line) and Spatial (bottom line) Vector Decay
Constants in Various Frames.} \label{covDecayConstantFig}
\end{figure}

Computations of the vector decay constant away from the weak binding
limit and the rest frame are displayed in Fig.
\ref{covDecayConstantFig}. One sees a reassuringly weak dependence
on the vector momentum $P$. There is, however, a 13\% difference in
the numerical value of the temporal and spatial decay constants,
which may be taken as a measure of the reliability of the method.

\section{Bottomonium}

The study of the dependence of bottomonium spectrum and the decay
constants on the parameters of the potential has been performed in a
similar way as for charmonia. We varied the parameters of the
potential and minimized the deviations of the calculated values of
six quantities from their experimental values. These six quantities
are: the masses of the first two excited vector meson states
relative to the mass of the ground state vector meson; the spin
average mass of the scalar ($0^{++}$), axial vector ($1^{++}$) and
tensor ($2^{++}$) meson states relative to the mass of the ground
state vector meson; the decay constants of the ground and the first
two excited vector meson states.

In order to find the best set of parameters we minimized $\chi^2$:
\begin{equation}
\chi^2=\frac{1}{N_{dof}}\sum_i\frac{(f^i-f^i_{exp})^2}{\sigma^2_i}
\end{equation}
where $N_{dof}$ is a number of degrees of freedom: $N_{dof}=N_f-N_p$
where $N_f$ is a number of observable quantities and $N_p$ is the
number of parameters we vary. In our case $N_f=6$ and $N_p=3$ (we
vary $\alpha$, $b$ and $m_b$) so $N_{dof}=3$. The standard
deviations of the experimental values $\sigma^2_i$ have been taken
from the Particle Data Group book. We calculate six quantities
$f^{i}$, and $f^i_{exp}$ are their experimental values.

We found that it is not possible to reproduce even five out of six
quantities within $10\%$ of their experimental value using standard
`Coulomb + linear' potential. However, it is possible with the
introduction of the momentum-dependent running coupling.

For the `Coulomb + linear' potential with logarithmic short range
behavior of running coupling we found that `the best fit' parameters
for the bottomonium are: $m_b=4.75$ GeV, $a_C=a_H=0.35$, $b=0.19$
GeV$^2$, and $\sigma=0.897$ GeV. Our results for the relative
differences of the calculated meson properties from the experimental
data are presented in Table \ref{dc_paramb} (the full spinor
structure has been used for calculation of our results in this
Table).

\begin{table}[!h]
\caption{Relative differences between the calculated and
experimental values in \%.}\label{dc_paramb}
\begin{center}
\begin{tabular}{c|ccc}

\hline

&C+L Rel &C+L log &C+L log\\
&&$\Lambda=0.4$ GeV &$\Lambda=0.25$ GeV\\
\hline \hline
$m_{\Upsilon'}-m_{\Upsilon}$     &-2.30 &-6.57 &-9.23 \\
$m_{\Upsilon''}-m_{\Upsilon}$    &0.01  &-3.12 &-5.13 \\
$m_{\chi}-m_{\Upsilon}$          &1.33  &-7.59 &-12.3 \\
$f_{\Upsilon}$                   &25.0  &1.13  &-6.07 \\
$f_{\Upsilon'}$                  &20.5  &2.70  &-1.45 \\
$f_{\Upsilon''}$                 &44.8  &24.9  &20.8  \\
\hline \hline
\end{tabular}
\end{center}
\end{table}

Predicted decay constants are presented here (Table \ref{Bdc} ). All
computations we performed as for charmonia.

One can see that agreement with available experimental data is
impressive for the model with running coupling (C+L log). We
conclude that the running coupling in C+L potential is needed to
reproduce right short-range behavior of the meson wave functions,
which is probed by the decay constants.

\begin{table}[!h]
\caption{Bottomonium Decay Constants (MeV).} \label{Bdc}
\begin{center}
\begin{tabular}{c|ccccc}
\hline
Meson & C+L NonRel & C+L Rel & C+L log  & C+L log & experiment \\
      &            &         & $\Lambda = 0.4$ GeV  & $\Lambda = 0.25$ GeV & \\
\hline \hline
$\eta_b$                       &979    &740   &638   &599     &\\
$\eta_b'$                      &644    &466   &423   &411     &\\
$\eta_b''$                     &559    &394   &362   &354     &\\
$\Upsilon$                     &963    &885   &716   &665     &$708\pm 8$\\
$\Upsilon'$                    &640    &581   &495   &475     &$482\pm 10$\\
$\Upsilon''$                   &555    &501   &432   &418     &$346\pm 50$\\
$\Upsilon'''$                  &512    &460   &400   &388     &$325\pm 60$\\
$\Upsilon^{(4)}$                &483    &431   &377   &367      &$369\pm 93$\\
$\Upsilon^{(5)}$                 &463    &412   &362   &351     &$240\pm 61$\\
$\chi_{b1}$                       &186    &150    &142    &136   &\\
$\chi_{b1}'$                      &205    &160   &152   &147  &\\
$\chi_{b1}''$                     &215    &164   &157   &152  &\\
\hline \hline
\end{tabular}
\end{center}
\end{table}

\clearpage

\section{Heavy-light meson decay constants}

The results of our calculations of heavy-light meson decay constants
in different models are presented in Table \ref{ewdc1}, and then
results of C+L model with running coupling are compared to the
experiment and other model calculations in Table \ref{ewdc}.

The first two columns of Table \ref{ewdc1} correspond to the
calculations with C+L potential with and without nonrelativistic
reduction of the quark spinors. One can see considerable difference
between these results not only for light mesons (as could have been
expected) but also for heavy mesons which are usually considered
being nonrelativistic (even for $B_c$ mesons the difference between
two columns is 30-50\% ).

Also, as for $c\bar{c}$ and $b\bar{b}$ mesons, introduction of
running coupling is needed to correctly describe short-range
behavior of heavy-light meson wave functions and bring decay
constant in better agreement with experiment.

Our results for all the meson decay constants (except $K$
pseudoscalar meson) agree quite well both with the experimental data
and other model calculations (where available). We would like to
point out that we used the same parameters of the potential for all
of our calculations in this section (global parameters:
$\alpha=0.594, b=0.162, \sigma=0.897$), they have not been refitted.

The fact that the decay constant of $K$ pseudoscalar meson is so
different both from its experimental value and other model
calculations lets us conclude that there are some effects, important
for light pseudoscalar mesons, that are missing in nonrelativistic
potential quark model. It might be related to the lack of the chiral
symmetry in this model, or maybe the absence of many-body effects.
Therefore it will be of interest in the future to perform the study
of light meson decay constants in Coulomb gauge model which takes
these effects into account.

\addtolength{\topmargin}{-2cm} \addtolength{\footskip}{3cm}

\begin{table}[h!]
\caption{\label{ewdc1} Decay constants of heavy-light mesons (MeV).
Global parameters have been used.}
\begin{center}
\begin{tabular}{|c|c|c|c|c|}

\hline \multicolumn{1}{|c|}{Meson}&
\multicolumn{4}{c|}{Nonrel wf}\\
\multicolumn{1}{|c|}{}& \multicolumn{2}{c|}{C+L potential}&
\multicolumn{2}{c|}{C+L log potential}\\
\multicolumn{1}{|c|}{}& \multicolumn{1}{c|}{nonrel}&
\multicolumn{1}{c|}{rel}& \multicolumn{1}{c|}{$0.4GeV$}&
\multicolumn{1}{c|}{$0.25GeV$}\\

\hline
&&&&\\
$K$                       &1116  &445  &425  &417   \\
$K^*$                     &332   &286  &261  &252   \\
$K_0^*$                   &97    &30   &30   &30    \\
&&&&\\
&&&&\\
$D$                       &489   &290  &269  &260   \\
$D^*$                     &318   &272  &240  &230   \\
$D_0$                     &221   &83   &82   &81    \\
&&&&\\
&&&&\\
$D_s$                     &627   &374  &337  &324   \\
$D_s^*$                   &447   &388  &324  &306   \\
$D_{s0}^*$                &174   &75   &73   &72    \\
&&&&\\
&&&&\\
$B$                       &267   &195  &175  &167   \\
$B^*$                     &232   &196  &169  &161   \\
$B_0$                     &207   &84   &83   &81    \\
&&&&\\
&&&&\\
$B_s$                     &394   &283  &242  &229   \\
$B_s^*$                   &349   &300  &241  &226   \\
$B_{s0}$                  &208   &98   &94   &92    \\
&&&&\\
&&&&\\
$B_c$                     &917   &623  &451  &415   \\
$B_c^*$                   &886   &779  &497  &450   \\
$B_{c0}$                  &174   &97   &86   &81    \\
\hline

\end{tabular}
\end{center}
\end{table}

\clearpage

\addtolength{\oddsidemargin}{-2.5cm}

\begin{table}[h!]
\caption{\label{ewdc} Decay constants of heavy-light mesons (MeV).
Global parameters have been used.}
\begin{center}
\begin{tabular}{c|c|c|c|c|c|c|c|c|l}

\hline \multicolumn{1}{|c|}{Meson}& \multicolumn{1}{c|}{}&
\multicolumn{2}{c|}{Lattice}& \multicolumn{1}{c|}{Light}&
\multicolumn{1}{c|}{QCD}& \multicolumn{2}{c|}{CQM}&
\multicolumn{1}{c|}{Bethe-}&
\multicolumn{1}{c|}{Experiment}\\
\multicolumn{1}{|c|}{}& \multicolumn{1}{c|}{C+L}&
\multicolumn{1}{c|}{quenched}& \multicolumn{1}{c|}{unquenched}&
\multicolumn{1}{c|}{cone}& \multicolumn{1}{c|}{sum rule}&
\multicolumn{2}{c|}{}& \multicolumn{1}{c|}{Salpeter}&
\multicolumn{1}{c|}{}\\
\multicolumn{1}{|c|}{}& \multicolumn{1}{c|}{log}&
\multicolumn{1}{c|}{\cite{lat1}}&
\multicolumn{1}{c|}{\cite{lat2}\cite{lat3}}&
\multicolumn{1}{c|}{\cite{LC}}&
\multicolumn{1}{c|}{\cite{SumRule1}\cite{SumRule2}}&
\multicolumn{1}{c|}{\cite{CQM1}} & \multicolumn{1}{c|}{\cite{CQM2}}&
\multicolumn{1}{c|}{\cite{Bethe1}\cite{Bethe2}}&
\multicolumn{1}{c|}{}\\

\hline \hline
&&&&&&&&&\\
$K$     &417 &152(6)(10) &152.0(6.1) & &            &155 &169 &157 &$153\pm 4$ ($e^+e^-$)\\
        &    &           &           & &            &    &    &&$158\pm 21$ ($\mu^+\mu^-$)\\
$K^*$   &252 &           &255.5(6.5) & &            &236 &    &&\\
$K_0^*$ &30  &           &           & &$427\pm 85$ &    &    &&\\
&&&&&&&&&\\
$D$     &260 &235(8)(14) &225(14)(14) &    &$205\pm 20$ &234 &234 &238         &$302\pm 94$ \cite{PDG}\\
        &    &           &            &    &            &    &    &            &$222.6\pm 16.7^{+2.8}_{-3.4}$ \cite{CLEO:DC}\\
$D^*$   &230 &           &            &216 &            &310 &    &$340\pm 23$ &\\
$D_0$   &81  &           &            &    &            &    &    &            &\\
&&&&&&&&&\\
$D_s$      &324   &266(10)(18)&$267(13)(17)(+10)$ &&$235\pm 24$ &268 &391 &241         &$246\pm 47$ ($\mu^+\mu^-$)\\
           &      &           &                   &&            &    &    &            &$281\pm 33$ ($\tau^+\tau^-$)\\
$D_s^*$    &306   &           &                   &&            &315 &    &$375\pm 24$ &\\
$D_{s0}^*$ &72    &           &                   &&            &    &    &            &\\
&&&&&&&&&\\
$B$   &167 &&216(9)(19)(4)(6) &150 &$203\pm 23$ &189 &191 &193         &\\
$B^*$ &161 &&                 &    &            &219 &    &$238\pm 18$ &\\
$B_0$ &81  &&                 &    &            &    &    &            &\\
&&&&&&&&&\\
$B_s$    &229 &&242(9)(34)(+38) & &$236\pm 30$ &218 &236 &195         &\\
$B_s^*$  &226 &&                & &            &251 &    &$272\pm 20$ &\\
$B_{s0}$ &92  &&                 &    &            &    &    &            &\\
&&&&&&&&&\\
$B_c$    &415 &&&&&421 &&&\\
$B_c^*$  &    &&                 &    &            &    &    &            &\\
$B_{c0}$ &    &&                 &    &            &    &    &            &\\
\hline \hline
\end{tabular}
\end{center}
\end{table}

\clearpage \addtolength{\oddsidemargin}{2.5cm}
\addtolength{\topmargin}{2cm} \addtolength{\footskip}{-3cm}

\chapter{FORM-FACTORS}\label{ff}
\section{Electromagnetic form-factors}
Single quark elastic and transition form factors for charmonia are
considered in the following sections. The agreement with recent
lattice computations is very good, but requires that the standard
nonrelativistic reduction of the current not be made and that the
running coupling described above be employed. As will be shown, this
obviates the need for the phenomenological $\kappa$ factor
introduced for electroweak decays in the ISGW model\cite{ISGW}.

\subsection{Charmonium single quark form factors}

Unfortunately elastic electromagnetic form factors are not
observables for charmonia; however this is an area where lattice
gauge theory can aid greatly in the development of models and
intuition.  In particular, a theorist can choose to couple the
external current to a single quark, thereby yielding a nontrivial
'pseudo-observable'. This has been done in Ref. \cite{JLlatt} and we
follow their lead here by considering the single-quark elastic
electromagnetic form factors for pseudoscalar, scalar, vector, and
axial vector charmonia.

\begin{figure}[h!]
\begin{center}
\includegraphics[angle=-90,width=15cm]{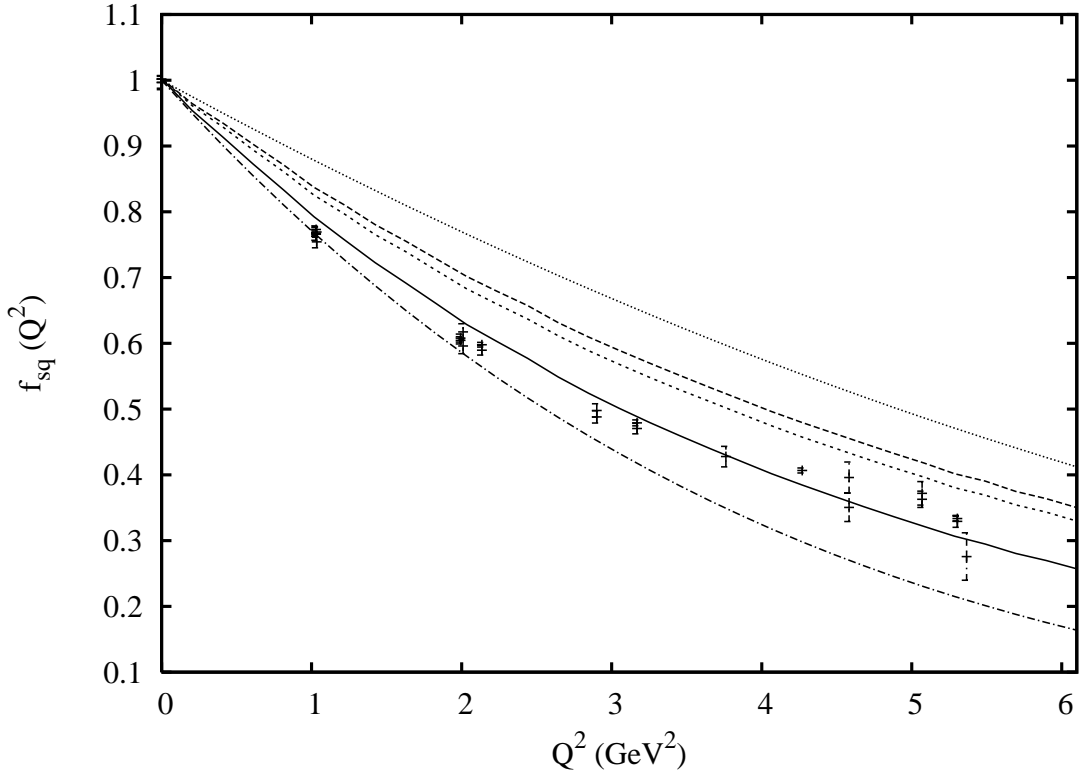}
\caption{The Single Quark $\eta_c$ Form-factor $f_{sq}(Q^2)$. From
top to bottom the curves are SHO, nonrelativistic BGS, relativistic
BGS, BGS log, and ISGW.} \label{etaFig}
\end{center}
\end{figure}

A variety of quark model computations of the $\eta_c$ single quark
elastic form factor are compared to lattice results in Fig.
\ref{etaFig}. It is common to use SHO wavefunctions when computing
complicated matrix elements. The dotted curve displays the
nonrelativistic form factor (Eq. \ref{ff0Eq}) with SHO wavefunctions
(the SHO scale is taken from Ref. \cite{cs}). Clearly the result is
too hard with respect to the lattice. This problem was noted by ISGW
and is the reason they introduce a suppression factor $\vec q \to
\vec q/\kappa$. ISGW set $\kappa = 0.7$ to obtain agreement with the
pion electromagnetic form factor. The same procedure yields the
dot-dashed curve in Fig. \ref{etaFig}. The results agrees well with
lattice for small $Q^2$; thus, somewhat surprisingly, the {\it ad
hoc} ISGW procedure appears to be successful for heavy quarks as
well as light quarks.

The upper dashed curve indicates that replacing SHO wavefunctions
with full Coulomb+linear wavefunctions gives a somewhat softer
nonrelativistic form factor. The same computation with the
relativistic expression (Eq. \ref{ffFullEq}), the lower dashed
curve, yields a slight additional improvement. Finally, the
relativistic BGS+log single quark elastic $\eta_c$ form factor is
shown as the solid line and is in remarkably good agreement with the
lattice (it is worth stressing that form factor data have not been
fit). It thus appears that the ISGW procedure is an {\it ad hoc}
procedure to account for relativistic dynamics and deviations of
simple SHO wavefunctions from Coulomb+linear+log wavefunctions.

\begin{figure}[h!]
\begin{center}
\includegraphics[angle=-90,width=15cm]{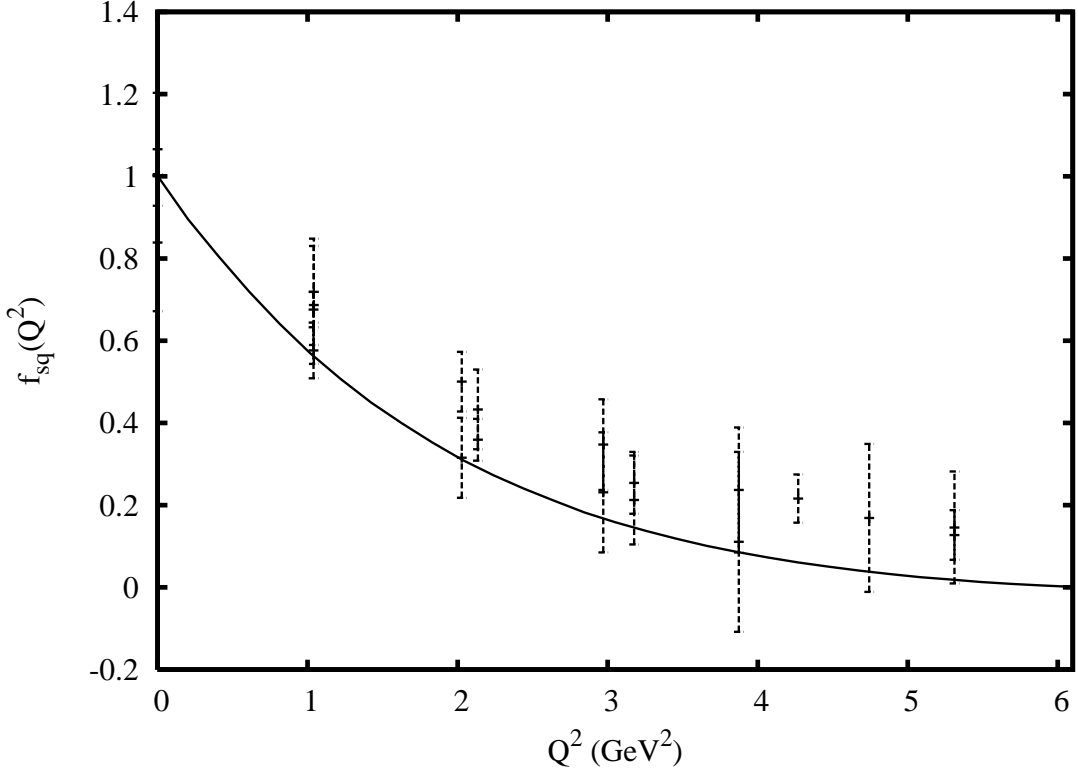}
\caption{The $\chi_{c0}$ Single Quark Form-factor $f_{sq}(Q^2)$.}
\label{chic0Fig}
\end{center}
\end{figure}

A similar procedure can be followed for the vector, scalar, and
axial elastic single quark form factors. The necessary Lorentz
decompositions and expressions for the form factors are given in
Appendix \ref{ffApp}. The single quark $\chi_{c0}$ elastic form
factor for the relativistic BGS+log case is shown in Fig.
\ref{chic0Fig}. The BGS model yields a very similar result and is
not shown. This appears to be generally true and hence most
subsequent figures will only display BGS+log results. As can be
seen, the agreement with the lattice data, although somewhat noisy,
is very good.

\begin{figure}[h!]
\includegraphics[angle=-90,width=8cm]{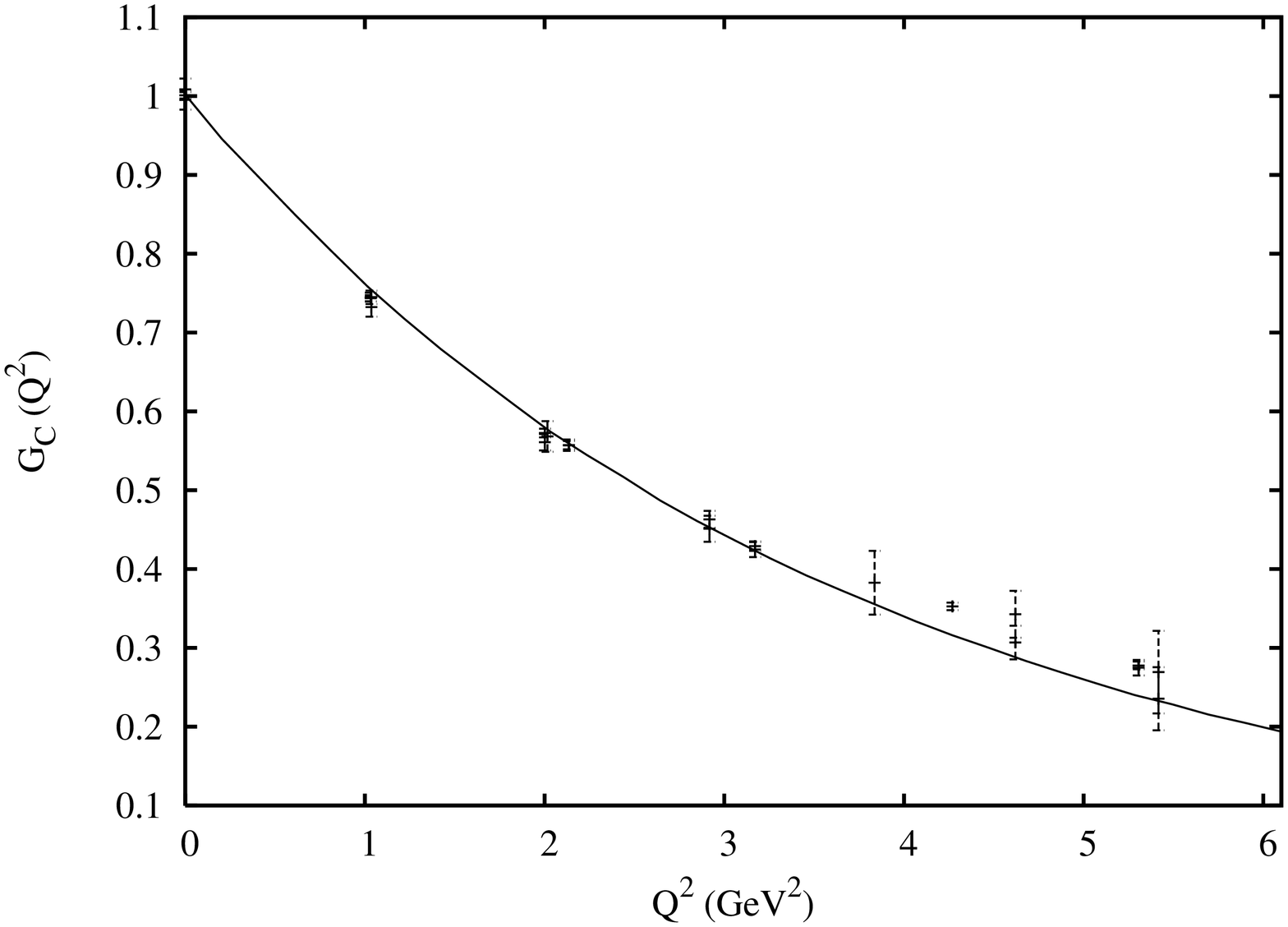}
\includegraphics[angle=-90,width=8cm]{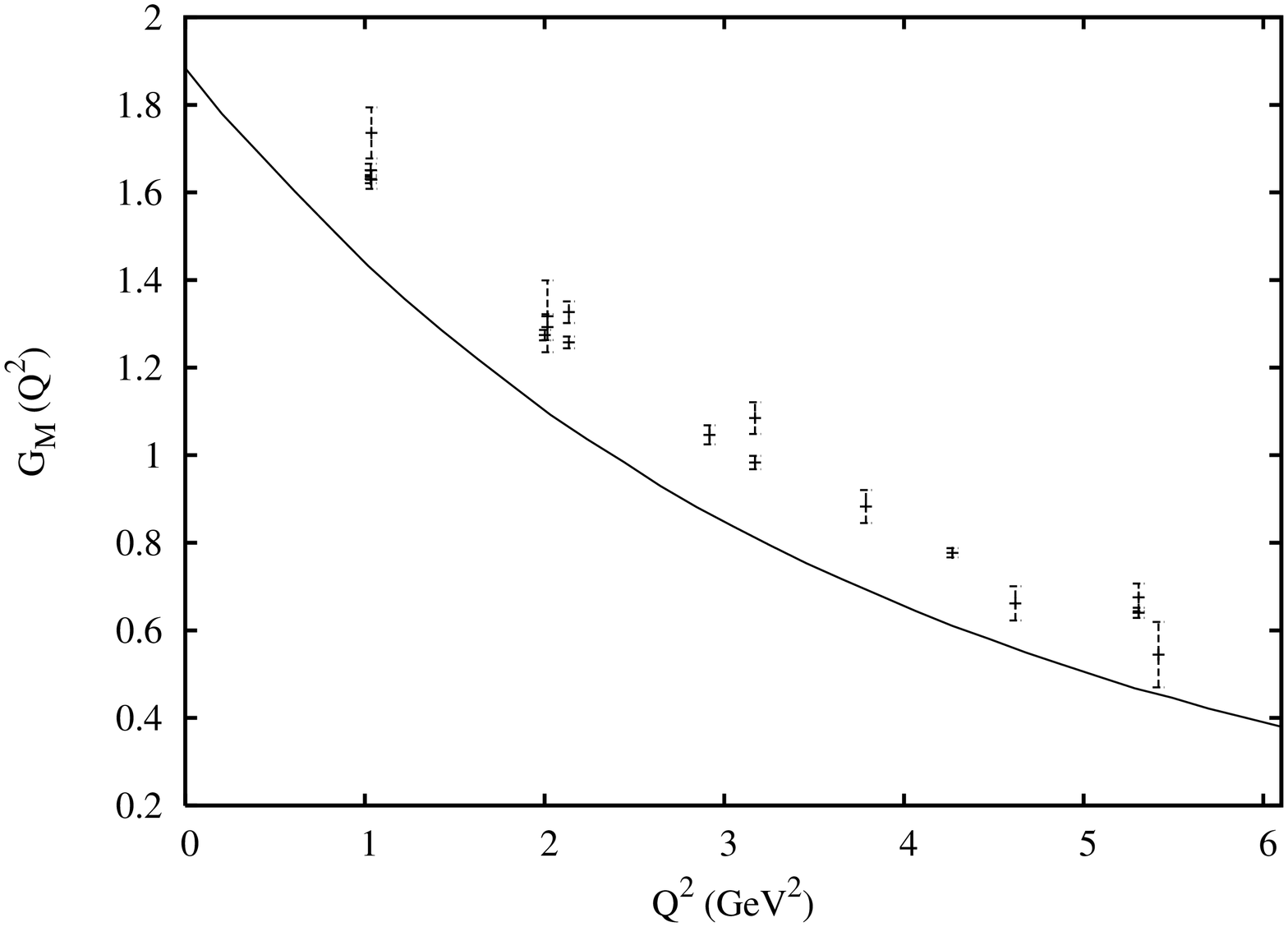}
\caption{Single Quark $J/\psi$ Form Factors $G_{sq}^C$ (left) and
$G_{sq}^M$ (right).} \label{psiFig}
\end{figure}

The left panel of Fig. \ref{psiFig} shows the single quark $J/\psi$
charge form factor. The agreement of the relativistic BGS+log model
with the lattice data is remarkable. The right panel of Fig.
\ref{psiFig} contains the magnetic dipole form factor (see Appendix
\ref{ffApp} for definitions). In this case the form factor at zero
recoil is model-dependent. In the nonrelativistic limit, Eq.
\ref{GMEq} implies that $G_M(\vec q = 0) = M_V/m \approx 2$. The
model prediction is approximately  10\% too small compared to the
lattice data. The lattice results have not been tuned to the
physical charmonium masses (charmonium masses are approximately 180
MeV too low); however it is unlikely that this is the source of the
discrepancy since the ratio $M/m$ is roughly constant when $M$ is
near the physical mass. Thus it appears that the problem lies in the
quark model. Reducing the quark mass provides a simple way to
improve the agreement; however the modifications to the spectrum due
to a 10\% reduction in the quark mass are difficult to overcome with
other parameters while maintaining the excellent agreement with
experiment.

\begin{figure}[h!]
\includegraphics[angle=-90,width=8cm]{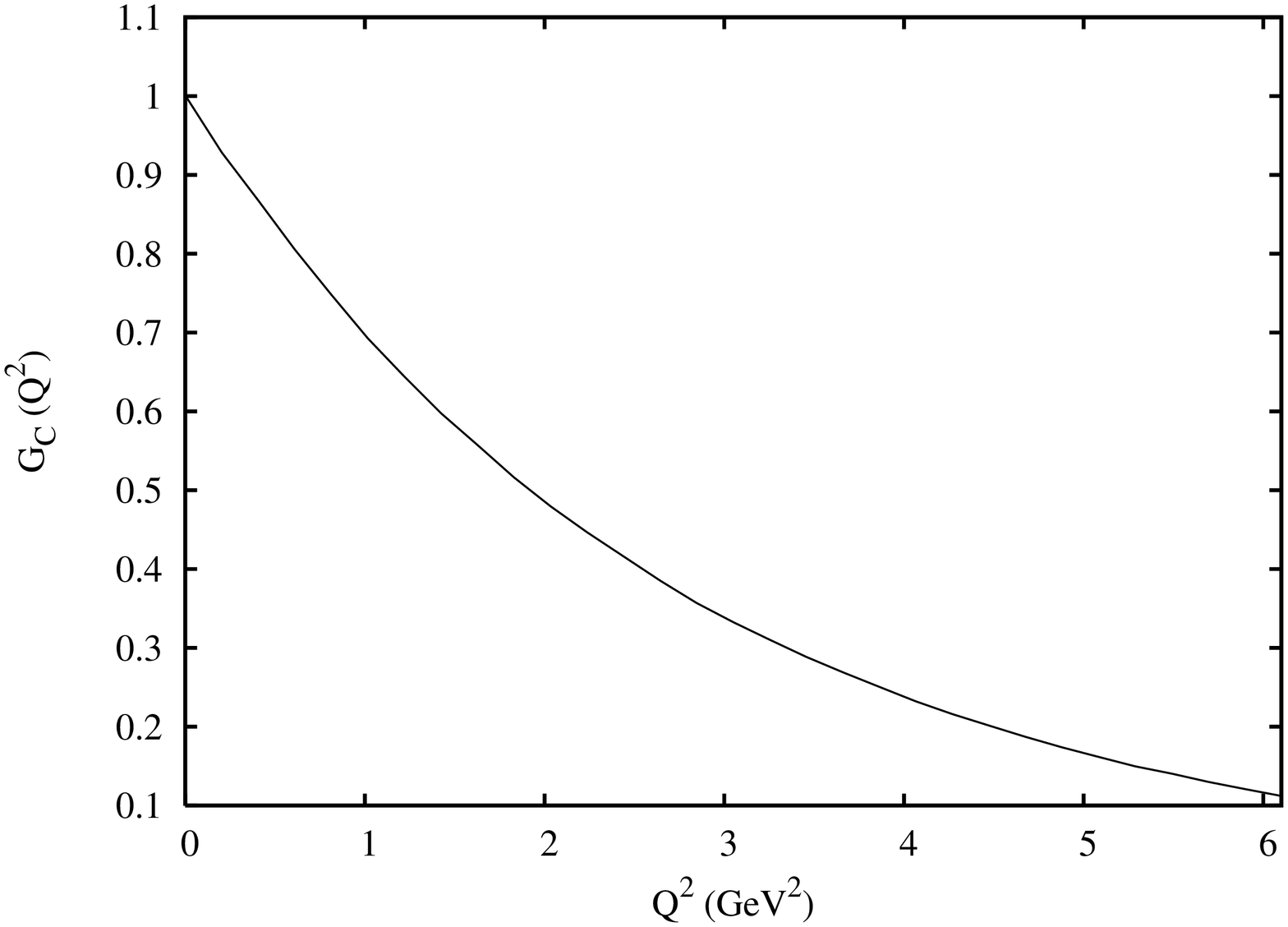}
\includegraphics[angle=-90,width=8cm]{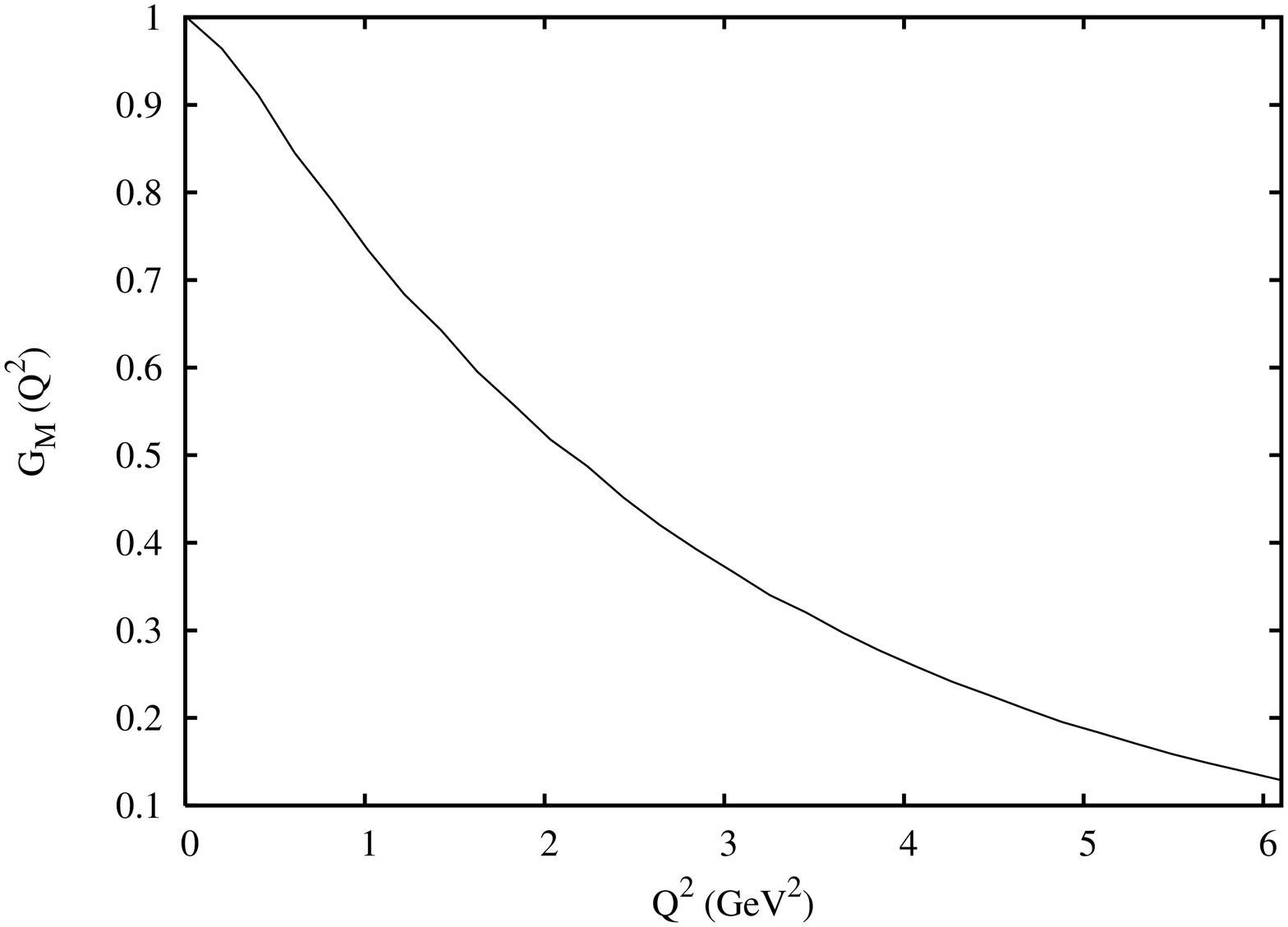}
\caption{Single Quark $h_c$ Form Factors $G_{sq}^C$ (left) and
$G_{sq}^M$ (right).} \label{SqhcFig}
\end{figure}

\begin{figure}[h!]
\includegraphics[angle=-90,width=8cm]{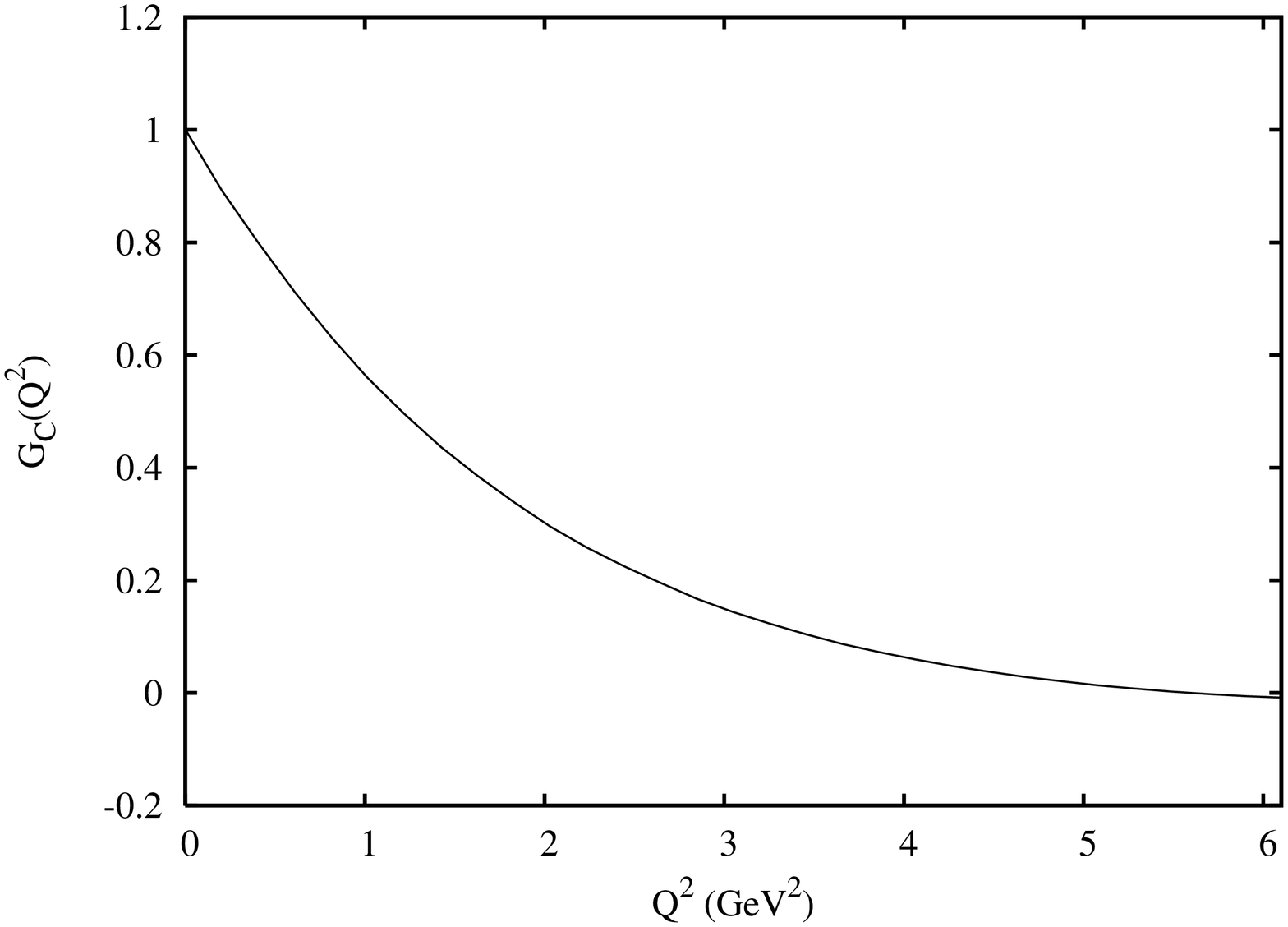} 
\includegraphics[angle=-90,width=8cm]{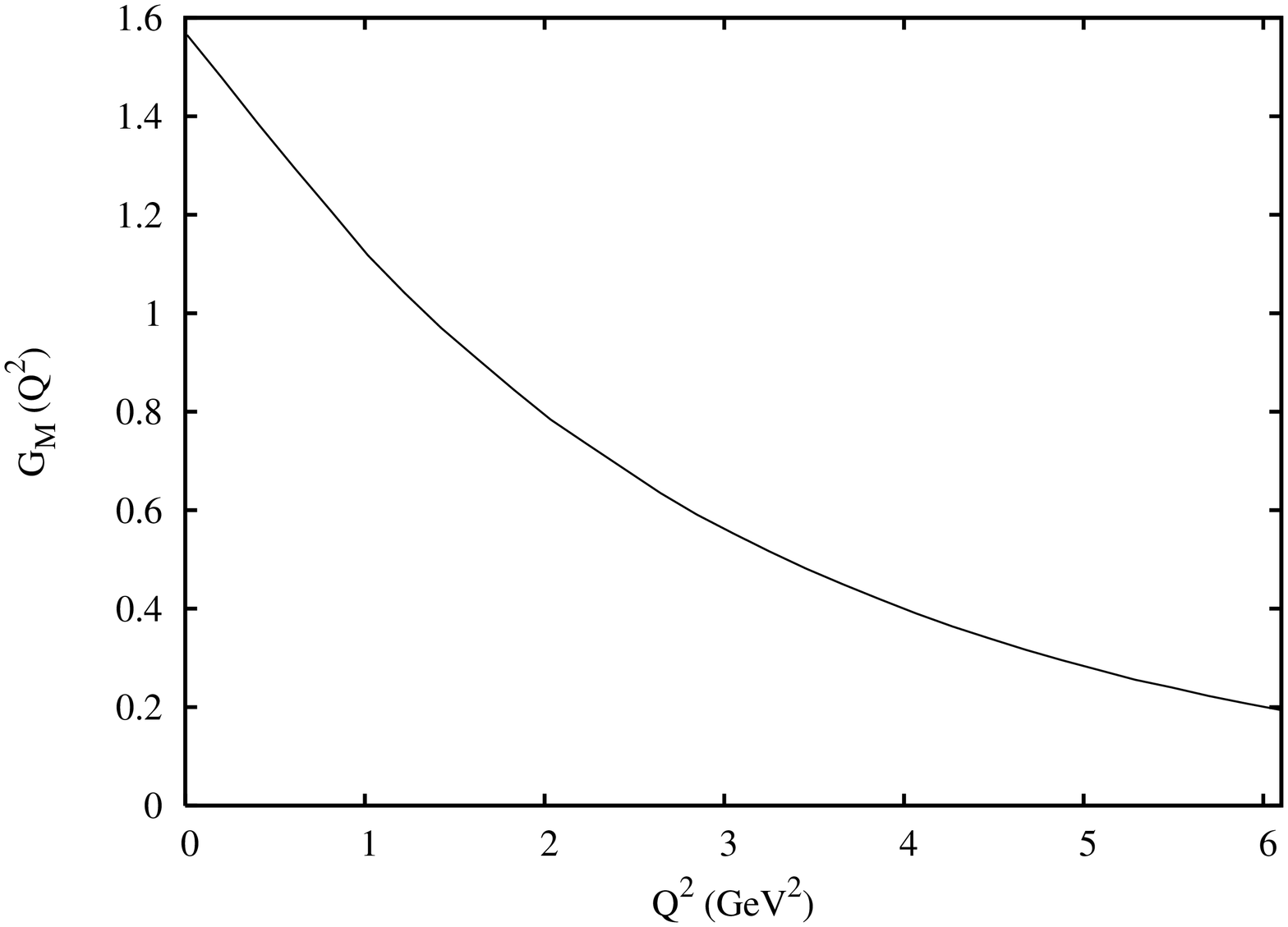}
\caption{Single Quark $\chi_{c1}$ Form Factors $G^C_{sq}$ (left) and
$G^M_{sq}$ (right).} \label{chic1SQFig}
\end{figure}

Predictions for the single quark elastic electromagnetic form
factors of the $h_c$ and $\chi_{c1}$ states are shown in Figs.
\ref{SqhcFig} and \ref{chic1SQFig}. As for the $J/\psi$, the charge
form factors are normalized at zero recoil, while the magnetic form
factors take on model-dependent values at zero recoil. In the
nonrelativistic limit these are $G^M_{sq}(\vec q =0) = M/(2m)$ for
the $h_c$ and $G^M_{sq}(\vec q = 0) = 3M/(4m)$ for the $\chi_{c1}$.

The presence of a kinematical variable in form factors makes them
more sensitive to covariance ambiguities than static properties such
as decay constants. In addition to  frame and current component
dependence, one also must deal with wavefunction boost effects that
become more pronounced as the recoil momentum increases. Presumably
it is preferable to employ a frame which minimizes wavefunction
boost effects since these are not implemented in the nonrelativistic
constituent quark model. Possible choices are (i) the initial meson
rest frame (ii) the final meson rest frame (iii) the Breit frame.
These frames correspond to different mappings of the three momentum
to the four momentum: $|\vec q|^2 = Q^2 (1 + \alpha)$ where $\alpha
= 0$ in the Breit frame and $\alpha = Q^2/4 M^2$ in the initial or
final rest frame (these expressions are for elastic form factors
with a meson of mass $M$). Furthermore, as with decay constants, it
is possible to compute the form factors by using different
components of the current.

We consider the $\eta_c$ elastic single quark form factor in greater
detail as an example. The form factor obtained from the temporal
component of the current in the initial meson rest frame is given in
Eqs. \ref{ffFullEq} and \ref{ff0Eq}. Computing with the spatial
components yields Eq. \ref{chiSpatialEq} with the nonrelativistic
limit

\begin{equation}
f_{sq}(Q^2) = \frac{2M}{m} \int {d^3 k \over (2\pi)^3}
\Phi(\vec{k})\Phi^*\left(\vec{k}+\frac{\vec{q}}{2}\right)\left(\vec
k + \frac{\vec q}{2}\right)\cdot \frac{\vec q}{q^2}
\end{equation}
This can be shown to be equivalent to

\begin{equation}
\frac{2M}{m} \frac{1}{4}\int d^3x |\Phi(x)|^2 {\rm e}^{-i \vec q
\cdot \vec x/2},
\end{equation}
which is Eq. \ref{ff0Eq} in the weak coupling limit. At zero recoil
this evaluates to $\frac{M}{2m}$, which is approximately 10\% too
small with respect to unity.  Once again, reducing the quark mass
presumably helps improve agreement.

\begin{figure}[h!]
\begin{center}
\includegraphics[angle=-90,width=15cm]{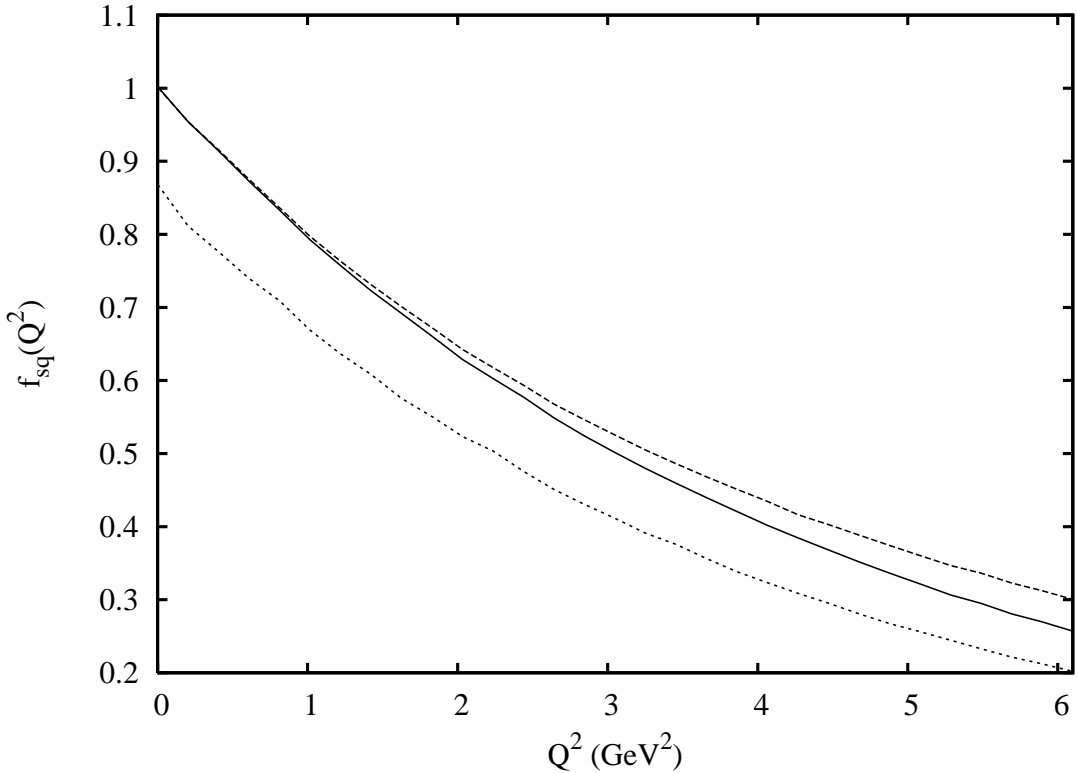}
\caption{Covariance Tests for the Single Quark $\eta_c$ Form
Factor.} \label{covTestFig}
\end{center}
\end{figure}

Fig. \ref{covTestFig} compares the various methods of computing the
$\eta_c$ single quark form factor. The solid line is the result of
Fig. \ref{etaFig}, computed in the initial rest frame with the
temporal component of the current. The dashed line is the
computation of the form factor in the Breit frame. The good
agreement is due to a cancelation between the different four-vector
mapping discussed above and the modifications induced by computing
the quark model form factor in the Breit frame. The lower dashed
line is the form factor computed from the spatial components of the
current (Eq. \ref{chiSpatialEq}). It is evidently too small compared
to the correctly normalized results by approximately a factor of
$2m/M$, indicating that the method is accurate at the 10\% level.

Finally, the large $Q^2$ behavior of pseudoscalar form factors is a
controversial topic. We do not presume to resolve the issues here;
rather we note that the preferred method for obtaining the form
factor yields an asymptotic behavior proportional to $\alpha_s(Q^2)
f_{Ps} M_{Ps} /Q^2$, which is similar,
 but not identical, to that expected in perturbative QCD\cite{FJ}.
Nevertheless, the model is not applicable in this regime and the
asymptotic scaling should not be taken seriously.

%

\subsection{Charmonium Transition Form Factors} 
\label{tffSect}

Transition form factors convolve differing wavefunctions and
therefore complement the information contained in single quark
elastic form factors. They also have the important benefit of being
experimental observables at $Q^2 = 0$.

The computation of transition form factors proceeds as for elastic
form factors, with the exception that the current is coupled to all
quarks. Lorentz decompositions and quark model expressions for a
variety of transitions are presented in App. \ref{ffApp}.  The
mapping between three-momentum and $Q^2$ is slightly different in
the case of transition form factors. In the Breit frame this is
\begin{equation}
|\vec q|^2 = Q^2 + {(m_2^2 - m_1^2)^2 \over Q^2 + 2m_1^2 + 2 m_2^2},
\end{equation}
while in the initial rest frame it is
\begin{equation}
|\vec q|^2 = {Q^4 + 2Q^2 (m_1^2+m_2^2) + (m_1^2 - m_2^2)^2\over 4
m_1^2}. \label{generalQsqEq}
\end{equation}
An analogous result holds for the final rest frame mapping.

\begin{figure}[h!]
\begin{center}
\includegraphics[angle=-90,width=15cm]{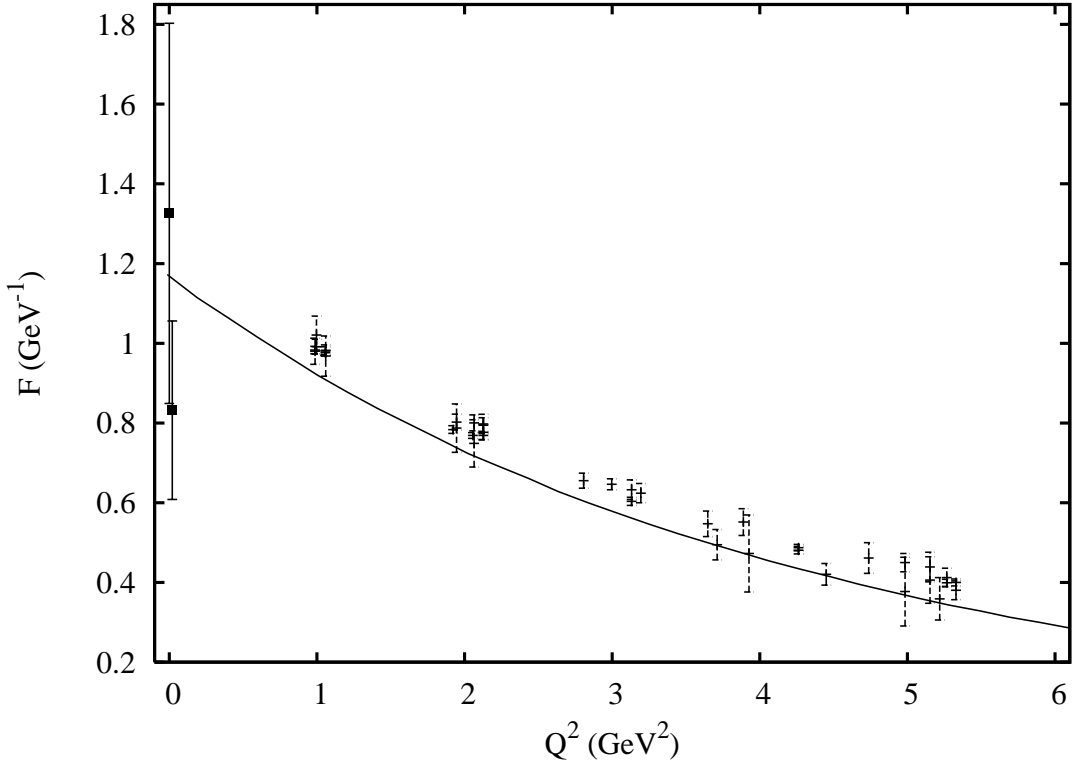}
\caption{Form Factor $F(Q^2)$ for $J/\psi\rightarrow \eta_c\gamma$.
Experimental points are indicated with squares.} \label{FpsiFig}
\end{center}
\end{figure}

\begin{figure}[h!]
\begin{center}
\includegraphics[angle=-90,width=15cm]{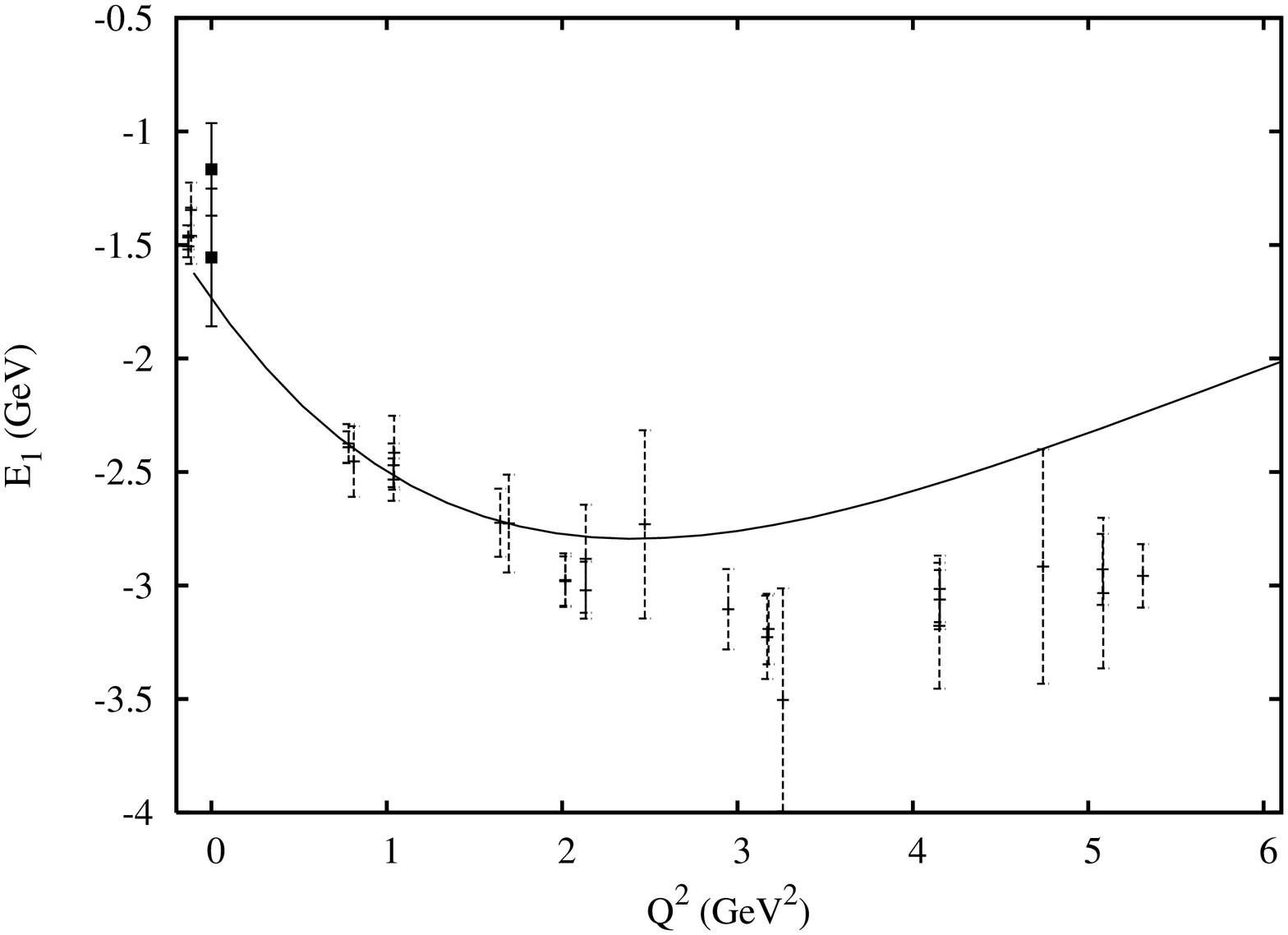} 
\caption{Form Factor $E_1(Q^2)$ for $\chi_{c0}\rightarrow
J/\psi\gamma$. Experimental points are indicated with squares.}
\label{E1Fig}
\includegraphics[angle=-90,width=15cm]{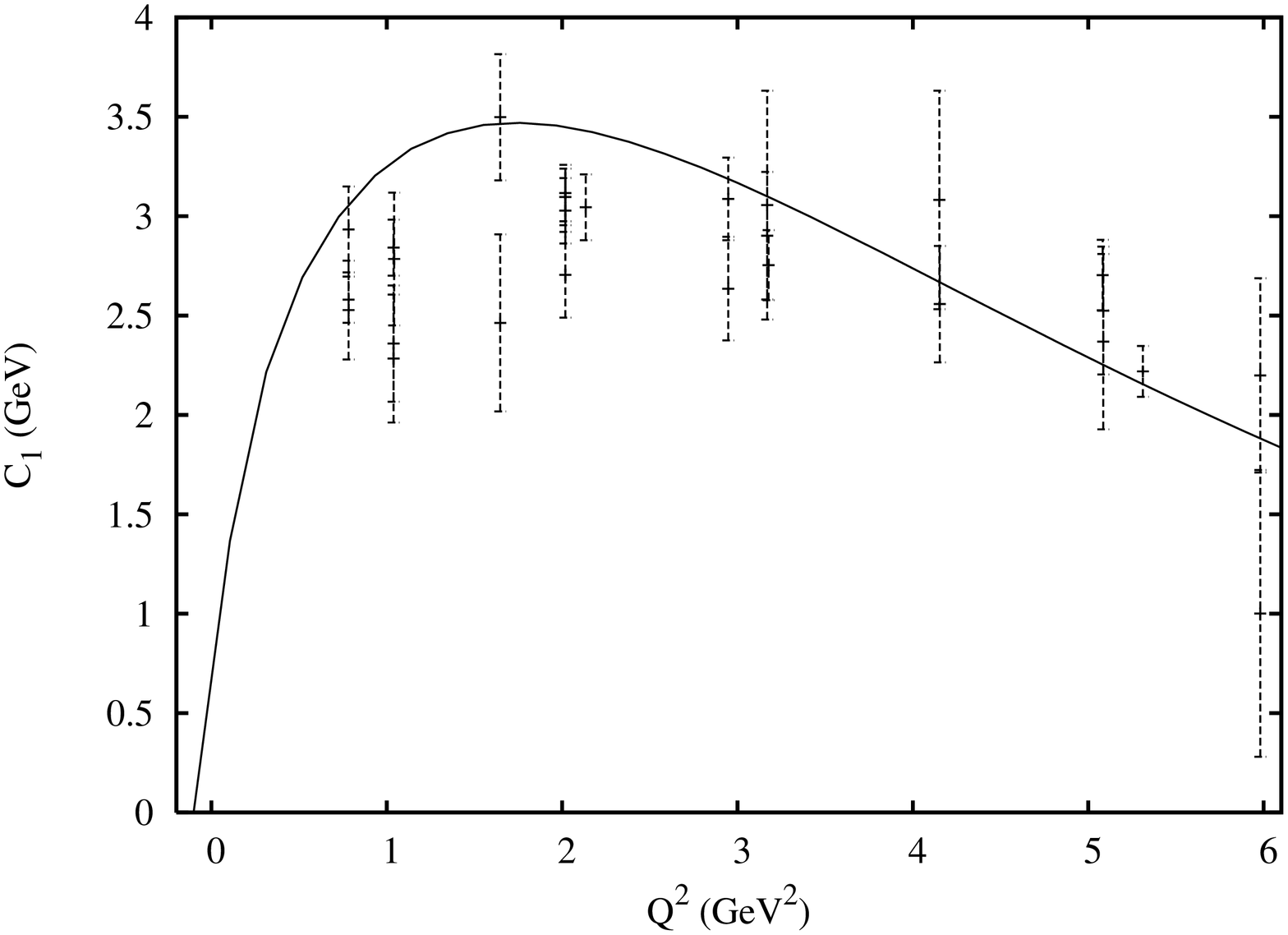}
\caption{Form Factor $C_1(Q^2)$ (right) for $\chi_{c0}\rightarrow
J/\psi\gamma$.} \label{C1Fig}
\end{center}
\end{figure}

\begin{figure}[h!]
\includegraphics[angle=-90,width=8cm]{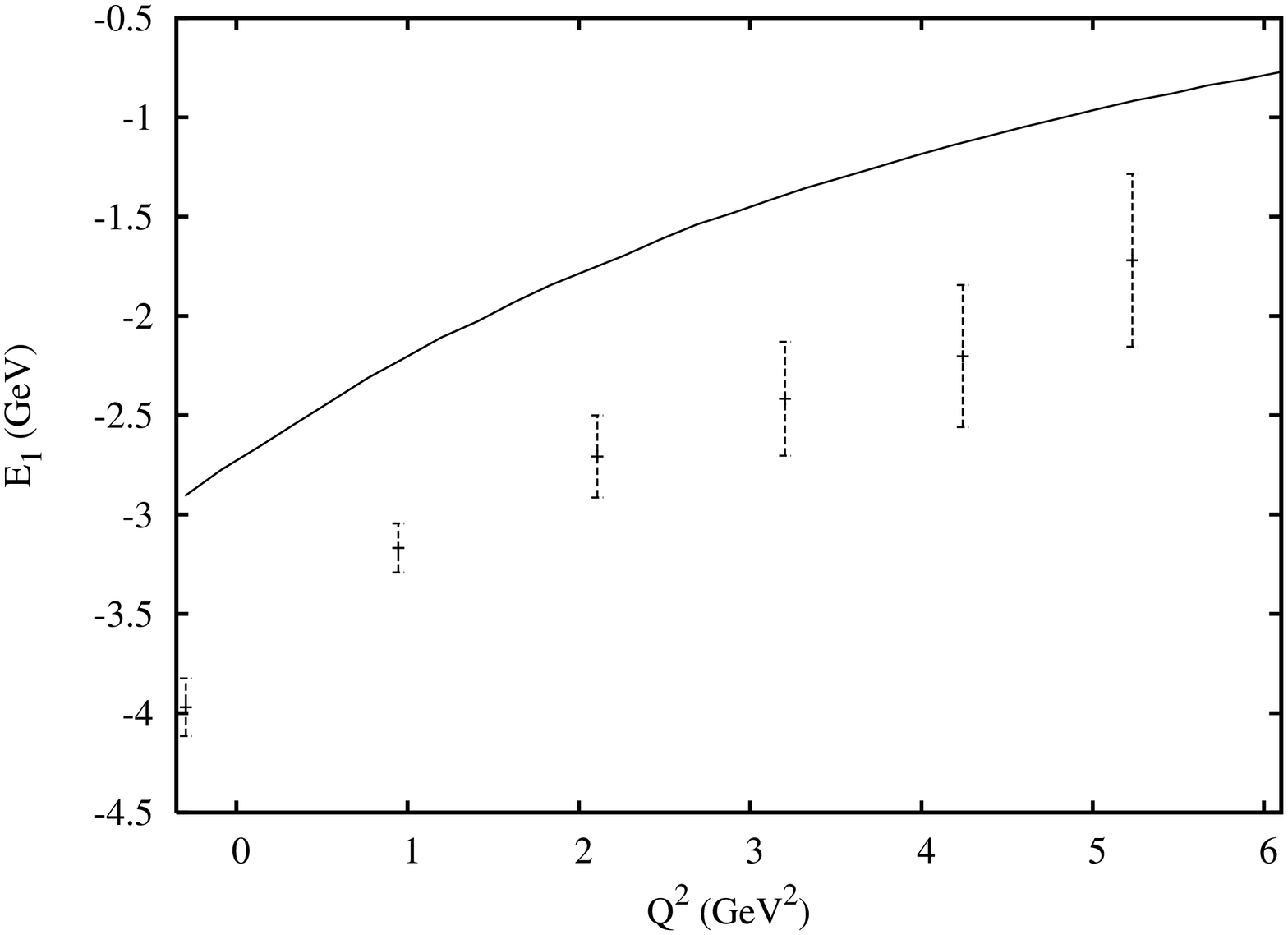} 
\includegraphics[angle=-90,width=8cm]{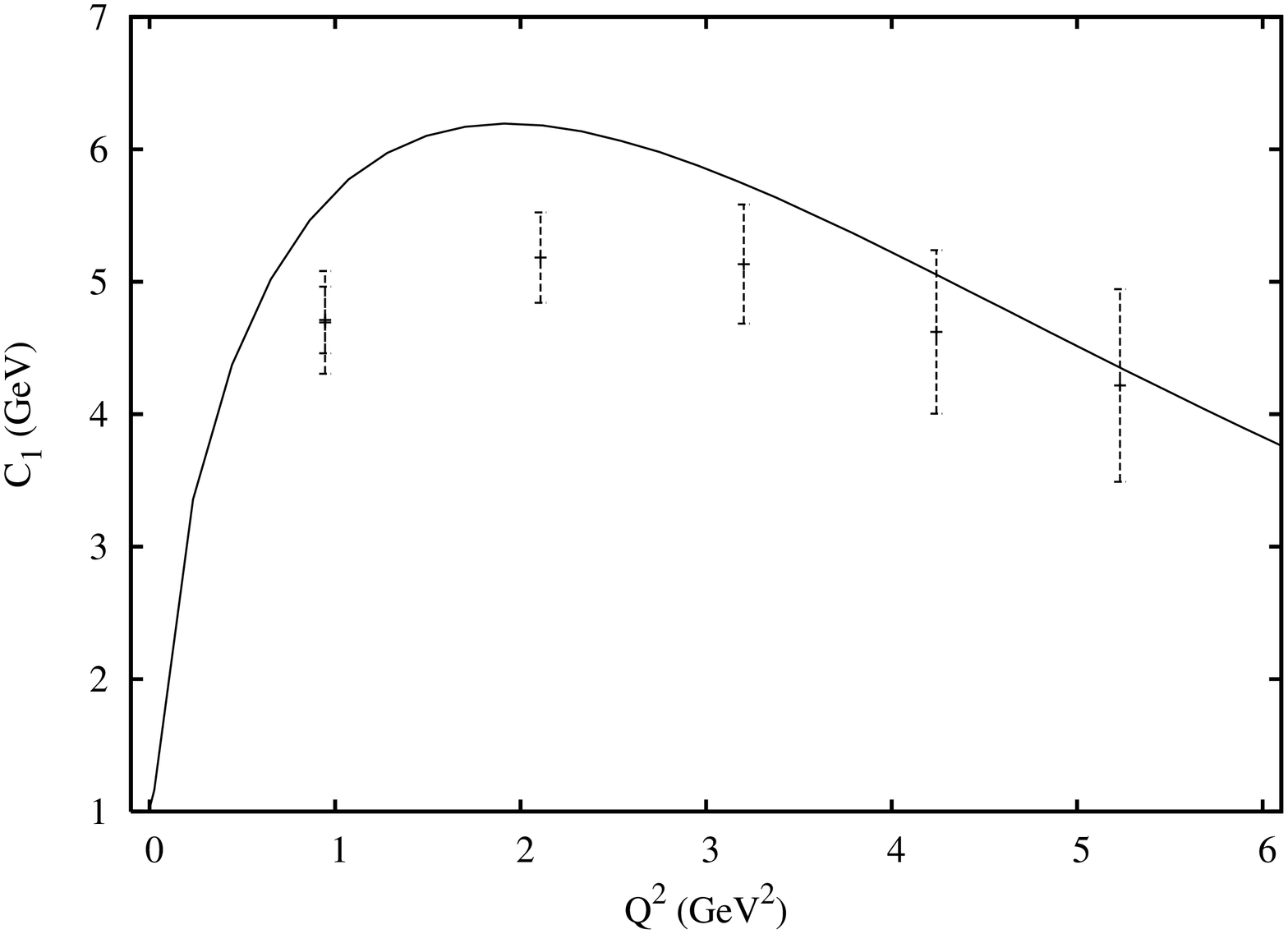}
\caption{Form Factors  $E_1(Q^2)$ (left) and $C_1(Q^2)$ (right) for
$h_c\rightarrow \eta_c\gamma$.} \label{hcFig}
\end{figure}

\begin{figure}[h!]
\includegraphics[angle=-90,width=15cm]{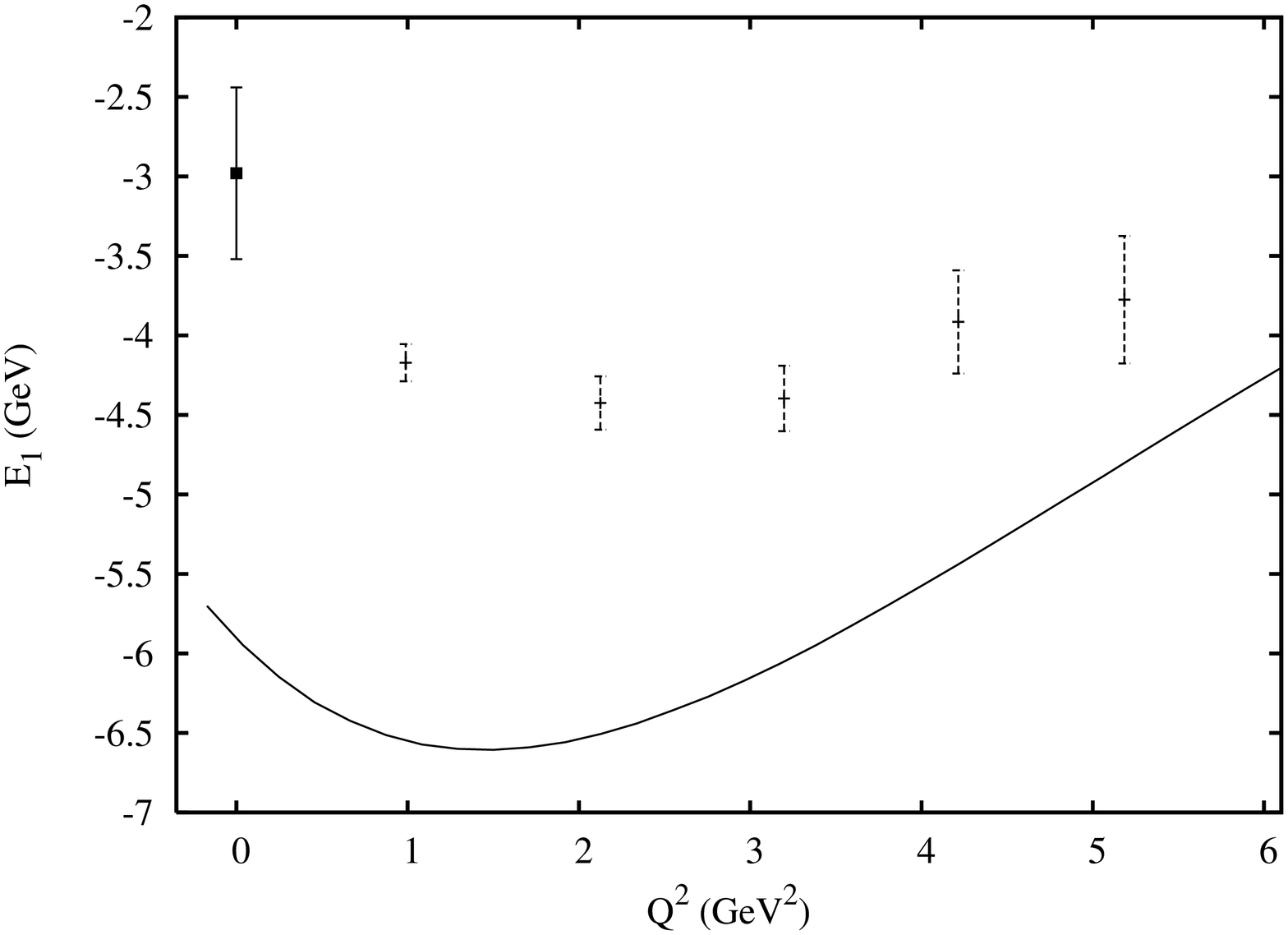}
\caption{Form Factor $E_1(Q^2)$ for $\chi_{c1} \to J/\psi\gamma$.
Experimental points are indicated with squares.}
\label{chi1psiE1Fig}
\end{figure}

\begin{figure}[h!]
\includegraphics[angle=-90,width=8cm]{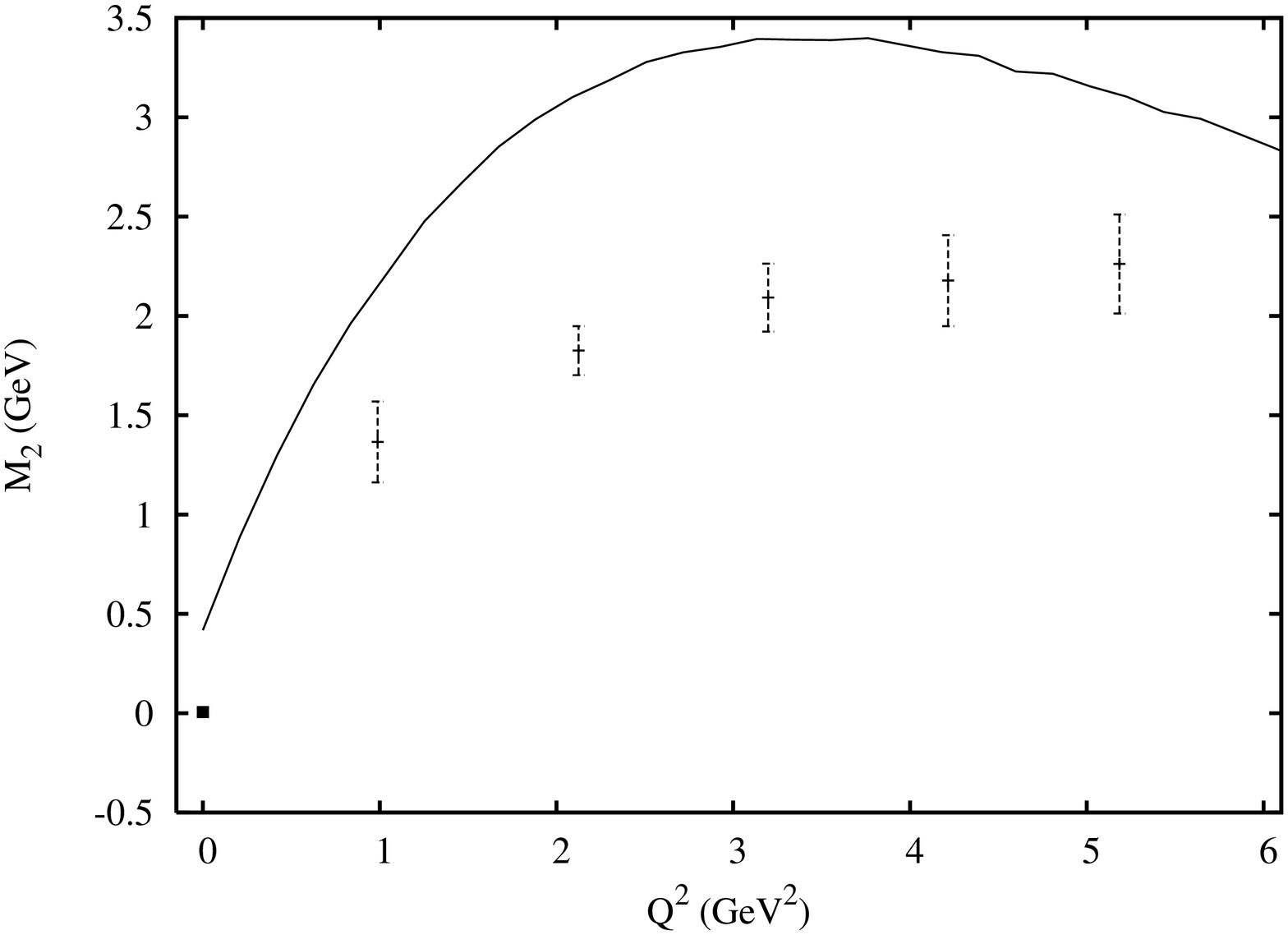} 
\includegraphics[angle=-90,width=8cm]{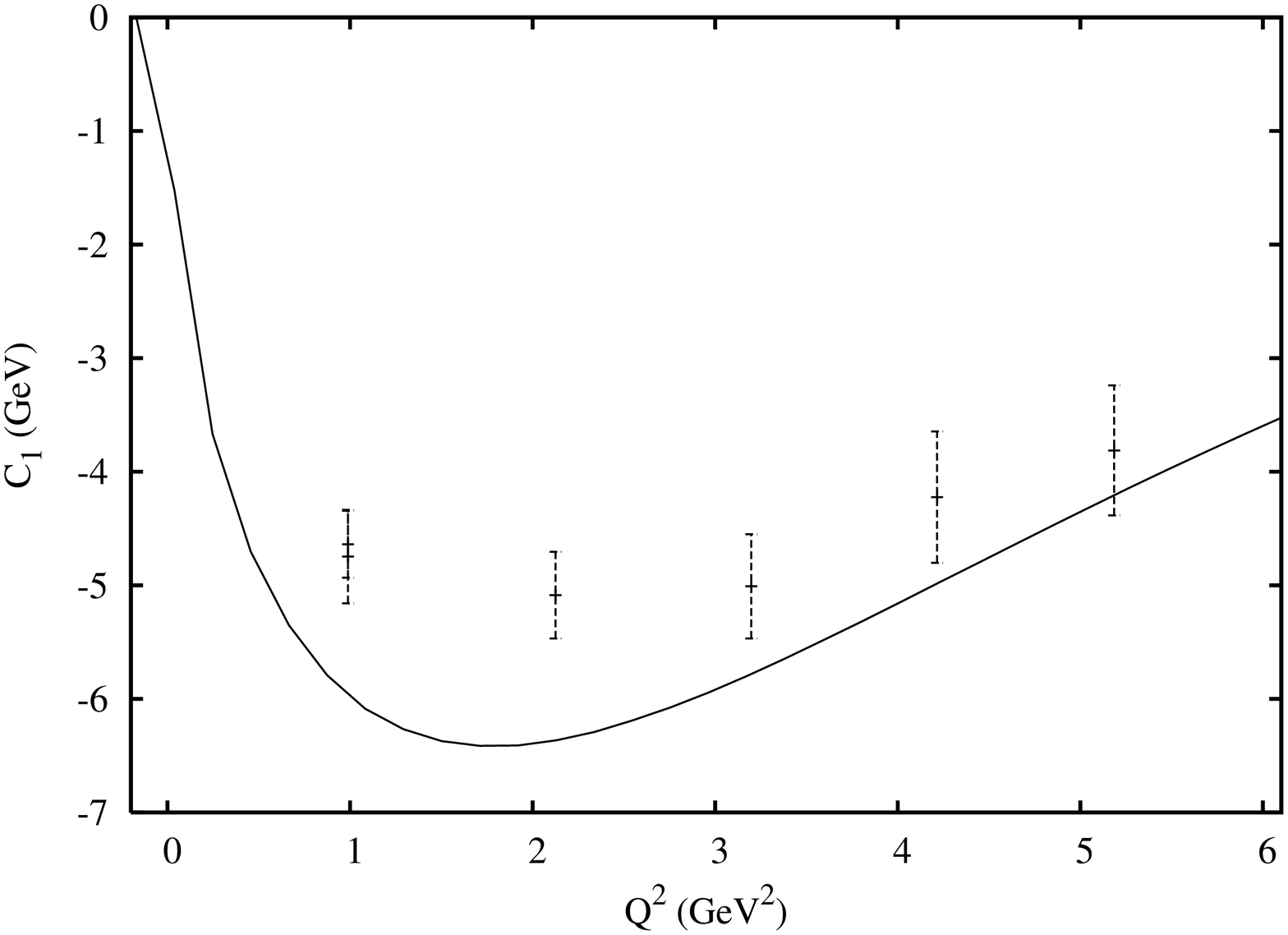}
\caption{Form Factors $M_2(Q^2)$ (left) and $C_1(Q^2)$ (right) for
$\chi_{c1}\rightarrow J/\psi\gamma$. Experimental points are
indicated with squares.} \label{chi1psiM2Fig}
\end{figure}

Computed form factors are compared to the lattice calculations of
Ref. \cite{JLlatt} and experiment (where available) in Figs.
\ref{FpsiFig} to \ref{chi1psiM2Fig}. Experimental measurements
(denoted by squares in the figures) have been determined as follows:
For $J/\psi \to \eta_c\gamma$ Crystal Barrel\cite{CB} measure
$\Gamma = 1.14 \pm 0.33$ keV. Another estimate of this rate may be
obtained by combining the  Belle measurement\cite{belle} of
$\Gamma(\eta_c \to \phi\phi)$ with the rate for $J/\psi \to
\eta_c\gamma \to \phi\phi\gamma$ reported in the PDG\cite{PDG}. One
obtains $\Gamma(J/\psi \to \eta_c\gamma) = 2.9 \pm 1.5$
keV\cite{JLlatt}. Both these data are displayed in Fig.
\ref{FpsiFig}.

Two experimental points for $\chi_{c0} \to J/\psi\gamma$ are
displayed in Fig. \ref{E1Fig}. These correspond to the PDG value
$\Gamma(\chi_{c0} \to J/\psi \gamma) = 115 \pm 14$ keV and a recent
result from CLEO\cite{cleo}: $\Gamma(\chi_{c0}\to J/\psi \gamma) =
204 \pm 31$ keV.

Finally, the experimental points for the $E_1$ and $M_2$ $\chi_{c1}
\to J/\psi \gamma$ multipoles (Fig. \ref{chi1psiM2Fig}) are
determined from the decay rate reported in the PDG and the ratio
$M_2/E_1 = 0.002 \pm 0.032$ determined by E835\cite{E835}.

Overall the agreement between the model, lattice, and experiment is
impressive.  The exception is the $E_1$ multipole for $\chi_{c1} \to
J/\psi \gamma$. We have no explanation for this discrepancy. Note
that the quenched lattice and quark model both neglect coupling to
higher Fock states, which could affect the observables. The
agreement with experiment indicates that such effects are small (or
can be effectively subsumed into quark model parameters and the
lattice scale), thereby justifying the use of the quenched
approximation and the simple valence quark model when applied to
these observables.

\begin{figure}[h!]
\includegraphics[angle=-90,width=8cm]{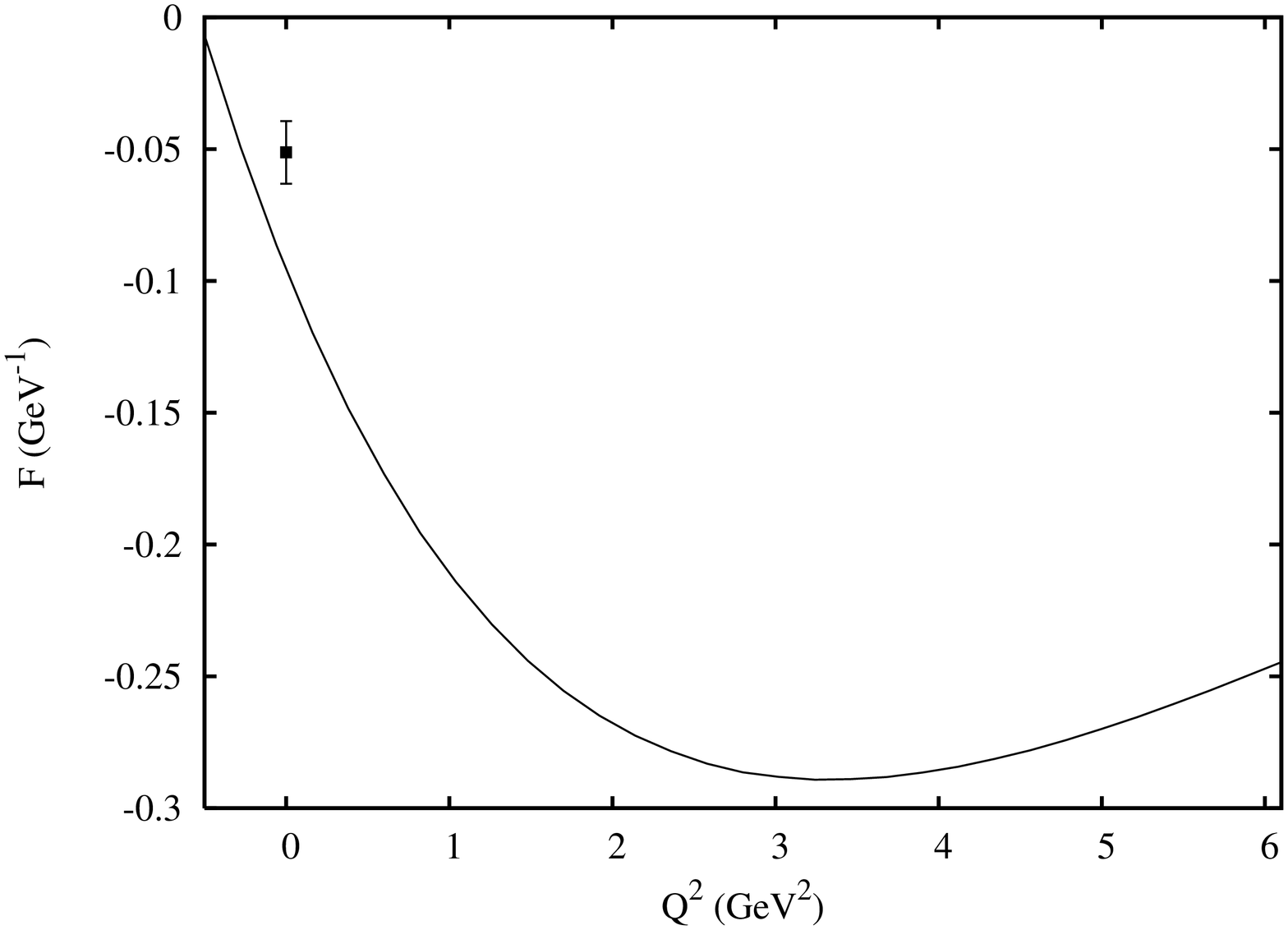} 
\includegraphics[angle=-90,width=8cm]{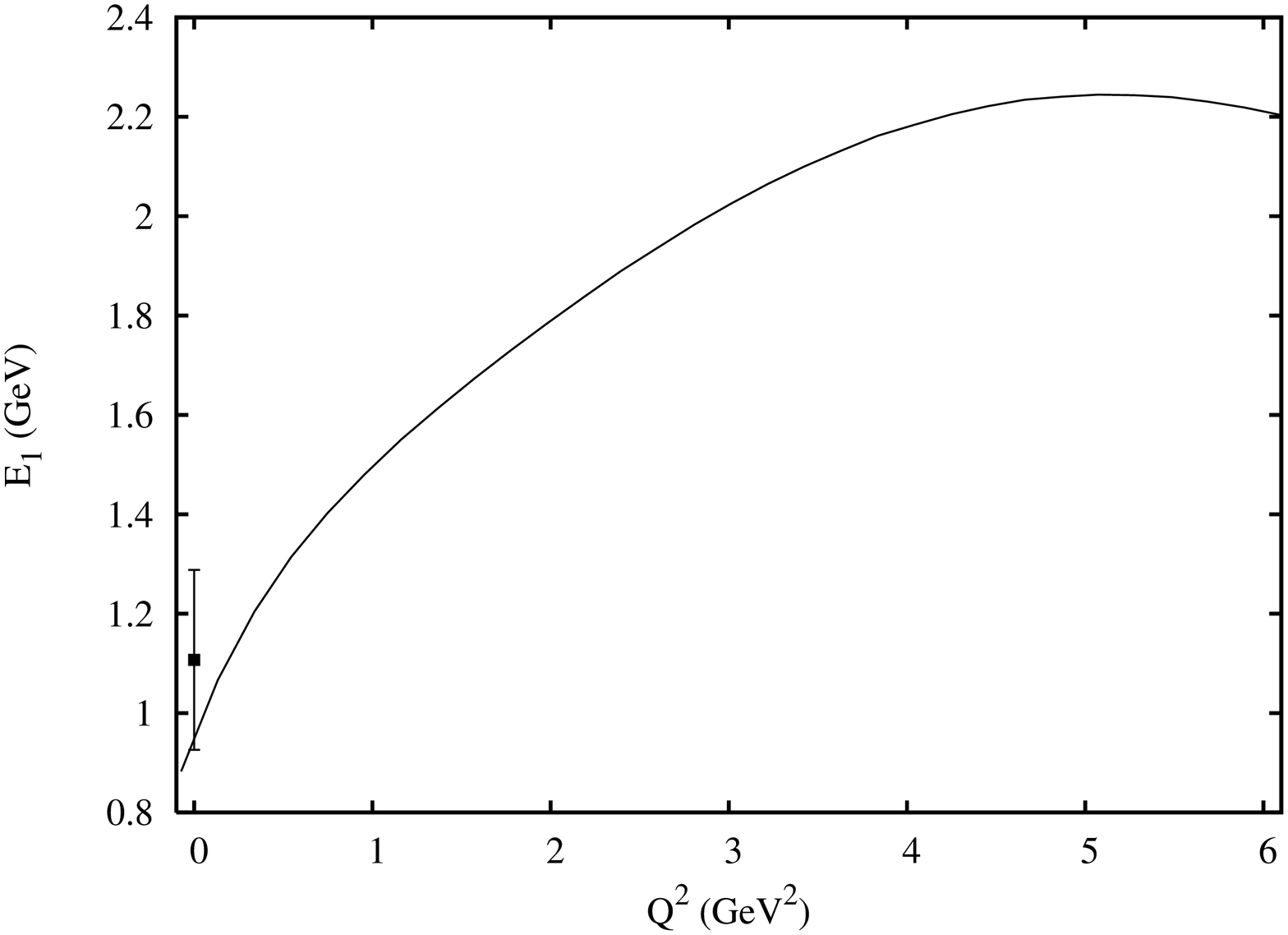}
\caption{Form Factor $F(Q^2)$ for $\psi(2S) \to \eta_c \gamma$
(left). Form factor $E_1(Q^2)$ for $\psi(2S) \to \chi_{c0}\gamma$
(right). Experimental points are indicated with squares.}
\label{psi2etaFig}
\end{figure}

Predictions for excited state form factors are simple to obtain in
the quark model (in contrast to lattice gauge theory, where
isolating excited states is computationally difficult). Two examples
are presented in Fig. \ref{psi2etaFig}. The agreement with
experiment (squares) is acceptable. \clearpage

\addtolength{\topmargin}{-2cm} \addtolength{\footskip}{3cm}

\clearpage \addtolength{\topmargin}{2cm}
\addtolength{\footskip}{-3cm}

\clearpage

\section{Electroweak form-factors}

Electroweak form-factors are measurable experimentally for a variety
of processes, so studying them in a particular model can help
greatly in improving the model and learning about its applicability.
In this section we present the results of our calculations of
electroweak transition form-factors and compare them to the
experiment (if available) and to other model calculations. The
details of the derivation of the expressions for the form-factors
are presented in the Appendix \ref{EWffApp}.

\begin{figure}[h]
\includegraphics[angle=-90,width=15cm]{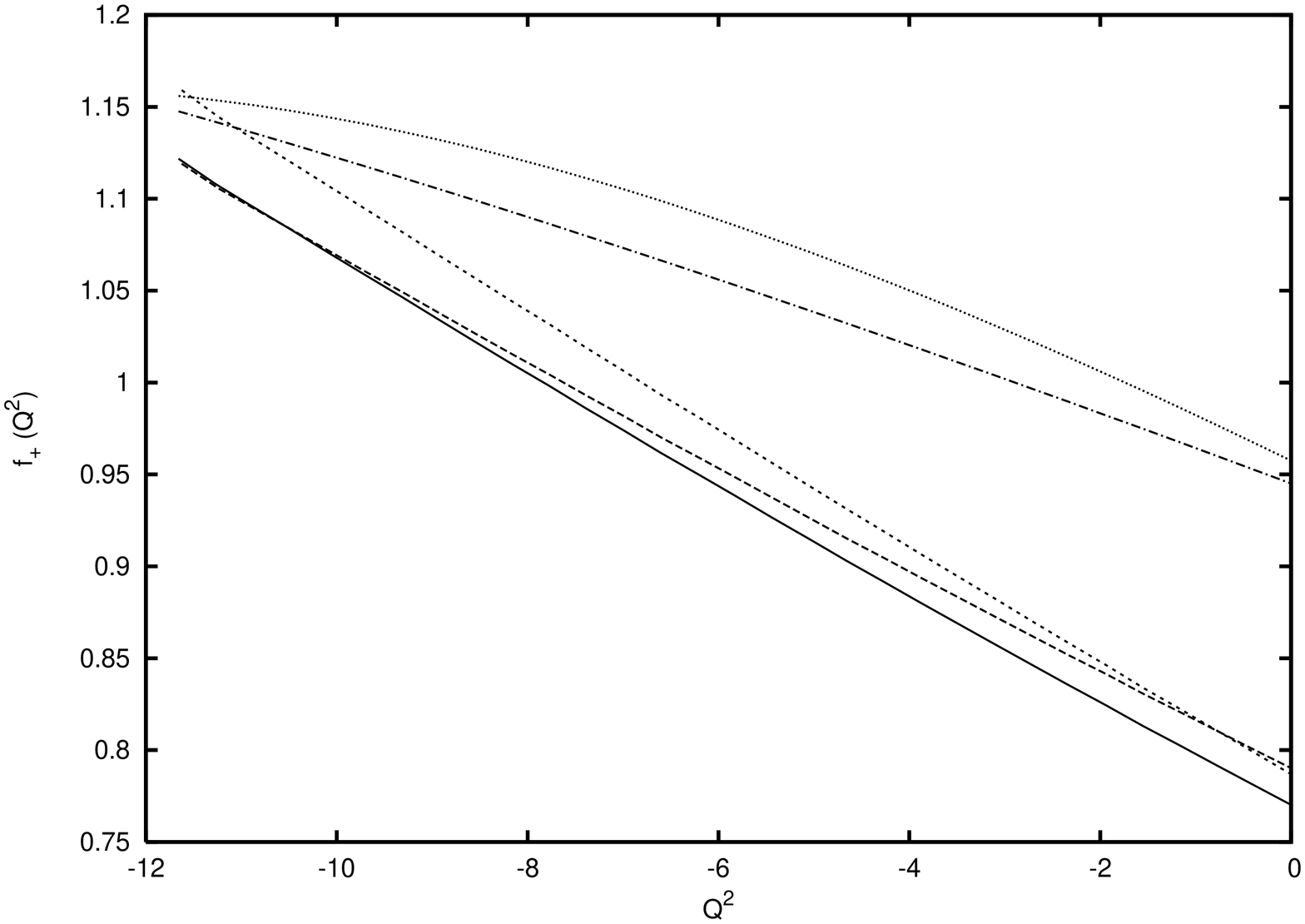}
\caption{\label{fplus}Form-factor $f_+(Q^2)$ of
$\bar{B}^0\rightarrow D^+$. From top to bottom at $Q^2=0$ the curves
are SHO, ISGW, relativistic C+L, nonrelativistic C+L and C+L log.}
\includegraphics[angle=-90,width=15cm]{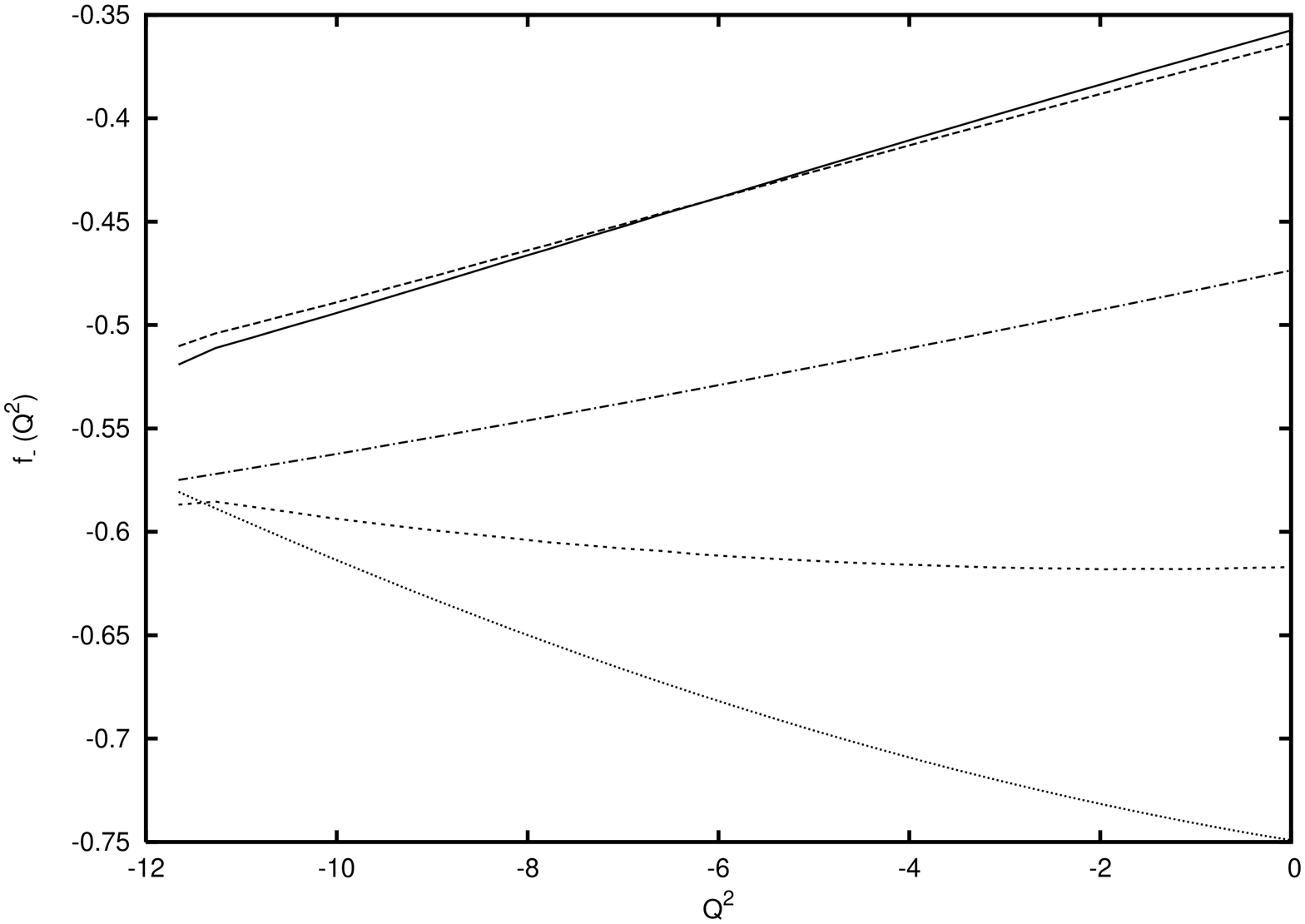}
\caption{\label{fminus}Form-factor $f_-(Q^2)$ of
$\bar{B}^0\rightarrow D^+$. From top to bottom at $Q^2=0$ the curves
are C+L log, relativistic C+L, ISGW, nonrelativistic C+L and SHO.}
\end{figure}

\begin{figure}[h]
\includegraphics[angle=-90,width=15cm]{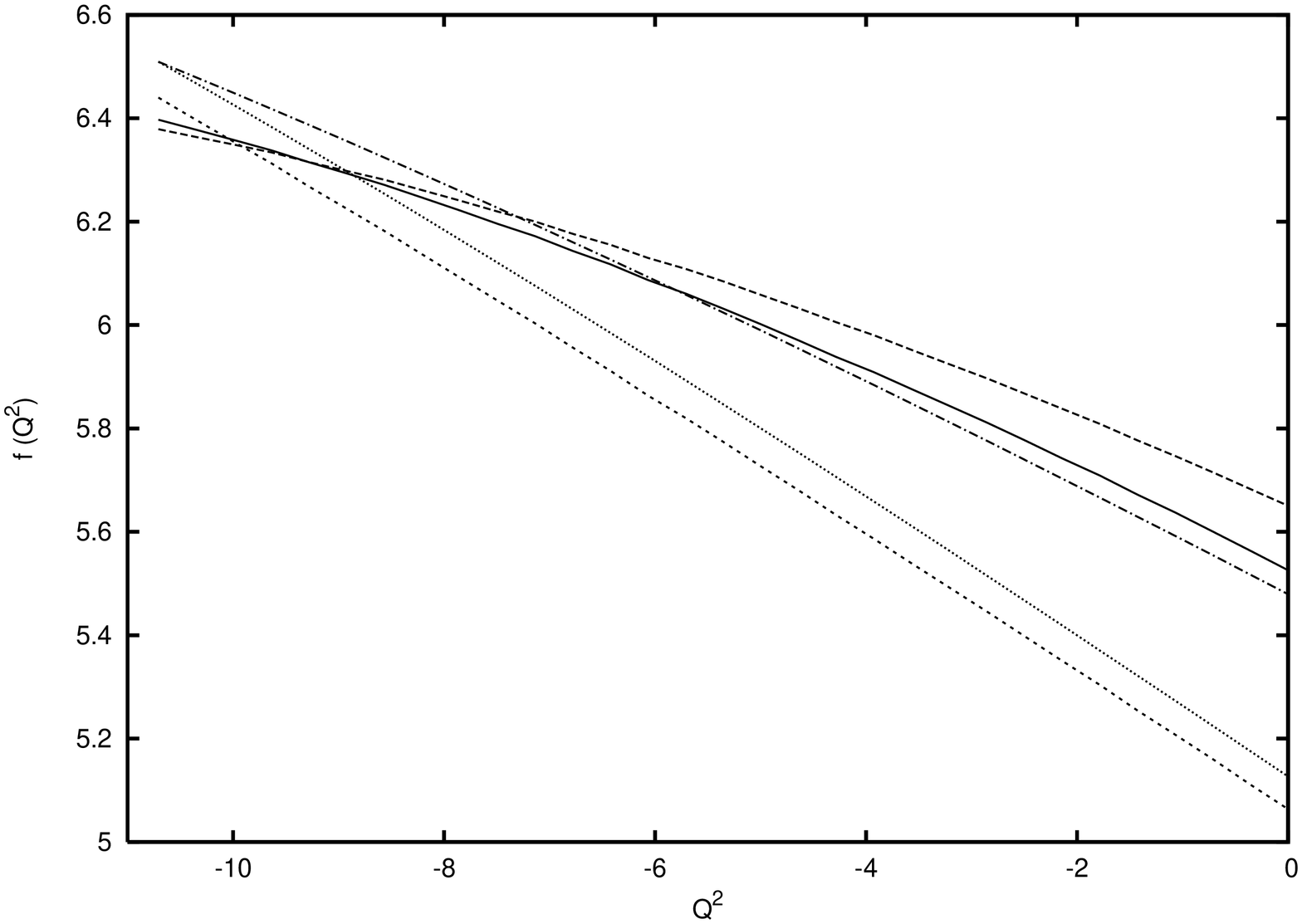}
\caption{\label{f}Form-factor $f(Q^2)$ of $\bar{B}^0\rightarrow
D^{*+}$. From top to bottom at $Q^2=0$ the curves are relativistic
C+L, C+L log, ISGW, SHO and nonrelativistic C+L.}
\includegraphics[angle=-90,width=15cm]{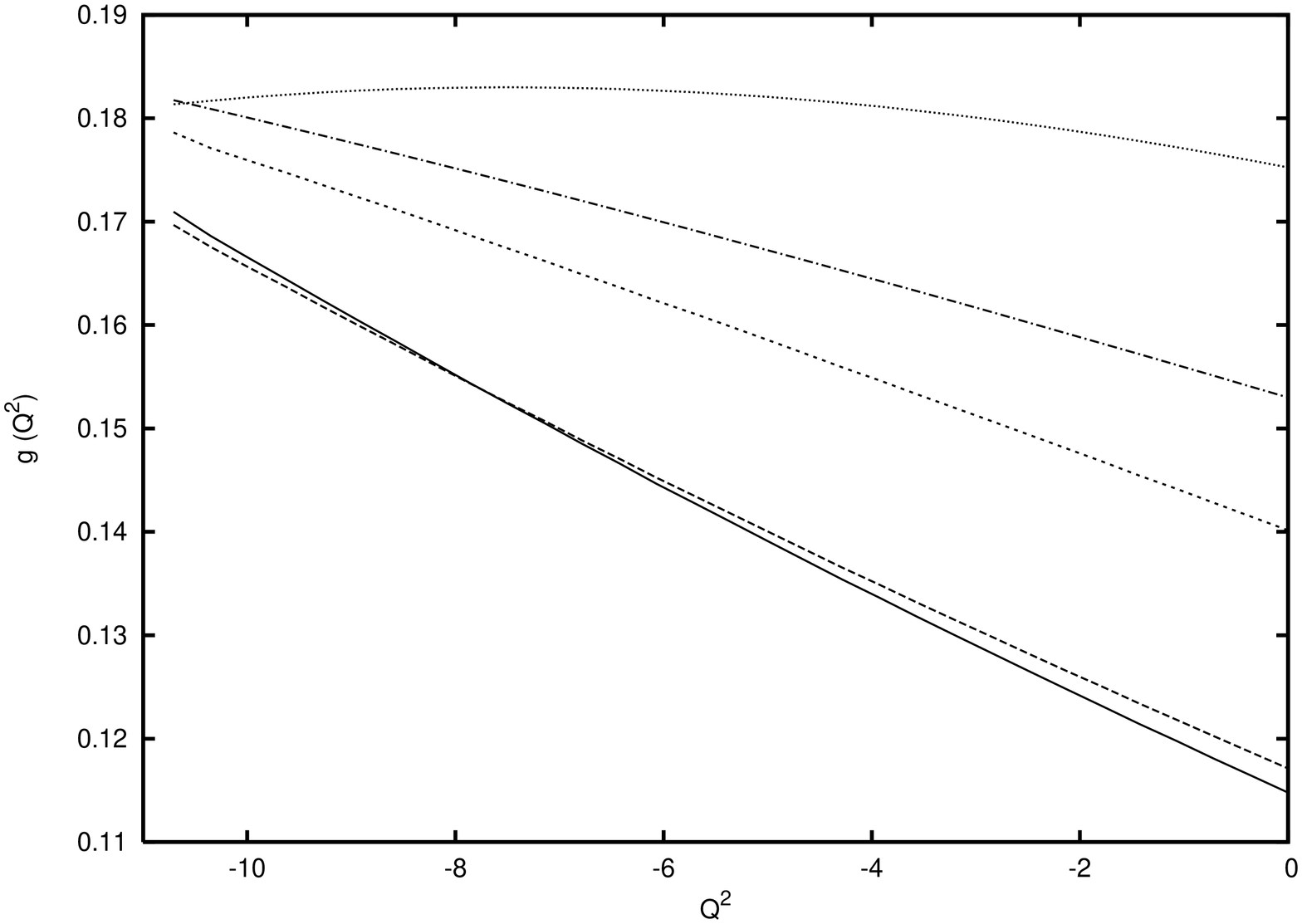}
\caption{\label{g}Form-factor $g(Q^2)$ of $\bar{B}^0\rightarrow
D^{*+}$. From top to bottom at $Q^2=0$ the curves are SHO, ISGW,
nonrelativistic C+L, relativistic C+L and C+L log.}
\end{figure}

\begin{figure}[h]
\includegraphics[angle=-90,width=15cm]{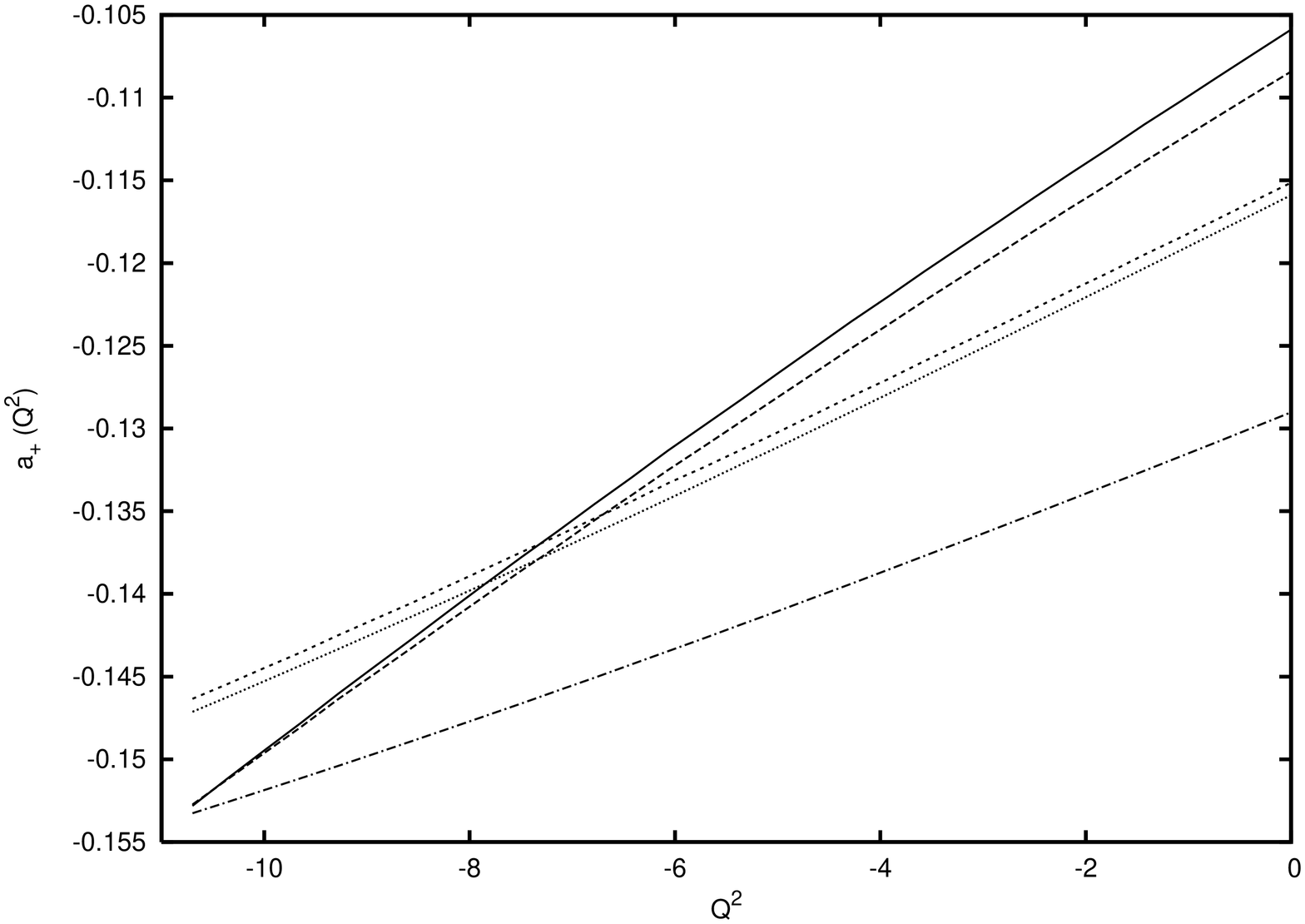}
\caption{\label{aplus}Form-factor $a_+(Q^2)$ of
$\bar{B}^0\rightarrow D^{*+}$. From top to bottom at $Q^2=0$ the
curves are C+L log, relativistic C+L, nonrelativistic C+L, SHO and
ISGW.}
\includegraphics[angle=-90,width=15cm]{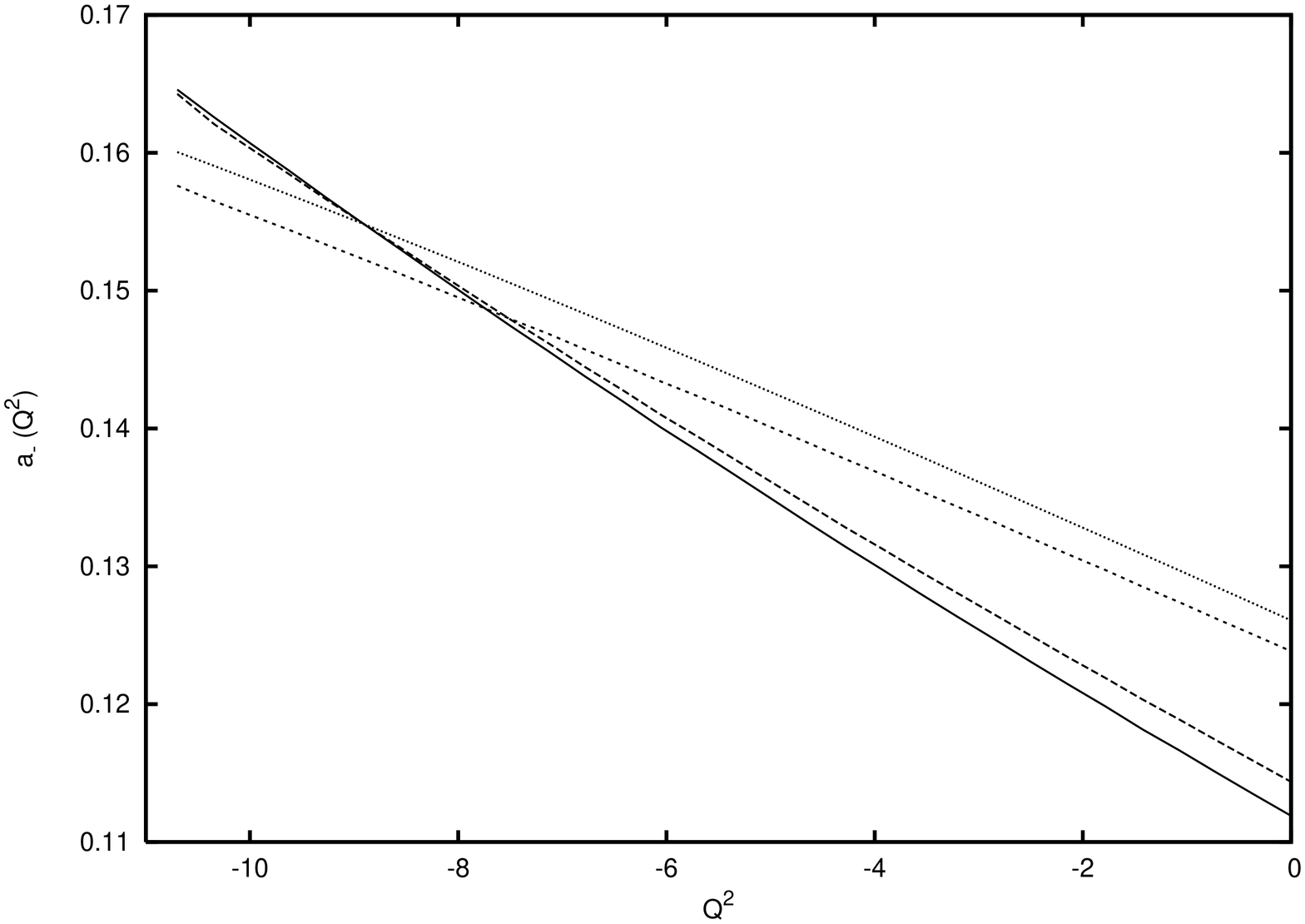}
\caption{\label{aminus}Form-factor $a_-(Q^2)$ of
$\bar{B}^0\rightarrow D^{*+}$. From top to bottom at $Q^2=0$ the
curves are SHO, nonrelativistic C+L, relativistic C+L and C+L log.}
\end{figure}

Results for the dependence of the form-factors on the momentum
transfer for $\bar{B}^0\rightarrow D^+$ and $\bar{B}^0\rightarrow
D^{*+}$ decays are presented in the figures
\ref{fplus}-\ref{aminus}. In the ISGW model SHO wave functions are
used as an approximation for the meson wave functions, and an
artificial factor $\kappa=0.7$ is introduced ($|\vec{q}|\rightarrow
|\vec{q}|/\kappa$). The formulae for this model are taken from their
paper \cite{ISGW}.

\begin{figure}[h]
\includegraphics[angle=-90,width=15cm]{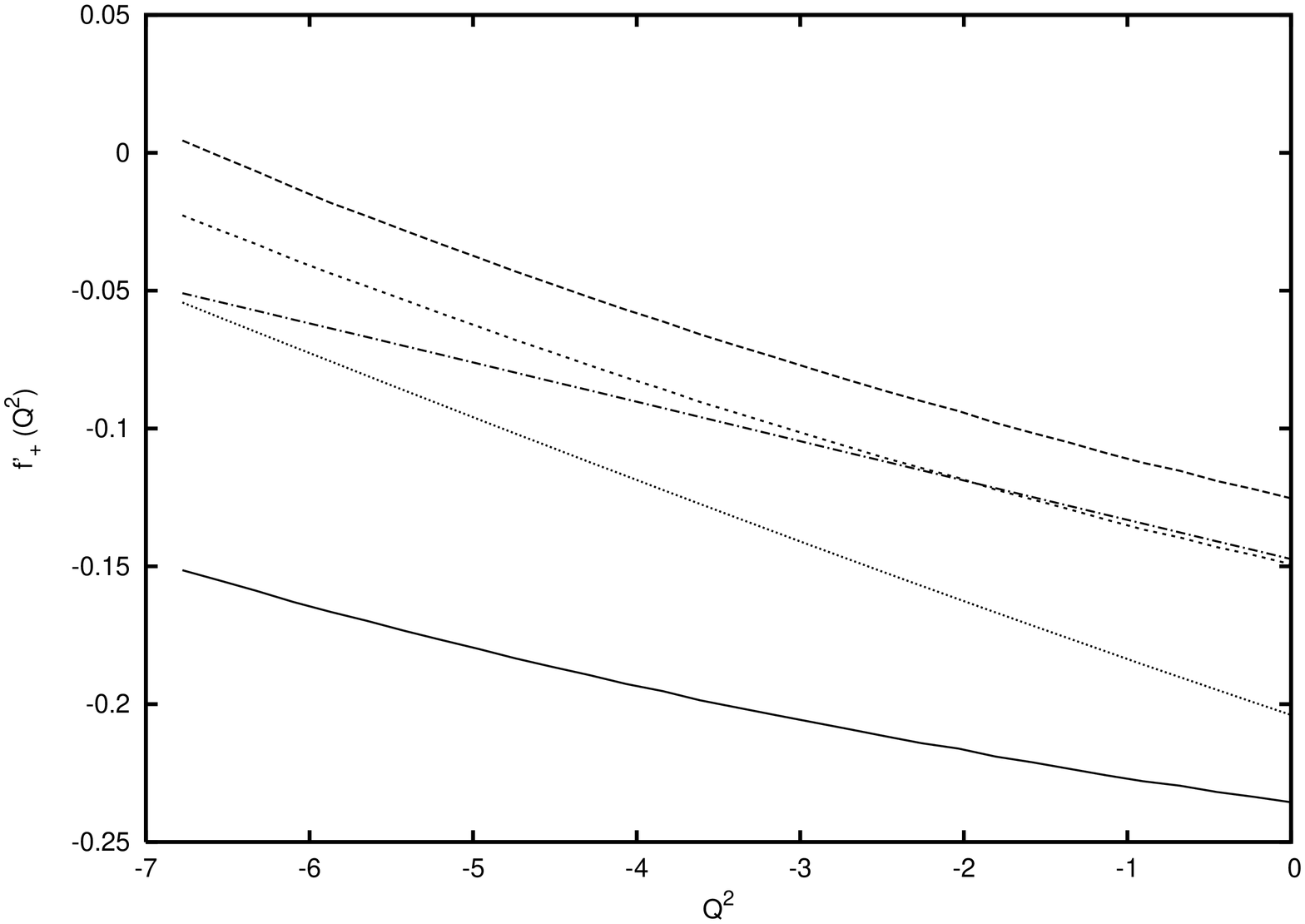}
\caption{\label{fplus'}Form-factor $f'_+(Q^2)$ of
$\bar{B}^0\rightarrow D^+(2S)$. From top to bottom at $Q^2=0$ the
curves are relativistic C+L, ISGW, nonrelativistic C+L, SHO and C+L
log.}
\includegraphics[angle=-90,width=15cm]{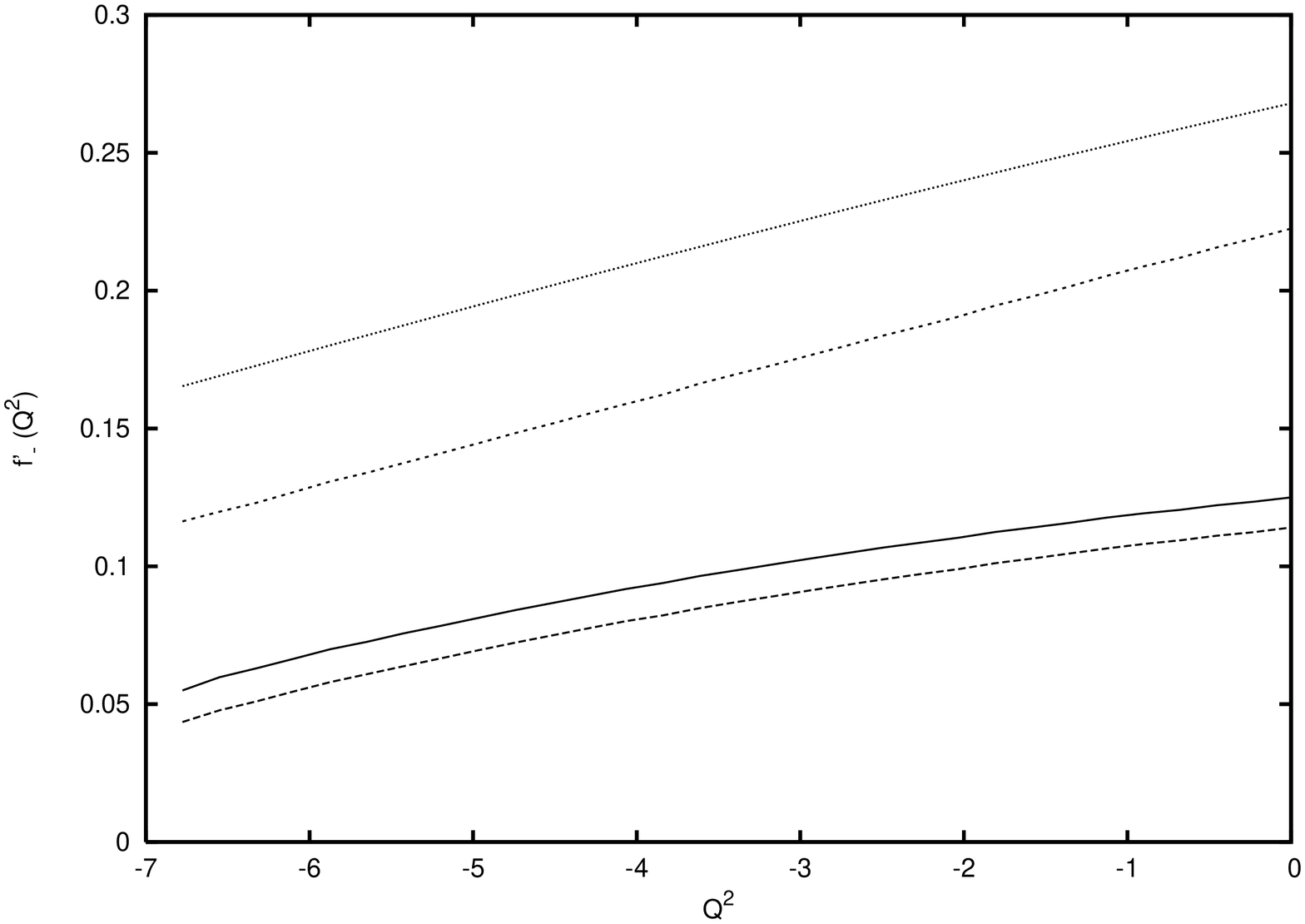}
\caption{\label{fminus'}Form-factor $f'_-(Q^2)$ of
$\bar{B}^0\rightarrow D^+(2S)$. From top to bottom at $Q^2=0$ the
curves are SHO, nonrelativistic C+L, C+L log and relativistic C+L.}
\end{figure}

\begin{figure}[h]
\includegraphics[angle=-90,width=15cm]{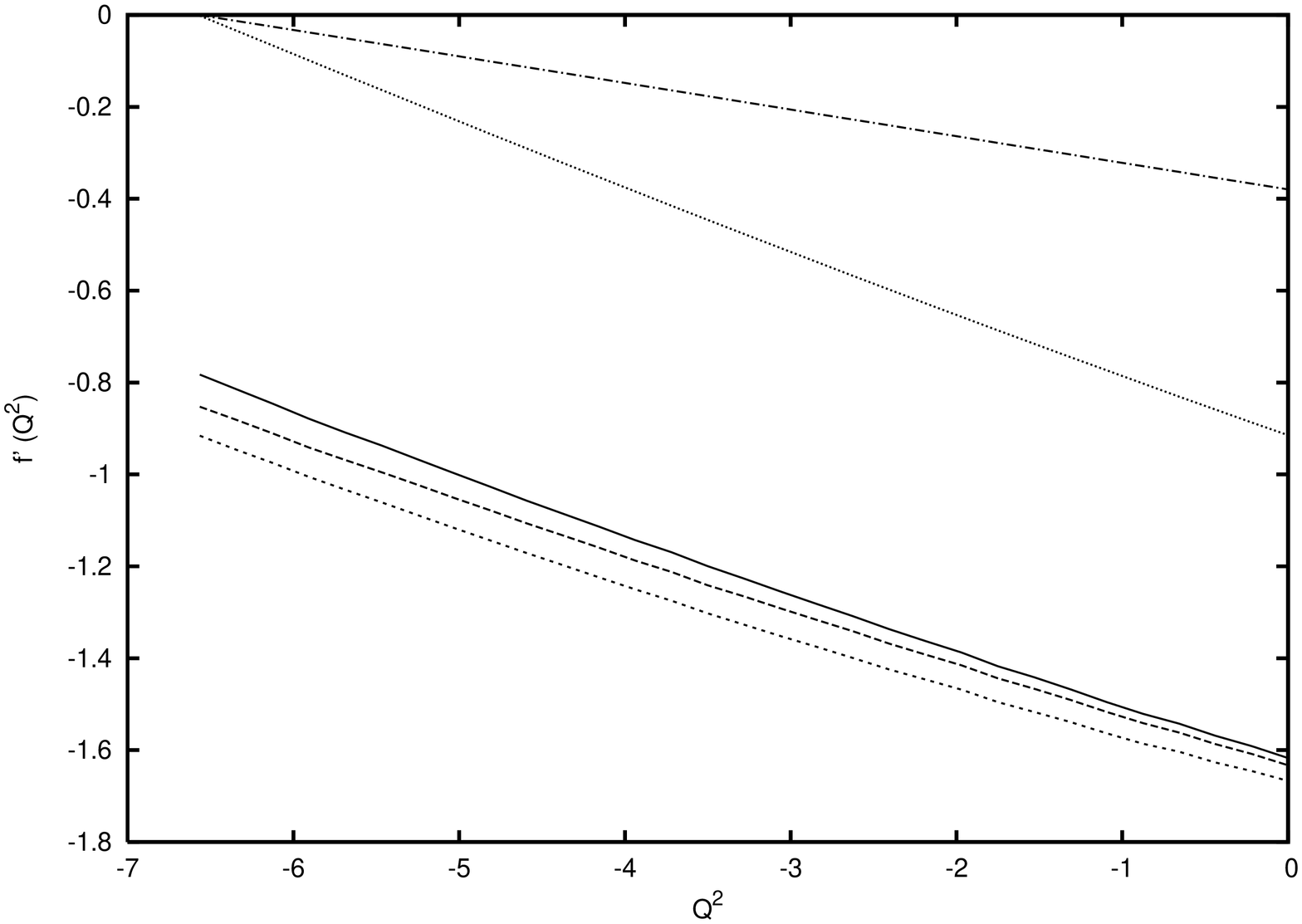}
\caption{\label{f'}Form-factor $f'(Q^2)$ of $\bar{B}^0\rightarrow
D^{*+}(2S)$. From top to bottom at $Q^2=0$ the curves are ISGW, SHO,
C+L log, relativistic C+L and nonrelativistic C+L.}
\includegraphics[angle=-90,width=15cm]{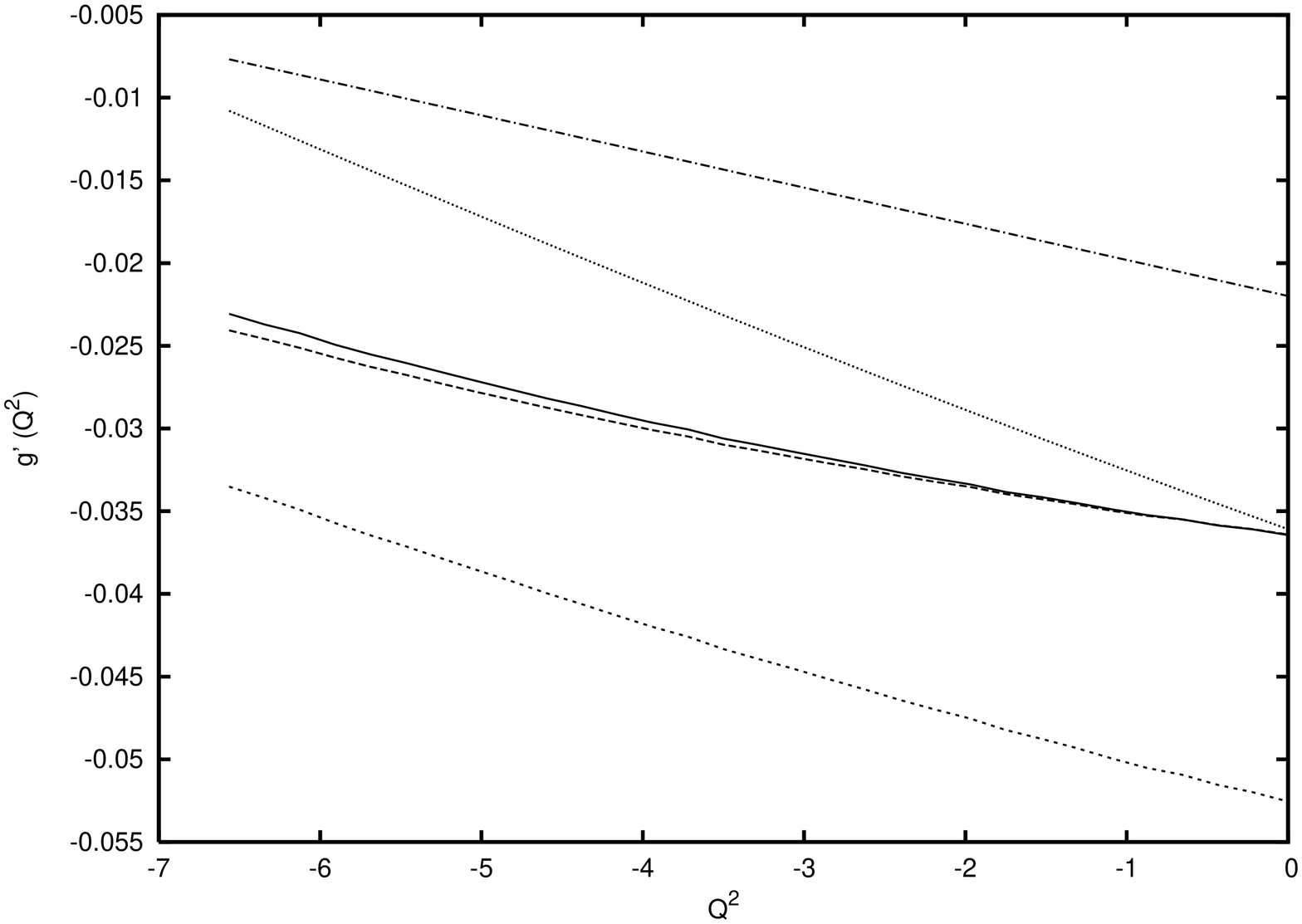}
\caption{\label{g'}Form-factor $g'(Q^2)$ of $\bar{B}^0\rightarrow
D^{*+}(2S)$. From top to bottom at $Q^2=0$ the curves are ISGW, SHO,
C+L log, relativistic C+L and nonrelativistic C+L.}
\end{figure}

\begin{figure}[h]
\includegraphics[angle=-90,width=15cm]{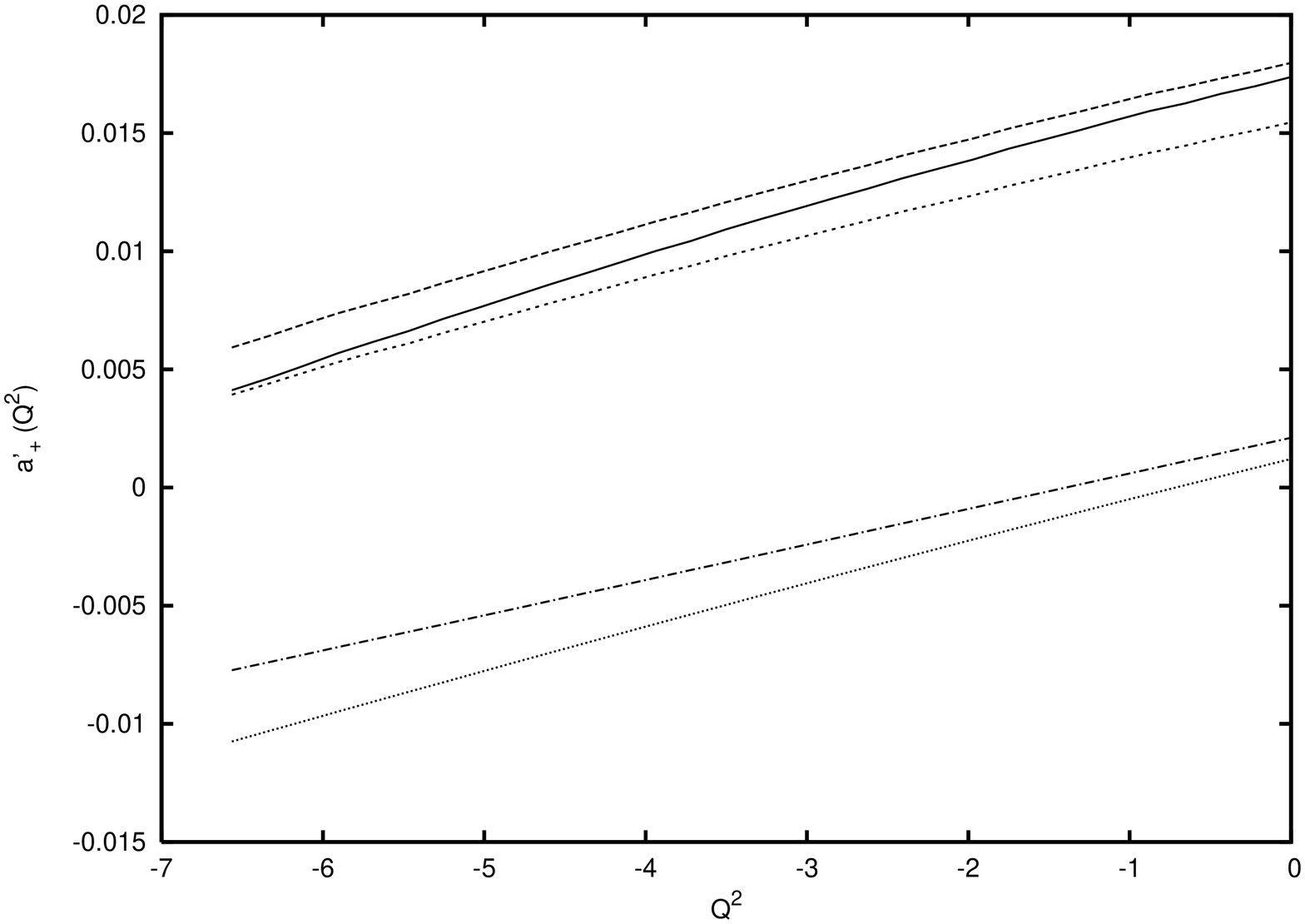}
\caption{\label{aplus'}Form-factor $a'_+(Q^2)$ of
$\bar{B}^0\rightarrow D^{*+}(2S)$. From top to bottom at $Q^2=0$ the
curves are relativistic C+L, C+L log, nonrelativistic C+L, ISGW and
SHO.}
\includegraphics[angle=-90,width=15cm]{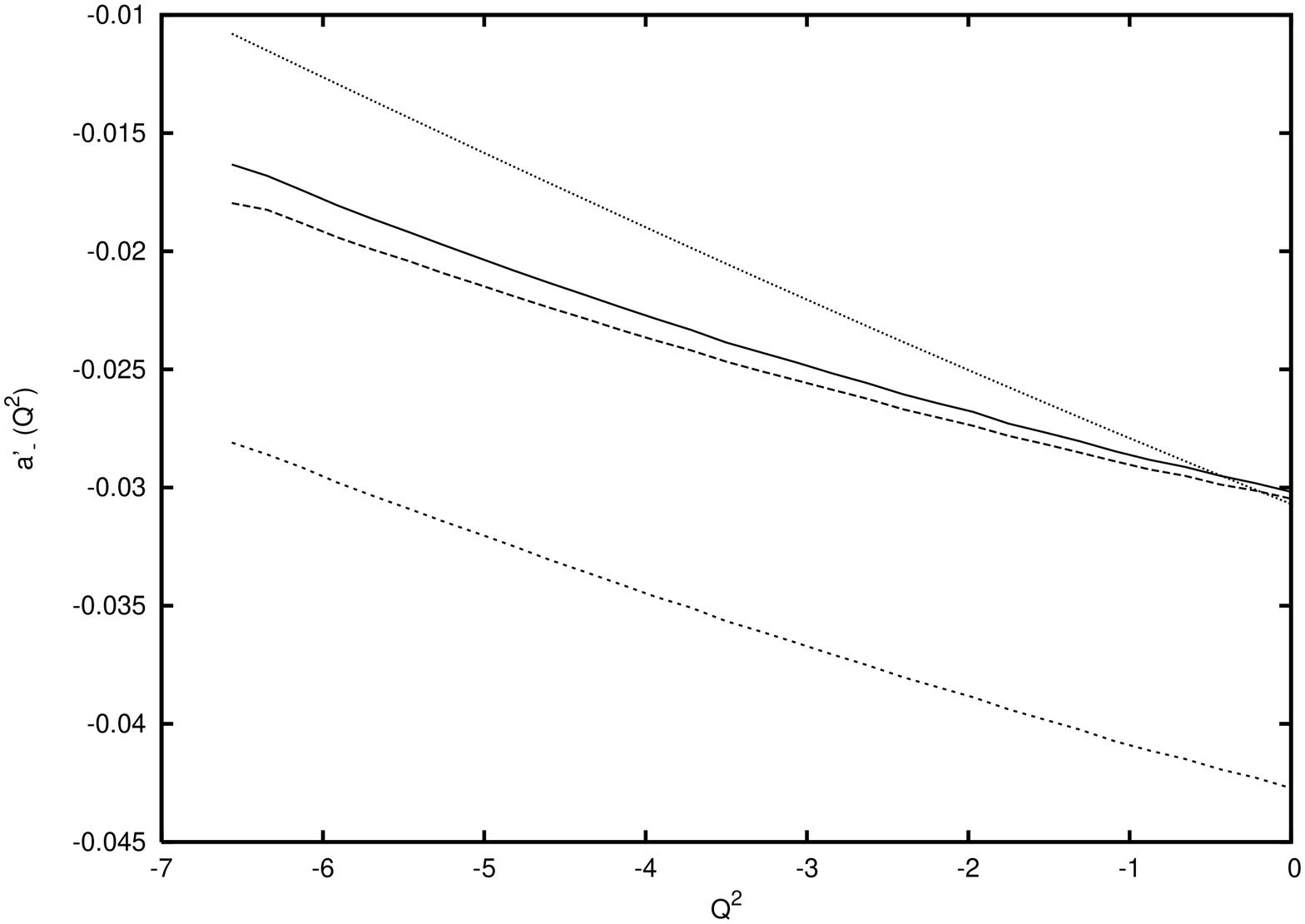}
\caption{\label{aminus'}Form-factor $a'_-(Q^2)$ of
$\bar{B}^0\rightarrow D^{*+}(2S)$. From top to bottom at $Q^2=0$ the
curves are C+L log, relativistic C+L, SHO and nonrelativistic C+L.}
\end{figure}

Results for the form-factors of the transitions to the excited
states $\bar{B}^0\rightarrow D^+ (2S)$ and $\bar{B}^0\rightarrow
D^{*+} (2S)$ are presented in the figures
\ref{fplus'}-\ref{aminus'}. As expected, for the transitions to the
excited states, form-factors for different models of the potential
are more different from each other than for ground state
transitions. It happens mostly because the wave functions start to
differ more between the models as we go to the higher states. Also,
in the SHO and ISGW models the pseudoscalar and vector meson wave
functions are the same, while the Coulomb+linear+hyperfine potential
model has a spin-dependent term that can distinguish between them,
and difference becomes even larger for the excited states. And as we
consider transitions to the excited vector meson states it becomes
very important to take that spin dependence into account as one
could see from our results for the form-factors: SHO and ISGW model
form-factors are significantly different from the
Coulomb+linear+hyperfine potential model.

\begin{figure}[h]
\includegraphics[angle=-90,width=15cm]{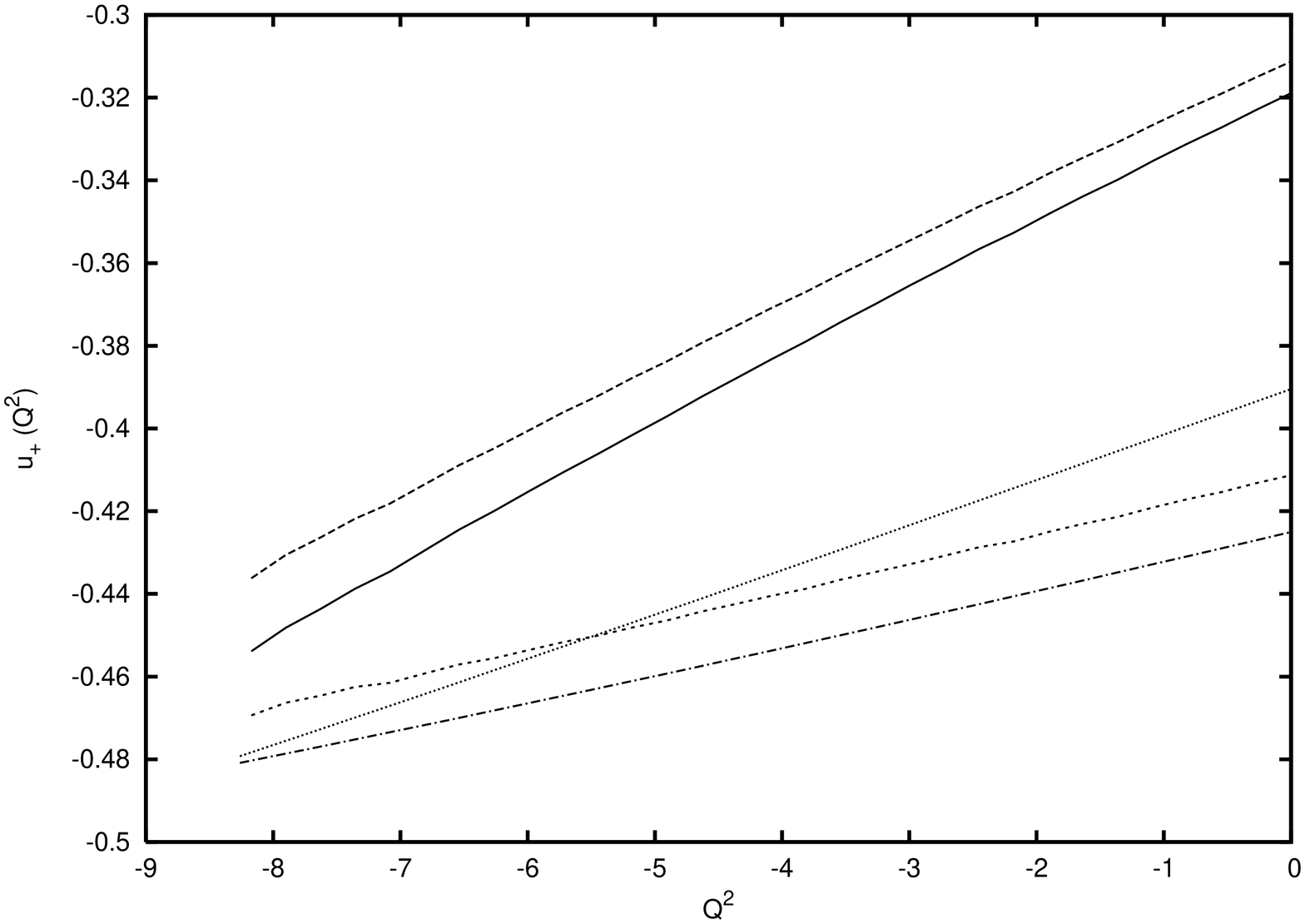}
\caption{\label{uplus}Form-factor $u_+(Q^2)$ of
$\bar{B}^0\rightarrow D_0^+$. From top to bottom at $Q^2=0$ the
curves are relativistic C+L, C+L log, SHO, nonrelativistic C+L and
ISGW.}
\includegraphics[angle=-90,width=15cm]{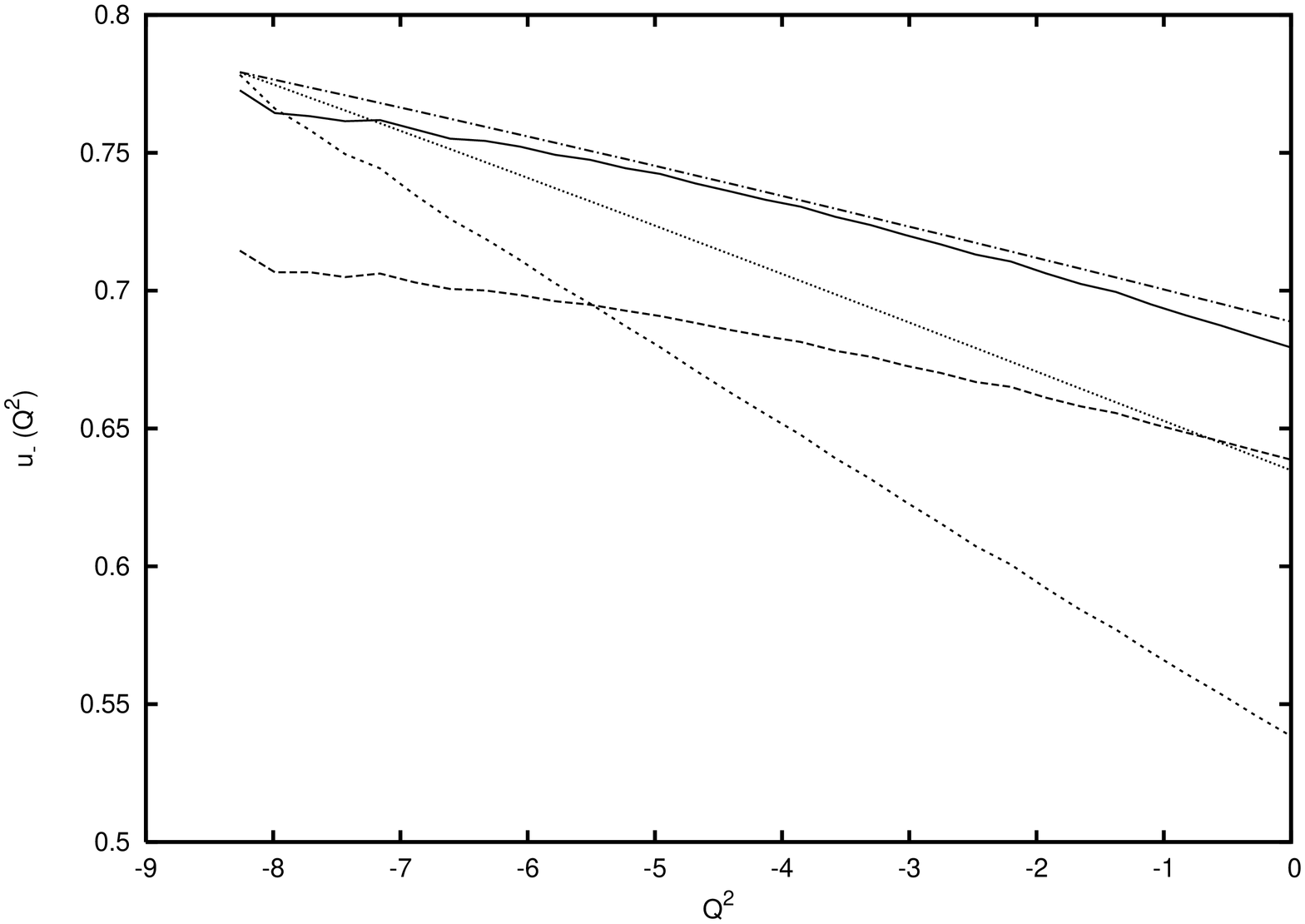}
\caption{\label{uminus}Form-factor $u_-(Q^2)$ of
$\bar{B}^0\rightarrow D_0^+$. From top to bottom at $Q^2=0$ the
curves are ISGW, C+L log, relativistic C+L, SHO and nonrelativistic
C+L.}
\end{figure}

Results for the pseudoscalar to scalar meson transition form-factors
are presented in the figures \ref{uplus} and \ref{uminus}. The
form-factors with the full relativistic expressions of the quark
spinors taken into account are quite different from the completely
nonrelativistic model results, so we conclude that the relativistic
corrections can be significant.

To compare to the experimental data presented in \cite{Rosner:HQET}
we have to calculate $F_V(w)$ (for $\bar{B}^0\rightarrow D^+$),
$F_A(w)$ and $F^*_V(w)$ (for $\bar{B}^0\rightarrow D^{*+}$) which
are defined by (these are decompositions of the matrix elements in
the heavy quark limit, so they are not the most general
expressions):
\begin{eqnarray}
V^{\mu}=\langle D(v')|V^{\mu}|B(v)\rangle &=&\sqrt{M_1M_2} F_V(w)(v+v')^{\mu},\nonumber\\
A^{\mu}=\langle D^*(v',\epsilon)|A^{\mu}|B(v)\rangle
&=&\sqrt{M_1M_2} F_A(w)
\Big[\left(\epsilon^{*}\right)^{\mu}(1+v\cdot v')-\epsilon^*\cdot v (v')^{\mu}\Big],\nonumber\\
(V^{*})^{\mu}=\langle D^*(v',\epsilon)|V^{\mu}|B(v)\rangle
&=&-i\sqrt{M_1M_2} F^*_V(w)
\epsilon^{\mu\nu\alpha\beta}\epsilon^*_{\nu}v_{\alpha}v'_{\beta}.
\end{eqnarray}
where $v=P_1/M_1,\,v'=P_2/M_2$ and
\begin{equation}
w=\frac{P_1\cdot P_2}{M_1 M_2}=\frac{M_1^2+M_2^2+q^2}{2M_1M_2}.
\end{equation}
In the $B(v)$ rest frame we have:
\begin{eqnarray}
V^0&=&\sqrt{M_1M_2}F_V(w)\left(1+\frac{E_2}{M_2}\right),\nonumber\\
\vec{V}&=&\vec{q}\sqrt{\frac{M_1}{M_2}}F_V(w),\nonumber\\
A^0&=&\left\{\begin{array}{cc}
             0 & if\,\,\,\, M_V=\pm 1,\\
             |\vec{q}|\sqrt{\frac{M_1}{M_2}}F_A(w) & if\,\,\,\, M_V=0,\\
            \end{array}\right.\nonumber\\
\vec{A}&=&\left\{\begin{array}{cc}
                \sqrt{M_1M_2}F_A(w)\left(1+\frac{E_2}{M_2}\right)\vec{\epsilon}^{\,*} & if\,\,\,\, M_V=\pm 1,\\
                \sqrt{M_1M_2}F_A(w)\left(1+\frac{E_2}{M_2}\right)\hat{e}_z & if\,\,\,\, M_V=0,\\
                \end{array}\right.\nonumber\\
(V^*)^0&=&0,\nonumber\\
\vec{V}^*&=&|\vec{q}|\sqrt{\frac{M_1}{M_2}}F^*_V(w)M_V\vec{\epsilon}^{\,*}.
\end{eqnarray}

Now there are two different expressions for $F_V(w)$ and $F_A(w)$
form-factors (one from the zero component of the matrix element and
one from the vector components):
\begin{eqnarray}
F_V(w)_0&=&\sqrt{\frac{M_2}{M_1}}\frac{V^0}{M_2+E_2},\nonumber\\
F_V(w)_{vec}&=&\sqrt{\frac{M_2}{M_1}}\frac{\vec{V}\cdot\vec{q}}{|\vec{q}|^2},\nonumber\\
F_A(w)_0&=&\sqrt{\frac{M_2}{M_1}}\frac{A^0}{|\vec{q}|},\,\,\,\,M_V=0,\nonumber\\
F_A(w)_{vec}&=&\left\{\begin{array}{cc}
                \sqrt{\frac{M_2}{M_1}}\frac{\vec{A}\cdot\vec{\epsilon}}{M_2+E_2},\,\,\,\, M_V=\pm 1,\\
                \sqrt{\frac{M_2}{M_1}}\frac{\vec{A}\cdot\hat{e}_z}{M_2+E_2},\,\,\,\, M_V=0,\\
                \end{array}\right.\nonumber\\
F^*_V(w)_{vec}&=&M_V\sqrt{\frac{M_2}{M_1}}\,\frac{\vec{V}^*\cdot\vec{\epsilon}}{|\vec{q}|}.
\end{eqnarray}
The two expressions for each of the form-factors $F_V(w)$ and
$F_A(w)$ should be equivalent to each other if our model is
covariant and the heavy quark approximation is good enough.

In the nonrelativistic approximation for SHO wave functions:
\begin{eqnarray}
F_V(w)_0&=&e^{-q^2\mu^2/4\beta^2},\nonumber\\
F_V(w)_{vec}&=&e^{-q^2\mu^2/4\beta^2}M_2\left(\frac{1}{m_2}-\frac{\mu}{2}\left(\frac{1}{m_1}+\frac{1}{m_2}\right)\right),\nonumber\\
F_A(w)_0&=&e^{-q^2\mu^2/4\beta^2}M_2\left(\frac{1}{m_2}-\frac{\mu}{2}\left(\frac{1}{m_1}+\frac{1}{m_2}\right)\right),\nonumber\\
F_A(w)_{vec}&=&e^{-q^2\mu^2/4\beta^2},\nonumber\\
F^*_V(w)_{vec}&=&e^{-q^2\mu^2/4\beta^2}M_2\left(\frac{1}{m_2}+\frac{\mu}{2}\left(\frac{1}{m_1}-\frac{1}{m_2}\right)\right).
\end{eqnarray}
It follows from the formulas above that $F_V(w)_0=F_V(w)_{vec}$ and
$F_A(w)_0=F_A(w)_{vec}$ if the heavy quark limit is satisfied:
$\bar{m}_1\ll m_1,\,\bar{m}_2\ll m_2$ and $M_2\approx m_2$.

\begin{figure}[h]
\begin{center}
\includegraphics[angle=-90,width=14cm]{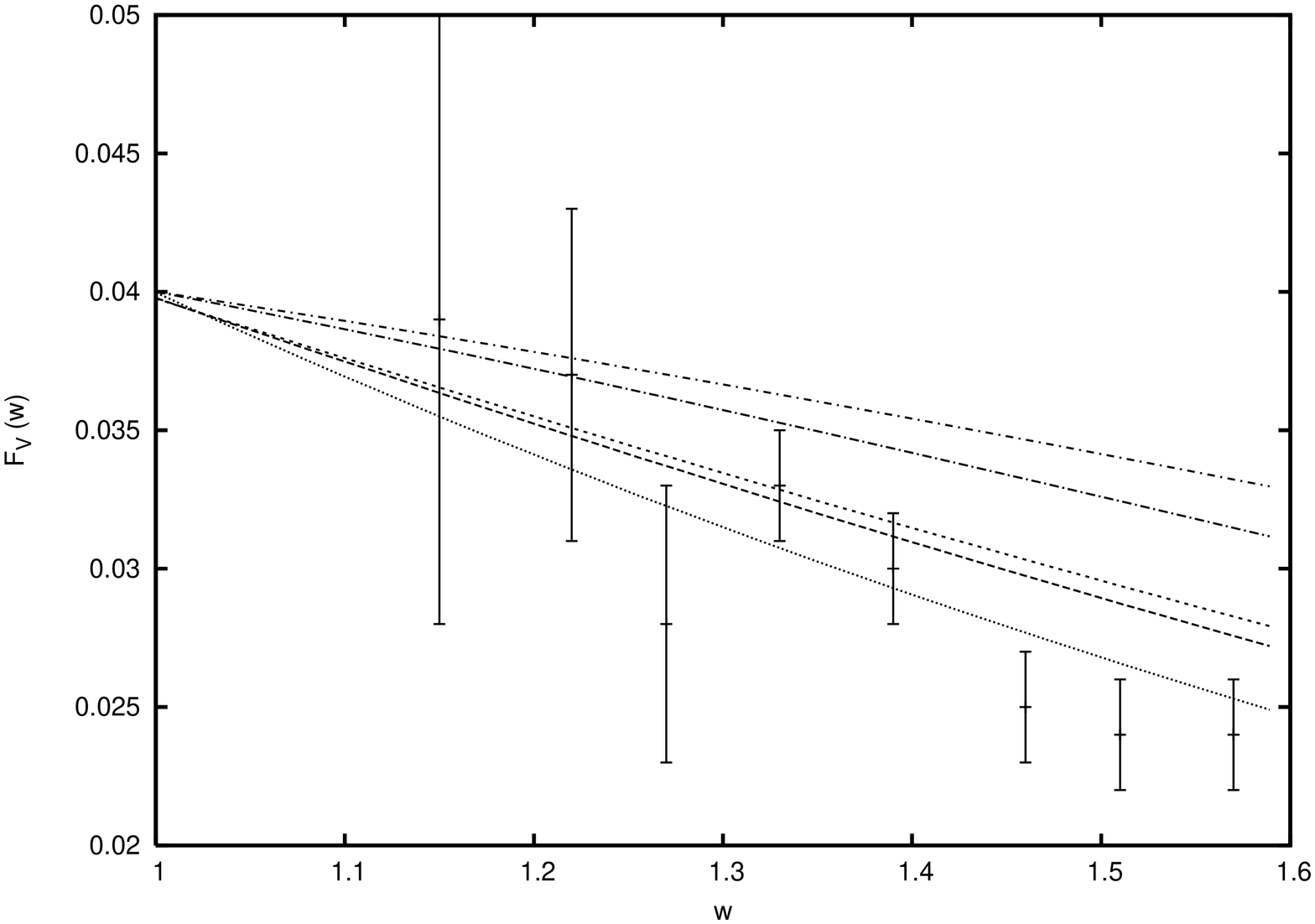}
\caption{\label{Fv}Form-factor $F_V(w)$ of $\bar{B}^0\rightarrow
D^+$. From top to bottom at $w=w_{max}$ the curves are ISGW, SHO,
relativistic C+L, C+L log and nonrelativistic C+L. Experimental data
is taken from \cite{Rosner:HQET}.}
\includegraphics[angle=-90,width=14cm]{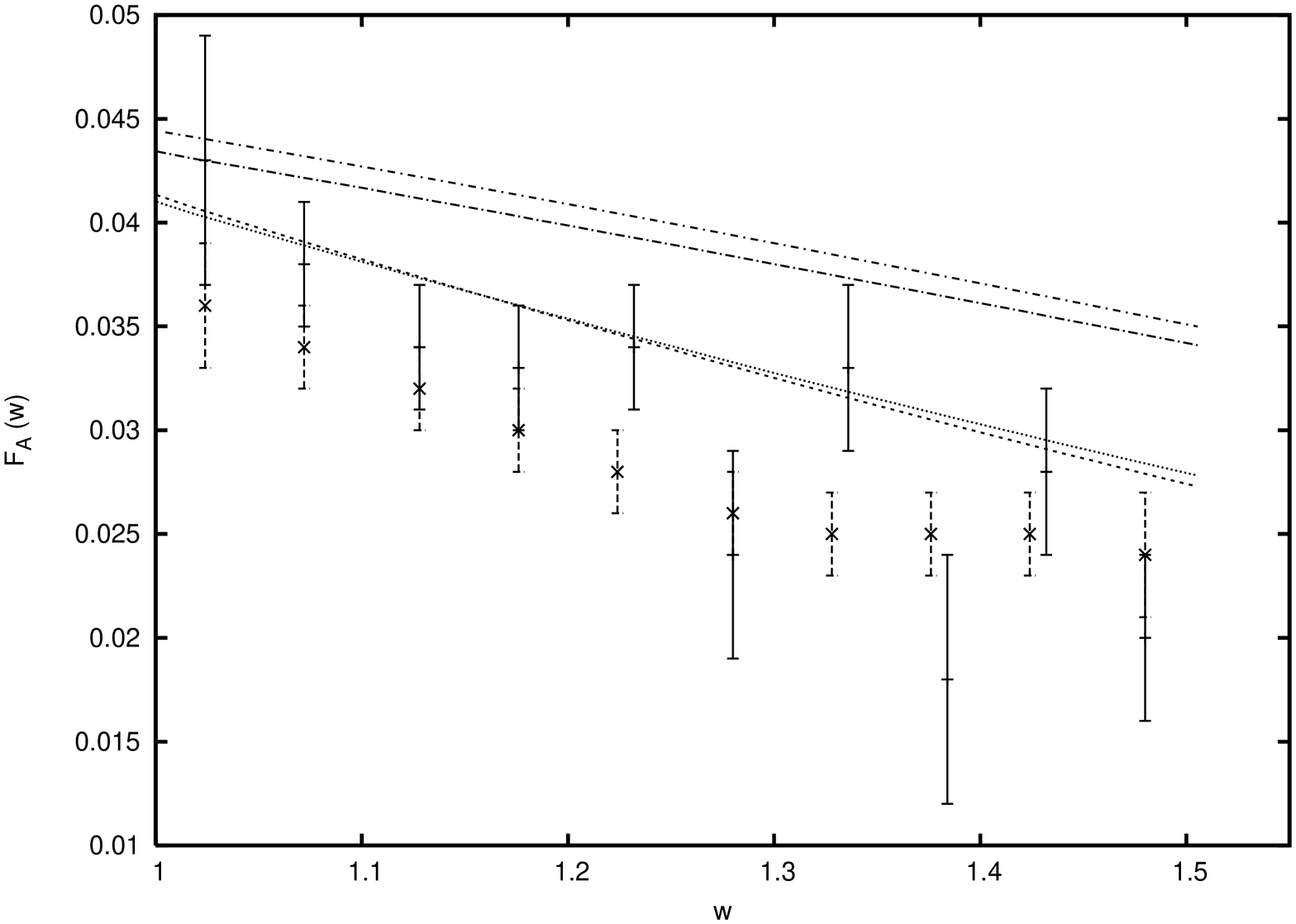}
\caption{\label{Fa}Form-factor $F_A(w)$ of $\bar{B}^0\rightarrow
D^{*+}$. From top to bottom at $w=w_{max}$ the curves are SHO,
nonrelativistic C+L, relativistic C+L and C+L log. Experimental data
is taken from \cite{Rosner:HQET}.}
\end{center}
\end{figure}

The results of our calculations of $F_V(w)$ and $F_A(w)$ for
different models of the potentials are compared to the experimental
data in figures \ref{Fv} and \ref{Fa}. Coulomb+linear+hyperfine
interaction potential model works better for both $F_V(w)$ and
$F_A(w)$ as expected. Also from $F_A(w)$ results it is obvious that
taking into account relativistic corrections for quark spinors is
important for consistency with the experimental data.

\begin{figure}[h]
\includegraphics[angle=-90,width=15cm]{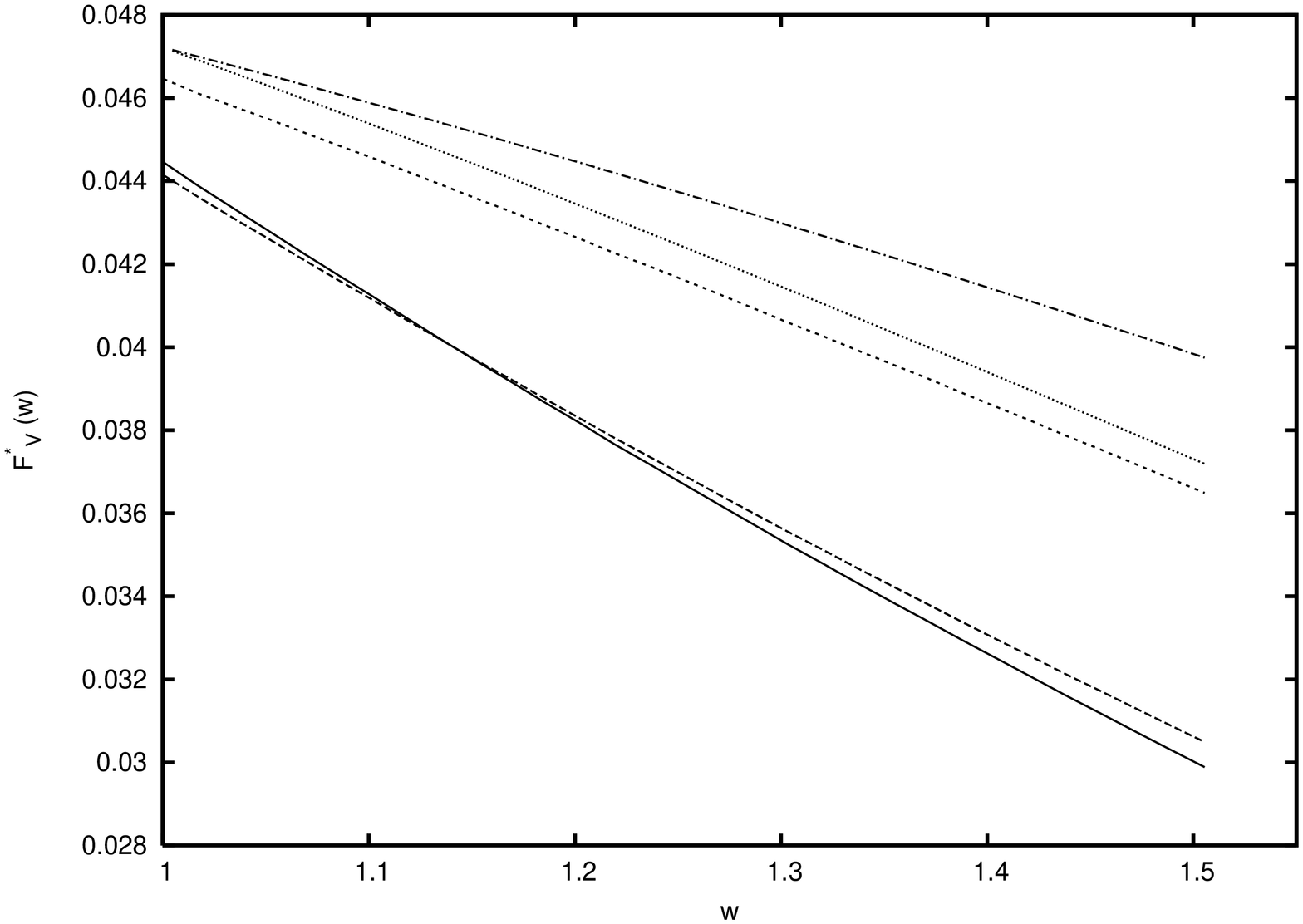}
\caption{\label{Fvv}Form-factor $F^*_V(w)$ of $\bar{B}^0\rightarrow
D^{*+}$. From top to bottom at $w=w_{max}$ the curves are SHO,
nonrelativistic C+L, relativistic C+L and C+L log.}
\end{figure}

Our results for $F^*_V(w)$ are presented in figure \ref{Fvv}.

We also want to compare our results to the heavy quark symmetry
calculations. In this context $h_{\pm}(w)$ form-factors are
introduced for $P_1(^1S_0)\rightarrow P_2(^1S_0)$ transition which
are related to the previously calculated form-factors:
\begin{eqnarray}
h_{\pm}(w)&=&\frac{M_1\pm M_2}{2\sqrt{M_1M_2}}f_+(Q^2)+\frac{M_1\mp
M_2}{2\sqrt{M_1M_2}}f_-(Q^2).
\end{eqnarray}

Our calculations give:
\begin{eqnarray}
h_{\pm}(w)=\frac{1}{2\sqrt{M_1M_2}}\left(V^0-(E_2\mp
M_2)\frac{\vec{V}\cdot\vec{q}}{|\vec{q}|^2}\right).
\end{eqnarray}
In the nonrelativistic approximation for SHO wave functions:
\begin{eqnarray}
h_+(w)&=&e^{-q^2\mu^2/4\beta^2},\\
h_-(w)&=&e^{-q^2\mu^2/4\beta^2}\left[1-M_2\left[\frac{1}{m_2}-\frac{\mu}{2}\left(\frac{1}{m_1}+\frac{1}{m_2}\right)\right]\right].
\end{eqnarray}

In the limit of infinitely heavy quark it follows from the heavy
quark symmetry that:
\begin{eqnarray}
h_+(w)&=&\xi(w),\\
h_-(w)&=&0,
\end{eqnarray}
where $\xi(w)$ is the Isgur-Wise function.

\begin{figure}[h]
\includegraphics[angle=-90,width=15cm]{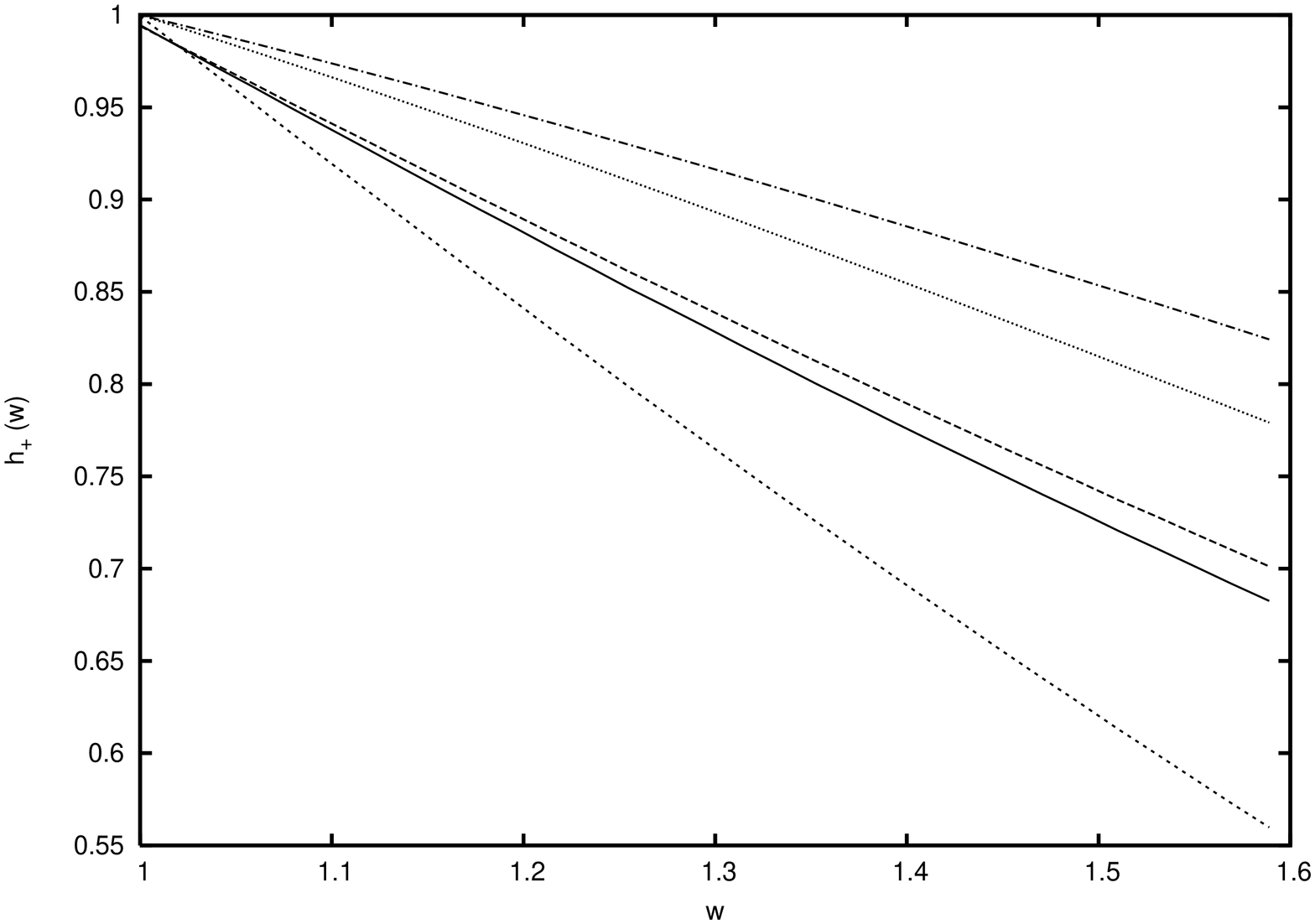}
\caption{\label{hplus}Form-factor $h_+(w)$ of $\bar{B}^0\rightarrow
D^+$. From top to bottom at $w=w_{max}$ the curves are ISGW, SHO,
relativistic C+L, C+L log and nonrelativistic C+L.}
\includegraphics[angle=-90,width=15cm]{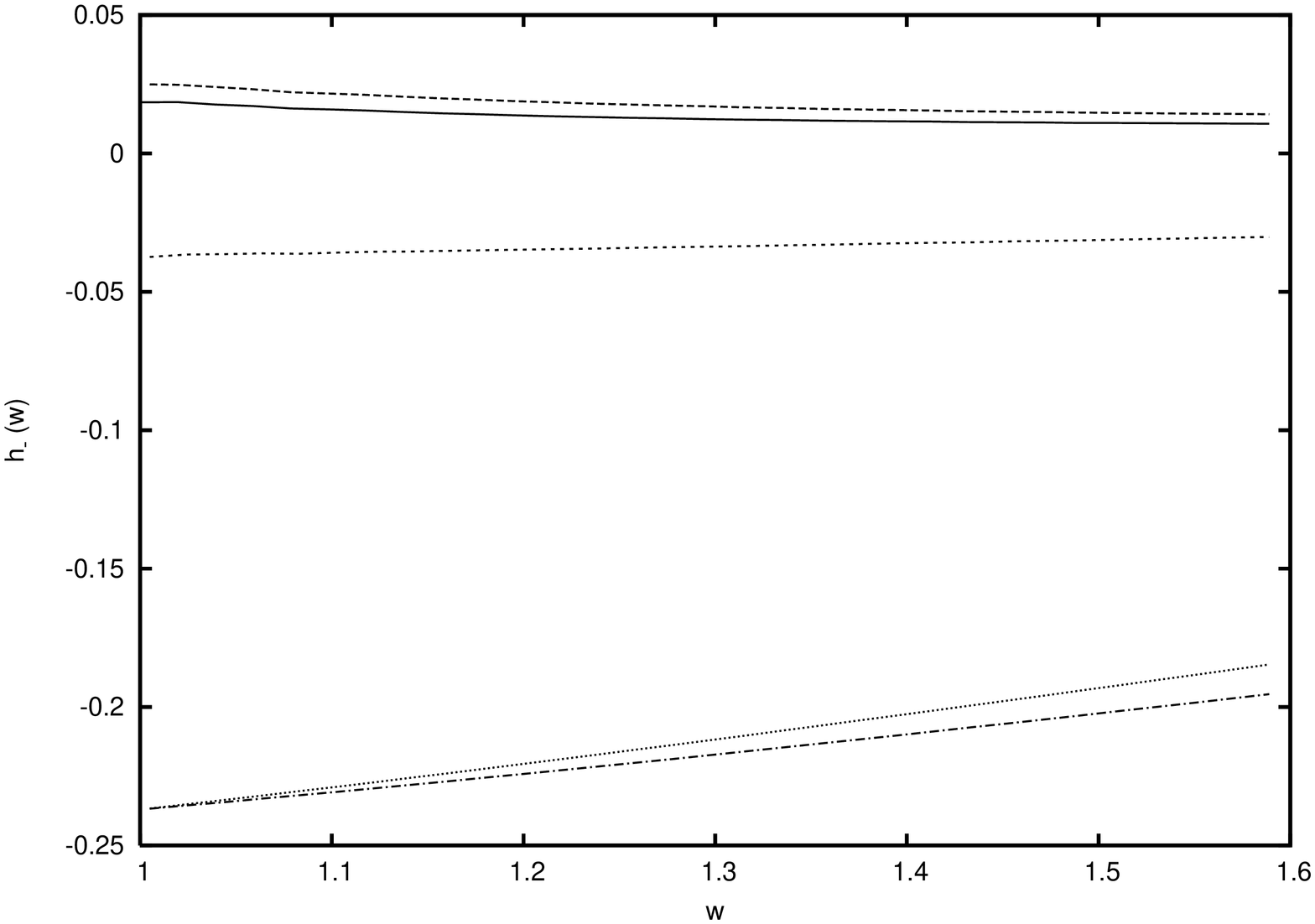}
\caption{\label{hminus}Form-factor $h_-(w)$ of $\bar{B}^0\rightarrow
D^+$. From top to bottom at $w=w_{max}$ the curves are relativistic
C+L, C+L log, nonrelativistic C+L, SHO and ISGW.}
\end{figure}

Our results for $h(w)$ form-factors for $\bar{B}^0\rightarrow D^+$
are presented in figures \ref{hplus}-\ref{hminus}. For all our
calculations $h_+(1)\approx 1$ just as it is supposed to be in the
heavy quark limit. $h_-(w)$ is consistent with zero in our
calculations using Coulomb + linear + hyperfine potential but not
SHO potential. In summary, our calculations are consistent with the
heavy quark EFT calculations of the form-factors for $B\rightarrow
D$ decays.

\begin{figure}[h]
\includegraphics[angle=-90,width=15cm]{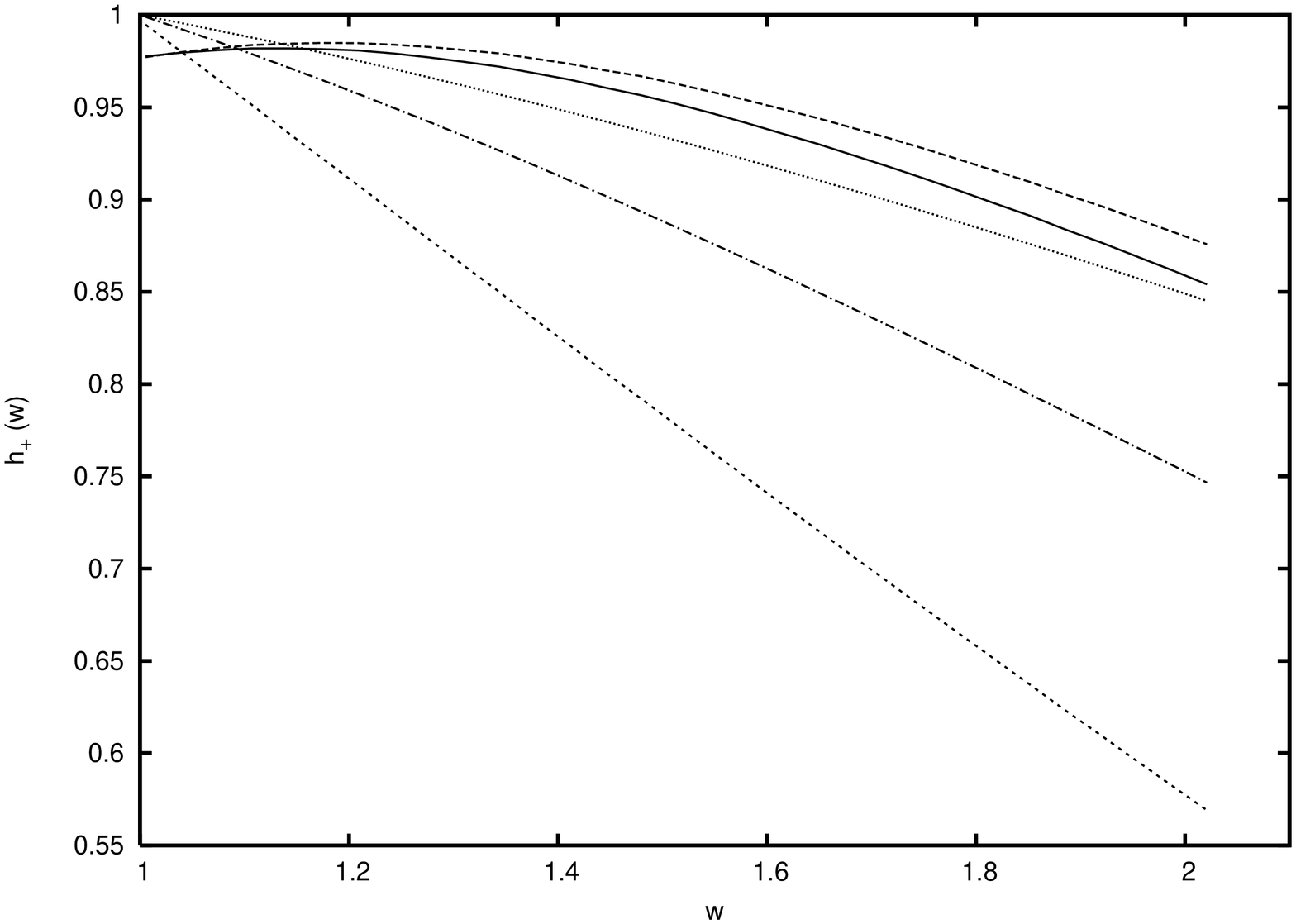}
\caption{\label{hplus_DK}Form-factor $h_+(w)$ of
$\bar{D}^0\rightarrow K^+$. From top to bottom at $w=w_{max}$ the
curves are relativistic C+L, C+L log, SHO, ISGW and nonrelativistic
C+L.}
\includegraphics[angle=-90,width=15cm]{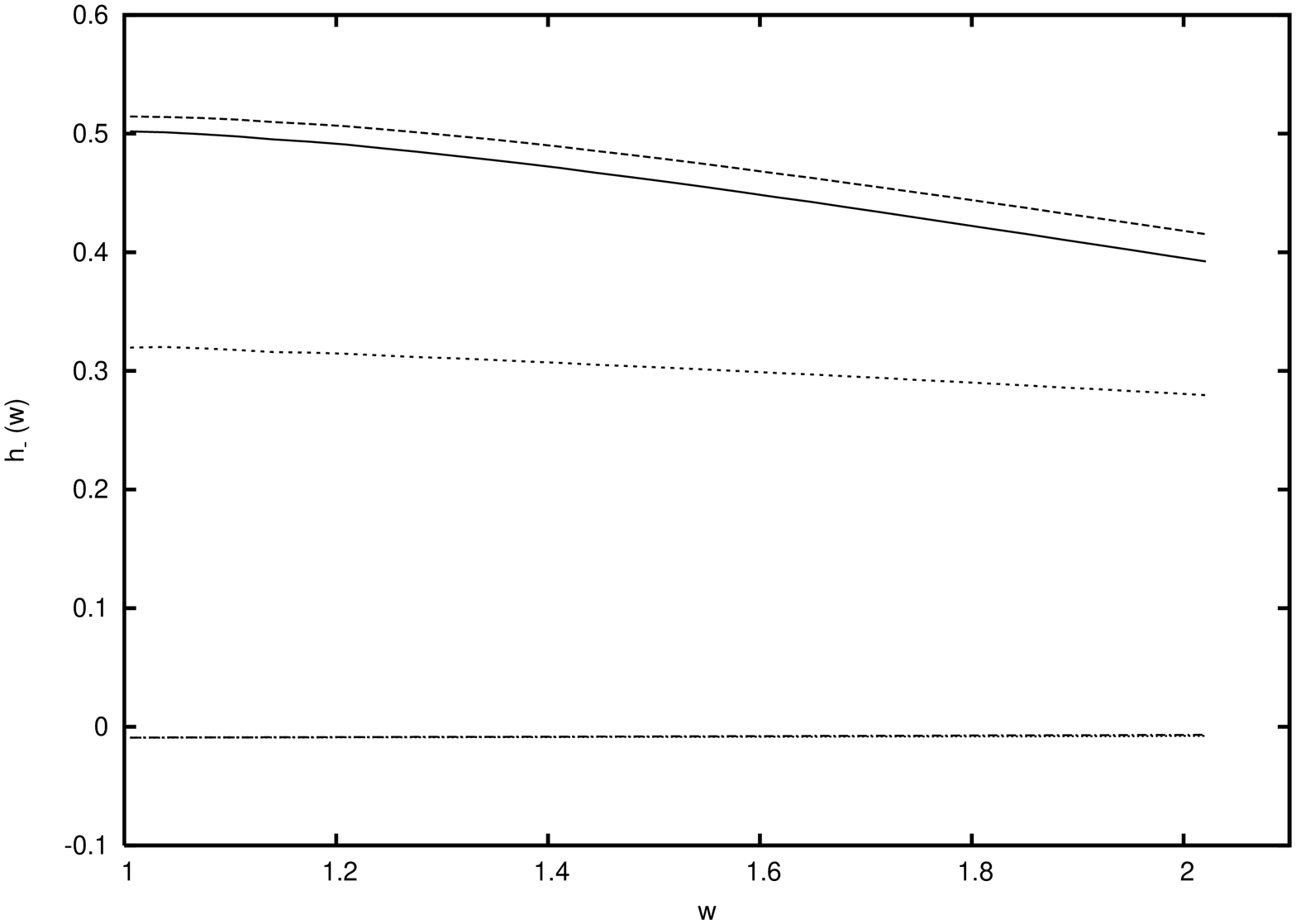}
\caption{\label{hminus_DK}Form-factor $h_-(w)$ of
$\bar{D}^0\rightarrow K^+$. From top to bottom at $w=w_{max}$ the
curves are relativistic C+L, C+L log, nonrelativistic C+L, SHO and
ISGW.}
\end{figure}

The results for $h(w)$ form-factors for $\bar{D}^0\rightarrow K^+$
are presented in figures \ref{hplus_DK}-\ref{hminus_DK}. Again we
get $h_+(1)\approx 1$ for all the models. But $h_-(w)$ is
significantly far from zero for Coulomb + linear + hyperfine
potential model, which means that finite mass corrections are
important for this case (as was expected since $K$ consists of light
quarks). It is interesting to note that for SHO and ISGW models
$h_-(w)$ is quite close to zero for $D\rightarrow K$ transitions
while being significantly different from zero for $B\rightarrow D$
transitions, and it should be opposite since B mesons should be
closer to the heavy quark limit than K mesons. It means that SHO
potential is not very good approximation for the quark interaction
since the form-factors calculated with SHO potential don't approach
heavy quark limit behavior as they should.

In the heavy quark symmetry limit the matrix elements for
$P(^1S_0)\rightarrow V(^3S_1)$ transition could be written as
\cite{Neubert}:
\begin{eqnarray}\label{HQL}
V^{\mu}&=&\langle V(\vec{P}_V)|\bar{q}\gamma^{\mu}q|P(\vec{P}_P)\rangle=i\sqrt{m_Pm_V}\epsilon^{\mu\nu\alpha\beta}\left(\epsilon^*_{M_V}\right)_{\nu} v'_{\alpha} v_{\beta} \xi(w),\\
A^{\mu}&=&\langle
V(\vec{P}_V)|\bar{q}\gamma^{\mu}\gamma^5q|P(\vec{P}_P)\rangle=\sqrt{m_Pm_V}\left[\left(\epsilon^{\,*}_{M_V}\right)^{\mu}(v\cdot
v'+1)-v'^{\mu}\left(\epsilon^*_{M_V}\cdot
v\right)\right]\xi(w).\nonumber
\end{eqnarray}

Comparing (\ref{PV}) with (\ref{HQL}) one finds that
\begin{eqnarray}
\lim_{m\rightarrow\infty}h_g(w)&=&\lim_{m\rightarrow\infty} \left\{2\sqrt{m_Pm_V}g(Q^2)\right\}=\xi(w),\nonumber\\
\lim_{m\rightarrow\infty}h_f(w)&=&\lim_{m\rightarrow\infty} \left\{\frac{f(Q^2)}{2\sqrt{m_Vm_P}}\right\}=\xi(w),\nonumber\\
\lim_{m\rightarrow\infty}h_{a-}(w)&=&\lim_{m\rightarrow\infty} \left\{-\sqrt{m_Pm_V}\left(a_+(Q^2)-a_-(Q^2)\right)\right\}=\xi(w),\nonumber\\
\lim_{m\rightarrow\infty}h_{a+}(w)&=&\lim_{m\rightarrow\infty}
\left\{a_+(Q^2)+a_-(Q^2)\right\}=0.
\end{eqnarray}

\begin{figure}[h]
\includegraphics[angle=-90,width=15cm]{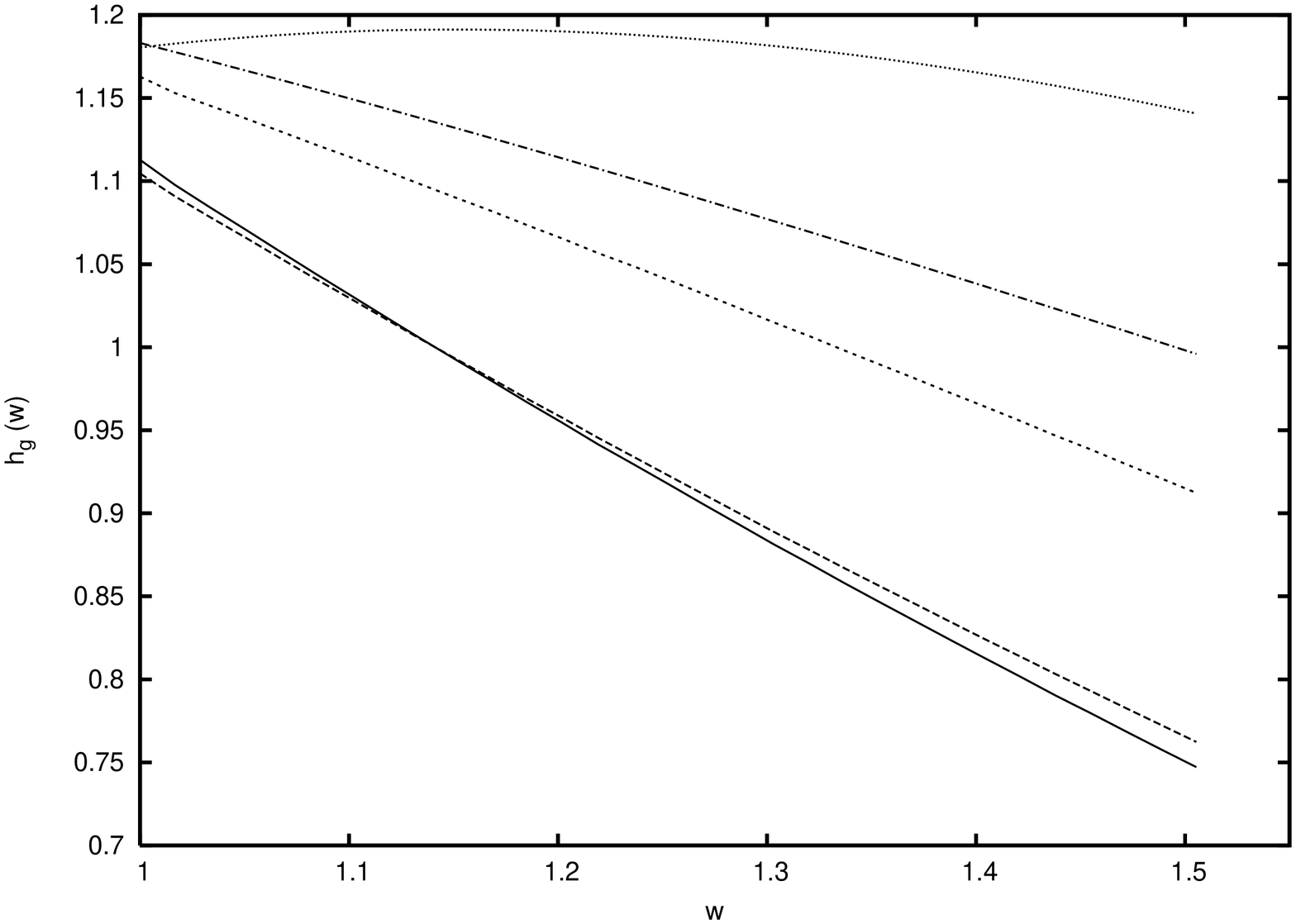}
\caption{\label{hg}Form-factor $h_g(w)$ of $\bar{B}^0\rightarrow
D^{*+}$. From top to bottom at $w=w_{max}$ the curves are SHO, ISGW,
nonrelativistic C+L, relativistic C+L and C+L log.}
\includegraphics[angle=-90,width=15cm]{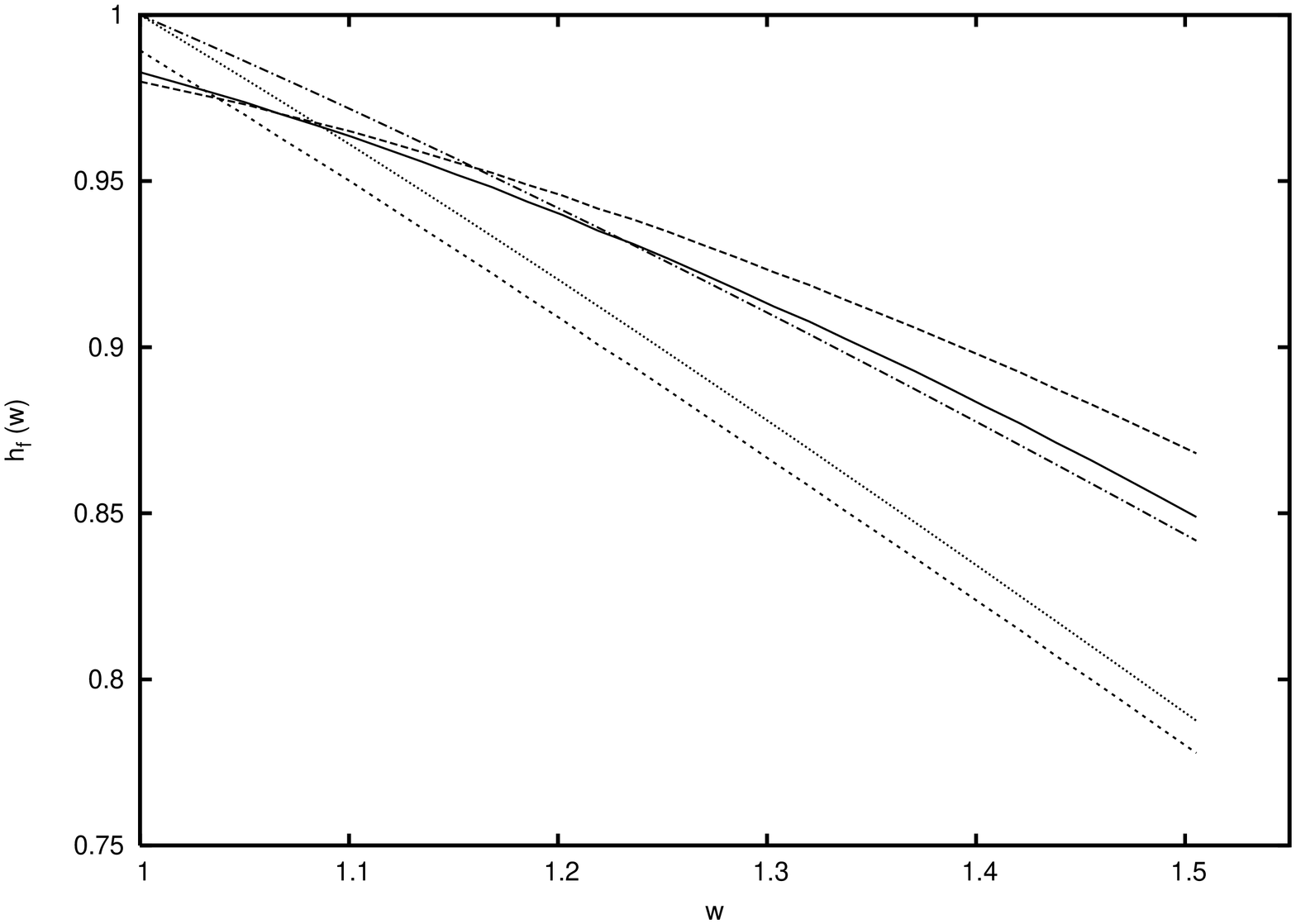}
\caption{\label{hf}Form-factor $h_f(w)$ of $\bar{B}^0\rightarrow
D^{*+}$. From top to bottom at $w=w_{max}$ the curves are
relativistic C+L, C+L log, ISGW, SHO and nonrelativistic C+L.}
\end{figure}

\begin{figure}[h]
\includegraphics[angle=-90,width=15cm]{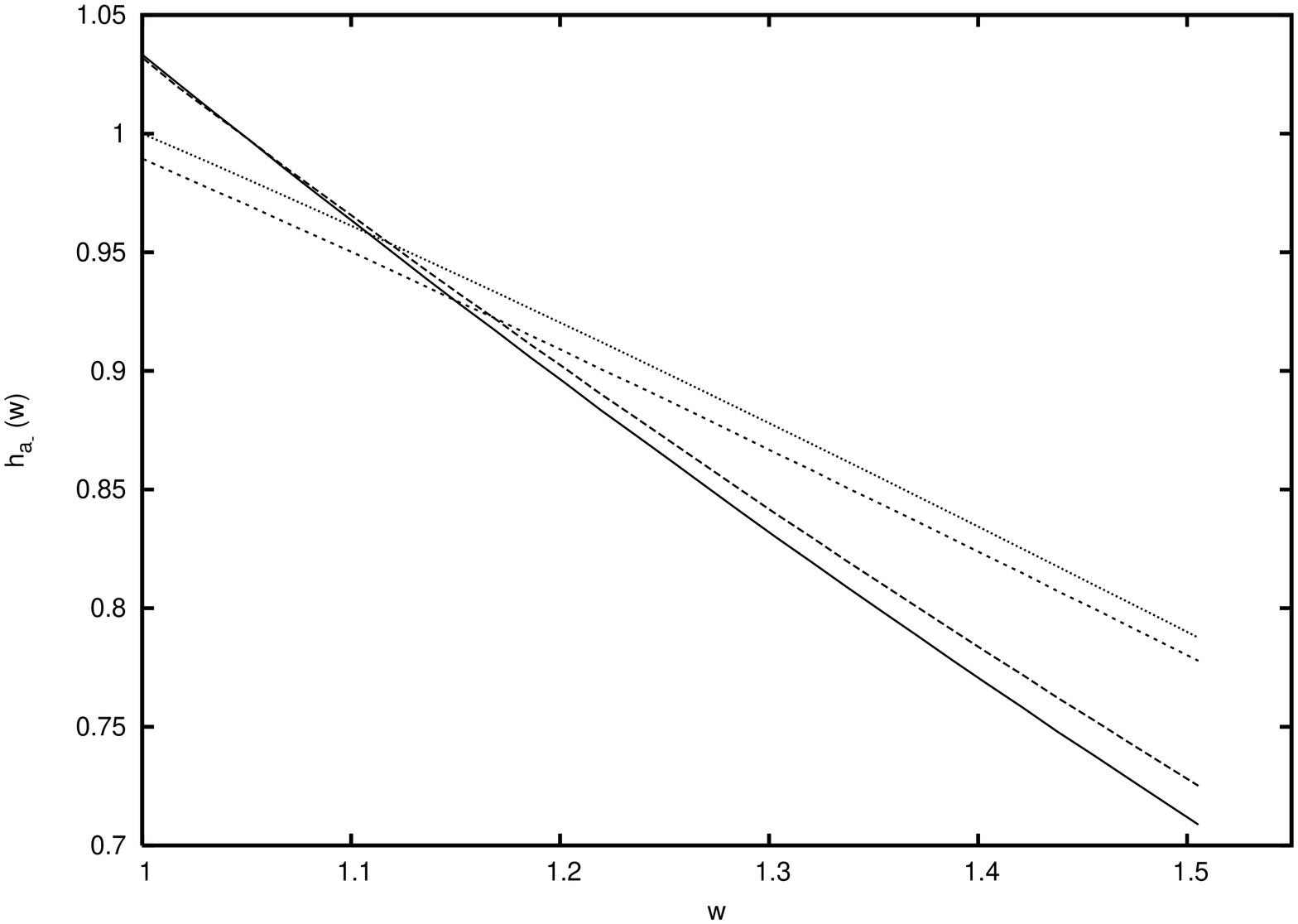}
\caption{\label{haminus}Form-factor $h_{a-}(w)$ of
$\bar{B}^0\rightarrow D^{*+}$. From top to bottom at $w=w_{max}$ the
curves are SHO, nonrelativistic C+L, relativistic C+L and C+L log.}
\includegraphics[angle=-90,width=15cm]{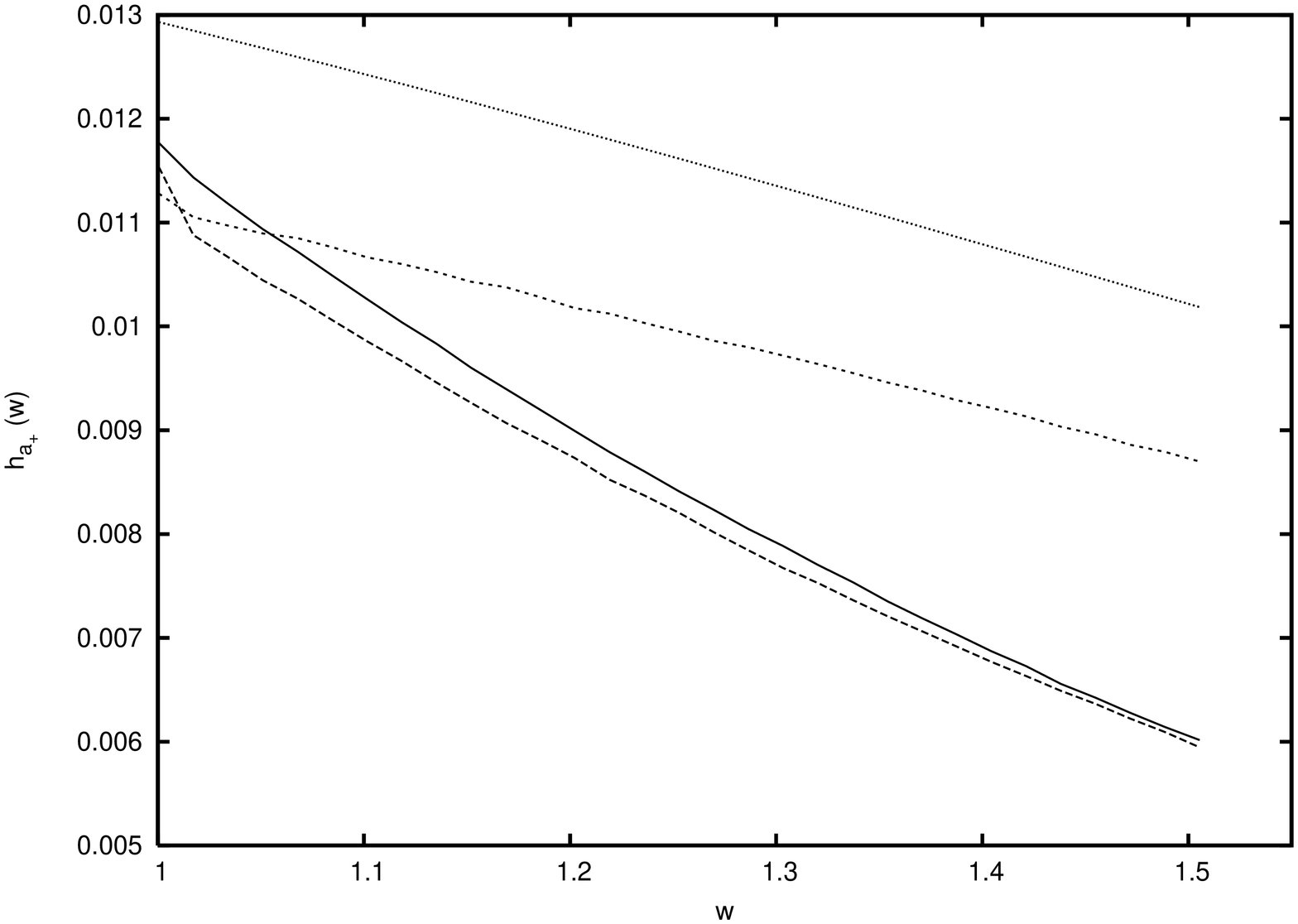}
\caption{\label{haplus}Form-factor $h_{a+}(w)$ of
$\bar{B}^0\rightarrow D^{*+}$. From top to bottom at $w=w_{max}$ the
curves are SHO, nonrelativistic C+L, C+L log and relativistic C+L.}
\end{figure}

Our results for $h_g(w)$, $h_f(w)$, $h_{a-}(w)$ and $h_{a+}(w)$ are
presented in figures \ref{hg}-\ref{haplus}.

\addtolength{\topmargin}{-2cm} \addtolength{\footskip}{3cm}

\clearpage \addtolength{\topmargin}{2cm}
\addtolength{\footskip}{-3cm}

\chapter{GAMMA-GAMMA TRANSITIONS}\label{gg_chapter}

In this chapter the details and the results of our calculations of
gamma-gamma transition decay rates are presented and discussed.

The general amplitude for two-photon decay of pseudoscalar
quarkonium can be written as

\begin{equation}
{\cal A}(\lambda_1 p_1; \lambda_2 p_2) =
\epsilon^*_\mu(\lambda_1,p_1) \epsilon^*_\nu(\lambda_2, p_2) {\cal
M}^{\mu\nu}
\end{equation}
with
\begin{equation}
{\cal M}^{\mu\nu}_{Ps} = i M_{Ps}(p_1^2, p_2^2, p_1\cdot p_2) \,
\epsilon^{\mu\nu\alpha\beta} \, p_{1\alpha}p_{2\beta}.
\end{equation}
The total decay rate is then $\Gamma(Ps \to \gamma\gamma) =
{m_{Ps}^3 \over 64 \pi} |M_{Ps}(0,0)|^2$.

Before moving on to the quark model computation, it is instructive
to evaluate the amplitude in an effective field theory that
incorporates pseudoscalars, vectors, and vector meson dominance. The
relevant Lagrangian density is

\begin{equation}
{\cal L} = -i Q m_V f_V V_\mu A^\mu - \frac{1}{2} Q F^{(V)} \eta
\tilde F_{\mu\nu} V^{\mu\nu}
\end{equation}
where $\tilde F^{\mu\nu} =
\frac{1}{2}\epsilon^{\mu\nu\alpha\beta}F_{\alpha\beta}$ and
$V^{\mu\nu} = \partial^\mu V^\nu - \partial^\nu V^\mu$, Q is the
charge of the quark \footnote{The vector meson dominance term is not
gauge invariant. Why this is not relevant here is discussed in Sect.
15 of Ref. \cite{RPF}.}. Evaluating the transition $Ps \to
\gamma\gamma$ yields

\begin{equation}
M_{Ps}(p_1^2,p_2^2) = \sum_V m_V f_V Q^2 \left({F^{(V)}(p_1^2)\over
p_2^2-m_V^2} + {F^{(V)}(p_2^2)\over p_1^2-m_V^2}\right).
\label{relMPsEq}
\end{equation}
Hence the pseudoscalar decay rate is

\begin{equation}
\Gamma(Ps \to \gamma\gamma) = {m_{Ps}^3 Q^4 \over 16 \pi} \left(
\sum_V {f_V F^{(V)}(0) \over m_V}\right)^2. \label{relRateEq}
\end{equation}
Notice that the desired cubic pseudoscalar mass dependence is
achieved in a simple manner in this approach (see the discussion in
section \ref{gg_theory}).

The application of this formula is complicated by well-known
ambiguities in the vector meson dominance model (namely, is $p_V^2 =
m_V^2$ or zero?). The time ordered perturbation theory of the quark
model suffers no such ambiguity (although, of course, it is not
covariant) and it is expedient to use the quark model to resolve the
ambiguity. We thus choose to evaluate the form factor at the
kinematical point $|\vec q| = m_{Ps}/2$, appropriate to $Ps \to
\gamma\gamma$ in the pseudoscalar rest frame. Applying  Eq.
\ref{generalQsqEq} to the virtual process $\eta_c \to J/\psi \gamma$
then implies that the argument of the form factor should be $Q^2 =
2.01$ GeV$^2$.

A simple estimate of the rate for $\eta_c \to \gamma\gamma$ can now
be obtained from Eq. \ref{relRateEq}, $f_{J/\psi} \approx 0.4$ GeV,
and $F^{(V)}(Q^2=2\ {\rm GeV}^2) \approx 0.7$ GeV$^{-1}$ (Fig.
\ref{FpsiFig}). The result is $\Gamma(\eta_c\to \gamma\gamma)
\approx 7.1$ keV, in reasonable agreement with experiment.

\begin{figure}[h!]
\includegraphics[angle=270,width=15cm]{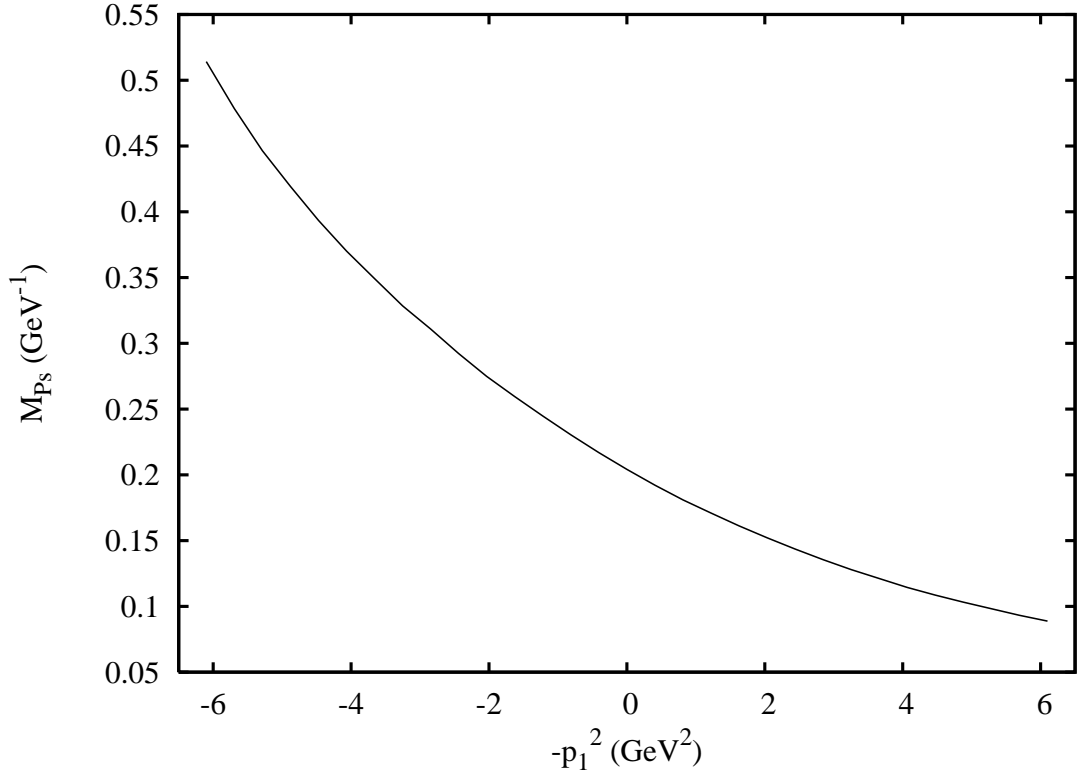}
\caption{The Two-photon Form Factor $M_{Ps}(p_1^2,p_2^2=0)$ for
$\eta_c \to \gamma\gamma$.} \label{MPsFig}
\end{figure}

Finally, the predicted form of the two-photon $\eta_c$ form factor
is shown in Fig. \ref{MPsFig} in the case that one photon is
on-shell. The result is a slightly distorted monopole (due to vector
resonances and the background term in Eq. \ref{relMPsEq}) that
disagrees strongly with naive factorization results. This prediction
have been successfully tested by lattice computations
\cite{Dudek:2006ut}, and this leads us to the conclusion that the
factorization model should be strongly refuted.

As motivated above, the microscopic description of the $\eta_c$
two-photon decay is best evaluated in bound state time ordered
perturbation theory. Thus one has

\begin{equation}
{\cal A}_{NR} = \sum_{\gamma,V}{\langle \gamma(\lambda_1,p_1)
\gamma(\lambda_2,p_2) | H | \gamma, V\rangle \, \langle \gamma, V |
H | Ps \rangle \over (m_{Ps} - E_{\gamma V}) } \label{microAEq}
\end{equation}
The second possible time ordering requires an extra vertex to permit
the transition $\langle Ps, V | \gamma\rangle$ and hence is higher
order in the Fock space expansion. Thus the second time ordering has
been neglected in Eq. \ref{microAEq}.

The amplitudes  can be written in terms of the relativistic
decompositions of the previous sections. One obtains the on-shell
amplitude
\begin{equation}
M_{Ps} = \sum_{V}Q^2 \sqrt{\frac{m_V}{E_V}} f_V
\frac{F^{(V)}(q)}{m_{Ps}-E_{\gamma V}(q)}. \label{MPsEq}
\end{equation}
We choose to label the momentum dependence with the nonrelativistic
$q=|\vec q|$ in these expressions \footnote{The naive application of
the method advocated here to light quarks will fail. In this case
the axial anomaly requires that $M_{Ps} = \frac{i \alpha}{\pi
f_\pi}$, which is clearly at odds with Eq. \ref{MPsEq}. The
resolution of this problem requires a formalism capable of
incorporating the effects of dynamical chiral symmetry breaking,
such as described in Refs. \cite{Norbert,rm}.}.

The total width is evaluated by summing over intermediate states,
integrating, and symmetrizing appropriately. Form factors and decay
constants are computed as described in the preceding sections. As
argued above, form factors are evaluated at the point $|\vec q|  =
m_{Ps}/2$. Table \ref{etaTab} shows the rapid convergence of the
amplitude in the vector principle quantum number $n$  for the
quantity $\frac{4\sqrt{2}}{Q\sqrt{m_{\eta_c}}}{\cal A}_{++}$.
Surprisingly, convergence is not so fast for the $\Upsilon$ system
and care must be taken in this case.
\begin{table}[!h]
\caption{Amplitude for $\eta_c\rightarrow\gamma\gamma$ ($10^{-3}$
GeV$^{-1}$).}
\begin{center}
\begin{tabular}{c|cc}
\hline
n  & BGS & BGS log \\
\hline \hline
1 & -211 &  -141 \\
2 &  -34 &  -30  \\
3 &  -10 &  -10  \\
\hline \hline
\end{tabular}
\end{center}
\label{etaTab}
\end{table}

Table \ref{ccggTab} presents the computed widths for the $\eta_c$,
$\eta_c'$, $\eta_c''$ and $\chi_{c0}$ mesons in a variety of models.
The second and third columns compare the predictions of the BGS
model with and without a running coupling. Use of the running
coupling reduces the predictions by approximately a factor of two,
bringing the model into good agreement with experiment. This is due,
in large part, to the more accurate vector decay constants provided
by the BGS+log model. In comparison, the results of Godfrey and
Isgur (labeled GI), which rely on naive factorization
 supplemented with the {\it ad hoc} pseudoscalar mass dependence discussed above,
 does not fare so well for the excited $\eta_c$ transition rate. Similarly a
computation using heavy quark spin symmetry (labeled HQ) finds a
large $\eta_c'$ rate. Columns 6 and 7 present results computed in
the factorization approach with nonrelativistic and relativistic
wavefunctions respectively.  Column 8 (Munz) also uses factorization
but computes with the Bethe-Salpeter formalism. The model of column
9 (CWV) employs factorization with wavefunctions determined by a
two-body Dirac equation. With the exception of the last model, it
appears that model variation in factorization approaches can
accommodate some, but never all, of the experimental data, in
contrast to the bound state perturbation theory result. However,
more and better data are required before this conclusion can be
firm. The experimental rate for $\eta_c'$ is obtained from Ref.
\cite{CLEO} and assumes that $Br(\eta_c \to K_SK\pi) = Br(\eta_c'
\to K_SK\pi)$. This assumption is supported by the measured rates
for $B \to K\eta_c$ and $K\eta_c'$ as explained in Ref. \cite{Song}.
Our predictions for the bottomonia are presented in Table
\ref{Bgamma2}.

\clearpage \addtolength{\oddsidemargin}{-2cm}
\begin{table}[h!]
\caption{Charmonium Two-photon Decay Rates (keV). For BGS log model
$\Lambda = 0.25$ GeV. }
\begin{tabular}{c|cc|cccccc|l}
\hline process & BGS & BGS log & G\&I\cite{GI} & HQ\cite{LP} &
A\&B\cite{AB} & EFG\cite{EFG} & Munz\cite{Munz}
& CWV\cite{CWV}& Experiment
 \\
\hline \hline $\eta_c  \rightarrow \gamma\gamma$ &14.2 &7.18 &6.76 &
7.46 & 4.8 & 5.5 & 3.5(4)
& 6.18 & $7.44\pm 2.8$\\
$\eta'_c \rightarrow \gamma\gamma$     &2.59   &1.71 &4.84 & 4.1 &
3.7  & 1.8 & 1.4(3)
& 1.95 & $1.3 \pm 0.6$ \\
$\eta''_c \rightarrow \gamma\gamma$     &1.78   &1.21 & -- & -- & --
& --  & 0.94(23)
& -- &  -- \\
$\chi_{c0} \rightarrow \gamma\gamma$   &5.77   &3.28 & -- & -- & --
& 2.9 & 1.39(16)
& 3.34 & $2.63\pm 0.5$ \\
\hline \hline
\end{tabular}
\label{ccggTab}
\end{table}

\clearpage \addtolength{\oddsidemargin}{2cm}
\begin{table}[h!]
\caption{Bottomonium Two-photon Decay Rates (keV). For BGS log model
$\Lambda = 0.25$ GeV.} \label{Bgamma2}
\begin{center}
\begin{tabular}{c|ccccccc|c}
\hline process & BGS & BGS log & G\&I\cite{GI} & Munz\cite{Munz} &
HQ\cite{LP1} &
A\&B\cite{AB} & EFG\cite{EFG}  & Experiment \\
\hline \hline
$\eta_b  \rightarrow \gamma\gamma$     &0.45    &0.23   &0.38 &0.22(0.04)   &0.56 &0.17 &0.35    & -- \\
$\eta_b' \rightarrow \gamma\gamma$     &0.11    &0.07   &--   &0.11(0.02)   &0.27 &--   &0.15    & --  \\
$\eta_b'' \rightarrow \gamma\gamma$    &0.063   &0.040  &--   &0.084(0.012) &0.21 &--   &0.10    & --  \\
$\chi_{b0} \rightarrow \gamma\gamma$   &0.126   &0.075  &--   &0.024(0.003) &--   &--   &0.038   & --  \\
\hline \hline
\end{tabular}
\end{center}
\end{table}

\chapter{RADIATIVE TRANSITIONS}\label{RadTrans}

Meson and baryon radiative transitions deserve a lot of
investigation since they are easily produced and the transition
operator is very well known. Since they belong to the
non-perturbative regime  of QCD they cannot be described from the
first principles. One of the theories which had a number of
successes in describing nonperturbative part of QCD is the
constituent quark model. In particular, quark models work quite well
for the meson spectrum, as was demonstrated in Chapter
\ref{spectrum}. But one needs to consider other observables (such as
electromagnetic transitions) since very different potentials can
lead to the similar mass spectra. Radiative transitions between
various mesons are very sensitive to the inter-quark potential, and
can provide significant help in testing various meson potentials and
wave functions and show us ways to improve the models.

In dealing with radiative transitions some typical approximations
are usually in use. Some of them are impulse approximation, dipole
approximation for E1 transitions \cite{cs,mcclary,barnes}, long wave
length approximation \cite{mcclary,cs,barnes}, non relativistic
approximation \cite{Cornell,cs,Bonnaz}. Also spherical harmonic
oscillator (SHO) wave functions are widely used to represent the
meson wave functions \cite{FM}. And almost in all the cases the
study of radiative transitions is performed only for the particular
sector of meson spectra (for example only heavy or only light
mesons) \cite{close,mcclary,barnes,book}.

Most of these approximations are taken from atomic and nuclear
physics where they describe radiative transitions rather well. But
when applied to mesons they are not always justified. For example,
long wave length approximation is defined by condition
$k_{\gamma}R\ll 1$ where $k_{\gamma}$ is the photon momentum, and
$R$ is the size of the source. For the meson radiative transitions
typically $k_{\gamma}=0.1\div0.5$ GeV and $R=0.5\div1$ fm
$=2.5\div5$ GeV$^{-1}$ so that the long wave length condition is not
always true. Also the long wave length approximation leads to
neglecting the recoil of the final meson, and in reality the
momentum of the final meson is often comparable to its mass. We
conclude that not only recoil should not be neglected but the
nonrelativistic approximation is not suitable in this case. To
preserve gauge invariance both transition operator and meson wave
functions should be relativistic. Some attempts have been made to
take the relativistic effects in radiative meson transitions into
account but other approximations have been used which can have a
larger effect on the result \cite{mcclary,Cornell,GI,Goity}.

The motivation for this work was to perform a detailed study of
meson radiative transitions and investigate the effects of different
approximations in the quark model. We used wave functions calculated
from the realistic potentials as well as SHO wave functions (for
comparison). Relativistic corrections in the transition operator as
well as in the wave functions have been taken into account. Higher
order diagrams (beyond impulse approximation) have been estimated.
Decays rates have been calculated for all the transitions for which
experimental data are available from the Particle Data Group book.

\section{Impulse approximation}
\subsection{Nonrelativistic constituent quark model}

In Tables \ref{cc_nonrel}, \ref{bb_nonrel} and \ref{light_nonrel}
results calculated in the nonrelativistic potential model for SHO
('Gaussian') and realistic Couolomb+linear+hyperfine potentials are
presented for $c\bar{c}$, $b\bar{b}$ and light mesons. Radiative
decay rates have been determined for both nonrelativistic
approximation and full relativistic expressions of quark spinors
(`nonrel' and `rel' columns) in the impulse approximation. Detailed
description of our method  and formulae for the decay rates are
presented in section \ref{imp}.

From the results in the tables \ref{cc_nonrel} and \ref{bb_nonrel}
one can conclude that $c\bar{c}$, and even $b\bar{b}$, mesons should
not be considered nonrelativistically as relativistic corrections
just in the transition operator make a big difference for the decay
rate. Taking into account relativistic corrections in the wave
functions will change the results even more.

Also, our results show that inter-quark potential has a considerable
effect on the decay rates, see for example
$\chi_{c0}\rightarrow\gamma J/\psi$ for SHO and Coulomb+linear
potentials.

In each case, for some of the transitions the results show agreement
with the experiment while for the other transitions they are far
off. This shows the importance of studying the whole range of
different mesons and quark-interquark potentials and also gives us a
hint that some important effects might be missing in a
nonrelativistic quark model.

\begin{table}[t!]
\caption{\label{cc_nonrel}$c\bar{c}$ meson radiative decay rates
(keV)}
\begin{center}
\begin{tabular}{c|c|c|c|c|c|c}

\hline \multicolumn{1}{c|}{}& \multicolumn{1}{c|}{}&
\multicolumn{2}{c|}{SHO}& \multicolumn{2}{c|}{Coulomb+linear}&
\multicolumn{1}{c}{Experiment}\\
\multicolumn{1}{c|}{}& \multicolumn{1}{c|}{$\gamma$(MeV)}&
\multicolumn{1}{c|}{nonrel}& \multicolumn{1}{c|}{rel}&
\multicolumn{1}{c|}{nonrel}& \multicolumn{1}{c|}{rel}&
\multicolumn{1}{c}{PDG \cite{PDG06}}\\

\hline \hline
$J/\psi\rightarrow\gamma\eta_c$        &115 &2.85 &2.52 &2.82 &2.11 &$1.21\pm 0.41$ \\
$\chi_{C0}\rightarrow\gamma J/\psi$    &303 &194  &167  &349  &276  &$135\pm 21$    \\
$\chi_{C1}\rightarrow\gamma J/\psi$    &389 &221  &193  &422  &325  &$317\pm 36$    \\
$\chi_{C2}\rightarrow\gamma J/\psi$    &430 &137  &114  &352  &260  &$416\pm 46$    \\
$\Psi (2S)\rightarrow\gamma\eta_c $    &639 &5.95 &3.21 &8.15 &1.41 &$0.88\pm 0.17$ \\
$\Psi (2S)\rightarrow\gamma \chi_{C0}$ &261 &23.4 &16.8 &13.6 &7.0  &$31.0\pm 2.6$  \\
$\Psi (2S)\rightarrow\gamma \chi_{C1}$ &171 &54.5 &40.3 &36.0 &20.4 &$29.3\pm 2.5$  \\
$\Psi (2S)\rightarrow\gamma \chi_{C2}$ &128 &77.4 &59.2 &55.4 &33.8 &$27.3\pm 2.5$  \\
$h_c\rightarrow\gamma\eta_c$           &496 &189  &162  &497  &363  & seen          \\
\hline
\end{tabular}
\end{center}
\end{table}

\begin{table}[t!]
\caption{\label{bb_nonrel}$b\bar{b}$-meson radiative decay rates
(keV). Parameters fitted to known bottomonium spectrum are employed
(see section \ref{bb-bar}).}
\begin{center}
\begin{tabular}{c|c|c|c|c|c|c}
\hline \multicolumn{1}{c|}{}& \multicolumn{1}{c|}{}&
\multicolumn{2}{c|}{SHO}& \multicolumn{2}{c|}{Coulomb+linear}&
\multicolumn{1}{c}{Experiment}\\
\multicolumn{1}{c|}{}& \multicolumn{1}{c|}{$\gamma$(MeV)}&
\multicolumn{1}{c|}{nonrel}& \multicolumn{1}{c|}{rel}&
\multicolumn{1}{c|}{nonrel}& \multicolumn{1}{c|}{rel}&
\multicolumn{1}{c}{PDG \cite{PDG06}}\\
\hline \hline
$\chi_{b0}(1P)\rightarrow\gamma\Upsilon(1S)$ &391 &11.2  &10.9  &33.7  &30.8    &seen           \\
$\chi_{b1}(1P)\rightarrow\gamma\Upsilon(1S)$ &423 &10.7  &10.4  &35.2  &32.1    &seen           \\
$\chi_{b2}(1P)\rightarrow\gamma\Upsilon(1S)$ &442 &7.79  &7.53  &33.5  &30.3    &seen           \\
$\Upsilon(2S)\rightarrow\gamma\eta_b(1S)$    &559 &0.028 &0.020 &0.005 &0.001   &$<0.016$ \\
$\Upsilon(2S)\rightarrow\gamma\chi_{b0}(1P)$ &162 &0.77  &0.73  &0.54  &0.43    &$1.22\pm 0.24$ \\
$\Upsilon(2S)\rightarrow\gamma\chi_{b1}(1P)$ &130 &1.98  &1.87  &1.45  &1.17    &$2.21\pm 0.32$ \\
$\Upsilon(2S)\rightarrow\gamma\chi_{b2}(1P)$ &110 &3.02  &2.87  &2.37  &1.95    &$2.29\pm 0.31$ \\
$\chi_{b0}(2P)\rightarrow\gamma\Upsilon(2S)$ &207 &8.47  &8.00  &16.0  &14.2    &seen           \\
$\chi_{b0}(2P)\rightarrow\gamma\Upsilon(1S)$ &743 &0.24  &0.18  &13.3  &10.6    &seen           \\
$\chi_{b1}(2P)\rightarrow\gamma\Upsilon(2S)$ &230 &9.03  &8.50  &17.4  &15.4    &seen           \\
$\chi_{b1}(2P)\rightarrow\gamma\Upsilon(1S)$ &764 &0.19  &0.15  &12.8  &10.2    &seen           \\
$\chi_{b2}(2P)\rightarrow\gamma\Upsilon(2S)$ &242 &8.53  &8.02  &17.5  &15.4    &seen           \\
$\chi_{b2}(2P)\rightarrow\gamma\Upsilon(1S)$ &777 &0.08  &0.07  &11.07 &8.77    &seen           \\
$\Upsilon(3S)\rightarrow\gamma\chi_{b0}(1P)$ &484 &0.025 &0.029 &0.196 &0.256   &$0.061\pm 0.030$ \\
$\Upsilon(3S)\rightarrow\gamma\chi_{b0}(2P)$ &122 &1.16  &1.06  &0.80  &0.64    &$1.20\pm 0.24$ \\
$\Upsilon(3S)\rightarrow\gamma\chi_{b1}(2P)$ &99  &3.03  &2.77  &2.16  &1.77    &$2.97\pm 0.56$ \\
$\Upsilon(3S)\rightarrow\gamma\chi_{b2}(2P)$ &86  &4.75  &4.37  &3.54  &2.92    &$3.00\pm 0.63$ \\
$\Upsilon(3S)\rightarrow\gamma\eta_{b}(1S)$  &867 &0.005 &0.003 &0.006 &0.001   &$<0.009$ \\
$\Upsilon(3S)\rightarrow\gamma\eta_{b}(2S)$  &343 &0.006 &0.003 &0.001 &0.000   &$<0.013$ \\
\hline
\end{tabular}
\end{center}
\end{table}

\begin{table}[t!]
\caption{\label{light_nonrel}Light meson radiative decay rates
(keV).}
\begin{center}
\begin{tabular}{c|c|c|c|c|c|c}
\hline \multicolumn{1}{c|}{}& \multicolumn{1}{c|}{}&
\multicolumn{2}{c|}{SHO}& \multicolumn{2}{c|}{Coulomb+linear}&
\multicolumn{1}{c}{Experiment}\\
\multicolumn{1}{c|}{}& \multicolumn{1}{c|}{$\gamma$(MeV)}&
\multicolumn{1}{c|}{nonrel}& \multicolumn{1}{c|}{rel}&
\multicolumn{1}{c|}{nonrel}& \multicolumn{1}{c|}{rel}&
\multicolumn{1}{c}{PDG \cite{PDG06}}\\
\hline \hline
$\rho^0\rightarrow\gamma\pi^0$         &376 &51.1  &20.9 &41.6  &13.1 &$87.8\pm 12.5$ \\
$\rho^{\pm}\rightarrow\gamma\pi^{\pm}$ &375 &50.9  &20.9 &41.5  &13.1 &$65.9\pm  7.8$ \\
$\rho\rightarrow\gamma\eta$            &195 &55.9  &26.1 &41.7  &14.9 &$43.2\pm  4.8$ \\
$w\rightarrow\gamma\pi^0$              &380 &470.  &192. &384.  &121. &$756\pm  30$   \\
$w\rightarrow\gamma\eta$               &200 &6.64  &3.09 &4.97  &1.78 &$4.16\pm 0.47$ \\
$\eta'\rightarrow\gamma \rho^0$        &165 &114.  &54.2 &84.5  &31.2 &$59.7\pm  6.7$ \\
$\eta'\rightarrow\gamma w$             &159 &11.5  &5.51 &8.55  &3.16 &$6.15\pm 1.16$ \\
$f_0(980)\rightarrow\gamma \rho^0$     &183 &518.  &233. &591.  &256. &               \\
$f_0(980)\rightarrow\gamma w$          &178 &55.8  &25.1 &63.8  &27.6 &               \\
$a_0(980)\rightarrow\gamma \rho$       &187 &59.3  &26.6 &67.4  &29.2 &               \\
$h_1\rightarrow\gamma a_0(980)$        &171 &28.3  &10.5 &28.4  &10.4 &               \\
$h_1\rightarrow\gamma f_0(980)$        &175 &3.35  &1.24 &3.36  &1.22 &               \\
$h_1\rightarrow\gamma \eta'$           &193 &24.2  &10.3 &42.8  &13.0 &               \\
$h_1\rightarrow\gamma \eta$            &457 &30.5  &11.1 &63.9  &17.0 &               \\
$h_1\rightarrow\gamma \pi^0$           &577 &459.  &152. &1097. &266. &               \\
$\phi\rightarrow\gamma\eta$            &363 &43.0  &27.1 &44.5  &21.4 &$55.4\pm  1.7$ \\
$b_1\rightarrow\gamma\pi^{\pm}$        &607 &50.5  &16.2 &124.5 &29.5 &$227\pm  75$ \\
$f_1(1285)\rightarrow\gamma\rho^0$     &406 &1066. &459. &1216. &489. &$1331\pm  389$ \\
$a_2(1320)\rightarrow\gamma\pi^{\pm}$  &652 &324.  &144. &93.4  &64.4 &$287\pm 30$ \\
\hline

\end{tabular}
\end{center}
\end{table}

\clearpage

\subsection{Coulomb gauge model}
In Tables \ref{cc_rel}, \ref{bb_rel} and \ref{light_rel} results
calculated in the Coulomb gauge model are presented for $c\bar{c}$,
$b\bar{b}$ and light mesons and compared to the experiment and
nonrelativistic potential model (column 2). Column 3 corresponds to
TDA approximation and column 4 is RPA approximation for the pion
wave function (for details on Coulomb gauge model see section
\ref{CoulombGauge}).

One can see a remarkable improvement in our result for the
transitions involving pion $\rho\rightarrow\gamma\pi$ and
$w\rightarrow\gamma\pi$ as we consider pion in RPA approximation of
the relativistic model.

Let us also point out better agreement with experiment for some of
the decays of $b\bar{b}$ mesons in the relativistic model. It is
quite unexpected to observe such a big difference in the decay rates
calculated in nonrelativistic and relativistic models for $b\bar{b}$
mesons as they are usually considered heavy. We conclude that
relativistic corrections and many-body effects are important even
for $b\bar{b}$ mesons.

For $c\bar{c}$ mesons we would like to point out that our results
for Coulomb gauge model differ substantially from nonrelativistic
potential model results, for some of the transitions agreement with
experiment is better and for some it is worse. The conclusion is
that effects taken into account in Coulomb gauge model are important
and require additional study.

\begin{table}[!h]
\caption{\label{cc_rel}$c\bar{c}$-meson radiative decay rates (keV)}
\begin{center}
\begin{tabular}{c|ccc}

\hline \multicolumn{1}{c|}{}& \multicolumn{1}{c}{Coulomb+Linear}&
\multicolumn{1}{c}{Coulomb gauge}&
\multicolumn{1}{c}{Experiment}\\
\multicolumn{1}{c|}{}& \multicolumn{1}{c}{potential}&
\multicolumn{1}{c}{TDA}&
\multicolumn{1}{c}{PDG \cite{PDG06}}\\

\hline \hline
$J/\psi\rightarrow\gamma\eta_c$                 &2.11 &4.15  &$1.21 \pm 0.41 $\\
$\chi_{C0}\rightarrow\gamma J/\psi$             &276  &358   &$135  \pm 21   $\\
$\chi_{C1}\rightarrow\gamma J/\psi$             &325  &412   &$317  \pm 36   $\\
$\chi_{C2}\rightarrow\gamma J/\psi$             &260  &278   &$416  \pm 46   $\\
$\Psi (2S)\rightarrow\gamma\eta_c $             &1.41 &0.92  &$0.88 \pm 0.17 $\\
$\Psi (2S)\rightarrow\gamma \chi_{C0}$          &7.0  &33.9  &$31.0 \pm 2.6  $\\
$\Psi (2S)\rightarrow\gamma \chi_{C1}$          &20.4 &67.8  &$29.3 \pm 2.5  $\\
$\Psi (2S)\rightarrow\gamma \chi_{C2}$          &33.8 &77.0  &$27.3 \pm 2.5  $\\
\hline
\end{tabular}
\end{center}
\end{table}

\begin{table}[!h]
\caption{\label{bb_rel}$b\bar{b}$-meson radiative decay rates (keV)}
\begin{center}
\begin{tabular}{c|ccc}

\hline \multicolumn{1}{c|}{}& \multicolumn{1}{c}{Coulomb+Linear}&
\multicolumn{1}{c}{Coulomb gauge}&
\multicolumn{1}{c}{Experiment}\\
\multicolumn{1}{c|}{}& \multicolumn{1}{c}{potential}&
\multicolumn{1}{c}{TDA}&
\multicolumn{1}{c}{PDG \cite{PDG06}}\\
\hline \hline

$\chi_{b0}(1P)\rightarrow\gamma\Upsilon(1S)$ &30.8 &26.9       &seen            \\
$\chi_{b1}(1P)\rightarrow\gamma\Upsilon(1S)$ &32.1 &27.0       &seen            \\
$\chi_{b2}(1P)\rightarrow\gamma\Upsilon(1S)$ &30.3 &23.6       &seen            \\
$\Upsilon(2S)\rightarrow\gamma\chi_{b0}(1P)$ &0.43 &1.27       &$1.22 \pm 0.24 $\\
$\Upsilon(2S)\rightarrow\gamma\chi_{b1}(1P)$ &1.17 &3.09       &$2.21 \pm 0.32 $\\
$\Upsilon(2S)\rightarrow\gamma\chi_{b2}(1P)$ &1.95 &4.16       &$2.29 \pm 0.31 $\\
$\chi_{b0}(2P)\rightarrow\gamma\Upsilon(2S)$ &14.2 &15.4       &seen            \\
$\chi_{b0}(2P)\rightarrow\gamma\Upsilon(1S)$ &10.6 &2.19       &seen            \\
$\chi_{b1}(2P)\rightarrow\gamma\Upsilon(2S)$ &15.4 &16.8       &seen            \\
$\chi_{b1}(2P)\rightarrow\gamma\Upsilon(1S)$ &10.2 &2.34       &seen            \\
$\chi_{b2}(2P)\rightarrow\gamma\Upsilon(2S)$ &15.4 &16.5       &seen            \\
$\chi_{b2}(2P)\rightarrow\gamma\Upsilon(1S)$ &8.77 &1.91       &seen            \\
$\Upsilon(3S)\rightarrow\gamma\chi_{b0}(2P)$ &0.64 &1.72       &$1.20 \pm 0.24 $\\
$\Upsilon(3S)\rightarrow\gamma\chi_{b1}(2P)$ &1.77 &4.05       &$2.97 \pm 0.56 $\\
$\Upsilon(3S)\rightarrow\gamma\chi_{b2}(2P)$ &2.92 &5.82       &$3.00 \pm 0.63 $\\

\hline

\end{tabular}
\end{center}
\end{table}

\begin{table}[!h]
\caption{\label{light_rel} Light meson radiative decay rates (keV)}
\begin{center}
\begin{tabular}{c|c|c|c|c}

\hline \multicolumn{1}{c|}{}& \multicolumn{1}{c|}{Coulomb+Linear}&
\multicolumn{2}{c|}{Coulomb gauge}&
\multicolumn{1}{c}{Experiment}\\
\multicolumn{1}{c|}{}& \multicolumn{1}{c|}{potential}&
\multicolumn{1}{c|}{TDA}& \multicolumn{1}{c|}{RPA}&
\multicolumn{1}{c}{PDG \cite{PDG06}}\\

\hline \hline
$\rho^0\rightarrow\gamma\pi^0$               &13.0 &39.3  &85.3 &$87.8 \pm 12.5 $\\
$\rho^\pm\rightarrow\gamma\pi^\pm$           &13.0 &39.3  &85.3 &$65.9 \pm 7.8  $\\
$\rho\rightarrow\gamma\eta$                  &14.2 &95.5  &     &$43.2 \pm 4.8  $\\
$w\rightarrow\gamma\pi^0$                    &121. &356.  &771. &$756  \pm 30   $\\
$w\rightarrow\gamma\eta$                     &1.77 &11.1  &     &$4.16 \pm 0.47 $\\
$\eta'\rightarrow\gamma \rho^0$              &33.4 &220.  &     &$59.7 \pm 6.7  $\\
$\eta'\rightarrow\gamma w$                   &3.18 &22.7  &     &$6.15 \pm 1.16 $\\
$f_0(980)\rightarrow\gamma \rho^0$           &263. &583.  &     &                \\
$f_0(980)\rightarrow\gamma w$                &27.7 &62.3  &     &                \\
$a_0(980)\rightarrow\gamma \rho$             &29.8 &66.6  &     &                \\
$h_1\rightarrow\gamma a_0(980)$              &11.3 &31.6  &     &                \\
$h_1\rightarrow\gamma f_0(980)$              &1.33 &3.69  &     &                \\
$h_1\rightarrow\gamma \eta'$                 &13.1 &14.9  &     &                \\
$h_1\rightarrow\gamma \eta$                  &17.0 &10.6  &     &                \\
$h_1\rightarrow\gamma \pi^0$                 &267. &121.  &173. &                \\
$b_1\rightarrow\gamma\pi^{\pm}$              &29.8 &264.  &508. &$227 \pm 75    $\\
$f_1(1285)\rightarrow\gamma\rho^0$           &492. &823.  &     &$1331\pm 389 $\\
$a_2\rightarrow\gamma\pi^{\pm}$              &63.4 &275.  &549. &$287 \pm 30  $\\
\hline

\end{tabular}
\end{center}
\end{table}
\clearpage

\section{Higher order diagrams}

We would now like to estimate the effect of higher order diagrams on
radiative transitions (see section \ref{HigherOrder} for definition
and description of higher order diagrams). We choose to describe
them in the time-ordered bound state perturbation theory. As an
example we consider radiative transition of $c\bar{c}$ vector meson
$J/\psi$ to pseudoscalar meson $\eta_c$.

In our approach impulse approximation is considered the leading
order in perturbation theory. The higher order terms take into
account the possibility of quark-antiquark pair creation (for
example, through $^3P_0$ model) and could be written as:
\begin{equation}\label{HO}
A_{HO}= \sum_{V} \frac{\langle\gamma\eta_c|H|V,\eta_c\rangle \langle
V,\eta_c |H|J/\psi\rangle}{m_{J/\psi}-E_{V \eta_c}}+ \sum_{S}
\frac{\langle \gamma\eta_c|H|S,\eta_c,\gamma\rangle \langle
S,\eta_c,\gamma |H|J/\psi\rangle}{m_{J/\psi}-E_{S \eta_c \gamma}}.
\end{equation}

The two terms in \ref{HO} describe different time-orderings of
$^3P_0$ and electromagnetic interactions and correspond to the two
diagrams in Fig. \ref{HOfig}. Here we would like to analyze the
first diagram; the second diagram could be calculated in a
completely analogous way.

The first diagrams has two stages:
\begin{enumerate}
  \item $J/\psi$ decays into $\eta_c$ and some other intermediate state
meson $V$. This corresponds to the matrix elements $\langle V,\eta_c
|H|J/\psi\rangle$ of the above formula.
  \item Intermediate state meson $V$ transforms into the photon, which
corresponds to $\langle\gamma\eta_c|H|V,\eta_c\rangle$.
\end{enumerate}

To calculate the amplitude of the process in perturbation theory we
multiply the amplitudes of two parts of the transition and then
divide by the energy denominator. The energy denominator is the
energy difference between the initial state meson $J/\psi$ in its
rest frame and the intermediate state consisting of $\eta_c$ and
$V$. We also have to sum over all the intermediate bound states $V$
which could be formed. Since in the second part of the process meson
$V$ transforms itself into the photon, it must have the same quantum
numbers as the photon (S=1, L=0). Only vector mesons have this set
of quantum numbers, so we sum over all possible vector mesons in the
intermediate state. In case of $c\bar{c}$ mesons these are ground
state $J/\psi$ and all its excited states.

The amplitudes for the first part of the diagram have been estimated
in the $^3P_0$ model using SHO wave functions. In this case they can
be calculated analytically:
\begin{eqnarray}
  \langle J/\psi,\eta_c|H_{^3P_0}|J/\psi\rangle &=&\gamma
  \left(\frac{4\pi}{\beta^2}\right)^{3/4}
  \frac{16q}{27}\,\,
  e^{-q^2/12\beta^2}
  \nonumber \\
  \langle \psi',\eta_c|H_{^3P_0}|J/\psi\rangle &=&\gamma
  \left(\frac{4\pi}{\beta^2}\right)^{3/4}
  \frac{32\sqrt{6}\,q}{243}\,
  \left( 1-\frac{q^2}{12\beta^2}\right) e^{-q^2/12\beta^2},
  \label{amp3P0}\\
  \langle \psi'',\eta_c|H_{^3P_0}|J/\psi\rangle &=&\gamma
  \left(\frac{4\pi}{\beta^2}\right)^{3/4}
  \frac{8\sqrt{30}\,q}{729}
  \left(1-\frac{4q^2}{15\beta^2}+\frac{q^4}{180\beta^4}\right)
  e^{-q^2/12\beta^2}.\nonumber
\end{eqnarray}
for the first three excitations of the intermediate vector meson
state. Here $q=|\vec{q}|$ is the magnitude of the vector meson (and
photon) momentum, $\beta$ is a parameter for SHO wave functions.

To find $x$ for the amplitudes above we first need to find the
intermediate vector meson momentum $\vec{q}$ from the energy
conservation law for the whole process:
\begin{equation*}
    E_{J/\psi}=E_{\eta}+E_{\gamma}.
\end{equation*}

In the initial state meson rest frame: $E_{J/\psi}=m_{J/\psi}$,
$E_{\gamma}=|\vec{q}|$, $E_{\eta}=\sqrt{m_{\eta}^2+|\vec{q}|^2}$,
and then:
\begin{equation*}
    |\vec{q}|=\frac{m_{J/\psi}^2-m_{\eta}^2}{2m_{J/\psi}}\approx
    0.115\,GeV.
\end{equation*}
The values of the parameters for $c\bar{c}$-mesons are:
$\gamma=0.35$,  $\beta=0.378$ GeV for $c\bar{c}$-mesons
\cite{HHreview}, then $x\approx 0.305$ (we also take
$\vec{x}\parallel OZ$), and the amplitudes of (\ref{amp3P0}) are:
\begin{eqnarray}
  \langle J/\psi,\eta_c|H_{^3P_0}|J/\psi\rangle &\approx& 0.681\,\,\text{GeV}^{-1/2},\nonumber \\
  \langle \psi',\eta_c|H_{^3P_0}|J/\psi\rangle &\approx& 0.368\,\,\text{GeV}^{-1/2},\nonumber\\
  \langle \psi'',\eta_c|H_{^3P_0}|J/\psi\rangle &\approx& 0.067\,\,\text{GeV}^{-1/2}.
\end{eqnarray}

The amplitude of second part of the process, transformation of the
vector meson into the photon, is proportional to the vector meson
decay constant (as it was defined in Section \ref{dcSection}):
\begin{equation}
\langle\gamma\eta_c|H|V,\eta_c\rangle =
\frac{e\,Q_q}{\sqrt{2m_V}}\frac{\left(\epsilon^*_{\lambda}\right)_{\mu}}{\sqrt{2q}}\,\langle
0|\bar{\Psi}\gamma^{\mu}\Psi|V\rangle = \frac{e\,Q_q}{2}
\sqrt{\frac{m_V}{q}} \, f_V \,
\epsilon_V^{\mu}\left(\epsilon^*_{\lambda}\right)_{\mu}=\frac{e\,Q_q}{2}
\sqrt{\frac{m_V}{q}}\, f_V\, \delta_{\lambda V}
\end{equation}
where $\epsilon^*_{\lambda}$ and $\epsilon_V$ are the photon and
vector meson polarization vectors.

To estimate the higher order diagram we take the experimental values
of $\psi$ decay constants:
\begin{eqnarray*}
  f_{J/\psi} &=& 0.411\,\,\text{GeV}, \\
  f_{\psi'} &=& 0.279\,\,\text{GeV}, \\
  f_{\psi''} &=& 0.174\,\,\text{GeV}.
\end{eqnarray*}

Then the first higher order diagram is (the first term of
eqn.\ref{HO}):
\begin{equation}\label{HO1}
A_{HO1}=A^{(0)}_{HO1}(1+a'+a''+...)
\end{equation}
where $A^{(0)}$ corresponds to the ground state vector meson
($J/\psi$) in the intermediate state:
\begin{equation}
    A^{(0)}_{HO1}=\frac{\langle\gamma\eta_c|H|J/\psi,\eta_c\rangle \langle
J/\psi,\eta_c |H|J/\psi\rangle}{m_{J/\psi}-E_{J/\psi \eta_c}}\approx
0.049\,\,\text{GeV}^{-1/2}
\end{equation}
and the coefficients are:
\begin{equation*}
  a' = \frac{\langle\gamma\eta_c|H|\psi',\eta_c\rangle}{\langle\gamma\eta_c|H|J/\psi,\eta_c\rangle}
  \frac{ \langle \psi',\eta_c |H|J/\psi\rangle}{ \langle J/\psi,\eta_c |H|J/\psi\rangle}
  \frac{m_{J/\psi}-E_{J/\psi \eta_c}}{m_{J/\psi}-E_{\psi' \eta_c}} \approx 0.334,
\end{equation*}

\begin{equation*}
  a'' = \frac{\langle\gamma\eta_c|H|\psi'',\eta_c\rangle}{\langle\gamma\eta_c|H|J/\psi,\eta_c\rangle}
  \frac{ \langle \psi'',\eta_c |H|J/\psi\rangle}{ \langle J/\psi,\eta_c |H|J/\psi\rangle}
  \frac{m_{J/\psi}-E_{J/\psi \eta_c}}{m_{J/\psi}-E_{\psi'' \eta_c}} \approx 0.036,
\end{equation*}

\begin{equation}
...
\end{equation}

One can see that the sum in (\ref{HO1}) converges rather fast,
mostly because of the decrease in the decay constants and increase
in the energy denominator for the excited states. So we have:
\begin{equation}
    A_{HO1}\approx
    A_{HO1}^{(0)}(1+0.334+0.036)=1.37\,A_{HO1}^{(0)}\approx 0.067\,\,\text{GeV}^{-1/2}.
\end{equation}

To compare to the leading order diagram (impulse approximation) we
need to compute higher order amplitude in the relativistic
convention:
\begin{equation}
    A^{(rel)}_{HO1}=\sqrt{2m_{J/\psi}}\,\sqrt{2E_{\eta}}\,\sqrt{2q}\,A_{HO1}=0.197\,\,\text{GeV}.
\end{equation}

The value of the leading order diagram is (see eqn. (\ref{impSHO})):
\begin{equation}
    A_{imp}=|\vec{q}|\sqrt{M_1E_2}
\frac{eQ_q+eQ_{\bar{q}}}{m_q} e^{-{q^2}/16\beta^2}\approx
0.095\,\,\text{GeV}.
\end{equation}

We conclude that higher order diagrams can be very significant (our
estimated value is larger than the impulse approximation amplitude)
and should be studied further.

%
%

\chapter{Summary}\label{Conclusions}

An investigation of meson properties in constituent quark models has
been reported in the present dissertation. The main goal of this
work was to study the limits of applicability of quark models, and
generate possible improvements. Typical approximations, widely used
in this type of models, are also analyzed.

The first chapter contains general introduction to QCD and its most
important properties including asymptotic freedom, confinement and
dynamical chiral symmetry breaking.

The rest of the dissertation is divided into two main parts:
`Theory' part (Chapter \ref{theory}), in which the detailed
description of our approach is given, and `Applications' part
(Chapters \ref{spectrum}, \ref{dc}, \ref{ff}, \ref{gg_chapter},
\ref{RadTrans}), in which our results are presented and discussed.

Chapter \ref{theory} explains the theory necessary for understanding
our approach. First, the nonrelativistic constituent quark model for
mesons is introduced. In this model a meson is approximated as a
bound state of a quark and an antiquark, and the presence of the
gluon is only taken into account through its effect on the
instantaneous interaction potential between the meson constituents.
The basic potential for quark-antiquark interaction consists of
three terms: Coulomb term is motivated by one-gluon exchange, linear
term represents a phenomenological model for color confinement and a
hyperfine term is spin dependent. This model of the potential can
describe the heavy meson spectrum with great accuracy, which means
that it contains all the features important for the masses of the
low lying states of heavy mesons. However, as we move away from
heavy meson spectroscopy to study other mesons or other meson
properties, the `Coulomb+linear' potential has to be modified as it
is not powerful enough.

We suggest two main modifications of the potential, which were
inspired by fundamental QCD properties and then verified by the
experiment. The first is making the interaction potential more
powerful in explaining complicated spin structure of hadrons by
adding terms calculated in perturbation theory with one-loop
corrections included. The second modification is including
momentum-dependence of the QCD coupling in the Coulomb term of the
potential.

The first modification was motivated by the puzzling heavy-light
meson spectroscopy which has been discovered in the last few years
and generated a lot of interesting ideas and new exotic models.
Until now, however, no model has been successful in giving complete
explanation of the phenomena. It has been shown, in particular, that
masses of some particles detected do not fit in the canonical
picture of nonrelativistic constituent quark model, with the usual
`Coulomb+linear' potential. We suggest that the problems in
explaining new states do not necessarily need new approach but could
be solved within the naive model by including spin-dependent terms
in the potential. These terms have been calculated in perturbation
theory with one-loop corrections included. It is demonstrated that
they do not destroy the agreement of experimentally known charmonium
and bottomonium spectra but can be especially important for mesons
with unequal quark masses. We show that the set of parameters for
the improved potential can be found to reproduce the masses of the
puzzling P-wave heavy-light meson states in Chapter \ref{spectrum}
on Spectroscopy.

Second modification of the potential (taking into account the
momentum dependence of the coupling) has been inspired by the
fundamental property of asymptotic freedom in QCD. According to this
property, for large energy scales the interaction between a quark
and an antiquark becomes weaker as the exchange momentum increases
(or distance between a quark and an antiquark decreases), and this
is not taken into account in the `Coulomb+linear' potential. The
fact that this naive expression works so well to explain the heavy
meson spectroscopy tells us that the momentum dependence of the
potential for small distances is not particularly important for the
meson masses. Of course, it can be important for other meson
properties. We show that some of the observables are very sensitive
to the introduction of running coupling, in particular, meson decay
constants and gamma-gamma transitions. We suggest that the behavior
of the running coupling should imitate the one of perturbative QCD
at small distances and saturate to a phenomenological value at large
distances. This assumption allows us to investigate meson properties
sensitive to the high energy scale and explain experimental data on
charmonium decay constants and gamma-gamma transitions. We perform
the global fit of the parameters for charmonium and bottomonium and
find that, only if the dependence of running coupling on the
momentum scale is taken into account, experimental values for the
vector meson decay constants can be reproduced.

Even though the constituent quark model works quite well for the
meson properties, it has its limits. In particular, explanation of
some of the QCD properties important to the light mesons, such as
dynamical chiral symmetry breaking or relativistic corrections for
the wave functions, is not possible in this model. Better
description can be achieved in the relativistic approach to the
mesons based on QCD Hamiltonian, which is introduced in section
\ref{CoulombGauge}. This approach takes into account many-body
effects and powerful enough to generate the description of dynamical
chiral symmetry breaking and the emergence of the pion as a
Goldstone boson in the theory.

Next, the main points of our approach to the meson transitions are
introduced. To describe strong meson decays a phenomenological
$^3P_0$ model have been used. Investigation of strong decays is not
the main goal of the present dissertation but the results of $^3P_0$
model calculations are important for the study of the effects of
pair creation on the electromagnetic transitions. For that reason, a
short introduction to $^3P_0$ model has been presented.

The main points of our description of electromagnetic and
electroweak transitions and definitions are presented in Chapter
\ref{theory} `Theory'. These transitions can provide us with
valuable information on the hadron structure since the transition
operator is very well known and much experimental data exists on the
subject. Still, the calculation of the observables is complicated
enough that numerous approximations are widely in use, which are
typically taken from nuclear physics and not justified to use for
hadrons. Our main motivation was to study the relevance of this
approach to hadrons by investigating meson transitions both with and
without making simplifying approximations. By comparing results one
can see the importance of the effects that have been neglected and
relevance of the effects to certain meson properties.

One of the approximations investigated is the nonrelativistic
approximation for quark spinors which is widely used especially for
heavy quarks. The important conclusions is that it is not justified
even for the heavy quarks, such as `charm' and `bottom'. Study of
charmonium and bottomonium meson properties have been performed
twice: 1)with nonrelativistic approximation for the quark spinors
and 2)with full relativistic expressions. The results of these two
cases are quite different (up to 50\% for charmonium decay constants
and up to 30\% for bottomonium). Relativistic corrections also make
a difference for the meson form-factors. Including relativistic
correction in our calculations changes the slope of the
form-factors. The same change in slope has been generated in ISGW
model by introducing the artificial factor $\kappa$, such that
$q\rightarrow q/\kappa$. This factor did not have any physical
interpretation and was introduced to achieve better agreement with
available experimental data. Our study gives a physical explanation
of the necessary change in the slope of the form-factor as the
effect of the relativistic corrections.

The impulse approximation is another simplification taken from the
nuclear physics, it completely ignores the possibility of
quark-antiquark pair creation and annihilation. The description of
the transitions in this approximation includes two diagrams
corresponding to the coupling of the external current to quark and
antiquark independently. We present the formalism for calculation of
the meson electromagnetic and electroweak form-factors in this
approximation. Our results have been compared to the quenched
lattice results for charmonium electromagnetic form-factors, and
they are in very good agreement. It leads us to believe that the
impulse approximation is a good description of the electromagnetic
transitions for charmonium.

One would like to study, however, other possible diagrams which
appear when we go beyond the impulse approximation and include the
possibility of the pair creation from the vacuum. Our description of
these diagrams is presented in section \ref{HigherOrder} of the
Chapter \ref{theory} `Theory'. Quark-antiquark pairs are assumed to
appear from the vacuum with $^3P_0$ quantum numbers (this model has
been quite successful for the description of strong meson decays).
They can interact with the constituents of the initial state meson
and might form bound states, which eventually transform to the final
state of the process. The time-ordered bound state perturbation
theory is used to calculate the amplitude of the transition. Our
estimation of the higher-order diagram for
$J/\psi\rightarrow\eta_c\gamma$ transition gives rather unexpected
result: the value of the amplitude for higher-order diagram is
larger than that of the impulse approximation amplitude. However, we
know that the impulse approximation works well for the charmonium
transitions from our study of the form-factors so including
higher-order diagram of comparable value might ruin the agreement
with lattice results. We also know that the $^3P_0$ model gives a
good description of the charmonium strong decays and the
time-ordered bound state perturbation theory is well justified. We
conclude that this situation has to be studied further and might
lead us to discovering some interesting phenomena not taken into
account in this approach.

Time-ordered bound state perturbation theory is important ingredient
in our approach to gamma-gamma transitions. Our method is quite
different from the perturbation theory calculations of this process
as it takes into account the infinite gluon exchange between quarks.
We find that this is rather important for a successful description
of gamma-gamma decays of charmonium, together with the momentum
dependence of the running coupling for short distances and
relativistic expressions of quark spinors. If we include all of this
effects then our results for gamma-gamma transitions of charmonium
states are in very good agreement with the experimental data. This
is the only approach that can explain all available experimental
results for gamma-gamma transitions of charmonium.

Finally, we explain the main differences of the nonrelativistic
constituent quark model and Coulomb gauge model in their application
to the study of radiative transitions. We find that the Coulomb
gauge model works particularly well for transitions involving pions.
This gives us hope that effects important for the pion behavior (and
absent from the nonrelativistic model) might have a reasonable
explanation in the Coulomb gauge model.

Overall, our results show that the quark model gives a satisfactory
description of meson properties. Modifications of the model,
suggested in this work, improve the limits of model applicability
and allow us to describe meson structure in a transparent way. It is
important to note that all the modifications have been motivated by
fundamental QCD properties and are not artificial adjustments of the
potential.

It has been demonstrated that the disagreement of the model
predictions with the experiment does not necessarily mean that the
formalism is wrong or the model is not applicable. It might be
possible that the important effects (which in principle can be
incorporated into the model) have been ignored. Investigation of
these effects and the ways they present themselves might give us
valuable information about fundamental QCD properties and hadron
structure in a simple framework.

The approach described in the present dissertation can be applied to
the investigation of the variety of interesting phenomena of low
energy QCD. Some of them can be studied by analyzing processes for
which experimental data is available, such as spectrum of excited
states of mesons and baryons, semi-leptonic and non-leptonic decays
of heavy-light mesons, hadron production and others. To investigate
the properties of light hadrons (and possibly the structure of the
nucleus) Coulomb gauge model can be applied. Hybrid hadron
properties can also be investigated after certain assumptions are
made about the hybrid structure in the model.

\clearpage
\appendix

\chapter{Decay constants} \label{DecayConstantsApp}

Decay constant definitions and quark model expressions for vector,
scalar, pseudoscalar, axial, and $^1P_1$ meson decay constants are
presented here.

\section{Vector Decay Constant}

The decay constant $f_V$ of the vector meson is defined as

\begin{equation}
m_Vf_V\epsilon^{\mu}=\langle0|\bar{\Psi}\gamma^{\mu}\Psi|V\rangle
\end{equation}
where $m_V$ is the vector meson mass, $\epsilon^{\mu}$ is its
polarization vector, $|V\rangle$ is the vector meson state.  The
decay constant has been extracted from leptonic decay rates with the
aid of the following:
\begin{equation}
\Gamma_{V\rightarrow e^{+}e^{-}}=\frac{e^4Q^2f^2_V}{12\pi
m_V}=\frac{4\pi\alpha^2}{3}\frac{Q^2f^2_V}{m_V}.
\end{equation}

Following the method described in the text yields the quark model
vector meson decay constant:
\begin{equation}\label{relfV}
f_V=\sqrt{\frac{3}{m_V}}\int\frac{d^3k}{(2\pi)^3} \Phi(\vec{k})
\sqrt{1+\frac{m_q}{E_k}} \sqrt{1+\frac{m_{\bar{q}}}{E_{\bar{k}}}}
\left(1+\frac{k^2}{3(E_k+m_q)(E_{\bar{k}}+m_{\bar{q}})}\right)
\end{equation}

The nonrelativistic limit of this yields the well-known
proportionality of the decay constant to the wavefunction at the
origin:
\begin{equation}
f_V=2\sqrt{\frac{3}{m_V}}\int\frac{d^3k}{(2\pi)^3} \Phi(\vec{k})
=2\sqrt{\frac{3}{m_V}} \tilde{\Phi}(r=0).
\end{equation}

\section{Pseudoscalar Decay Constant}

The decay constant $f_P$ of a pseudoscalar meson is defined by

\begin{equation}
p^{\mu}f_P= i \langle0|\bar{\Psi}\gamma^{\mu}\gamma^5\Psi|P\rangle
\end{equation}
where $p^{\mu}$ is the meson momentum and $|P\rangle$ is the
pseudoscalar meson state. The pseudoscalar decay rate is then
\begin{equation}
\Gamma_{P\rightarrow l^{+}{\nu}_l}=\frac{G_F^2}{8\pi}
|V_{q\bar{q}}|^2 f_P^2 m_l^2 m_P
\left(1-\frac{m_l^2}{m_P^2}\right)^2.
\end{equation}

The quark model expression for the decay constant is
\begin{equation}
f_P=\sqrt{\frac{3}{m_P}}\int\frac{d^3k}{(2\pi)^3}
\sqrt{1+\frac{m_q}{E_k}}\sqrt{1+\frac{m_{\bar{q}}}{E_{\bar{k}}}}
\left(1- \frac{k^2}{(E_k+m_q)(E_{\bar{k}}+m_{\bar{q}})} \right)
\Phi(\vec{k}).
\end{equation}

In the nonrelativistic limit this reduces to the same expression as
the vector decay constant.

\section{Scalar Decay Constant}

The decay constant $f_S$ of the scalar meson is defined by

\begin{equation}
p^{\mu}f_S=\langle 0|\bar{\Psi}\gamma^{\mu}\Psi|S\rangle,
\end{equation}
which yields the quark model result:
\begin{equation}
f_S=\sqrt{\frac{3}{m_S}}\frac{\sqrt{4\pi}}{(2\pi)^3} \int k^3 dk\,
\sqrt{1+\frac{m_q}{E_k}}\sqrt{1+\frac{m_{\bar{q}}}{E_{\bar{k}}}}
\left(\frac{1}{E_{\bar{k}}+m_{\bar{q}}}-\frac{1}{E_k+m_q} \right)
R(k).
\end{equation}
Here and in the following,  $R$ is the radial wavefunction defined
by $\Phi(k) = Y_{lm}R(k)$ with $\int \frac{d^3k}{(2\pi)^3} |\Phi|^2
= 1$.


\section{Axial Vector Decay Constant}

The decay constant $f_A$ of the axial vector meson is defined as

\begin{equation}
\epsilon^{\mu}f_Am_A=\langle
0|\bar{\Psi}\gamma^{\mu}\gamma^5\Psi|A\rangle
\end{equation}
where $\epsilon^{\mu}$ is the meson polarization vector, $m_A$ is
its mass and $|A\rangle$ is the axial vector meson state. The quark
model decay constant is thus
\begin{equation}
f_A=-\sqrt{\frac{2}{m_A}}\frac{\sqrt{4\pi}}{(2\pi)^3} \int k^3
dk\,\sqrt{1+\frac{m_q}{E_k}}\sqrt{1+\frac{m_{\bar{q}}}{E_{\bar{k}}}}
\left(\frac{1}{E_{\bar{k}}+m_{\bar{q}}}+\frac{1}{E_k+m_q} \right)
R(k).
\end{equation}

\section{$h_c$ Decay Constant}

The decay constant $f_{A'}$ of the $^1P_1$ state meson is defined
by:

\begin{equation}
\epsilon^{\mu}f_{A'}m_{A'}=\langle
0|\bar{\Psi}\gamma^{\mu}\gamma^5\Psi|^1P_1\rangle
\end{equation}
where $\epsilon^{\mu}$ is the meson polarization vector, $m_{A'}$ is
its mass and $|^1P_1\rangle$ is its state.
 The resulting quark model decay constant is given by
\begin{equation}
f_{A'}=\frac{1}{\sqrt{m_{A'}}}\frac{\sqrt{4\pi}}{(2\pi)^3} \int k^3
dk
\,\sqrt{1+\frac{m_q}{E_k}}\sqrt{1+\frac{m_{\bar{q}}}{E_{\bar{k}}}}
\left(\frac{1}{E_{\bar{k}}+m_{\bar{q}}}-\frac{1}{E_k+m_q} \right)
R(k).
\end{equation}

\appendix
\chapter{Electromagnetic Form Factors} \label{ffApp}

A variety of Lorentz invariant multipole decompositions (see Ref.
\cite{JLlatt}) and quark model expressions for these multipoles are
presented in the following.

Each transition form-factor is normally a sum of two terms
corresponding to the coupling of the external current to the quark
and antiquark. For quarkonium these two terms are equal to each
other, so in the following we only present formulas corresponding to
the single quark coupling. In general both terms have to be
calculated.

\section{Pseudoscalar Form Factor}

The most general Lorentz covariant decomposition for the
electromagnetic transition matrix element between two pseudoscalars
is:
\begin{equation}
\langle
P_2(p_2)|\bar{\Psi}\gamma^{\mu}\Psi|P_1(p_1)\rangle=f(Q^2)(p_2+p_1)^{\mu}+g(Q^2)(p_2-p_1)^{\mu}
\end{equation}

To satisfy time-reversal invariance the form-factors $f(Q^2)$ and
$g(Q^2)$ have to be real. The requirement that the vector current is
locally conserved gives a relation between two form-factors:
\begin{equation}
g(Q^2)=f(Q^2)\frac{M^2_2-M^2_1}{Q^2}.
\end{equation}
Thus the matrix element can be written as:
\begin{equation}
\langle
P_2(p_2)|\bar{\Psi}\gamma^{\mu}\Psi|P_1(p_1)\rangle=f(Q^2)\left((p_2+p_1)^{\mu}-\frac{M^2_2-M^2_1}{q^2}(p_2-p_1)^{\mu}\right)
\end{equation}
In case of two identical pseudoscalars the second term vanishes.

Computing with the temporal component of the current in the quark
model formalism yields (for quarkonium)
\begin{eqnarray}
f(Q^2)&=&\frac{\sqrt{M_1E_2}}{(E_2+M_1)-\frac{M^2_2-M^2_1}{q^2}(E_2-M_1)}\\
&\times&\int \frac{d^3k}{(2\pi)^3} \Phi(\vec{k})
\Phi^*\left(\vec{k}+\frac{\vec{q}}{2}\right)
\sqrt{1+\frac{m_q}{E_k}}\sqrt{1+\frac{m_q}{E_{k+q}}}
\left(1+\frac{(\vec{k}+\vec{q})\cdot\vec{k}}
{(E_k+m_q)(E_{k+q}+m_q)}\right).\nonumber
\end{eqnarray}

In case of identical pseudoscalars in the non-relativistic
approximation the formula above simplifies to
\begin{equation}\label{ff0}
f(Q^2)=\frac{2\sqrt{M_1E_2}}{E_2+M_1} \int \frac{d^3k}{(2\pi)^3}
\Phi(\vec{k}) \Phi^*\left(\vec{k}+\frac{\vec{q}}{2}\right).
\end{equation}

Similar expressions occur when the computation is made with the
spatial components of the electromagnetic current:
\begin{eqnarray}
f(Q^2)\!\!\!&=&\!\!\!\frac{\sqrt{M_1E_2}}{1-\frac{M^2_2-M^2_1}{q^2}}
\,\frac{\vec{q}}{|\vec{q}|^2}\cdot\! \int \frac{d^3k}{(2\pi)^3}
\Phi(\vec{k}) \Phi^*\left(\vec{k}+\frac{\vec{q}}{2}\right)
\sqrt{1+\frac{m_q}{E_k}}\sqrt{1+\frac{m_q}{E_{k+q}}}
\left(\frac{\vec{k}}{E_k+m_q}+\frac{\vec{k}+\vec{q}}
{E_{k+q}+m_q}\right).\nonumber\\
\label{chiSpatialEq}
\end{eqnarray}

In this case the nonrelativistic approximation for the single quark
form factor is
\begin{eqnarray}
f(Q^2)=\frac{\sqrt{M_1E_2}}{m|\vec{q}|^2} \vec{q}\cdot\int
\frac{d^3k}{(2\pi)^3} \Phi(\vec{k})
\Phi^*\left(\vec{k}+\frac{\vec{q}}{2}\right)(2\vec{k}+\vec{q}).
\end{eqnarray}

%
%
Covariance requires the same expression for the temporal and spatial
form factors. Comparing the formula above to the expression for the
temporal form factor (\ref{ff0}) shows that covariance is recovered
in the nonrelativistic and weak coupling limits (where $M_1+M_2 \to
4m$).

\section{Vector Form Factors}

The most general Lorentz covariant decomposition for the
electromagnetic transition matrix element between two identical
vectors is:
\begin{eqnarray}
\langle V(p_2)|\bar{\Psi}\gamma^{\mu}\Psi|V(p_1)\rangle=
-(p_1+p_2)^{\mu}\left[G_1(Q^2) (\epsilon^*_2 \cdot \epsilon_1)
+ \frac{G_3(Q^2)}{2m^2_V} (\epsilon^*_2 \cdot p_1) (\epsilon_1 \cdot p_2) \right]\nonumber\\
+G_2(Q^2)\left[\epsilon^{\mu}_1 (\epsilon^*_2 \cdot p_1) +
\epsilon^{\mu*}_2 (\epsilon_1 \cdot p_2) \right]
\end{eqnarray}
These form-factors are related to the standard charge, magnetic
dipole and quadrupole multipoles by
\begin{eqnarray}
G_C&=&\left(1+\frac{2}{3}\eta\right)G_1-\frac{2}{3}\eta G_2+\frac{2}{3}\eta(1+\eta)G_3\nonumber\\
G_M&=&G_2\nonumber\\
G_Q&=&G_1-G_2+(1+\eta)G_3 \label{G3Eq}
\end{eqnarray}
where $\eta=\frac{Q^2}{4m^2_q}$.

Quark model expressions for these are:
\begin{eqnarray}
G_2(Q^2)=-\frac{\sqrt{m_VE_2}}{|\vec{q}|^2} \int
\frac{d^3k}{(2\pi)^3} \Phi(\vec{k})
\Phi^*\left(\vec{k}+\frac{\vec{q}}{2}\right)
\sqrt{1+\frac{m_q}{E_k}}\sqrt{1+\frac{m_q}{E_{k+q}}} \left(
 \frac{\vec{k}\cdot\vec{q}}{E_k+m_q}-\frac{\vec{k}\cdot\vec{q}+|\vec{q}|^2}{E_{k+q}+m_q}\right)
\label{GMEq}
\end{eqnarray}
and
\begin{equation}
G_1(Q^2)=\frac{\sqrt{m_VE_2}}{m_V+E_2} \int \frac{d^3k}{(2\pi)^3}
\Phi(\vec{k}) \Phi^*\left(\vec{k}+\frac{\vec{q}}{2}\right)
\sqrt{1+\frac{m_q}{E_k}}\sqrt{1+\frac{m_q}{E_{k+q}}}
\left(1+\frac{(\vec{k}+\vec{q})\cdot\vec{k}}
{(E_k+m_q)(E_{k+q}+m_q)}\right)
\end{equation}
or
\begin{equation}
G_1(Q^2)=\frac{\sqrt{m_VE_2}}{|\vec{q}|^2} \int
\frac{d^3k}{(2\pi)^3} \Phi(\vec{k})
\Phi^*\left(\vec{k}+\frac{\vec{q}}{2}\right)
\sqrt{1+\frac{m_q}{E_k}}\sqrt{1+\frac{m_q}{E_{k+q}}}
\left(\frac{\vec{k}\cdot\vec{q}}{E_k+m_q}+\frac{\vec{k}\cdot\vec{q}+|\vec{q}|^2}{E_{k+q}+m_q}\right).
\end{equation}

$G_3$ can be expressed in terms of $G_1$ and $G_2$ in two different
ways:
\begin{equation}
G_3=\frac{2m^2_V}{|\vec{q}|^2}\left(1-\frac{E_2}{m_V}\right)G_1+\frac{2m_V}{E_2+m_V}G_2
\end{equation}
or
\begin{equation}
G_3=\frac{2m_V(m_V-E_2)}{|\vec{q}|^2}(G_1-G_2).
\end{equation}
One can establish that $G_3\rightarrow G_2-G_1$ as
$|\vec{q}|\rightarrow 0$ from either equation.

\section{Scalar Form Factor}

The most general Lorentz covariant decomposition for the
electromagnetic transition matrix element between two scalars is:
\begin{equation}
\langle
S_2(p_2)|\bar{\Psi}\gamma^{\mu}\Psi|S_1(p_1)\rangle=f(Q^2)(p_2+p_1)^{\mu}+g(Q^2)(p_2-p_1)^{\mu}.
\end{equation}

As with pseudoscalars, this can be written as
\begin{equation}
\langle
S_2(p_2)|\bar{\Psi}\gamma^{\mu}\Psi|S_1(p_1)\rangle=f(Q^2)\left((p_2+p_1)^{\mu}-\frac{M^2_2-M^2_1}{q^2}(p_2-p_1)^{\mu}\right).
\end{equation}

In the case of identical scalars the quark model calculation gives
\begin{eqnarray}
f(Q^2)=\frac{\sqrt{M_1E_2}}{E_2+M_1} \int \frac{d^3k}{(2\pi)^3}
\Phi(\vec{k}) \Phi^*\left(\vec{k}+\frac{\vec{q}}{2}\right)
\sqrt{1+\frac{m_q}{E_k}}\sqrt{1+\frac{m_q}{E_{k+q}}}
\left(1+\frac{(\vec{k}+\vec{q})\cdot\vec{k}}{(E_k+m_q)(E_{k+q}+m_q)}\right).
\end{eqnarray}

In the nonrelativistic limit this reduces to

\begin{equation}
f(Q^2)= \int \frac{d^3k}{(2\pi)^3} \Phi(\vec{k})
\Phi^*\left(\vec{k}+\frac{\vec{q}}{2}\right).
\end{equation}

%

\section{Vector-Pseudoscalar Transition Form Factor}

The most general Lorentz covariant decomposition for the
electromagnetic transition matrix element between vector and
pseudoscalar is:
\begin{equation}
\langle
P(p_P)|\bar{\Psi}\gamma^{\mu}\Psi|V(p_V)\rangle=iF(Q^2)\epsilon^{\mu\nu\alpha\beta}(\epsilon_{M_V})_{\nu}(p_V)_{\alpha}(p_P)_{\beta}.
\end{equation}
Computing with the spatial components of the current then gives

\begin{equation}
F(Q^2) = -\sqrt{\frac{E_P}{m_V}} \,\frac{1}{|\vec{q}|^2} \int
\frac{d^3k}{(2\pi)^3} \Phi_V(\vec{k})
\Phi_P^*\left(\vec{k}+\frac{\vec{q}}{2}\right)
\sqrt{1+\frac{m_q}{E_k}}\sqrt{1+\frac{m_q}{E_{k+q}}}
\left(\frac{\vec{k}\cdot\vec{q}}{E_k+m_q}-\frac{\vec{k}\cdot\vec{q}+|\vec{q}|^2}
{E_{k+q}+m_q}\right).
\end{equation}

In the nonrelativistic approximation in zero recoil limit
$\vec{q}\rightarrow 0$ this reduces to
\begin{equation}
F(Q^2)|_{\vec{q}\rightarrow 0}=\frac{1}{m_q}\sqrt{\frac{m_P}{m_V}}.
\end{equation}

\section{Scalar-Vector Transition Form Factors}

The most general Lorentz covariant decomposition for the
electromagnetic transition matrix element between scalar ($^3P_0$)
meson state and vector ($^3S_1$) is
\begin{eqnarray}
\langle V(p_V)|\bar{\Psi}\gamma^{\mu}\Psi|S(p_S)\rangle=
\Omega^{-1}(Q^2)\Bigg( E_1(Q^2)\left[
\Omega(Q^2)\epsilon^{*\mu}_{M_V}-\epsilon^*_{M_V}\cdot p_S
(p^{\mu}_V p_V\cdot p_S-m^2_V p^{\mu}_S)\right]\Bigg.\nonumber\\
\left.+\frac{C_1(Q^2)}{\sqrt{Q^2}}m_V \epsilon^*_{M_V}\cdot
p_S\left[p_V\cdot p_S(p_V+p_S)^{\mu}-m^2_S p^{\mu}_V-m^2_V
p^{\mu}_S\right]\right)
\end{eqnarray}
where $\Omega(Q^2) \equiv (p_V\cdot
p_S)^2-m^2_Vm^2_S=\frac{1}{4}\left[(m_V-m_S)^2-Q^2\right]\left[(m_V+m_S)^2-Q^2\right]$,
and takes the simple value $m^2_s|\vec{q}|^2$ in the rest frame of a
decaying scalar.

$E_1$ contributes to the amplitude only in the case of transverse
photons, while $C_1$ contributes only for longitudinal photons.
Quark model expressions for the multipole form factors are

\begin{eqnarray}
C_1(Q^2) = -2\frac{\sqrt{Q^2}}{|\vec{q}|}\frac{\sqrt{E_Vm_S}}{4\pi}
&&\!\!\!\!\!\!\!\!\int \frac{d^3k}{(2\pi)^3}
R_S(\vec{k})R_V\left(\vec{k}+\frac{\vec{q}}{2}\right)
\sqrt{1+\frac{m_q}{E_k}}\sqrt{1+\frac{m_q}{E_{k+q}}}\nonumber\\
&&\times\left(\cos{\Theta}+\frac{k^2+|\vec{k}|\cdot|\vec{q}|}
{(E_k+m_q)(E_{k+q}+m_q)}\right)
\end{eqnarray}

\begin{eqnarray}
C_1(Q^2) = 2\frac{\sqrt{E_Vm_S}}{4\pi}\frac{\sqrt{Q^2}}{|\vec{q}|}
&&\!\!\!\!\!\!\!\!\int \frac{d^3k}{(2\pi)^3} R_S(\vec{k})
R_V\left(\vec{k}+\frac{\vec{q}}{2}\right)
\sqrt{1+\frac{m_q}{E_k}}\sqrt{1+\frac{m_q}{E_{k+q}}}\nonumber\\
&&\times\left(\frac{k}{E_k+m_q}+\frac{q\cos{\Theta}}{E_{k+q}+m_q}
+\frac{k\cos{2\Theta}}{E_{k+q}+m_q}\right).
\end{eqnarray}
The first(second) expression for $C_1(Q^2)$ is calculated from the
temporal(spatial) matrix element of the current.

\begin{eqnarray}
E_1(Q^2)=-2\frac{\sqrt{E_Vm_S}}{4\pi} \int \frac{d^3k}{(2\pi)^3}
R_S(\vec{k}) R_V\left(\vec{k}+\frac{\vec{q}}{2}\right)
\sqrt{1+\frac{m_q}{E_k}}\sqrt{1+\frac{m_q}{E_{k+q}}}\left[\frac{k}{E_k+m_q}
-\frac{k\cos{\Theta}+q}{E_{k+q}+m_q}\right]\nonumber.
\end{eqnarray}

\section{$h_{c}$-Pseudoscalar Transition Form Factor}

The most general Lorentz covariant decomposition for the
electromagnetic transition matrix element between $^1P_1$ meson
state and pseudoscalar ($^1S_0$) is

\begin{eqnarray}
\langle
P(p_P)|\bar{\Psi}\gamma^{\mu}\Psi|A(p_A)\rangle=\Omega^{-1}(Q^2)\Bigg(
E_1(Q^2)\left[ \Omega(Q^2)\epsilon^{\mu}_{M_L}-\epsilon_{M_L}\cdot
p_P(p^{\mu}_A p_A\cdot p_P-m^2_A p^{\mu}_P)
\right]\Bigg.\nonumber\\
\left.+\frac{C_1(Q^2)}{\sqrt{Q^2}}m_A \epsilon_{M_L}\cdot
p_P\left[p_A\cdot p_P(p_A+p_P)^{\mu}-m^2_P p^{\mu}_A-m^2_A
p^{\mu}_P\right]\right).
\end{eqnarray}

Quark model expressions for the form factors are
\begin{eqnarray}
E_1(Q^2)=\frac{\sqrt{3m_AE_P}}{8\pi}\, \int \frac{d^3k}{(2\pi)^3}
R_A(\vec{k}) R_P\left(\vec{k}+\frac{\vec{q}}{2}\right)
&&\!\!\!\!\!\!\!\!\sqrt{1+\frac{m_q}{E_k}}\sqrt{1+\frac{m_q}{E_{k+q}}}\\
&&\times k\sin^2{\Theta}
\left(\frac{1}{E_k+m_q}+\frac{1}{E_{k+q}+m_q}\right)\nonumber
\end{eqnarray}
and

\begin{eqnarray}
C_1(Q^2) =
-\frac{\sqrt{3m_AE_P}}{4\pi}\frac{\sqrt{Q^2}}{|\vec{q}|}\, \int
\frac{d^3k}{(2\pi)^3} R_A(\vec{k})
R_P\left(\vec{k}+\frac{\vec{q}}{2}\right)
&&\!\!\!\!\!\!\!\!\sqrt{1+\frac{m_1}{E_k}}\sqrt{1+\frac{m_2}{E_{k+q}}}\\
&&\times\cos{\Theta}
\left(1+\frac{k^2+kq\cos{\Theta}}{(E_k+m_q)(E_{k+q}+m_q)}\right)\nonumber.
\end{eqnarray}

\section{Axial Vector - Vector Transition Form Factor}

The most general Lorentz covariant decomposition for the
electromagnetic transition matrix element between axial vector
($^3P_1$) meson state and vector ($^3S_1$) is

\begin{eqnarray}
&&\langle V(p_V)|\bar{\Psi}\gamma^{\mu}\Psi|A(p_A)\rangle=\frac{i}{4\sqrt{2}\Omega(Q^2)}\epsilon^{\mu\nu\rho\sigma}(p_A-p_V)_{\sigma}\times\nonumber\\
&\times& \Bigg[
E_1(Q^2)(p_A+p_V)_{\rho}\bigg(2m_A[\epsilon_{M_A}\cdot
p_V](\epsilon^*_{M_V})_{\nu}
+2m_V[\epsilon^*_{M_V}\cdot p_A](\epsilon_{M_A})_{\nu}\bigg)\nonumber\\
&&+M_2(Q^2)(p_A+p_V)_{\rho}\bigg(2m_A[\epsilon_{M_A}\cdot
p_V](\epsilon^*_{M_V})_{\nu}
-2m_V[\epsilon^*_{M_V}\cdot p_A](\epsilon_{M_A})_{\nu}\bigg)\nonumber\\
&&+\frac{C_1(Q^2)}{\sqrt{Q^2}}\bigg(-4\Omega(Q^2)(\epsilon_{M_A})_{\nu} (\epsilon^*_{M_V})_{\rho}\\
&&+(p_A+p_V)_{\rho}\bigg[ (m^2_A-m^2_V+Q^2)[\epsilon_{M_A}\cdot
p_V](\epsilon^*_{M_V})_{\nu}+
(m^2_A-m^2_V-Q^2)[\epsilon^*_{M_V}\cdot p_A](\epsilon_{M_A})_{\nu}
\bigg]\bigg)\Bigg].\nonumber
\end{eqnarray}

Quark model expressions for the form factors are
\begin{eqnarray}
E_1(Q^2)=-\frac{\sqrt{3m_AE_V}}{8\pi}\, \int \frac{d^3k}{(2\pi)^3}
R_A(\vec{k})
&&\!\!\!\!\!\!\!\!\!\!R_V\left(\vec{k}+\frac{\vec{q}}{2}\right)
\sqrt{1+\frac{m_q}{E_k}}\sqrt{1+\frac{m_q}{E_{k+q}}}\\
&&\times\left(\frac{k(3-\cos^2{\Theta})}{E_k+m_q}+\frac{k(1-3\cos^2{\Theta})-2q\cos{\Theta}}{E_{k+q}+m_q}\right),\nonumber
\end{eqnarray}

\begin{eqnarray}
M_2(Q^2)=-\frac{\sqrt{3m_AE_V}}{8\pi}\, \int \frac{d^3k}{(2\pi)^3}
R_A(\vec{k})
&&\!\!\!\!\!\!\!\!\!\!R_V\left(\vec{k}+\frac{\vec{q}}{2}\right)
\sqrt{1+\frac{m_q}{E_k}}\sqrt{1+\frac{m_q}{E_{k+q}}}\\
&&\times\left(\frac{k(1-3\cos^2{\Theta})}{E_k+m_q}-\frac{k(1-3\cos^2{\Theta})+2q\cos{\Theta}}{E_{k+q}+m_q}\right)\nonumber
\end{eqnarray}

and

\begin{eqnarray}
C_1(Q^2) = \frac{\sqrt{3m_AE_V}}{2\pi}\frac{\sqrt{Q^2}}{|\vec{q}|}\,
\int \frac{d^3k}{(2\pi)^3} R_A(\vec{k})
R_V\left(\vec{k}+\frac{\vec{q}}{2}\right)
&&\!\!\!\!\!\!\!\!\!\!\sqrt{1+\frac{m_q}{E_k}}\sqrt{1+\frac{m_q}{E_{k+q}}}\\
&&\times\left(\cos{\Theta}+\frac{k^2\cos{\Theta}+\frac{1}{2}kq(1+\cos^2{\Theta})}
{(E_k+m_q)(E_{k+q}+m_q)}\right).\nonumber
\end{eqnarray}

\appendix
\chapter{Electroweak form-factors}\label{EWffApp}

\section{Pseudoscalar-pseudoscalar transition}

The most general Lorentz covariant decomposition for the electroweak
transition matrix element between two pseudoscalars is:
\begin{equation}
V^{\mu}-A^{\mu}=\langle
P_2(\vec{p}_2)|\bar{\Psi}\gamma^{\mu}(1-\gamma^5)\Psi|P_1(\vec{p}_1)\rangle=f_+(Q^2)(p_1+p_2)^{\mu}+f_-(Q^2)(p_1-p_2)^{\mu}.
\end{equation}
Here $P_1$ is the initial state meson with the mass $M_1$ which
consists of a quark with the mass $m_1$ and an antiquark with the
mass $\bar{m}_1$. Similarly, final state meson $P_2$ has the mass
$M_2$ and consists of a quark and an antiquark with the masses $m_2$
and $\bar{m}_2$.

The matrix element is parity invariant. To satisfy time-reversal
invariance the form-factors $f_+(Q^2)$ and $f_-(Q^2)$ have to be
real.

Axial matrix element is equal to zero for this case:
\begin{equation}
A^{\mu}=\langle
P_2(\vec{p}_2)|\bar{\Psi}\gamma^{\mu}\gamma^5\Psi|P_1(\vec{p}_1)
\rangle=0,
\end{equation}
so
\begin{equation}
V^{\mu}=\langle
P_2(\vec{p}_2)|\bar{\Psi}\gamma^{\mu}\Psi|P_1(\vec{p}_1)\rangle=f_+(Q^2)(p_1+p_2)^{\mu}+f_-(Q^2)(p_1-p_2)^{\mu}.
\end{equation}

In the $P_1$ rest frame we have:
$p_1=(M_1,0,0,0),\,p_2=(E_2,0,0,|\vec{q}|)$ and then:
\begin{eqnarray}
V^0&=&f_+(Q^2)(M_1+E_2)+f_-(Q^2)(M_1-E_2),\\
\vec{V}&=&\vec{q}\left(f_+(Q^2)-f_-(Q^2)\right).
\end{eqnarray}

Now we can express the form-factors in terms of $V^0$ and $\vec{V}$:
\begin{eqnarray}
f_+(Q^2)=\frac{V^0}{2M_1}-\frac{E_2-M_1}{2M_1}\frac{\vec{V}\cdot\vec{q}}{|\vec{q}|^2},\\
f_-(Q^2)=\frac{V^0}{2M_1}-\frac{E_2+M_1}{2M_1}\frac{\vec{V}\cdot\vec{q}}{|\vec{q}|^2}.
\end{eqnarray}

Matrix elements $V^0$ and $\vec{V}$ could be calculated in the quark
model:
\begin{eqnarray}
V^0=\sqrt{M_1E_2} \int \frac{d^3k}{(2\pi)^3} \Phi_1(k)
\Phi_2^*\left(\vec{k}+\vec{q}\frac{\bar{m}_2}{m_2+\bar{m}_2}\right)
&&\!\!\!\!\!\!\!\!\!\!\sqrt{1+\frac{m_1}{E_k}}\sqrt{1+\frac{m_2}{E_{k+q}}}\\
&&\times\left(1+\frac{(\vec{k}+\vec{q})\cdot\vec{k}}
{(E_k+m_1)(E_{k+q}+m_2)}\right),\nonumber\\
\vec{V}=\sqrt{M_1E_2} \int \frac{d^3k}{(2\pi)^3} \Phi_1(k)
\Phi_2^*\left(\vec{k}+\vec{q}\frac{\bar{m}_2}{m_2+\bar{m}_2}\right)
&&\!\!\!\!\!\!\!\!\!\!\sqrt{1+\frac{m_1}{E_k}}\sqrt{1+\frac{m_2}{E_{k+q}}}\\
&&\times\left(\frac{\vec{k}}{E_k+m_1}+\frac{\vec{k}+\vec{q}}
{E_{k+q}+m_2}\right).\nonumber
\end{eqnarray}

Then the general expressions for $f_+(Q^2)$ and $f_-(Q^2)$ are:
\begin{eqnarray}
&&f_{\pm}(Q^2)=\frac{1}{2}\sqrt{\frac{E_2}{M_1}} \int
\frac{d^3k}{(2\pi)^3} \Phi_1(k)
\Phi_2^*\left(\vec{k}+\vec{q}\frac{\bar{m}_2}{m_2+\bar{m}_2}\right)
\sqrt{1+\frac{m_1}{E_k}}\sqrt{1+\frac{m_2}{E_{k+q}}}\\
&&\times\left(1+\frac{(\vec{k}+\vec{q})\cdot\vec{k}}
{(E_k+m_1)(E_{k+q}+m_2)}-\frac{(E_2\mp
M_1)(\vec{k}\cdot\vec{q})}{|\vec{q}|^2}\left(\frac{1}{E_k+m_1}+\frac{1}
{E_{k+q}+m_2}\right)-\frac{E_2-M_1}{E_{k+q}+m_2}\right).\nonumber
\end{eqnarray}

In the nonrelativistic approximation $m/E_k \approx 1$:
\begin{eqnarray}
V^0&=&2\sqrt{M_1M_2} \int \frac{d^3k}{(2\pi)^3} \Phi_1(k)
\Phi_2^*\left(\vec{k}+\vec{q}\frac{\bar{m}_2}{m_2+\bar{m}_2}\right),\\
\vec{V}&=&\sqrt{M_1M_2} \int \frac{d^3k}{(2\pi)^3} \Phi_1(k)
\Phi_2^*\left(\vec{k}+\vec{q}\frac{\bar{m}_2}{m_2+\bar{m}_2}\right)
\left(\frac{\vec{k}}{m_1}+\frac{\vec{k}+\vec{q}} {m_2}\right).
\end{eqnarray}

If we use SHO wave functions as an approximation for the meson wave
functions, then form-factors could be calculated analytically. The
SHO wave function for a ground state pseudoscalar meson is:
\begin{equation}
\Phi(k)=\left(\frac{4\pi}{\beta^2}\right)^{3/4}e^{-k^2/2\beta^2}.
\end{equation}
Then the matrix elements are:
\begin{eqnarray}
V^0&=&2\sqrt{M_1M_2}e^{-q^2\mu^2/4\beta^2},\\
\vec{V}&=&\vec{q}\sqrt{M_1M_2}e^{-q^2\mu^2/4\beta^2}\left[\frac{1}{m_2}-\frac{\mu}{2}\left(\frac{1}{m_1}+\frac{1}{m_2}\right)\right]
\end{eqnarray}
and the form-factors are:
\begin{eqnarray}
f_+=\sqrt{\frac{M_2}{M_1}}e^{-q^2\mu^2/4\beta^2}\left[1-\frac{M_2-M_1}{2}\left[\frac{1}{m_2}-\frac{\mu}{2}\left(\frac{1}{m_1}+\frac{1}{m_2}\right)\right]\right],\\
f_-=\sqrt{\frac{M_2}{M_1}}e^{-q^2\mu^2/4\beta^2}\left[1-\frac{M_2+M_1}{2}\left[\frac{1}{m_2}-\frac{\mu}{2}\left(\frac{1}{m_1}+\frac{1}{m_2}\right)\right]\right],
\end{eqnarray}
where
\begin{equation}
\mu=\frac{\bar{m}_2}{m_2+\bar{m}_2}.
\end{equation}

If we consider transition of the ground state pseudoscalar meson to
the first excited state pseudoscalar meson then the decomposition of
the current matrix elements will of course be the same. The only
difference will be the wave function of the final state meson. In
the SHO basis the wave function of the first excited state is:
\begin{equation}
\Phi(k)=\left(\frac{4\pi}{\beta^2}\right)^{3/4} \sqrt{\frac{3}{2}}
\left( 1-\frac{2k^2}{3\beta^2}\right) e^{-k^2/2\beta^2}.
\end{equation}

Then the matrix elements are:
\begin{eqnarray}
V^0&=&-\sqrt{\frac{M_1M_2}{6}}\,\frac{\mu^2|\vec{q}|^2}{\beta^2}\,e^{-q^2\mu^2/4\beta^2},\\
\vec{V}&=&-\vec{q}\sqrt{\frac{M_1M_2}{6}}e^{-q^2\mu^2/4\beta^2}\left[\mu\left(\frac{1}{m_1}+\frac{1}{m_2}\right)+\frac{\mu^2|\vec{q}|^2}{2m_2\beta^2}\right]
\end{eqnarray}
and the form-factors are:
\begin{eqnarray}
f'_+=\frac{1}{2}\sqrt{\frac{M_2}{6M_1}}e^{-q^2\mu^2/4\beta^2}\left[\mu(M_2-M_1)\left(\frac{1}{m_1}+\frac{1}{m_2}\right)-\frac{\mu^2|\vec{q}|^2}{\beta^2}\left(1-\frac{M_2-M_1}{2m_2}\right)\right],\\
f'_-=\frac{1}{2}\sqrt{\frac{M_2}{6M_1}}e^{-q^2\mu^2/4\beta^2}\left[\mu(M_2+M_1)\left(\frac{1}{m_1}+\frac{1}{m_2}\right)-\frac{\mu^2|\vec{q}|^2}{\beta^2}\left(1-\frac{M_2+M_1}{2m_2}\right)\right].
\end{eqnarray}

\section{Pseudoscalar-vector transition}

The most general Lorentz covariant decompositions for the
electroweak transition matrix elements between a pseudoscalar and a
vector are:
\begin{eqnarray}\label{PV}
V^{\mu}&=&\langle V(\vec{P}_V)|\bar{q}\gamma^{\mu}q|P(\vec{P}_P)\rangle=ig(Q^2)\epsilon^{\mu\nu\alpha\beta}\left(\epsilon^*_{M_V}\right)_{\nu} \left(P_P+P_V\right)_{\alpha} \left(P_P-P_V\right)_{\beta},\\
A^{\mu}&=&\langle
V(\vec{P}_V)|\bar{q}\gamma^{\mu}\gamma^5q|P(\vec{P}_P)\rangle=f(Q^2)\left(\epsilon^{*}_{M_V}\right)^{\mu}+a_+
(\epsilon^*_{M_V}\cdot P_P)(P_P+P_V)^{\mu}+a_-
(\epsilon^*_{M_V}\cdot P_P)(P_P-P_V)^{\mu}\nonumber.
\end{eqnarray}

In the rest frame of the decaying pseudoscalar:
$P_P=(m_P,0,0,0),\,\, P_V=(E_V,0,0,|\vec{q}|)$.

If $M_V=\pm 1$ then:
\begin{eqnarray}
\epsilon_{M_V}^*=\epsilon^*_{\pm
1}=\left(0,\mp\frac{1}{\sqrt{2}},\frac{i}{\sqrt{2}},0\right)\nonumber
\end{eqnarray}
and
\begin{eqnarray}
V^0&=&A^0=0,\label{V0A0}\\
\vec{V}&=&2M_Vg(Q^2)|\vec{q}|m_P \vec{\epsilon}^{\,*}_{M_V},\\
\vec{A}&=&f(Q^2)\vec{\epsilon}_{M_V}^{\,*},
\end{eqnarray}
so
\begin{eqnarray}
g(Q^2)&=&M_V\frac{\vec{V}\cdot\vec{\epsilon}_{M_V}}{2|\vec{q}|m_P},\\
f(Q^2)&=&\vec{A}\cdot\vec{\epsilon}_{M_V}.
\end{eqnarray}

If $M_V=0$ then:
\begin{eqnarray}
\epsilon^*_{M_V}=\epsilon^*_0=\left(\frac{|\vec{q}|}{m_V},0,0,\frac{E_V}{m_V}\right)\nonumber
\end{eqnarray}
and
\begin{eqnarray}
V^0&=&\vec{V}=0,\label{V0Vvec}\\
A^0&=&f(Q^2)\frac{|\vec{q}|}{m_V}+a_+(Q^2)|\vec{q}|\frac{m_P(m_P+E_V)}{m_V}+a_-(Q^2)|\vec{q}|\frac{m_P(m_P-E_V)}{m_V},\\
\vec{A}&=&f(Q^2)\frac{E_V}{m_V}\vec{\epsilon}_0+a_+(Q^2)|\vec{q}|\frac{m_P}{m_V}\vec{q}-a_-(Q^2)|\vec{q}|\frac{m_P}{m_V}\vec{q},
\end{eqnarray}
\clearpage \addtolength{\oddsidemargin}{-1cm} so
\begin{eqnarray}
a_-&=&A^0\frac{m_V}{2m_P^2|\vec{q}|}-\frac{f(Q^2)}{2m_P^2}-\frac{(m_P+E_V)m_V}{2m_P^2|\vec{q}|^2}\left(\frac{\vec{A}\cdot\vec{q}}{|\vec{q}|}-f(Q^2)\frac{E_V}{m_V}\right),\\
a_+&=&A^0\frac{m_V}{2m_P^2|\vec{q}|}-\frac{f(Q^2)}{2m_P^2}+\frac{(m_P-E_V)m_V}{2m_P^2|\vec{q}|^2}\left(\frac{\vec{A}\cdot\vec{q}}{|\vec{q}|}-f(Q^2)\frac{E_V}{m_V}\right).
\end{eqnarray}

In the quark model:

\begin{equation}
V^0\!\!=\!\!\sqrt{E_Vm_P} \int \frac{d^3k}{(2\pi)^3} \Phi_P(\vec{k})
\Phi^*_V\left(\vec{k}+\vec{q}\frac{\bar{m}_2}{\bar{m}_2+m_2}\right)
\sqrt{1+\frac{m_1}{E_k}}\sqrt{1+\frac{m_2}{E_{k+q}}}
\left(\frac{M_V\left(\vec{k}\cdot\vec{\epsilon}^{\,*}_{M_V}\right)|\vec{q}|}
{(E_k+m_1)(E_{k+q}+m_2)}\right)\!\!=\!\!0,\nonumber
\end{equation}

\begin{equation}
A^0\!\!=\!\!\sqrt{E_Vm_P} \!\!\int \frac{d^3k}{(2\pi)^3}
\Phi_P(\vec{k})
\Phi^*_V\left(\vec{k}+\vec{q}\frac{\bar{m}_2}{\bar{m}_2+m_2}\right)
\sqrt{1+\frac{m_1}{E_k}}\sqrt{1+\frac{m_2}{E_{k+q}}}
\left(\frac{\vec{k}}{E_k+m_1}+\frac{\vec{k}+\vec{q}}{E_{k+q}+m_2}\right)\vec{\epsilon}^{\,*}_{M_V},\nonumber
\end{equation}
which is consistent with (\ref{V0A0}) and (\ref{V0Vvec}), and

\begin{equation}
\vec{V}\!\!=\!i\sqrt{E_Vm_P} \!\!\int\!\! \frac{d^3k}{(2\pi)^3}
\Phi_P(\vec{k})
\Phi_V^*\!\left(\vec{k}+\vec{q}\frac{\bar{m}_2}{\bar{m}_2+m_2}\right)\!
\sqrt{1+\frac{m_1}{E_k}}\sqrt{1+\frac{m_2}{E_{k+q}}}
\left(\frac{\vec{k}}{E_k+m_1}-\frac{\vec{k}+\vec{q}}{E_{k+q}+m_2}\right)\times\vec{\epsilon}^{\,*}_{M_V},\nonumber
\end{equation}

\begin{eqnarray}
\vec{A}=\!\!\sqrt{E_Vm_P}\! \int \!\frac{d^3k}{(2\pi)^3}
\Phi_P(\vec{k})
&&\!\!\!\!\!\!\!\!\Phi_V^*\left(\vec{k}+\vec{q}\frac{\bar{m}_2}{\bar{m}_2+m_2}\right)
\!\sqrt{1+\frac{m_1}{E_k}}\!\sqrt{1+\frac{m_2}{E_{k+q}}}\\
&&\times\left(\vec{\epsilon}^{\,*}_{M_V}+\frac{\left((\vec{k}+\vec{q})\cdot\vec{\epsilon}^{\,*}_{M_V}\right)\vec{k}+\left(\vec{k}\cdot\vec{\epsilon}^{\,*}_{M_V}\right)\vec{q}-(\vec{k}\cdot\vec{q})\vec{\epsilon}^{\,*}_{M_V}}{(E_k+m_1)(E_{k+q}+m_2)}\right).\nonumber
\end{eqnarray}

Since $\vec{q}$ is the only vector in the first integral above,
$\vec{V}$ is proportional to
$(\vec{q}\times\vec{\epsilon}^{\,*}_{M_V})$. Then for
$\vec{q}\parallel OZ$ we have $\vec{V}=0$ if $M_V=0$, which is
consistent with (\ref{V0Vvec}).

We can now write down expressions for the form-factors in the quark
model:
\begin{eqnarray}
&&g(Q^2)=M_V\frac{\vec{V}\cdot\vec{\epsilon}_{M_V}}{2|\vec{q}|m_P}\nonumber\\
&=&\!\!M_V\frac{i\vec{\epsilon}_{M_V}}{2|\vec{q}|}
\sqrt{\frac{E_V}{m_P}} \!\int \!\!\frac{d^3k}{(2\pi)^3}
\Phi_P(\vec{k})
\Phi_V^*\left(\vec{k}+\vec{q}\frac{\bar{m}_2}{\bar{m}_2+m_2}\right)
\!\!\sqrt{1+\frac{m_1}{E_k}}\!\sqrt{1+\frac{m_2}{E_{k+q}}}
\!\left(\frac{\vec{k}}{E_k+m_1}-\frac{\vec{k}+\vec{q}}{E_{k+q}+m_2}\right)\times\vec{\epsilon}^{\,*}_{M_V}\nonumber\\
&=&\!\!-\frac{1}{2|\vec{q}|} \sqrt{\frac{E_V}{m_P}} \!\int
\!\!\frac{d^3k}{(2\pi)^3} \Phi_P(\vec{k})
\Phi_V^*\left(\vec{k}+\vec{q}\frac{\bar{m}_2}{\bar{m}_2+m_2}\right)
\!\!\sqrt{1+\frac{m_1}{E_k}}\!\sqrt{1+\frac{m_2}{E_{k+q}}}
\!\left(\frac{k\cos{\Theta_k}}{E_k+m_1}-\frac{k\cos{\Theta_k}+|\vec{q}|}{E_{k+q}+m_2}\right),
\end{eqnarray}

\begin{eqnarray}
f(Q^2)\!\!=\!\!\sqrt{E_Vm_P}\! \int \!\!\frac{d^3k}{(2\pi)^3}
\Phi_P(\vec{k})
\Phi_V^*\!\left(\vec{k}+\vec{q}\frac{\bar{m}_2}{\bar{m}_2+m_2}\right)
\!\!\sqrt{1+\frac{m_1}{E_k}}\!\sqrt{1+\frac{m_2}{E_{k+q}}}
\!\left(\!1+\frac{\frac{1}{2}k^2\sin^2{\Theta_k}-k|\vec{q}|\cos{\Theta_k}}{(E_k+m_1)(E_{k+q}+m_2)}\right),
\end{eqnarray}

\begin{eqnarray}
a_+&=&A^0\frac{m_V}{2m_P^2|\vec{q}|}-\frac{f(Q^2)}{2m_P^2}+\frac{(m_P-E_V)m_V}{2m_P^2|\vec{q}|^2}\left(\frac{\vec{A}\cdot\vec{q}}{|\vec{q}|}-f(Q^2)\frac{E_V}{m_V}\right),\\
a_-&=&A^0\frac{m_V}{2m_P^2|\vec{q}|}-\frac{f(Q^2)}{2m_P^2}-\frac{(m_P+E_V)m_V}{2m_P^2|\vec{q}|^2}\left(\frac{\vec{A}\cdot\vec{q}}{|\vec{q}|}-f(Q^2)\frac{E_V}{m_V}\right).
\end{eqnarray}

In the nonrelativistic approximation with SHO wave functions we
have:
\begin{eqnarray}
V^0&=&0,\nonumber\\
\vec{V}&=&|\vec{q}|\sqrt{m_Vm_P}\,\,e^{-q^2\mu^2/4\beta^2}\left(\frac{1}{m_2}+\frac{\mu}{2}\left(\frac{1}{m_1}-\frac{1}{m_2}\right)\right)M_V\vec{\epsilon}^{\,*}_{M_V},\nonumber\\
A^0&=&|\vec{q}|\sqrt{m_Vm_P}\,\,e^{-q^2\mu^2/4\beta^2}\left(\frac{1}{m_2}-\frac{\mu}{2}\left(\frac{1}{m_1}+\frac{1}{m_2}\right)\right)\delta_{M_V0},\nonumber\\
\vec{A}&=&2\sqrt{m_Vm_P}\,\,e^{-q^2\mu^2/4\beta^2}\vec{\epsilon}^{\,*}_{M_V},
\end{eqnarray}
and then
\begin{eqnarray}
g(Q^2)&=&\frac{1}{2}\sqrt{\frac{m_V}{m_P}}\,\,e^{-q^2\mu^2/4\beta^2}\left(\frac{1}{m_2}+\frac{\mu}{2}\left(\frac{1}{m_1}-\frac{1}{m_2}\right)\right),\nonumber\\
f(Q^2)&=&2\sqrt{m_Vm_P}\,\,e^{-q^2\mu^2/4\beta^2},\nonumber\\
a_+(Q^2)&=&-\frac{1}{2\sqrt{m_Vm_P}}e^{-q^2\mu^2/4\beta^2}\left[1 + \frac{m_V}{m_P} - \frac{m^2_V}{m_P}\left[\frac{1}{m_2}-\frac{\mu}{2}\left(\frac{1}{m_1}+\frac{1}{m_2}\right)\right]\right],\nonumber\\
a_-(Q^2)&=&\frac{1}{2\sqrt{m_Vm_P}}e^{-q^2\mu^2/4\beta^2}\left[1 -
\frac{m_V}{m_P} +
\frac{m^2_V}{m_P}\left[\frac{1}{m_2}-\frac{\mu}{2}\left(\frac{1}{m_1}+\frac{1}{m_2}\right)\right]\right].
\end{eqnarray}

For the transition to the first excited state of the vector meson
the matrix elements are:
\begin{eqnarray}
V^0&=&0,\nonumber\\
\vec{V}&=&|\vec{q}|\sqrt{\frac{m_Vm_P}{6}}\,\,e^{-q^2\mu^2/4\beta^2}\left[\mu\left(\frac{1}{m_1}-\frac{1}{m_2}\right)-\frac{\mu^2|\vec{q}|^2}{2m_2\beta^2}\right]M_V\vec{\epsilon}^{\,*}_{M_V},\nonumber\\
A^0&=&-|\vec{q}|\sqrt{m_Vm_P}\,\,e^{-q^2\mu^2/4\beta^2}\left[\mu\left(\frac{1}{m_1}+\frac{1}{m_2}\right)+\frac{\mu^2|\vec{q}|^2}{2m_2\beta^2}\right]\delta_{M_V0},\nonumber\\
\vec{A}&=&-\sqrt{\frac{m_Vm_P}{6}}\,\,\frac{\mu^2|\vec{q}|^2}{\beta^2}\,\,e^{-q^2\mu^2/4\beta^2}\vec{\epsilon}^{\,*}_{M_V},
\end{eqnarray}
\clearpage \addtolength{\oddsidemargin}{1cm} and then
\begin{eqnarray}
g(Q^2)&=&\frac{1}{2}\sqrt{\frac{m_V}{6m_P}}\,\,e^{-q^2\mu^2/4\beta^2}\left[\mu\left(\frac{1}{m_1}-\frac{1}{m_2}\right)-\frac{\mu^2|\vec{q}|^2}{2m_2\beta^2}\right],\nonumber\\
f(Q^2)&=&-\sqrt{\frac{m_Vm_P}{6}}\,\,\frac{\mu^2|\vec{q}|^2}{\beta^2}\,\,e^{-q^2\mu^2/4\beta^2},\\
a_+(Q^2)&=&\frac{1}{2\sqrt{6m_Vm_P}}e^{-q^2\mu^2/4\beta^2}\left[\frac{\mu^2|\vec{q}|^2}{2\beta^2}\left(1 + \frac{m_V}{m_P} - \frac{m^2_V}{m_2m_P}\right)-\frac{m^2_V\mu}{m_P}\left(\frac{1}{m_1}+\frac{1}{m_2}\right)\right],\nonumber\\
a_-(Q^2)&=&-\frac{1}{2\sqrt{6m_Vm_P}}e^{-q^2\mu^2/4\beta^2}\left[\frac{\mu^2|\vec{q}|^2}{2\beta^2}\left(1
- \frac{m_V}{m_P} +
\frac{m^2_V}{m_2m_P}\right)+\frac{m^2_V\mu}{m_P}\left(\frac{1}{m_1}+\frac{1}{m_2}\right)\right].\nonumber
\end{eqnarray}

\section{Pseudoscalar-scalar transition}

The vector matrix element vanishes for pseudoscalar to scalar
transition:
\begin{equation}
\langle
S(\vec{p}_2)|\bar{\Psi}\gamma^{\mu}\Psi|P(\vec{p}_1)\rangle=0,
\end{equation}
and the most general Lorentz covariant decomposition for the axial
matrix element is:
\begin{equation}
\langle
S(\vec{p}_2)|\bar{\Psi}\gamma^{\mu}\gamma^5\Psi|P(\vec{p}_1)\rangle=u_+(Q^2)(p_1+p_2)^{\mu}+u_-(Q^2)(p_1-p_2)^{\mu}.
\end{equation}
Here $P$ is the initial state meson with the mass $M_1$ which
consists of a quark with the mass $m_1$ and an antiquark with the
mass $\bar{m}_1$. Similarly, final state meson $S$ has the mass
$M_2$ and consists of a quark and an antiquark with the masses $m_2$
and $\bar{m}_2$.

The matrix element is parity invariant. To satisfy time-reversal
invariance the form-factors $u_+(Q^2)$ and $u_-(Q^2)$ have to be
real.

In the $P$ rest frame we have:
$p_1=(M_1,0,0,0),\,p_2=(E_2,0,0,|\vec{q}|)$ and then:
\begin{eqnarray}
A^0&=&u_+(Q^2)(M_1+E_2)+u_-(Q^2)(M_1-E_2),\\
\vec{A}&=&\vec{q}\left(u_+(Q^2)-u_-(Q^2)\right).
\end{eqnarray}

Now we can express the form-factors in terms of $A^0$ and $\vec{A}$:
\begin{eqnarray}
u_+(Q^2)=\frac{A^0}{2M_1}-\frac{E_2-M_1}{2M_1}\frac{\vec{A}\cdot\vec{q}}{|\vec{q}|^2},\\
u_-(Q^2)=\frac{A^0}{2M_1}-\frac{E_2+M_1}{2M_1}\frac{\vec{A}\cdot\vec{q}}{|\vec{q}|^2}.
\end{eqnarray}

Matrix elements $A^0$ and $\vec{A}$ could be calculated in the quark
model:
\begin{eqnarray}
A^0=\sqrt{M_1E_2} \sum_{M_LM_S}\!\!\!\langle 00|1M_L1M_S\rangle \int
\frac{d^3k}{(2\pi)^3} \Phi_1(k)
\Phi_2^*&&\!\!\!\!\!\!\!\!\!\left(\vec{k}+\vec{q}\frac{\bar{m}_2}{m_2+\bar{m}_2}\right)
\sqrt{1+\frac{m_1}{E_k}}\sqrt{1+\frac{m_2}{E_{k+q}}}\nonumber\\
&&\times\left(\frac{\vec{\epsilon}^{\,*}_{M_S}\cdot\vec{k}}{E_k+m_1}+\frac{\vec{\epsilon}^{\,*}_{M_S}\cdot(\vec{k}+\vec{q})}{E_{k+q}+m_2}\right),
\end{eqnarray}

\begin{eqnarray}
\vec{A}=\sqrt{M_1E_2} \sum_{M_LM_S}\langle 00|1M_L1M_S\rangle \int
\frac{d^3k}{(2\pi)^3} \Phi_1(k)
&&\!\!\!\!\!\!\!\!\!\Phi_2^*\left(\vec{k}+\vec{q}\frac{\bar{m}_2}{m_2+\bar{m}_2}\right)
\sqrt{1+\frac{m_1}{E_k}}\!\sqrt{1+\frac{m_2}{E_{k+q}}}\\
&&\times\left(\vec{\epsilon}^{\,*}_{M_S} +
\frac{\vec{k}\left(\vec{\epsilon}^{\,*}_{M_S}\cdot(\vec{k}+\vec{q})\right)+\left(\vec{k}\times(\vec{q}\times\vec{\epsilon}^{\,*}_{M_S})\right)}{(E_k+m_1)(E_{k+q}+m_2)}\right),\nonumber
\end{eqnarray}

where
\begin{equation}
\vec{\epsilon}^{\,*}_{M_S}=\left\{ \begin{array}{cc}
                 (0,0,1) & M_S=0,\\
                 \left(-\frac{1}{\sqrt{2}},\frac{i}{\sqrt{2}},0 \right) & M_S=1,\\
                 \left(\frac{1}{\sqrt{2}},\frac{i}{\sqrt{2}},0 \right) & M_S=-1.
                 \end{array}\right.\,\,\,\,\,\,\,\,
\end{equation}

In the nonrelativistic approximation $m/E_k \approx 1$:
\begin{eqnarray}
A^0&=&\sqrt{M_1M_2} \sum_{M_LM_S}\langle 00|1M_L1M_S\rangle \int
\frac{d^3k}{(2\pi)^3} \Phi_1(k)
\Phi_2^*\left(\vec{k}+\vec{q}\frac{\bar{m}_2}{m_2+\bar{m}_2}\right)
\left(\frac{\vec{k}}{m_1}+\frac{\vec{k}+\vec{q}}
{m_2}\right)\cdot\vec{\epsilon}^{\,*}_{M_S},\nonumber\\
\vec{A}&=&2\sqrt{M_1M_2} \sum_{M_LM_S}\langle 00|1M_L1M_S\rangle
\int \frac{d^3k}{(2\pi)^3} \Phi_1(k)
\Phi_2^*\left(\vec{k}+\vec{q}\frac{\bar{m}_2}{m_2+\bar{m}_2}\right)
\vec{\epsilon}^{\,*}_{M_S}.
\end{eqnarray}

If we use SHO wave functions as an approximation for the meson wave
functions, then form-factors could be calculated analytically. The
SHO wave function for a ground state pseudoscalar meson is:
\begin{equation}
\Phi(k)=\left(\frac{4\pi}{\beta^2}\right)^{3/4}e^{-k^2/2\beta^2},
\end{equation}
and for a ground state scalar meson it is:
\begin{equation}
\Phi(k)=\left(\frac{4\pi}{\beta^2}\right)^{5/4} \sqrt{\frac{2}{3}}
ke^{-k^2/2\beta^2},
\end{equation}

Then the matrix elements are:
\begin{eqnarray}
A^0&=&-\beta\sqrt{\frac{3M_1M_2}{2}}\left(\frac{1}{m_1}+\frac{1}{m_2}\right)
\left[1-\frac{\mu^2|\vec{q}|^2}{6\beta^2}+\frac{\mu|\vec{q}|^2}{\beta^2}\frac{m_1}{m_1+m_2}\right]e^{-q^2\mu^2/4\beta^2},\nonumber\\
\vec{A}&=&-\sqrt{\frac{2M_1M_2}{3}}\,\frac{\mu
|\vec{q}|}{\beta}\,e^{-q^2\mu^2/4\beta^2}\,\vec{\epsilon}_0\,\delta_{M_S0}.
\end{eqnarray}
and the form-factors are:
\begin{eqnarray}
u_+=-\frac{1}{2}\sqrt{\frac{3M_2}{2M_1}}e^{-q^2\mu^2/4\beta^2}\left[\beta\left(\frac{1}{m_1}+\frac{1}{m_2}\right)\left[1-\frac{\mu^2|\vec{q}|^2}{6\beta^2}+\frac{\mu|\vec{q}|^2}{\beta^2}\frac{m_1}{m_1+m_2}\right]-(M_2-M_1)\frac{2\mu}{3\beta}\right],\nonumber\\
u_-=-\frac{1}{2}\sqrt{\frac{3M_2}{2M_1}}e^{-q^2\mu^2/4\beta^2}\left[\beta\left(\frac{1}{m_1}+\frac{1}{m_2}\right)\left[1-\frac{\mu^2|\vec{q}|^2}{6\beta^2}+\frac{\mu|\vec{q}|^2}{\beta^2}\frac{m_1}{m_1+m_2}\right]-(M_2+M_1)\frac{2\mu}{3\beta}\right],\nonumber
\end{eqnarray}
where
\begin{equation}
\mu=\frac{\bar{m}_2}{m_2+\bar{m}_2}.\nonumber
\end{equation}


\begin{thebibliography}{99}
\bibitem{gamma}
 Several conventions for  are currently in use, which typically differ in the overall normalization
 of equation \ref{Ham3P0}.
\end{thebibliography}
\end{document}